\documentclass[12pt,a4paper]{book}
\usepackage{amsfonts}    % Sets equations to a math font
\usepackage{amsmath}     % Used to write equations
\usepackage{amssymb}     % Allows maths symbols to be used
\usepackage{array}       % Makes tables and align equations
\usepackage{blindtext}   % Inputs large paragraphs with no meaning to test layout (lorem ipsum)
\usepackage{calligra}    % For cursive font on the declaration page
\usepackage{enumitem}    % Allows dot points to be customised
\usepackage{fancyhdr}    % Gives more control of headers and footnotes
\usepackage{float}       % Used to align graphics
\usepackage[T1]{fontenc} % Adds extra characters, such as \dh
\usepackage{footmisc}    % Better footnote handling
\usepackage{graphbox}    % For aligning figures in minipages
\usepackage{graphicx}    % Used to add graphics
\usepackage{indentfirst} % Indents first paragraph in a section
\usepackage{lmodern}     % Allows arbitrary font sizing. Useful for footnotes/etc
\usepackage{longtable}   % Allows multi-page tables
\usepackage{mathtools}   % Makes amsmath equations prettier
\usepackage{microtype}   % Makes large blocks of text look nicer by subtly changing some `micro-typographic' elements
\usepackage{multirow}    % Allows cells in tables to take multiple rows
\usepackage{newtxmath}   % Adds some fonts
\usepackage{pdfpages}    % Allows .pdf files to be inserted
\usepackage{physics}     % Adds extra functionality to the equations package (e.g. partial differentials)
\usepackage{ragged2e}    % Justifies text in the table footnotes
\usepackage{subfig}      % Allows subfigures to be made and referenced
\usepackage{tikz}        % Can be used to create diagrams
\usepackage{titlesec}    % Makes chapter titles look better
\usepackage{url}         % Allows urls to be placed into the document
\usepackage{xcolor}      % Defines colours and allows new colours to be created

\usepackage[english]{babel}                        % Allows characters from a given language to compile correctly (e.g. just English here)
\usepackage[justification=centering]{caption}      % Used to add captions to figures and forces a centre justification
\usepackage[utf8]{inputenc}                        % Allows extra characters to compile correctly, commonly found in references
\usepackage[scr=boondoxo,scrscaled=1.05]{mathalfa} % Allows lowercase caligraphic math font, but is called with \mathscr instead of overwritting \mathcal
\usepackage[percent]{overpic}                      % Allows text to be placed over a figure
\usepackage[section,below]{placeins}               % Prevents figures from being placed in the following section

\usepackage[
bindingoffset=20mm,     % Extra spacing on the inner edge of the pages
top=25mm,               % Extra space added for header
bottom=15mm,
left=15mm,
right=15mm,
]{geometry}
\usepackage[
backend=bibtex,       % Fixes an issue where some characters aren't recognised
natbib=true,          % Sets compatibility with natbib commands (like \citep)
style=authoryear,     % Style of the bibliography
citestyle=authoryear, % Style of the inline citations
maxbibnames=3,        % Maximum number of authors in the bibliography entry
minbibnames=3,        % biblatex will truncate bibliography author lists exceeding maxcitenames to one author plus "et al." (mincitenames=1). This forces the minimum to a different value
maxcitenames=2,       % Maximum number of names in the inline citation
uniquelist=false,     % In case the author published multiple papers in the same year
dashed=false,         % Prevents authors referenced multiple times from being replaced with dashes
url=false,
doi=false,
isbn=false,
eprint=false
]{biblatex}

\usepackage{csquotes} % Ensures biblatex compiles references with quotes correctly
\usepackage{hyperref} % Makes references clickable

\usetikzlibrary{arrows.meta}
\usetikzlibrary{positioning,arrows}
\usetikzlibrary{decorations.pathmorphing}
\usetikzlibrary{decorations.markings}
\usetikzlibrary{shapes.misc}
\tikzset{
photon/.style={decorate, decoration={snake}, draw=black},
particle/.style={draw=black, postaction={decorate},
    decoration={markings,mark=at position .5 with {\arrow[draw=black]{>}}}},
pics/carc/.style args={#1:#2:#3}{
    code={\draw[pic actions] (#1:#3) arc(#1:#2:#3);}},
cross/.style={cross out, draw, 
    minimum size=2*(#1-\pgflinewidth), 
    inner sep=0pt, outer sep=0pt}
}

\newcommand{\horizontalLine}{\rule{\linewidth}{0.5mm}}

\newcommand{\graya}{\mbox{$\gamma$-ray}}   % Gamma ray as a compound adjective
\newcommand{\grayn}{\mbox{$\gamma$~ray}}   % Gamma ray as a noun
\newcommand{\grays}{$\gamma$~rays}         % Gamma ray as a pluralised noun
\newcommand{\pp}{\mbox{$p$--$p$}~collision}
\newcommand{\pps}{\mbox{$p$--$p$}~collisions}

\newcommand{\hi}{H\,{\scshape i}}
\newcommand{\htwo}{H$_2$}
\newcommand{\hii}{H\,{\scshape ii}}

\newcommand{\GP}{\textsc{Galprop}}
\newcommand{\galdef}{\textsc{galdef}}
\newcommand{\sclass}{\textsc{sourceClass}}
\newcommand{\hpx}{{HEALPix}}

\newcommand{\tibet}{Tibet~AS${\gamma}$}
\newcommand{\fermi}{\mbox{\textit{Fermi}--LAT}}
\newcommand{\argo}{\mbox{{Argo}--{YBJ}}}
\newcommand{\hess}{{H.E.S.S.}} % Technically it should be {H.E.S.S.\@}; however, LaTeX handles this automatically

\newcommand\blindfootnote[1]{%
  \begingroup
  \renewcommand\thefootnote{}\footnote{#1}%
  \addtocounter{footnote}{-1}% Subtract one to the counter to ensure all other footnotes continue counting correctly
  \endgroup
}

\title{High-Energy Cosmic-Ray Propagation in the Milky Way and the Associated Diffuse Gamma-Ray Emission} % Thesis title
\author{Peter~Marinos}      % Author name
\date{November, 2023}          % Date of submission

\xdefinecolor{UofA-Navy}{RGB}{16, 37, 53}
\xdefinecolor{UofA-Navy70}{RGB}{89, 102, 114}
\xdefinecolor{UofA-Navy50}{RGB}{136, 146, 154}
\xdefinecolor{UofA-Navy30}{RGB}{183, 190, 193}
\xdefinecolor{UofA-Red}{RGB}{212, 0, 0}
\xdefinecolor{UofA-MidBlue}{RGB}{0, 90, 156}
\xdefinecolor{UofA-MidBlue70}{RGB}{76, 140, 186}
\xdefinecolor{UofA-MidBlue50}{RGB}{128, 173, 206}
\xdefinecolor{UofA-MidBlue30}{RGB}{179, 206, 255}
\xdefinecolor{UofA-Black}{RGB}{0, 0, 0}
\xdefinecolor{UofA-Black70}{RGB}{77, 77, 77}
\xdefinecolor{UofA-Black50}{RGB}{128, 128, 128}
\xdefinecolor{UofA-Black30}{RGB}{179, 179, 179}
\xdefinecolor{UofA-DigMidBlue}{RGB}{0, 75, 131}
\xdefinecolor{UofA-DigGrey}{RGB}{241, 241, 241}
\xdefinecolor{UofA-DigPaleBlue}{RGB}{184, 220, 235}
\xdefinecolor{UofA-Gold}{RGB}{133, 97, 2}
\xdefinecolor{UofA-Gold50}{RGB}{163, 117, 11}

\newcommand{\apj}{{ApJ}}%                                    % Astrophysical Journal
\newcommand{\mnras}{{MNRAS}}%                                % Monthly Notices of the Royal Astronomical Society
\titleformat{\chapter}[display]{\normalfont\Large\bfseries}{\chaptertitlename\ \thechapter}{20pt}{\Huge}
\titlespacing*{\chapter}{0pt}{-40pt}{40pt} % This decreases the amount above while keeping the amount below the same
\titlespacing*{\section}{0pt}{3em}{2em} % Sets the space above to 3em and the space below to 2em

\graphicspath{{figures/}}

\bibliography{ref.bib}

\renewbibmacro{in:}{}
\AtEveryBibitem{\clearlist{language}}
\hypersetup{
  colorlinks, % Removes borders around links
  citecolor = UofA-DigMidBlue,  % Changes colour of \cite commands
  linkcolor = UofA-DigMidBlue,  % Changes colour of \ref commands
  urlcolor  = UofA-MidBlue,     % Changes colour of \url command
}

\DeclareFieldFormat{journaltitle}{%
  \iffieldundef{url}
    {\iffieldundef{doi}
      {#1} % If both DOI and URL are missing, display the title as is
      {\href{https://doi.org/\thefield{doi}}{#1}}} % If url is missing but doi is present, use the doi as a link
    {\href{\thefield{url}}{#1}} % If url is present, use it as a link
}
\DeclareFieldFormat{eid}{%
  \iffieldundef{eid}
    {#1}
    {\href{https://arxiv.org/abs/\thefield{eid}}{#1}}
}
\DeclareCiteCommand{\parencite}[\mkbibparens]
  {\usebibmacro{prenote}}
  {\usebibmacro{citeindex}\printtext[bibhyperref]{\usebibmacro{cite}}}
  {\multicitedelim}
  {\usebibmacro{postnote}}
\DeclareCiteCommand{\textcite}
  {\boolfalse{cbx:parens}}
  {\usebibmacro{citeindex}\printtext[bibhyperref]{\usebibmacro{textcite}\ifbool{cbx:parens}{\bibcloseparen\global\boolfalse{cbx:parens}}{}}}
  {\multicitedelim}
  {\usebibmacro{textcite:postnote}}
\DeclareCiteCommand{\citealt}
  {}
  {\printtext[bibhyperref]{\printnames{labelname}\setunit{\addspace}\printfield{year}}}
  {\multicitedelim}
  {\usebibmacro{textcite:postnote}}

\AtBeginDocument{}%
\AtBeginDocument{}%
\AtBeginDocument{}%
\AtBeginDocument{}%

\let\mtcleardoublepage\cleardoublepage
\renewcommand{\cleardoublepage}{\clearpage{\pagestyle{empty}\mtcleardoublepage}}
\makeatletter
\def\@endpart{\vfil\newpage
              \if@twoside
                  \if@openright
                    \null
                    \thispagestyle{empty}%
                    \newpage
                  \fi
              \fi
              \if@tempswa
                \twocolumn
              \fi}
\makeatother

\assignpagestyle{\chapter}{empty}

\makeatletter
\renewcommand{\chaptermark}[1]{%
    \markboth{%
    \MakeUppercase{\ifnum\c@secnumdepth>\m@ne \@chapapp \ \thechapter. \ \fi #1}%
    }{%
    \MakeUppercase{\ifnum\c@secnumdepth>\m@ne \@chapapp \ \thechapter. \ \fi #1}}%
}
\makeatother

\begin{document}
    
    % Insert the title page. This creates two pages; however, by default this will have page numbers. Begin by setting page style to empty so the second page is completely blank
    \pagestyle{empty}       % Sets all pages to not have numbers
    % This is the title page
\begin{titlepage}
    
    % We want everything on the title page to be centred
    \begin{center}
        
        % Insert the University of Adelaide logo at the top of the page
        % \includegraphics[width=0.45\textwidth]{figures/UofA_logo.eps} \\ \vspace{1.2cm}
        \includegraphics[width=0.45\textwidth]{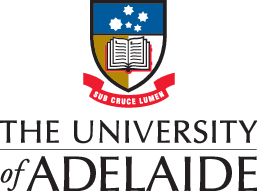} \\ \vspace{1.2cm}
        % Faculty
        \textsc{\LARGE Faculty of Sciences} \\ \vspace{0.4cm}
        % School
        \textsc{\Large School of Physical Sciences} \\ \vspace{0.7cm}
        
        \makeatletter % Allows the use of the @ symbol, required in inserting author/etc
        
        % Title
        \horizontalLine \\ \vspace{0.6cm}
        \huge \textbf{\LARGE \@title} \\ \vspace{0.3cm}
        \horizontalLine \vspace{1.2cm}
        
        % Insert the Author
        % \Large \emph{PhD Candidate:} \\
        \LARGE \textsc{\@author} \vspace{1.2cm}
        
        % Principal supervisor
        \large \emph{Principal Supervisor:} \\
        Prof.~Gavin~\textsc{Rowell} %\vspace{0.5cm} % Adjust this depending on how long the title is
        
        % Co-supervisor
        \large \emph{Co-Supervisor:} \\
        Dr.~Sabrina~\textsc{Einecke}
        
        % Add the date
        \vfill % Push date to the bottom of the page
        \large Submitted: November, 2023 \\
        \large Accepted: January, 2024 \\
        \large Revised: March, 2025
        
        \makeatother
        
    \end{center}
    
\end{titlepage}   % Input the title page
    
    % Begin the documents front matter
    \frontmatter
    
    % Table of Contents
    {
      \hypersetup{hidelinks} % Hide links only within this block
      \tableofcontents
    }
    
    % Abstract
    % Set up a new page as the abstract. Add it to the table of contents (toc) as a chapter, but make it a section as that is the preferred page setup
\newpage
\phantomsection
\addcontentsline{toc}{chapter}{Abstract}
\section*{Abstract}

Cosmic rays~(CRs) are electrons, protons, and other nuclei that have been accelerated to energies ranging from 1\,MeV to 100\,EeV~($10^{6}$~to~$10^{20}$\,eV). As CRs travel through the interstellar medium~(ISM) they are deflected by the magnetic fields that permeate interstellar space, propagating for up to millions of years and across distances as large as thousands of parsecs.

While propagating, CRs collide with gas particles and interact with the infrared radiation from stars and the cosmic microwave background~(CMB). These interactions create gamma-ray photons that can be observed at Earth. Observations of gamma~rays show a large-scale background component of emission along the plane of the Milky Way. Even after a century since CRs were discovered there is no widespread consensus on the locations where Galactic CRs are accelerated, their composition, or which processes create most observed gamma~rays.

Simulations of Galactic CR transport were performed with the software \GP, with the resulting gamma-ray flux calculated up to the PeV regime. The impact of altering parameters such as the number and distribution of CR sources, the distribution of infrared radiation between stars, and the distribution and strength of the Galactic magnetic field~(GMF), were investigated. For the first time the modelling variation in the TeV predictions due to uncertainties in the Galactic distributions was quantified. Additionally, the modelling variation from considering a stochastic placement of the CR sources was quantified up to 1\,PeV.

The simulation results were compared to the most detailed Galactic TeV gamma-ray survey: the \hess{} Galactic plane survey~(HGPS). The \GP{} predictions were broadly compatible with the large-scale emission from the HGPS after accounting for both the catalogued sources and estimates of the unresolved source fraction. At 1\,TeV the gamma-ray emission from CR electrons was found to contribute $\sim$50\% to the large-scale emission, with this leptonic component increasing further as energy increased. The GMF was found to be an important modelling consideration above 1\,TeV as it impacted the large-scale emission by approximately a factor of two. Additionally, the CR electron flux at Earth above 1\,TeV was found to vary by over a factor of ten over a period of a few million years due to the short cooling times and diffusion distances of the TeV CR electrons. The \GP{} models were found to agree with observations of the diffuse gamma~rays in the TeV regime by \hess, and the PeV regime by LHAASO, extending the demonstrated accuracy of \GP{} into the TeV--PeV regime.

The results in this thesis will aid in discovering the origin of the diffuse gamma-ray emission and allows constraints to be placed on how the CR sources are distributed in the MW. The results will also inform the next generation of experiments, such as the Cherenkov telescope array~(CTA), on possible observation strategies and background considerations. It was also found that the proposed CTA Galactic plane survey will be sensitive enough to observe the large-scale diffuse gamma-ray emission in the TeV regime.

    % Declaration of Originality
    % Set up a new page as the declaration of originality. Add it to the table of contents (toc) as a chapter, but make it a section as that is the preferred page setup
\newpage
\phantomsection
\addcontentsline{toc}{chapter}{Declaration of Originality}
\section*{Declaration of Originality}

\noindent
I certify that this work contains no material which has been accepted for the award of any other degree or diploma in my name in any university or other tertiary institution and, to the best of my knowledge and belief, contains no material previously published or written by another person, except where due reference has been made in the text. In addition, I certify that no part of this work will, in the future, be used in a submission in my name for any other degree or diploma in any university or other tertiary institution without the prior approval of the University of Adelaide and where applicable, any partner institution responsible for the joint award of this degree.

\noindent
The author acknowledges that the copyright of published works contained within this thesis resides with the copyright holder(s) of those works.

\noindent
I give permission for the digital version of my thesis to be made available on the web, via the University's digital research repository, the Library Search, and through web search engines, unless permission has been granted by the University to restrict access for a period of time.

\noindent
I acknowledge the support I have received for my research through the provision of an Australian Government Research Training Program
Scholarship.

\noindent
This thesis was submitted and accepted with a draft version of the paper ``On the Temporal Variability of the Galactic Multi-TeV Interstellar Emission''. This revised edition has been updated to include the published version of the manuscript.

% Add some vertical space so that it looks a little nicer. Play with the amount before submission
\vspace{2em}
\noindent
\begin{figure}[H]
    \centering
    \includegraphics[width=\textwidth]{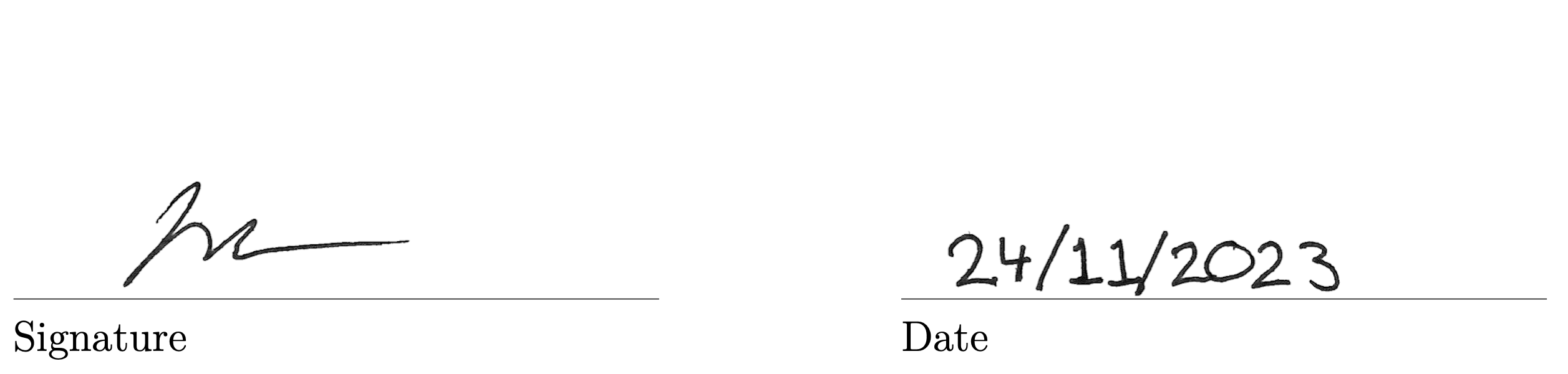}
\end{figure}

% % Original version, before signature
% \textbf{Signature: \hspace{9cm} Date:}
% Create a table to draw lines on the page for the signature and date fields
% \linewidth is the width of the column
% @{} = nothing on left and right border
% p{\hspace{0.15\textwidth}} adds horizontal space between the columns, where the p{} controls the column width
% \begin{tabular}{@{}p{0.4\textwidth}@{\hspace{0.15\textwidth}}p{0.4\textwidth}@{}}
%     {\Large \calligra Peter~Marinos} & {\LARGE \calligra 03\,/\,03\,/\,2023} \\[-0.9em]
%     \rule{\linewidth}{0.25pt} & \rule{\linewidth}{0.25pt} \\
%     Name & Date
% \end{tabular}
% \begin{tabular}{@{}p{0.4\textwidth}@{\hspace{0.15\textwidth}}p{0.4\textwidth}@{}}
%      &  \\[-0.9em]
%     \rule{\linewidth}{0.25pt} & \rule{\linewidth}{0.25pt} \\
%     Signature & Date
% \end{tabular}

\vfill
\noindent
\textsc{List of my published papers within this thesis:}
\begin{itemize}
    \item Paper 1: The Steady-State Multi-TeV Diffuse Gamma-Ray Emission Predicted with \GP{} and Prospects for the Cherenkov Telescope Array. Published in \mnras{} in February~2023. \nocite{MarinosP.2023}
    \item Paper 2: On the Temporal Variability of the Galactic Multi-TeV Interstellar Emissions. Published in \apj{} in March~2025. \nocite{MarinosP.2025}
\end{itemize}
    
    % Acknowledgements
    % Set up a new page as the acknowledgements. Add it to the table of contents (toc) as a chapter, but make it a section as that is the preferred page setup
\newpage
\phantomsection
\addcontentsline{toc}{chapter}{Acknowledgements}
\section*{Acknowledgements}

I acknowledge the support I have received for my research through the provision of an Australian Government Research Training Program Scholarship. Travel grants were provided by both the Astronomical Society of Australia and the Australian Institute of Physics.

This thesis makes use of the software \GP~\citep{MoskalenkoI.1998,StrongA.1998} which can be found at~\url{https://galprop.stanford.edu/}. \GP{} development is partially funded via NASA grants 80NSSC22K0477, 80NSSC22K0718, and 80NSSC23K0169. The Python packages utilised extensively throughout this work are: \verb|Numpy|~\citep{HarrisC.2020}, \verb|matplotlib|~\citep{HunterJ.2007}, \verb|SciPy|~\citep{VirtanenP.2020}, \verb|Astropy|~\citep{RobitailleT.2013,PriceWhelanA.2018}, and \verb|gammapy|~\citep{DeilC.2017}. The software \verb|HEALPix|~\citep{GorskiK.2005}, and the related Python package, is used for some results in this thesis.
The public data from the \hess{} Galactic plane survey~\citep{AbdallaH.2018a} is used for some results presented in this thesis and can be found at~\url{https://www.mpi-hd.mpg.de/hfm/HESS/hgps/}.
This thesis also makes use of NASA's Astrophysics Data System~(\url{https://ui.adsabs.harvard.edu/}).
The services provided by the Phoenix HPC service at the University of Adelaide supported all computational requirements.

On a more personal note, I would like to thank Prof.~Gavin Rowell and Dr.~Troy Porter. I would have never been able to finish without their support and indelible teachings. And thank you to Dr.~Fabien Voisin for his tireless work on UofA's HPC service.
I would also like to thank Dr.~Sabrina Einecke, A.~Prof.~Gu{\dh}laugur J{\'o}hannesson, and Dr.~Igor Moskalenko for their support and input on my manuscripts.
For my PhD colleagues, I would like to thank Tiffany Collins, Simon Lee, Olaf K{\"o}nig, and Rami Alsulami.

I would also like to thank my beautiful partner, Dr.~Hannah Twidale, for I fear the repercussions of omitting her mention.
Thank you to Brad Michelbach and Wayne Maloney for all the coffee and beer, respectively.
I must also thank my Mum, who listened to my endless ramblings and provided unremitting support throughout my PhD, and my Father, who was always there to join me at the Clipsal.
    
    % Insert the bulk of the thesis in the main matter
    \mainmatter
    \pagestyle{fancy}
    \fancyhead[RE]{\nouppercase{\leftmark}}
    \fancyhead[LE]{\thepage}
    \fancyhead[LO]{\nouppercase{\rightmark}}
    \fancyhead[RO]{\thepage}
    \fancyhead[CE,CO]{\ }
    \fancyfoot[RE,LE,RO,LO,CE,CO]{\ }
    \chapter{Introduction} \label{chap:intro}

In the early nineteen hundreds, ionising radiation was thought to be caused by the decay of radioactive materials in Earth's crust. The radiation would then travel upwards, ionising the atmosphere. A famous balloon flight by Viktor Hess in 1912 found that the radiation increased with altitude, showing that this ionisation radiation was from the cosmos -- hence the name: cosmic rays~\citep[CRs; as coined by][]{MillikanR.1928}. After another balloon flight during a solar eclipse, it was shown that this radiation was not from the Sun and must therefore come from outside the Solar system~\citep{HessV.1912}. Viktor Hess was later awarded a Nobel prize for these discoveries. Later, George Lema{\^i}tre and Manuel Vallarta measured an anisotropy in the CR arrival direction by utilising two arrays of Geiger counters separated by a cloud chamber. The anisotropy is caused by the magnetic field of the Earth, with positive charges coming from the West. Lema{\^i}tre and Vallarta observed that the majority of CRs came from the West, implying that most CRs had positive charges~\citep{LemaitreG.1933}.

CRs are highly energetic~(relativistic) particles, such as protons, anti-protons, atomic nuclei, electrons, positrons, and even neutrinos and anti-neutrinos, though usually only the charged particles are considered. CRs travel through space with velocities close to that of the speed of light -- so close that it becomes more useful to discuss CRs in terms of their energy. CR energies are typically measured in eV, or electron-Volts, and can be in the range of 1\,MeV to 100\,EeV~($10^{6}$~to~$10^{20}$\,eV), or one billion to one hundred quintillion electron-Volts -- enough energy to power a 60-Watt incandescent lightbulb for just over two seconds contained within a particle a trillion times smaller than a grain of sand~\citep{ClayR.1997}.

The study of CRs led to the discovery of many particles, from the muon by~\citet{AndersonC.1936} and its confirmation as a new particle by~\citet{StreetJ.1937}. Both the pion and kaon were discovered through CR observations by~\citet{LattesC.1947}. CRs are an integral part of the Milky Way~(MW) and the interstellar medium~(ISM). As CRs apply pressure to the interstellar gas they are essential to the formation and evolution of Galaxies and their stars~\citep{SemenovV.2021}. CRs also affect the chemistry in the ISM~\citep{BayetE.2011} and Earth's atmosphere~\citep{NicoletM.1975} and are responsible for the creation of certain elements such as boron.

The CR energy density in the Galactic plane is approximately 1\,eV\,cm$^{-3}$, which is similar to the energy density of the Galactic magnetic field~(GMF), the turbulent motions of the interstellar gas, and even that of the interstellar radiation field~\citep[i.e.~all light emitted by stars, the CMB, and thermal emission from dust combined;][]{WebberW.1998}. Therefore, CRs represent an important dynamical component of the MW. However, after more than a century since their discovery, many open questions remain. The mechanisms that accelerate CRs to these extreme energies, and even the mechanisms behind how they travel away from their sources, are not fully understood.

The locations where CRs are accelerated are referred to as CR sources. These sources are thought to be shell-type supernova remnants~(SNRs), pulsar wind nebulae~(PWNe), stellar clusters, gamma-ray~(\graya) binaries, or even the supermassive black holes at the centres of galaxies. There are a multitude of processes that can theoretically accelerate the CRs to the energies we observe, with each source type possibly employing more than one mechanism. The energy spectrum of CRs provides valuable insight into these acceleration processes, as each possible mechanism results in a different spectrum of CRs being injected into the ISM. However, as CRs are charged particles, their paths are deflected by the large-scale~(>100\,pc) and small-scale~(<100\,pc) magnetic fields throughout the MW.

Both the large-scale and small-scale magnetic fields in the MW have some randomness due to the ionised gases in the ISM creating their own localised magnetic fields. The CRs will scatter off the magnetic fields, performing a random walk through the ISM as they spread away from their acceleration sites. This process is known as diffusion.
CR diffusion in the MW is a complex physical process -- while CRs propagate they interact with the ISM, losing energy and creating \grays{} through collisions with the ISM gas, bremsstrahlung interactions with the ISM nuclei, and inverse Compton scattering on low-energy photons. As \grays{} are not deflected by the ambient medium they can be used to probe the underlying CR physics surrounding the CR accelerators and provide critical information on particle interactions.

\begin{figure}[t]
    \centering
    \includegraphics[width=\textwidth]{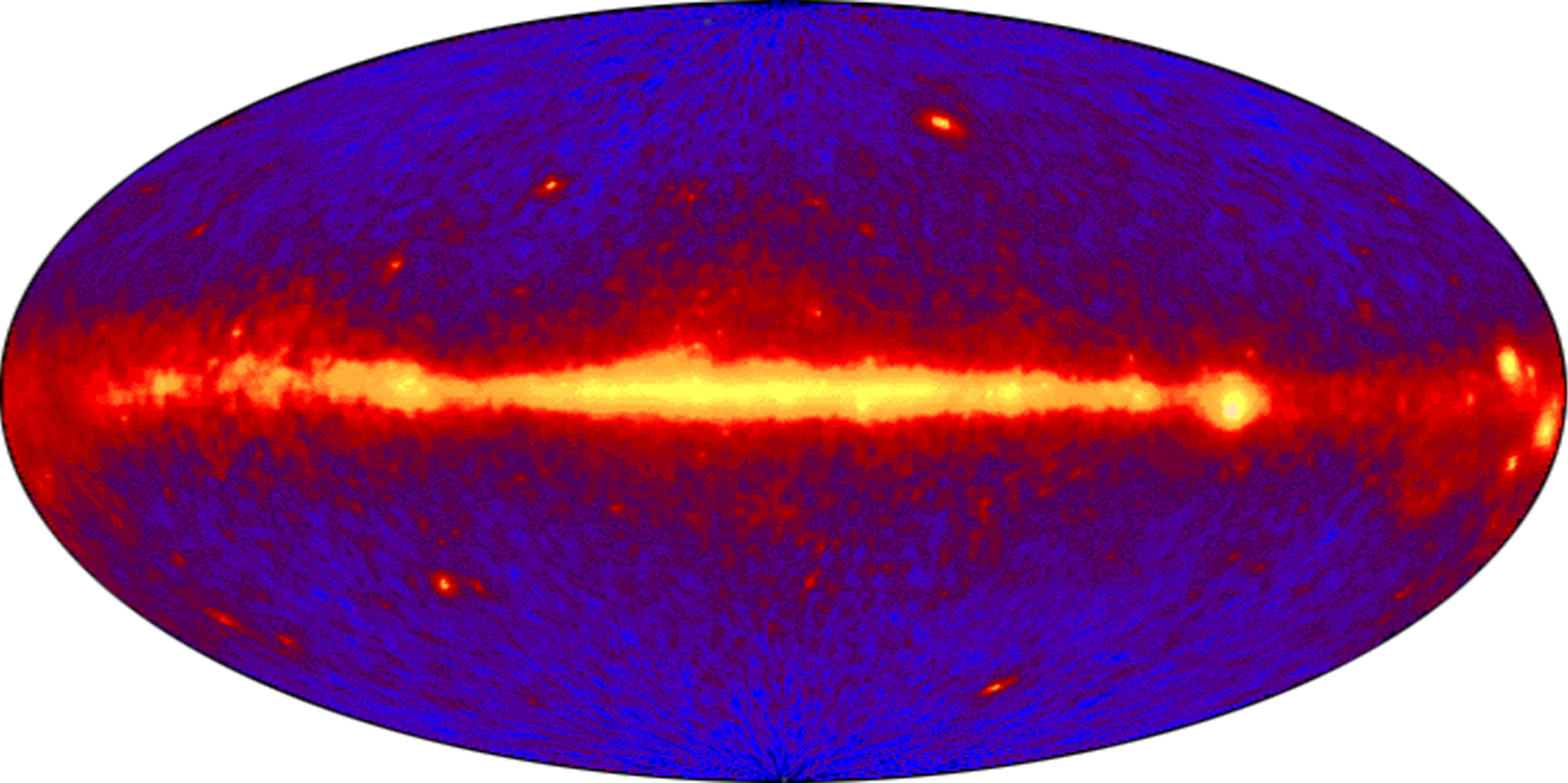}
    \includegraphics[width=\textwidth]{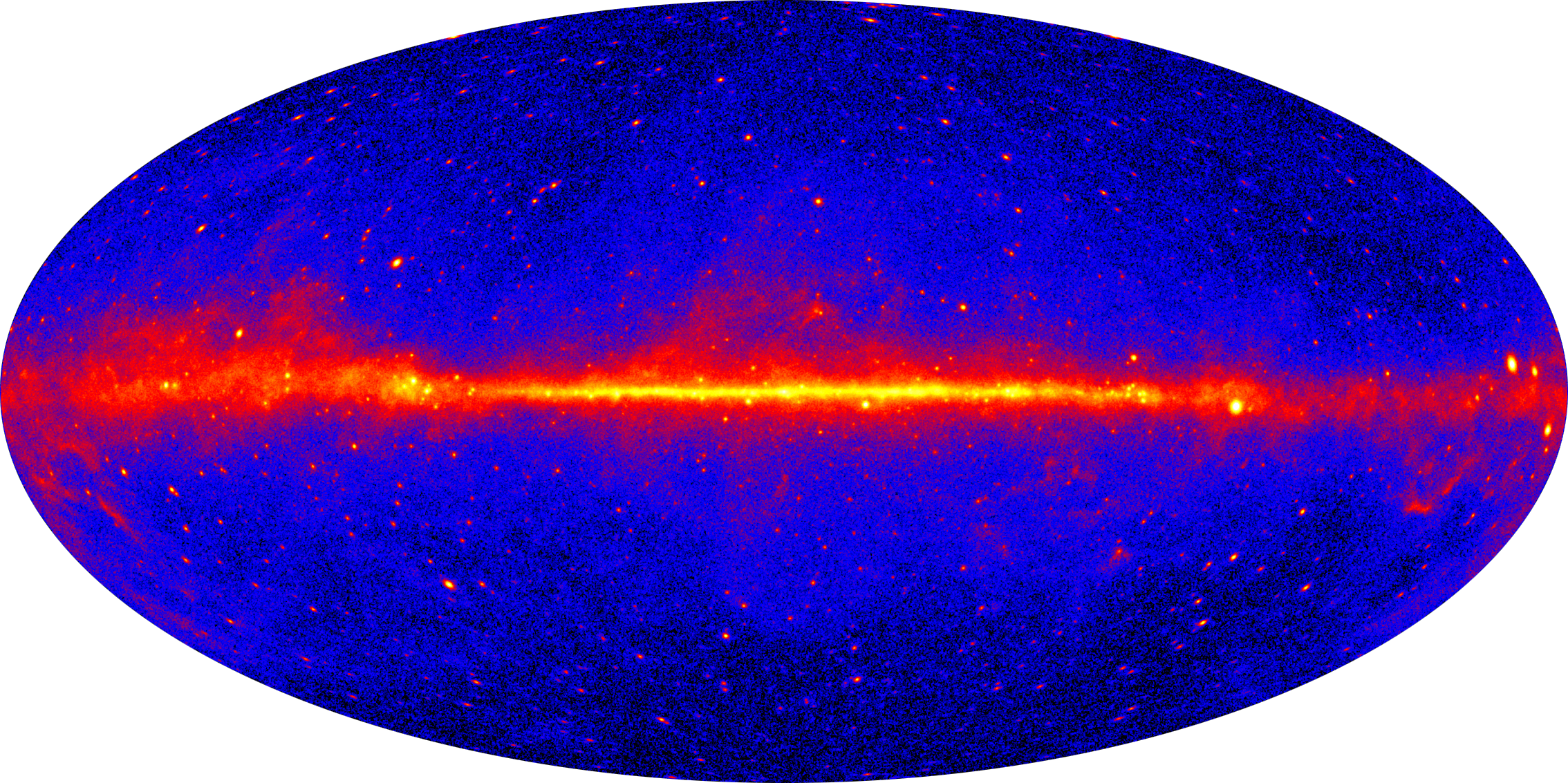}
    \caption{The EGRET \graya{} flux integrated above 100\,MeV~\citep[top image;][]{HartmanR.1999}, and the \fermi{} \graya{} flux integrated above 1\,GeV for the first five years of operation~\citep[bottom image;][]{AckermannM.2012}.}
    \label{fig:fermi image}
\end{figure}

The OSO--3~\citep{KraushaarW.1972} mission was a satellite launched in~1967 that observed X-rays and \grays. OSO--3 observed a large structure of \grays{} along the Galactic plane of the MW in the MeV to GeV energy range. Later, COS--B~\citep{BignamiG.1983} was able to distinguish discrete \graya{} sources from the large-scale structure. Following these observatories, the EGRET~\citep{KanbachG.1988} mission was able to resolve over~200 \graya{} sources and provided the first detailed view of the large-scale \graya{} emission for energies between 30\,MeV and 30\,GeV~\citep{HunterS.1997}. With the launch of \fermi~\citep{AtwoodW.2009}, which is still in operation, there are now over~5,000 identified MeV to GeV \graya{} sources in the MW, with the total Galactic MeV and GeV \graya{} emission being dominated by a large-scale structure~\citep{AbdollahiS.2020}.
Maps of the all-sky EGRET and \fermi{} \graya{} emission can be seen in \autoref{fig:fermi image}. Both the EGRET and \fermi{} skymaps show discrete sources and a large structure along the Galactic plane. Difference in large-scale \graya{} structure can also be seen between 100\,MeV and 1\,GeV skymaps.
For the TeV energy regime the dominant emission changes -- the \hess{} collaboration, in the \hess{} Galactic plane survey~(HGPS), found that the individual \graya{} sources dominated over the diffuse, large-scale \graya{} emission~\citep{AbdallaH.2018a}. Despite five decades of observations of the large-scale \graya{} emission, the origin of the diffuse emission has not yet reached a consensus.

The large-scale \graya{} emission is defined as the \graya{} structure along the Galactic plane. It covers all Galactic longitudes and spans up to 10$^{\circ}$ in Galactic latitude for MeV \grays. The large-scale \graya{} emission includes both the emission from unresolved \graya{} sources and the \graya{} emission from diffuse CRs. The large-scale \textit{diffuse} \graya{} emission is that from only diffuse CRs~(i.e.~CRs that have travelled~>100\,pc from their initial site of acceleration). Recently the large-scale emission has been observed in the TeV energy range by \argo~\citep{BartoliB.2015}, the PeV energy range by \tibet~\citep{AmenomoriM.2021} and LHAASO~\citep{SciascioG.2016,CaoZ.2023}, and will likely soon be observed by HAWC~\citep{AlbertA.2022a}. The Cherenkov telescope array~\citep[CTA;][]{AcharyaB.2018} is currently under construction and will likely observe the diffuse emission at TeV energies within its first decade~\citep{AceroF.2013,AcharyaB.2018}.

The diffuse emission can overpower low surface-brightness \graya{} sources, especially in the GeV--TeV energy regime. Hence, it is critical that accurate and precise models of the background diffuse emission be constructed. For example, the \fermi{} collaboration employed some of the emissions predicted by the CR propagation software \GP{} as a background interstellar emissions model. Using \GP{} as a background model enabled the \fermi{} collaboration to extract \graya{} source characteristics and properties from the survey~\citep{AckermannM.2012}. With more sensitive TeV observatories, the dimmest \graya{} sources will only be observable if employing a well-motivated and accurate diffuse model as a background interstellar emissions model. Furthermore, diffuse emission estimates will improve the accuracy and precision of \graya{} source analyses at GeV--PeV energies. Improved diffuse emission estimates may even help confirm the existence of dark matter through the observations of CRs and \grays{} from dark matter particle self-annihilations in the Galactic centre~\citep{ConradJ.2017}.

Creating accurate models of the MW's \graya{} emission across the entire energy range requires modelling of multiple interacting systems and processes, such as: the gas density and its ionisation, atomic, and chemical composition, the distribution of CR sources across the Milky Way, the interstellar radiation field~(ISRF), and the Galactic magnetic field~(GMF). These processes and distributions must be modelled simultaneously. In this thesis the CR propagation and \graya{} production code \GP{} is used to perform this modelling.

Although the HGPS has been compared to various other TeV \graya{} facilities~\citep[e.g.][]{NeronovA.2020,AbdallaH.2021}, there have been no comparisons with the simulated TeV \graya{} flux from CR propagation codes. The primary goal of this thesis is then to compare the large-scale TeV \graya{} emission~(and estimates of the diffuse emission) from \hess{} to the simulated diffuse TeV \graya{} emission from \GP. Characterising the precision of the models at TeV energies is integral to understanding the validity of using said models as backgrounds and will aid in constraining the origin of the diffuse \graya{} emission.

\vspace{2em}
\noindent
To achieve these goals, this thesis will begin by describing the \GP{} models in \autoref{chap:galprop} -- from how CR diffusion is simulated, to how the \graya{} emission is calculated. A description of the Galactic distributions used throughout the modelling is also included. \autoref{chap:hess} will then focus on \graya{} observations, particularly the HGPS, discussing limitations and important considerations for estimates of the TeV large-scale diffuse emission. \autoref{chap:comparing galprop and hess} will bring the simulations and observations together, detailing how a fair comparison between the predictions and observations was ensured. The first published work is included as \autoref{chap:paper 1} and contains the results from the comparisons between \GP{} and the HGPS.

\noindent
The \GP{} simulations are then extended to use a time-dependent solution to the diffusion equation. The motivations are discussed, and the results scrutinised, in \autoref{chap:time dep}. The effects of using a time-dependent solution on the local CR results, the Galaxy-wide \graya{} results, and the implications for comparisons to the HGPS, are included in the second published work in \autoref{chap:paper 2}.

\noindent
Future observations from CTA present a leap forward for TeV \graya{} measurements. Using the \graya{} predictions from \GP{} and the simulated performance of CTA, it will be determined whether future observations by CTA will be able to observe the diffuse \graya{} emission in \autoref{chap:CTA and gammapy}.

\noindent
Finally, \autoref{chap:summary} summarises the work and findings from this thesis and discusses future extensions.

    \chapter{Modelling Cosmic-Ray Transport with GALPROP} \label{chap:galprop}

Modelling the broad-band non-thermal \graya{} emissions within the Milky Way~(MW) requires simultaneously accounting for the CR energy losses and non-thermal emissions, as all interactions are governed by the same physics.
There are multiple software packages to perform these simulations, such as: \GP~\citep{MoskalenkoI.1998,StrongA.1998}, DRAGON~\citep{EvoliC.2008}, and PICARD~\citep{KissmannR.2014}.
This project uses the \GP{} framework for modelling CR transport throughout the MW.

\GP{} builds on the approach described by \citet{StrongA.1993,StrongA.1995,StrongA.1996b} and began with the goal of constraining CR production and propagation in the MW.
\GP{} has an extensive history of successfully reproducing observations: from the local CR spectra~\citep[e.g.][]{MoskalenkoI.2002,StrongA.2004}, the Galactic synchrotron emission~\citep[e.g.][]{StrongA.2011}, and the Galactic diffuse \graya{} emission from MeV to PeV \graya{} energies~\citep[e.g.][]{StrongA.2004,PorterT.2008,MarinosP.2023,MarinosP.2025}.
The modelled interstellar non-thermal emission from \GP{} has been used extensively as a background model by the \fermi{} collaboration~\citep[e.g.][]{AbdoA.2010,AceroF.2016a,AceroF.2016b,AjelloM.2016}, and systematic studies using CR propagation models such as \GP{} have been employed for extracting source properties for various supernova remnants~\citep[SNRs;][]{AceroF.2016b}.
In the near future, \GP{} will aid CTA in determining source properties for faint sources such as PWNe.
In this work \GP{} version~57 is utilised -- for a full description of features before version~57 see~\citet{StrongA.2013}, and for the additions to version~57 see~\citet{PorterT.2022}.

\GP{} numerically solves the transport equation to calculate the propagation of charged CR particles moving between 3D pixels~(also named voxels, and are referred to as the propagation cells in this thesis). \GP{} then uses these CR densities to calculate the all-sky \graya{} emission. \GP{} takes advantage of a wide variety of multi-messenger astronomical datasets, from CR flux measurements to radio observations, from both ground and space-based experiments. Data sources used by \GP{} include Ulysses, both of the \textit{Voyager} probes, Bell Labs, \fermi, PAMELA, AMS, and many others.

This chapter begins with an overview of CR propagation and \graya{} production, with a focus on the distances and time scales involved in modelling diffusion throughout the MW.
The \GP{} input distributions will also be discussed, including: the CR source distribution, the ISM gas distribution, the interstellar radiation field~(ISRF), and the Galactic magnetic field~(GMF). The diffusion parameters, such as the CR flux spectral shapes, CR normalisations, etc., will also be discussed. A brief explanation on applying the information in this chapter to set up a physically informative \GP{} simulation is given in \autoref{chap:running galprop}.

\section{Cosmic-Ray Transport} \label{sect:CR diffusion}

CRs are relativistic particles such as electrons~($e^{-}$), positrons~($e^{+}$), anti-protons~($\bar{p}$), protons/hydrogen nuclei~($p$), and heavier nuclei\footnote{Although neutrinos~($\nu$) can also be classified as CRs, this thesis focuses on charged particles.}. CRs can have energies as low as a 1\,MeV~($10^{6}$\,eV) or as high as 100\,EeV~($10^{20}$\,eV). This thesis focuses on CRs in the 0.1--$10^{3}$\,TeV~($10^{11}$--$10^{15}$\,eV) energy regime. These CRs have rest masses significantly smaller than their kinetic energies; hence, this energy regime is known as ultra-relativistic. The kinetic energy of a relativistic particle is given by:

\begin{align}
    E_{k} &= (\gamma_{\mathrm{CR}} - 1)m_{\mathrm{CR}} c^2 \\
    \gamma_{\mathrm{CR}} &= \frac{1}{\sqrt{1 - \beta_{\mathrm{CR}}^{2}}} \label{eq:lorentz factor}\\
    \beta_{\mathrm{CR}} &= \mathrm{v}_{\mathrm{CR}} / c
\end{align}

\noindent
where the subscript `CR' denotes that the variable can be applied to any CR particle, $\gamma_{\mathrm{CR}}$ is the Lorentz factor, $m_{\mathrm{CR}}$ is the rest mass, $c$ is the speed of light in a vacuum, and $\mathrm{v}_{\mathrm{CR}}$ is the total velocity of the CR. The CR species can be any charged particle -- from a hydrogen nucleus/proton to an iron nucleus, to an electron or positron. For the ultra-relativistic regime, where the kinetic energy is much larger than the rest mass of the CR~(i.e.~$\gamma_{\mathrm{CR}} \gg 1$), the above equations simplify to:

\begin{align}
    E_{k} &\approx \gamma_{\mathrm{CR}} m_{\mathrm{CR}}c^{2} \\
    \therefore \gamma_{\mathrm{CR}} &\approx \frac{E_{k}}{m_{\mathrm{CR}}c^{2}}
\end{align}

\noindent
where all variables have been defined previously. For a 1\,TeV proton, which has a rest-mass energy of $0.938 \times 10^{9}$\,eV, the ultra-relativistic approximation is accurate to 0.1\%.

As CRs have an electric charge they are deflected by the magnetic fields that permeate interstellar space. The CRs will travel in a curved path along the magnetic field lines, where the radius of the curved path is known as the gyro-radius. The gyro-radius of a charged particle in a magnetic field is given by:

\begin{align}
    r_{g} &= \frac{m_{\mathrm{CR}} \mathrm{v}_{\perp}}{\abs{q_{\mathrm{CR}}} B} \label{eq:rg}
\end{align}

\noindent
where $B$ is the strength of the magnetic field, $\mathrm{v}_{\perp}$ is the velocity of the CR perpendicular to the magnetic field, and $q_{\mathrm{CR}}$ is the charge of the CR species. As the rest mass of an ultra-relativistic particle makes a negligible contribution to the total energy, the gyro-radius can be simplified as:

\begin{align}
    E^{2} &= (p c)^{2} + (m_{\mathrm{CR}} c^{2} )^{2} \label{eq:energy} \\
    \therefore E_{k}^{2} &\approx (pc)^{2} \label{eq:rel approx} \\
    \Rightarrow r_g &= \frac{E_{k}}{c \abs{q_{\mathrm{CR}}} B} \label{eq:rg(E)}
\end{align}

\noindent
where $p$ is the momentum of the CR perpendicular to the direction of the magnetic field.
As the gyro-radius of a CR is proportional to the kinetic energy, lower energy CRs will be deflected more than higher energy CRs. Additionally, as the gyro-radius is inversely proportional to the magnetic field strength, CRs travelling through a stronger magnetic field will be deflected more than CRs travelling through a weaker magnetic field.
For the Milky~Way~(MW), the magnetic field strength varies between 3--10\,$\mu$G~\citep{RobitailleT.2012}. Therefore, CRs with energies below approximately $10^{15}$\,eV~($10^{3}$\,TeV) will have gyro-radii of approximately 110--360\,pc.
As the Galactic disk of the MW has a thickness of approximately 200--300\,pc, CRs with energies lower than 1\,PeV are unlikely to escape the MW, resulting in a diffuse `sea' of very-high-energy~(VHE) CRs dispersed throughout the Galaxy.

As the process through which the CRs propagate throughout the Galaxy depends on the magnetic field, it is useful to define a particle rigidity, $\varrho_{\mathrm{CR}}$. The magnetic rigidity is the resistance of a charged particle against deflection in a magnetic field/Lorentz force. The rigidity of a CR~($\varrho_{\mathrm{CR}}$) is typically measured in Volts and is given by:

\begin{equation}
    \begin{aligned}
        \varrho_{\mathrm{CR}} &= B r_{g} \\
        &= \frac{E_{k}}{c \abs{q_{\mathrm{CR}}}} \label{eq:rigidity}
    \end{aligned}
\end{equation}

\noindent
where all variables have been defined previously. Equations \ref{eq:rg(E)} and \ref{eq:rigidity} show that higher energy CRs~(or CRs with lower electric charges) are more resilient against deflection by a magnetic field. Within the MW there are two components to the magnetic field; the regular field created by the global structure of the MW~(e.g.~the spiral arms), and the irregular field caused by local~(i.e.~<100\,pc) variations caused by the ionised gas, stars, or other processes and objects. The regular field leads to a slow-moving bulk flow of CRs, while the irregular component introduces stochastic diffusion\footnote{For a recent review of CR diffusion, see~\citet{BeckerTjusJ.2020}.} as the CRs interact with~(and are deflected by) random magnetohydrodynamic waves.
Given the strength of the Galactic magnetic field~\citep[GMF; $\sim$3\,$\mu$G;][]{RobitailleT.2012} and the $\sim$kpc distances travelled by VHE CRs, the trajectories of the VHE CRs within the MW are dominated by diffusive processes.

As CRs diffuse they undergo a variety of interactions with the ISM -- from proton--proton collisions~(\pps) with the ISM gas, bremsstrahlung interactions with the ISM gas nuclei, inverse Compton~(IC) scattering on the ISRF, and synchrotron interactions with magnetic fields. These processes transfer the CR's energy to \grays{} or other lower-energy photons and particles, and will be discussed in more depth in \autoref{ssect:hadronic emission} and \autoref{ssect:leptonic emission}. The time it takes for a CR to lose all of its kinetic energy via a particular process is known as the cooling time, and is given by:

\begin{align}
    t_{\mathrm{cool}}(E_{k}) &= \int_{0}^{E_{k}} \frac{\mathrm{d}E}{\Dot{E}_{k}} \label{eq:cooling time}
\end{align}

\noindent
where $\Dot{E}_{k}$ is the rate that the CR loses kinetic energy. If $\Dot{E}_{k}$, then the cooling time can be simplified to $t_{\mathrm{cool}} = E_{k} / \Dot{E_{k}}$.

CRs in the TeV regime travel in approximately straight lines between scattering events.
During the scattering, the kinetic energy of the CR is transferred into photons.
After the scattering event the CR will begin travelling in a different, random direction. This process is repeated until all kinetic energy is lost, and is referred to as a random-walk process.
The diffusion distance is found by averaging over many scattering events.
For a given diffusion time, $t(E_{k})$, and by relating the random-walk theory to CR diffusion in three dimensions~\citep{ChandrasekharS.1943}, the root-mean-square distance that a CR will diffuse is described by the well-known equation:

\begin{align}
    d(E_{k}) &= \sqrt{6 D(E_{k}) t(E_{k})} \label{eq:diffusion distance}
\end{align}

\noindent
where $D(E_{k})$ represents the diffusion coefficient.
It is worth noting that the average distance a CR will diffuse before losing all of its kinetic energy~(i.e.~the cooling distance) can be found by substituting the cooling time from~\autoref{eq:cooling time} into \autoref{eq:diffusion distance}.

Throughout the literature it is common to see the diffusion coefficient defined as a power law in energy, and sometimes as a function of both energy and magnetic field strength. The diffusion coefficient\footnote{In general, $D_{xx}$ and $D_{pp}$ are diffusion tensors, allowing the diffusion speed to vary with direction. However, as \GP{} has not yet implemented anisotropic diffusion, they are referred to as diffusion coefficients throughout this work.} is given as a function of CR rigidity~\citep{SeoE.1994} with the form:

\begin{align}
    D_{xx}(\varrho_{\mathrm{CR}}) &= \beta_{\mathrm{CR}} D_{xx,0} \left( \frac{\varrho_{\mathrm{CR}}}{\varrho_{0}} \right)^{\delta} \label{eq:diffusion coefficient}
\end{align}

\noindent
where $\delta$ is the spectral index of the diffusion coefficient~(often called the diffusion index), and $\varrho_{0}=4$\,GV is the rigidity at which the diffusion coefficient is normalised. The normalisation diffusion coefficient, $D_{xx,0}$ varies with the chosen CR source model~(see \autoref{ssect:parameter optimisation}). The diffusion index is typically given by $\delta=1/3$ for a Kolmogorov spectrum of the ISM turbulence~\citep{KolmogorovA.1941} and $\delta=1/2$ for Iroshnikov-Kraichnan turbulence~\citep{IroshnikovP.1964,KraichnanR.1965}. In \autoref{ssect:parameter optimisation} the diffusion index is treated as a parameter that is fit to observational CR data.

For a more complete picture of CR diffusion, the partial differential equation known as the transport equation must be used. \GP{} uses the differential transport equation in three spatial dimensions over time to propagate CRs through different volume elements, hereafter referred to as the `propagation cells', or simply as cells. The calculation is performed for each nuclear species, where the partial differential transport equation~\citep{GinzburgV.1964,BerezinskiiV.1990} is given by:

\begin{equation}
    \begin{aligned}
        \pdv{\psi}{t} &= Q(\Vec{r},\,p) \\
        &\quad + \Vec{\nabla} \cdot (D_{xx} \Vec{\nabla} \psi - \Vec{v}_{\mathrm{conv}} \psi) \\
        &\quad + \pdv{}{p}p^{2}D_{pp}\pdv{}{p}\frac{1}{p^{2}}\psi \\
        &\quad - \pdv{}{p} \left( \Dot{p}\psi - \frac{p}{3}(\Vec{\nabla} \cdot \Vec{v}_{\mathrm{conv}})\psi \right) \\
        &\quad - \frac{\psi}{T_{f}} - \frac{\psi}{T_{r}} \label{eq:Transport Equation}
    \end{aligned}
\end{equation}

\noindent
where $\psi = \psi(\Vec{r},\,p,\,t)$ is the CR density per unit of particle momentum at the coordinate $\Vec{r}$ and time $t$, $\Dot{p}$ is the change in the CR momentum over time, $D_{xx}$ and $D_{pp}$ are the spatial and momentum-space diffusion coefficients, respectively, and are assumed to be independent of the CR species, $\Vec{v}_{\mathrm{conv}}$ is the convection velocity of the Galactic wind, and $T_{f}$ and $T_{r}$ are the time scales for nuclear fragmentation and radioactive decay, respectively. The source term, $Q(\Vec{r},\,p)$, captures all CRs being injected into the MW. The CRs being created as secondary particles will be discussed in \autoref{sssect:secondary creation}, and the CRs being injected into the MW by CR accelerators will be discussed in \autoref{sect:source dist.}.

For the individual terms in the transport equation: $\Vec{\nabla} D_{xx} \Vec{\nabla} \psi$ accounts for the spatial diffusion of the CRs, $\Vec{\nabla} \cdot \Vec{v}_{\mathrm{conv}} \psi$ accounts for advection of the CRs due to the motion of the ISM gas, $\pdv{}{p}p^{2}D_{pp}\pdv{}{p}\frac{1}{p^{2}}\psi$ accounts for the reacceleration of the CRs due to interactions with magnetic fields, $\pdv{}{p} \Dot{p}\psi$ accounts for radiative loses of the CRs, and $\pdv{}{p} \frac{p}{3}(\Vec{\nabla} \cdot \Vec{v}_{\mathrm{conv}})\psi$ accounts for adiabatic expansion (and compression) of the ISM gas which decreases (or increases) the CR density within the volume of gas.
The transport equation must be solved numerically as the combination of spatially and temporally varying differential terms make it impossible to find a closed analytical expression.

To numerically solve \autoref{eq:Transport Equation} \GP{} has a variety of solution methods that can be chosen by the user.
The Crank-Nicolson finite differencing technique~\citep{PressW.1992} is used in this thesis, which is second-order accurate in time and unconditionally stable independent of the timestep size~\citep{PressW.1992}. The Crank-Nicolson technique calculates the CR energy density moving from one spatial cell to another, where adjacent cells in all three spatial dimensions influence one another.
The time-updating step for the Crank-Nicholson solution method is given by:

\begin{align}
    \psi_{\mathscr{j}+1} &= \psi_{\mathscr{j}} + \frac{1}{2} \left[ \left( \pdv{\psi}{t} \right)_{\mathscr{j}} + \left( \pdv{\psi}{t} \right)_{\mathscr{j}+1} \right] \Delta t \label{eq:Crank-Nicholson}
\end{align}

\noindent
where $\mathscr{j}$ denotes the current timestep and $\Delta t$ is the size of the timestep.
A description of the full finite differencing technique in 3D, including descriptions of the boundary conditions for the various spatial axes, can be found in~\citet{StrongA.2013}.
For accurate, meaningful results, the minimum timestep must be shorter than the fastest cooling time of any particle included in the simulation~(\autoref{eq:cooling time}). Similarly, the size of the cells must be smaller than the shortest cooling distance of any particle in the simulation~(found by substituting \autoref{eq:cooling time} into \autoref{eq:diffusion distance}).

\section{Cosmic-Ray Energy Losses and Gamma-Ray Production} \label{sect:gamma-ray production}

As CRs travel through the Galaxy they can interact with the ISM gas, the ISRF, and the GMF. These interactions will transfer the kinetic energy of the CRs into photons~(including \grays). The process of CRs converting their kinetic energy into other forms is often referred to as cooling. The rate of cooling depends on the process and the CR species, limiting the length of time and distance that the CRs can diffuse.

This section provides broad explanations of how \graya{} photons are produced through the various processes in the MW. References to the full formalism used by \GP{} will be included; however, the equations provided here will be approximations.
The included approximations allow the cooling times and cooling distances~(i.e.~maximum diffusion distances) for various processes to be calculated to first order.
The main assumptions made in this section will be for large-scale~(>100\,pc), Galaxy-wide diffusion, and for ultra-relativistic~($E_{k}>0.01$\,TeV) CRs.

\subsection{Hadronic Processes} \label{ssect:hadronic emission}

CRs can collide with the atoms in the ambient gas in the ISM, with these collisions being energetic enough to produce subatomic particles and \grays. This interaction applies to CR protons and other heavier CR nuclei.
In general, the hadronic emission is due to nuclei--nuclei~($N$--$N$) collisions.

The formalism used by \GP{} to calculate the hadronic emission can be found in~\citet{MoskalenkoI.1998}, which uses equations from~\citet{DermerC.1986a} and~\citet{DermerC.1986b}.
\GP{} calculates the full $N$--$N$~collision interaction cross section for any CR nuclei from $^{1}$H to $^{64}$Ni onto a variety of gas species~\citep{StrongA.1998,StrongA.2007,StrongA.2009}. The total inelastic cross sections for $p$, $d$~(deuteron), $^{3}$He, and $^{4}$He are parameterised for any arbitrary target, with \GP{} using the formalism from~\citet{FerrandoP.1988,BarashenkovV.1994,WellischH.1996}. \GP{} also includes the formalism from~\citet{TripathiR.1996}, with corrections given in~\citet{PorterT.2022}.
As 90\% of the ISM gas is hydrogen~(see \autoref{sect:ISM gas}), and 90\% of hadronic CRs are protons~\citep{MewaldtR.1994}, only proton--proton collisions (\pps) will be discussed in depth here.

\subsubsection{Proton--Proton Collisions} \label{sssect:pp collisions}

% As the interaction does not actually create 'either a neutral or charged', but creates all three in random amounts, show as one interaction
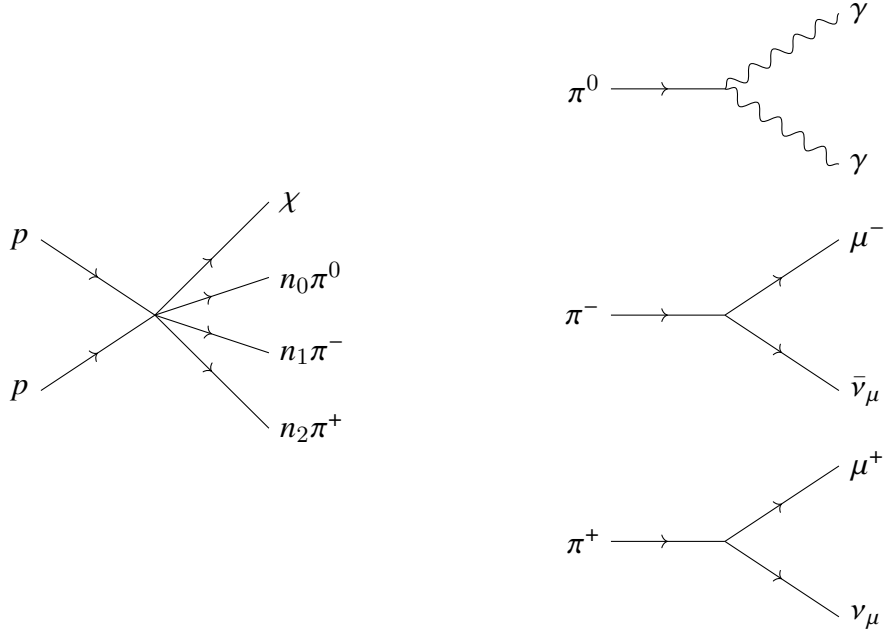
\begin{figure}[t]

    \centering

    \begin{tikzpicture}[node distance=1cm and 1.5cm]

        \coordinate[label=left:$p$] (p1);                    % p1
        \coordinate[below right=of p1] (int1);               % ineraction1
        \coordinate[below left=of int1,label=left:$p$] (p2); % p2

        \draw[particle] (p1) -- (int1);   % proton1 to interaction1
        \draw[particle] (p2) -- (int1);   % proton2 to interaction1

        \coordinate[right=of int1] (temp); % temporary point

        \coordinate[above=0.5cm of temp, label=right:$n_{0} \pi^{0}$] (pi0); % pi-
        \coordinate[above=of pi0, label=right:$\chi$] (2ps); % protons

        \coordinate[below=0.5cm of temp, label=right:$n_{1} \pi^{-}$] (pi-); % pi0
        \coordinate[below=of pi-, label=right:$n_{2} \pi^{+}$] (pi+); % pi+

        \draw[particle] (int1) -- (2ps);  % protons leaving interaction1
        \draw[particle] (int1) -- (pi0);  % pi0 leaving interaction1
        \draw[particle] (int1) -- (pi-);  % pi- leaving interaction1
        \draw[particle] (int1) -- (pi+);  % pi+ leaving interaction1

        \coordinate[right=of temp] (spacer0);
        \coordinate[right=of spacer0] (spacer1);

        % pi- decay

        \coordinate[right=of spacer1, label=left:$\pi^{-}$] (pi-1);
        \coordinate[right=of pi-1] (int4);
        \coordinate[above right=of int4, label=right:$\mu^{-}$] (mu-);
        \coordinate[below right=of int4, label=right:$\bar{\nu}_{\mu}$] (antinmu);

        \draw[particle] (pi-1) -- (int4);
        \draw[particle] (int4) -- (mu-);
        \draw[particle] (int4) -- (antinmu);

        % pi0 decay

        \coordinate[above=of mu-, label=right:$\gamma$] (g2); % gamma ray
        \coordinate[above left=of g2] (int2);
        \coordinate[above right=of int2, label=right:$\gamma$] (g1); % gamma ray
        \coordinate[left=of int2, label=left:$\pi^{0}$] (pi01);

        \draw[particle] (pi01) -- (int2);
        \draw[photon] (int2) -- (g1); % pi0 decaying to gamma
        \draw[photon] (int2) -- (g2); % pi0 decaying to gamma

        % pi+ decay

        \coordinate[below=of antinmu, label=right:$\mu^{+}$] (mu+);
        \coordinate[below left=of mu+] (int3);
        \coordinate[left=of int3, label=left:$\pi^{+}$] (pi+1);
        \coordinate[below right=of int3, label=right:$\nu_{\mu}$] (nmu);

        \draw[particle] (pi+1) -- (int3);
        \draw[particle] (int3) -- (mu+);
        \draw[particle] (int3) -- (nmu);

    \end{tikzpicture}

    \captionsetup{type=figure,justification=centering,format=plain}
    \captionof{figure}{A \pp{} creating neutral pions ($\pi^{0}$) and charged pions ($\pi^{\pm}$) is shown on the left, with $\chi$ representing any additional particles created within the interaction to conserve baryon number and electric charge.
    The interaction has the equation $p + p \rightarrow \chi + n_{0} \pi^{0} + n_{1} \pi^{-} + n_{2} \pi^{+}$, where the number of neutral, negative, and positive pions produced in the collision are given by $n_{0}$, $n_{1}$, and $n_{2}$, respectively.
    The three pion decay pathways are shown on the right.
    The top pathway shows a neutral pion decaying into two \grays{} ($\gamma$) via the equation $\pi^{0} \rightarrow 2\gamma$.
    The central pathway shows a negatively charged pion decaying into a negatively charged muon ($\mu^{-}$) and a muon antineutrino~($\bar{\nu}_{\mu}$) via the equation $\pi^{-} \rightarrow \mu^{-} + \bar{\nu}_{\mu}$.
    The bottom pathway shows a positively charged pion decaying into a positively charged muon ($\mu^{+}$) and a muon neutrino~($\nu_{\mu}$) via the equation $\pi^{+} \rightarrow \mu^{+} + \nu_{\mu}$.
    Although it is not shown, the charged muons will decay further into electrons, positrons, muon antineutrinos, and electron antineutrinos~($\Bar{\nu}_{e}$) via the equations $\mu^{-} \rightarrow e^{-} + \nu_{\mu} + \Bar{\nu}_{e}$ and $\mu^{+} \rightarrow e^{+} + \Bar{\nu}_{\mu} + \nu_{e}$.}
    \label{fig:pp diagrams}
\end{figure}

Neutral, positive, and negative pions~($\pi$) can be produced when a CR proton collides with another proton. To create a pion, the CR proton must have a total energy~(rest mass plus kinetic energy) greater than approximately 1.22\,GeV~\citep{KelnerS.2006}. The pions decay quickly -- charged pions have lifetimes of $2.6 \times 10^{-8}$\,s and neutral pions have lifetimes of $8.5 \times 10^{-17}$\,s~\citep{ZylaP.2020}. Depending on the charge of pion, the decay produces \grays, muons~($\mu$), or neutrinos~($\nu$).
Diagrams of \pps{} and the various pion pathways, along with the interaction equations, are shown in \autoref{fig:pp diagrams}.

The energy loss rate of CR protons due to collisions with the ISM hydrogen gas~\citep{AharonianF.1996} is given by:

\begin{align}
    \left[ \frac{\mathrm{d} E_{p}}{\mathrm{d}t} \right]_{\mathrm{pp}} &= \sigma_{\mathrm{pp}} n_{\mathrm{H}} f c E_{p}
\end{align}

\noindent
where $c$ is the speed of light in a vacuum, $n_{\mathrm{H}}$ is the number density of the ambient hydrogen gas~(see \autoref{sect:ISM gas}), $E_{p}$ is the kinetic energy of the CR proton. The elasticity of the collision is given by $f$ and is a function of both the target gas species and the kinetic energy of the CR~\citep{GaisserT.1990}. For CR protons in the ultra-relativistic regime colliding with the ambient hydrogen gas, the elasticity is often approximated by $f=0.5$.
$\sigma_{\mathrm{pp}}$ is the cross section for a \pp{} and is calculated explicitly by \GP{} by using a function fit to cross-section data from~\citep{DermerC.1986a}. For a first-order approximation, the \pp{} cross section from 10\,GeV to 1\,PeV can be given by $\sigma_{\mathrm{pp}} \approx 40 \times 10^{-31}$\,m$^{2}$ following~\citet{AharonianF.1996}. The approximate cooling time for CR protons in the ISM is then given by:

\begin{align}
    t_{\mathrm{pp}} &= 5.3 \times 10^{7} \left( \frac{n_{\mathrm{H}}}{\mathrm{cm}^{-3}} \right)^{-1}\,\mathrm{yr} \label{eq:cool pp}
\end{align}

\noindent
where all variables have been defined previously. With the above assumptions, the cooling time for the CR protons undergoing \pp{} interactions is proportional only to the density of the medium that the CR proton is travelling through.

\subsubsection{Secondary Particle Creation} \label{sssect:secondary creation}

As shown in \autoref{fig:pp diagrams}, \pps{} may create secondary particles. In a similar fashion, $N$--$N$~collisions are likely to create many secondary particles including other atoms with atomic numbers lower than the initial CR or ISM gas particle.
\GP{} considers the following reactions for the calculation of secondaries: $p$--$p$, $p$--$N$, $N$--$p$, and $N$--$N$.
The cross sections for these interactions can be found in~\citet{TanL.1983}.
Calculations for $\bar{p}$ production and propagation are given by~\citet{MoskalenkoI.2002,MoskalenkoI.2003,KachelriessM.2015,KachelriessM.2019}, and the production of the neutral mesons, secondary electrons, and positrons is calculated using the formalism from~\citet{DermerC.1986a,DermerC.1986a}. \GP{} can also use the more recent formalism from~\citet{KamaeT.2006,KachelriessM.2012,KachelriessM.2014,KachelriessM.2019}.

The secondary particles can be considered as an additional source of CRs. The source term~($Q(\Vec{r},\,p)$) from \autoref{eq:Transport Equation} has a component due to secondary particle creation given by:

\begin{align}
    Q(\Vec{r},\,p^{\prime}) &= \beta_{\mathrm{CR}} c n_{H}(\Vec{r}) \int \left( \psi_{\mathrm{CR}}(\Vec{r},\,p) \frac{\mathrm{d}\sigma(p,\,p^{\prime})}{\mathrm{d}p^{\prime}} \right)\,\mathrm{d}p \label{eq:secondary source term}
\end{align}

\noindent
where $\psi$ is the CR density~(from \autoref{eq:Transport Equation}), $\Vec{r}$ is the Galactic position of the secondary production, and $\beta_{\mathrm{CR}}$ is given by \autoref{eq:lorentz factor}. $p^{\prime}$ is the momentum of the secondary particle and $\mathrm{d}\sigma(p,\,p^{\prime})/\mathrm{d}p^{\prime}$ is the differential cross section of the secondary particle interaction, which is different for each process that can create secondaries. The full description of how the secondary source term is implemented in \GP{} can be found in~\citet{MoskalenkoI.1998}. The secondary source function implemented in v57 of \GP{} is defined for primary particles up to and including $^{64}$Ni and includes secondary particle creation from nuclear fragmentation and radioactive decay of all isotopes~\citep{PorterT.2022}.

\subsection{Leptonic Processes} \label{ssect:leptonic emission}

When a charged particle is decelerated it radiates its energy in the form of a photon. This emission mechanism disproportionately impacts low-mass particles, such as electrons, as they are decelerated more easily than higher-mass particles such as protons.
In extreme environments such as active galactic nuclei~(AGN), protons and heavier nuclei can be decelerated significantly and emit \grays{} via the processes detailed in this section. However, \grays{} are unlikely to be emitted by hadronic CRs via these processes for the cases discussed in this thesis.
Hence this section will only discuss the CR electrons\footnote{The leptonic emission processes also impact the CR positrons, and all equations are identical. Only the CR electrons are discussed here for simplicity.}.

There are three main processes that cause CRs to be decelerated: bremsstrahlung, synchrotron radiation, and IC scattering. Bremsstrahlung and IC interactions in the $E>1$\,TeV energy regime are catastrophic processes, converting all kinetic energy of the electron into a single \graya{} photon. Synchrotron interactions do not often create \grays{} and are more likely to produce radiation from radio to X-ray photons. However, a discussion of the synchrotron energy losses is included below as they are an important energy-loss mechanism for the CR electrons.

\subsubsection{Bremsstrahlung} \label{sssect:Bremsstrahlung}

\begin{figure}
    \centering
    \begin{tikzpicture}[node distance=1cm and 1.5cm]

        \coordinate[label=below:$N$] (N);   % centre
        \path (N) +(0,+0.25) coordinate (ac);   % above centre
        \path (N) +(-0.177,0.177) coordinate (lc);
        \path (N) +(+0.177,0.177) coordinate (rc);
        \coordinate[below left=of lc, label=left:$e^{-}$] (bl);  % bottom left
        \coordinate[below right=of rc, label=right:$e^{-\prime}$] (br); % bottom right
        \coordinate[right=of ac, label=right:$\gamma$] (mr); % top right

        \draw[particle] (bl) -- (lc); % bottom left electron
        \draw[photon] (mr) -- (ac); % centre photon
        \draw[particle] (rc) -- (br); % bottom right electron
        
        \node[circle,fill=black,inner sep=0pt, minimum size=0.2cm] (a) at (0,0) {};
        
        \draw (0,0) pic{carc=45:135:0.25};

    \end{tikzpicture}
    \captionsetup{type=figure,justification=centering,format=plain}
    \captionof{figure}{A Bremsstrahlung interaction, where an electron ($e^{-}$) is deflected by the strong electric fields around an atomic nucleus~(shown by the black circle), giving the final states $e^{-\prime}$ and $\gamma$. The equation for a Bremsstrahlung interaction is given by: $e^{-} + N \rightarrow e^{-\prime} + \gamma + N$.}
    \label{fig:bremsstrahlung diagram}
\end{figure}
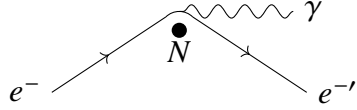

Bremsstrahlung radiation is emitted when a charged particle with some velocity is rapidly slowed down by the charged nucleus of another atom. As the charged particle is decelerated, the kinetic energy is transferred into a photon.
Hence, CRs diffusing through the ISM gas can undergo bremsstrahlung interactions with the nuclei of the ambient ISM gas particles.
In the relativistic regime a large fraction of the kinetic energy of the CR is converted into a high-energy \grayn. For an ultra-relativistic CR almost all of the kinetic energy is converted into a \grayn.
A diagram of a bremsstrahlung interaction, along with the interaction equation, is shown in \autoref{fig:bremsstrahlung diagram}.

\GP{} calculates the full bremsstrahlung energy loss rate and bremsstrahlung interaction cross sections for a range of ambient ISM gas species for any CR energy. \GP{} calculates bremsstrahlung radiation from both leptonic and hadronic CR interactions with the ISM gas. The full formalism used by \GP{} for the bremsstrahlung calculations can be found in~\citet{StrongA.2000}.

For CR electrons, which produce the majority of bremsstrahlung radiation, the energy loss rate due to bremsstrahlung interactions is given by:

\begin{align}
    \left[ \frac{\mathrm{d}E_{e}}{\mathrm{d}t} \right]_{\mathrm{brem}} &= c \beta_{e} \sum_{s} \left[ n_{s}(\Vec{r}) \int \left( E_{\gamma} \frac{\mathrm{d} \sigma_{\mathrm{brem}} (E_{e},\,E_{\gamma},\,s)}{\mathrm{d} E_{\gamma}} \right)\,\mathrm{d}E_{\gamma} \right]
\end{align}

\noindent
where $E_{e}$ is the kinetic energy of the electron, $E_{\gamma}$ is the energy of the resulting \grayn, and $n_{s}(\Vec{r})$ is the density of the ISM gas at the location $\Vec{r}$, where the subscript $s$ denotes the atomic species of the gas~(e.g.~$n_{\mathrm{H}}$ is the density of the hydrogen gas, see \autoref{sect:ISM gas}). $\mathrm{d} \sigma_{\mathrm{brem}}(E_{e},\,E_{\gamma},\,s) / \mathrm{d} E_{\gamma}$ is the differential cross section of a bremsstrahlung interaction on said ISM gas species, and is given by:

\begin{align}
    \frac{\mathrm{d} \sigma_{\mathrm{brem}} (E_{e},\,E_{\gamma},\,s)}{\mathrm{d} E_{\gamma}} &= \frac{\alpha r_{e}^{2}}{E_{\gamma}} \left( \left[ 1 + \left( \frac{\gamma_{f}}{\gamma_{e}} \right)^{2} \right] \Phi_{s,1} (E_{e},\,E_{\gamma},\,s) - \frac{2}{3} \frac{\gamma_{f}}{\gamma_{e}} \Phi_{s,2} (E_{e},\,E_{\gamma},\,s) \right) \\
    \gamma_{f} &= \frac{E_{e} - E_{\gamma}}{m_{e} c^{2}}
\end{align}

\noindent
where $\alpha$ is the fine structure constant, $\gamma_{e}$ is the Lorentz factor for the electron, $\gamma_{f}$ is the final Lorentz factor for the electron~(i.e.~the Lorentz factor of the electron after the interaction) and $r_{e} \approx 2.817 \times 10^{-15}$\,m is the classical radius of the electron. The functions $\Phi_{s,1}$ and $\Phi_{s,2}$ are calculated individually for every species of gas in the ISM via the Schiff formula~\citep{KochH.1959} and represent the electron scattering.

The bremsstrahlung in the ISM can be approximated using similar assumptions as for the \pps~(see \autoref{sssect:pp collisions}). The ISM gas consists of 90\% hydrogen~\citep[][see \autoref{sect:ISM gas}]{FerriereK.2001}.
The most simplistic case is that for ionised hydrogen~(\hii) as it is an unshielded charge~(i.e.~it has no bound electron). Hence, the charge distribution of \hii{} is due simply to the single proton in the nucleus. Using the Schiff formula for ionised hydrogen, it can be found that $\Phi_{\mathrm{H}\,\textsc{ii},1}=\Phi_{\mathrm{H}\,\textsc{ii},2}=\Phi_{\mathrm{H}\,\textsc{ii}}$, which is given by:

\begin{align}
    \Phi_{\mathrm{H}\,\textsc{ii}} = 4 \left[ \ln \left( \frac{2 E_{e}}{m_{e} c^{2}} \frac{E_{e} - E_{\gamma}}{E_{\gamma}} \right) - \frac{1}{2} \right]
\end{align}

\noindent
where all variables have been defined previously. Following~\citet{AharonianF.2004b}, further assumptions can be made. For CR electrons with energies~$E_{e} > 2$\,MeV the CR electron is unlikely to be captured by the nucleus. For this energy range the electron losses are a significant fraction of the total kinetic energy of the electron. Additionally, as the average ISM gas density is low~(varying between $10^{1}$ and $10^{3}$\,atoms\,cm$^{-3}$) a bremsstrahlung interaction is unlikely to produce a cascade of many collisions through the ISM gas. Hence, bremsstrahlung interactions can be viewed as individual collisions. These three assumptions allow the rate of energy lost via bremsstrahlung interactions at large scales~(i.e.~>100\,pc) to be simplified down to the form:

\begin{align}
    \left[ \frac{\mathrm{d}E_{e}}{\mathrm{d}t} \right]_{\mathrm{brem}} &= \frac{9 n_{\mathrm{H}} \sigma_{\mathrm{brem}} c}{7} E_{e}
\end{align}

\noindent
where $\sigma_{\mathrm{brem}} \approx 1.88 \times 10^{-30}$\,m$^{-2}$ is the integrated cross section for the bremsstrahlung interaction on \hii{} for ultra-relativistic CR electrons. With these assumptions and simplifications, the final form for the cooling time~\citet{AharonianF.2004b} is given by:

\begin{align}
    t_{\mathrm{brem}} &= 4.0 \times 10^{7} \left( \frac{n_{\mathrm{H}}}{\mathrm{cm}^{-3}} \right)^{-1}\,\mathrm{yr} \label{eq:cool brem}
\end{align}

\noindent
where all variables have been defined previously. The cooling time for bremsstrahlung interactions at scales >100\,pc depends only on the density of the medium through which the CR electron travels. Similarly to \pps, gas clouds are expected to emit bremsstrahlung \grays, and the kinetic energy of the CR electrons is transferred into \grays{} faster as the ISM gas density increases.

\subsubsection{Synchrotron Radiation} \label{sssect:Synchrotron}

CR electrons travelling through a magnetic field are decelerated, emitting synchrotron photons in the process. Synchrotron radiation is typically made up of non-thermal, low-energy photons such as radio waves; however, it can also account for X-rays in extreme environments~(e.g.~around pulsars). Although the synchrotron photon flux is not calculated in this thesis, the synchrotron energy losses can become significant for Galactic regions with intense magnetic fields. The synchrotron losses impact the CR electron density throughout the MW.
The GMF models considered in this work will be discussed later in \autoref{sect:GMF}.
A diagram of a synchrotron interaction, along with the interaction equation, is shown in \autoref{fig:synchotron diagram}.

\begin{figure}
    \centering
    \begin{tikzpicture}[node distance=1cm and 1.5cm]

        \coordinate[] (centre);   % centre
        \path (centre) +(0,+0.25) coordinate (ac);   % above centre
        \path (centre) +(-0.177,0.177) coordinate (lc);
        \path (centre) +(+0.177,0.177) coordinate (rc);
        \coordinate[below left=of lc, label=left:$e^{-}$] (bl);  % bottom left
        \coordinate[below right=of rc, label=right:$e^{-\prime}$] (br); % bottom right
        \coordinate[right=of ac, label=right:$\gamma$] (mr); % top right

        \draw[particle] (bl) -- (lc); % bottom left electron
        \draw[photon] (mr) -- (ac); % centre photon
        \draw[particle] (rc) -- (br); % bottom right electron
        
        \draw (0,-0.05) node[cross=0.15cm] {};
        \node[label=below:$\Vec{B}$] (b) at (0,0) {};
        
        \draw (0,0) pic{carc=45:135:0.25};

    \end{tikzpicture}
    \captionsetup{type=figure,justification=centering,format=plain}
    \captionof{figure}{A synchrotron event, where an electron ($e^{-}$) is deflected by a magnetic field~($\Vec{B}$, directed into the page) giving the final states $e^{-\prime}$ and a lower-energy photon $\gamma$. The equation for synchrotron interactions is given by: $e^{-} + \Vec{B} \rightarrow e^{-\prime} + \gamma$.}
    \label{fig:synchotron diagram}
\end{figure}

The amount of energy an electron loses via synchrotron depends on the angle between the electron and the magnetic field lines~\citep{RybickiG.1985}.
For the regular component of the GMF \GP{} uses the formalism for the synchrotron energy losses~(and the synchrotron photon flux) from~\citet{RybickiG.1985}. For the irregular/random component, \GP{} uses the formalism from~\citet{GhiselliniG.1988,WaelkensA.2009}.
For calculating the cooling times to first order, the interaction can be averaged over all possible angles. This assumption is fair if the velocities of the electrons are randomly oriented, which is valid if the electrons are truly diffuse. Hence, the energy loss rate~\citep{RybickiG.1985} can be approximated by:

\begin{align}
    \left[ \frac{\mathrm{d}E_{e}}{\mathrm{d}t} \right]_{\mathrm{syn}} &= \frac{32 \pi r_{e}^{2}}{9} c \beta_{e}^{2} \gamma_{e}^{2} U_{B} \\
    U_{B} &= \frac{|\Vec{B}|^{2}}{8 \pi}
\end{align}

\noindent
where $\beta_{e}$ is perpendicular to the magnetic field, $r_{e} \approx 2.817 \times 10^{-15}$\,m is the classical radius of the electron, $|\Vec{B}|^{2}$ is the magnitude of magnetic field vector, and $U_{B}$ is the energy density of the magnetic field.
For diffusion across large distances~(>100\,pc) it is important to consider the Galactic magnetic field, which will be discussed in more depth in \autoref{sect:GMF}.
Assuming the CRs are in the relativistic regime, and focusing on large distances, the cooling time can be simplified~\citep{AharonianF.2004b} giving:

\begin{align}
    t_{\mathrm{syn}}(E_{e}) &= 12 \times 10^{6} \left( \frac{B}{\mathrm{\mu G}} \right)^{-2} \left( \frac{E_{e}}{\mathrm{TeV}} \right)^{-1}\,\mathrm{yr} \label{eq:cool sync}
\end{align}

\noindent
where all variables have been defined previously. The synchrotron cooling time is inversely proportional to the square of the magnetic field strength.
The dependence of the \graya{} emission on the GMF, and the results on the impact of altering the GMF, will be discussed later in \autoref{chap:paper 1}.

\subsubsection{Inverse Compton Scattering} \label{sssect:IC}

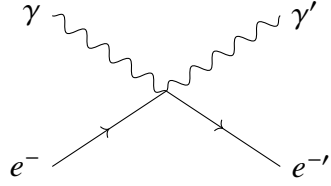
\begin{figure}
    \centering
    \begin{tikzpicture}[node distance=1cm and 1.5cm]

        \coordinate[label=left:$\gamma$] (tl);   % Top left
        \coordinate[below right=of tl] (centre); % centre
        \coordinate[below left=of centre, label=left:$e^{-}$] (bl);  % bottom left
        \coordinate[below right=of centre, label=right:$e^{- \prime}$] (br); % bottom right
        \coordinate[above right=of centre, label=right:$\gamma^{\prime}$] (tr); % top right

        \draw[photon] (tl) -- (centre); % top left photon
        \draw[photon] (centre) -- (tr); % top right photon
        \draw[particle] (bl) -- (centre); % bottom left electron
        \draw[particle] (centre) -- (br); % bottom right electron

    \end{tikzpicture}
    \captionsetup{type=figure,justification=centering,format=plain}
    \captionof{figure}{An IC scattering event, where a photon ($\gamma$) collides with an electron ($e^{-}$). The electron will impart some energy to the photon, giving the final states $e^{-\prime}$ and $\gamma^{\prime}$. The equation for the IC scattering interaction is given by: $e^{-} + \gamma \rightarrow e^{- \prime} + \gamma^{\prime}$.}
    \label{fig:IC scattering diagram}
\end{figure}

Inverse Compton~(IC) scattering is the process in which a CR electron interacts with a photon with lower energy, such as a CMB photon, infrared~(IR), optical, ultra-violet, or X-ray photon. For some \graya{} sources, such as blazars, IC scattering can also occur on the radio photons emitted via synchrotron~\citep[synchrotron self-Compton;~e.g.][]{JonesT.1974,BloomS.1996}. The electron transfers some fraction of its kinetic energy to the photon, upscattering the photon to higher energies. In the relativistic regime the electron transfers a significant fraction of its kinetic energy to the photon, with the process being catastrophic in the ultra-relativistic regime. If the electron-photon system has enough total energy, then the photons can potentially be upscattered into the \graya{} regime, with this process significantly contributing to the \graya{} flux observed at Earth~\citep{AharonianF.1997}. A diagram of an IC scattering interaction, along with the interaction equation, is shown in \autoref{fig:IC scattering diagram}.

The amount of energy transferred, and even the likelihood of the interaction, depends on the initial energies of both the electron and low-energy photon, as well as the relative angles between the electron and the photon. \GP{} calculates the anisotropic IC scattering, which considers the effects of the electron and photon angles relative to one another, with the full formalism for the anisotropic IC scattering cross sections found in~\citet{MoskalenkoI.2000}, with details on the full directionality of the ISRF is given in~\citet{MoskalenkoI.2006}. The two most recent models for the background photon field used by \GP{} are given by~\citet{PorterT.2017} and are detailed in \autoref{sect:ISRF}. For calculating the cooling time, the amount of energy transferred can be approximated to first order by averaging over all possible electron and background photon relative angles. Hence, the energy lost to IC scattering events~\citep{AharonianF.2004b} is given by:

\begin{align}
    \left[ \frac{\mathrm{d}E_{e}}{\mathrm{d}t} \right]_{\mathrm{IC}} = \frac{4 \sigma}{3} c \gamma_{e}^{2} U_{\mathrm{rad}}
\end{align}

\noindent
where $U_{\mathrm{rad}}$ is the energy density of the radiation field, which includes contributions from the cosmic microwave background~\citep[CMB; 0.26\,eV\,cm$^{-3}$;][]{AharonianF.2004b,FixsenD.2009} and the interstellar radiation field~(discussed later in \autoref{sect:ISRF}). The IC scattering interaction cross section between the electron and photon is given by $\sigma$.
The energy ranges of the electron-photon system where the recoil force on the electron is negligible is named the Thomson regime.
For the Thomson regime the interaction cross section is constant, where the Thomson cross section~\citep{ThomsonJ.1933} is given by:

\begin{equation}
    \begin{aligned}
        \sigma_{T} &= \frac{8\pi}{3} r_{e}^{2} \label{eq:thomson cross section} \\
        &\approx 6.652 \times 10^{-29}\,\mathrm{m}^{2}
    \end{aligned}
\end{equation}

\noindent
where $r_{e} \approx 2.817 \times 10^{-15}$\,m is the classical radius of the electron. As the recoil force on the electron becomes significant, the IC scattering interaction enters the Klein-Nishina~(KN) regime. The interaction cross section in the KN regime is smaller than in the non-relativistic limit, with the IC scattering interaction becoming less likely. Furthermore, in the KN regime the fraction of kinetic energy transfer approaches unity~\citep[i.e.~the electron transfers all of its kinetic energy to the photon;][]{BlumenthalG.1970}. The recoil force is significant when $\Gamma \gg 1$, where $\Gamma$ is a unitless quantity given by:

\begin{align}
    \Gamma &= \frac{4 E_{e} E_{b}}{(m_{e} c^{2})^{2}}
\end{align}

\noindent
where $E_{b}$ is the energy of the `background'~(low energy) photon. Rearranging for the electron energy and substituting various photon energies, the KN regime begins at $E_{e} \approx 100$\,TeV for CMB photons and $E_{e} \approx 5$\,TeV for IR photons, with both the CMB and IR fields accounted for in \GP~(as described in \autoref{sect:ISRF}). The KN cross section depends on the relative angle between the CR electron and background photon. When discussing diffuse electrons, the average over all angles can be taken. It is common throughout the literature to write the KN cross section in terms of the scattered photon energy~($E_{\gamma}$), or as a ratio of the scattered photon energy to the initial kinetic energy of the electron. Using the former convention, and recalling that the electron gives almost all its kinetic energy to the photon in the KN regime, the differential KN cross section is given by:

\begin{align}
    \frac{\mathrm{d}\sigma_{\mathrm{KN}}(E_{e},\,E_{b},\,E_{\gamma})}{\mathrm{d}E_{\gamma}} &= 2 \pi r_{e}^{2} \frac{(m_{e} c^2)^2}{E_{e}^{2} E_{b}} G(\Gamma,\,\xi) \label{eq:KN cross section} \\
    G(\Gamma,\,\xi) &= 2 \xi \ln (\xi) + ( 1 + 2 \xi ) ( 1 - \xi ) + \frac{( \xi \Gamma )^{2} ( 1 - \xi )}{2 ( 1 + \xi \Gamma )} \\
    \xi &= \frac{E_{\gamma}}{\Gamma E_{e} ( 1 - E_{\gamma} / E_{e} )} \\
    E_{\gamma} &\approx ( \gamma_{e} - 1 ) m_{e} c^{2} + E_{b}
\end{align}

\noindent
where $E_{\gamma}$ is the energy of the resulting \grayn, $G(\Gamma,\,\xi)$ and $\xi$ are unitless functions, and $E_{b} \ll m_{e} c^{2} \ll E_{e}$. The assumption that $E_{b} \ll m_{e} c^{2}$ is true for both the CMB and IR photon fields. For electron energies $E_{e} \approx 10$\,TeV and for incident IR photons, it is found that $E_{e} E_{b} \approx (m_{e} c^{2})^{2}$. Taking this assumption, and further assuming that $E_{e} \approx E_{\gamma}$, the integral KN cross section simplifies to $\sigma_{\mathrm{KN}} = 2 \pi r_{e}^{2} G(\Gamma,\xi)$. The latter form is common throughout the literature as IR photons account for the majority of the IC scattering. However, it should be noted that this approximation lies on the edge of the KN regime, and the condition $\Gamma \gg 1$ is only marginally satisfied.

In the KN regime the IC cross section decreases and the synchrotron losses begin to dominate over the IC losses. For $\Gamma < 10^{4}$, which is valid for $E_{e} \leq 100$\,TeV and $E_{b} \leq 10^{-1}$\,eV, the ratio of the energy radiated into synchrotron versus IC is given by:

\begin{align}
    \frac{\left[ \mathrm{d}E_{e} / \mathrm{d}t \right]_{\mathrm{IC}}}{\left[ \mathrm{d}E_{e} / \mathrm{d}t \right]_{\mathrm{syn}}} &= \frac{U_{\mathrm{rad}}}{U_{B}} \frac{1}{( 1 + \Gamma )^{1.5}}
\end{align}

\noindent
where all variables have been defined previously. For the Galactic ISRF~(CMB, IR, optical, and ultra-violet photons), and a weaker Galactic magnetic field strength~(inter-spiral-arm regions; 3\,$\mu$G; see \autoref{sect:GMF}) where the IC only just dominates over the synchrotron, the ratio of IC to synchrotron remains approximately constant for electron energies $E_{e} < 100$\,TeV despite the KN cross section decreasing~\citep{ModerskiR.2005}. For stronger Galactic magnetic fields~(intra-spiral-arm regions; 10\,$\mu$G; see \autoref{sect:GMF}), the synchrotron losses can dominate across all electron energies. Assuming electron energies $E_{e} < 100$\,TeV the cooling time for the IC losses~\citep{AharonianF.2004b,ModerskiR.2005} is given by:

\begin{align}
    t_{\mathrm{IC}}(E_{e}) &= 3.1 \times 10^{8} \left( \frac{U_{\mathrm{rad}}}{\mathrm{eV} \ \mathrm{cm}^{-3}} \right)^{-1} \left( \frac{E_{e}}{\mathrm{GeV}} \right)^{-1}\,\mathrm{yr} \label{eq:cool ic}
\end{align}

\noindent
where all variables have been defined previously.
Although this equation is accurate for electron kinetic energies $E_{e} < 100$\,TeV, it over-estimates the electron IC scattering losses on the IR component of the ISRF for $E_{e} > 100$\,TeV.
For the CMB component, the equation is valid up to $E_{e} > 1$\,PeV.
Above these energies the electrons become more likely to lose energy via synchrotron interactions, and \autoref{eq:cool ic} becomes an overestimate of the IC scattering cooling time.

The IC scattering cooling time is inversely proportional to $U_{\mathrm{rad}}$, which is isotropic for the CMB photon field and follows the distribution of stars in the MW for IR, optical, and UV photons.
The dependence of IC scattering on the ISRF, as well as the impacts on the \graya{} emission caused by altering the ISRF, will be discussed later in \autoref{chap:paper 1}.

\subsection{Gamma-Ray Flux} \label{ssect:GP flux calculation}

The \graya{} emissivity represents the magnitude of \graya{} radiation emitted by some volume. The emissivity is a function of the \graya{} energy and is described in \GP{} as a function of position. The \graya{} emissivity is given by:

\begin{align}
    \varepsilon(\Vec{r},\,E_{\gamma}) &= \frac{1}{V} \int \left( \frac{\mathrm{d}\sigma(E_{k},\,E_{\gamma})}{\mathrm{d}E_{\gamma}} \psi(\Vec{r},\,E_{k}) \right)\,\mathrm{d}E_{k} \label{eq:emissivity}
\end{align}

\noindent
where $V$ is the volume of the propagation cell and $\mathrm{d}\sigma(E_{k},\,E_{\gamma}) / \mathrm{d}E_{\gamma}$ is the differential cross section of an interaction with a CR with kinetic energy $E_{k}$ creating a \grayn{} with energy $E_{\gamma}$.
As the synchrotron emission is not calculated in this thesis, the equations for the synchrotron emissivity~\citep[][Equations~51--57]{StrongA.2013} are not included here. The \graya{} flux~($J$, provided by \GP{} in units of MeV$^{-1}$\,cm$^{-2}$\,s$^{-1}$\,sr$^{-1}$) is the integral over the line of sight~($x$) of the emissivity. The IC, bremsstrahlung, and pion-decay \graya{} fluxes, respectively, are given by:

\begin{align}
    J_{\mathrm{IC}}(\mathscr{l},\,\mathscr{b},\,E_{\gamma}) &= \frac{1}{4 \pi} \int \varepsilon_{\mathrm{ic}}(\Vec{r},\,E_{\gamma})\,\mathrm{d}x \label{eq:ic intensity from emissivity} \\
    J_{\mathrm{brem}}(\mathscr{l},\,\mathscr{b},\,E_{\gamma}) &= \frac{1}{4 \pi} \sum_{s} n_{s}(\Vec{r}) \int \varepsilon_{\mathrm{brem}}(\Vec{r},\,E_{\gamma})\,\mathrm{d}x \label{eq:brem intensity from emissivity} \\
    J_{\pi^{0}}(\mathscr{l},\,\mathscr{b},\,E_{\gamma}) &= \frac{1}{4 \pi} \sum_{s} n_{s}(\Vec{r}) \int \varepsilon_{\pi^{0}}(\Vec{r},\,E_{\gamma})\,\mathrm{d}x \label{eq:pion intensity from emissivity}
\end{align}

\noindent
where $(\mathscr{l},\,\mathscr{b})$ are the Galactic sky coordinates and $n_{s}(\Vec{r})$ is the density of the ISM gas at the location $\Vec{r}$. For bremsstrahlung emission and pion-decay emission the \graya{} emissivity is defined per gas species~($s$). Therefore, a sum must be performed over the ISM gas species when calculating the pion-decay and bremsstrahlung \graya{} flux integrals.

\subsubsection{Pair Absorption} \label{sssect:pair absorption}

As \grays{} travel through the ISM they may collide with a background photon from the CMB or ISRF~(see \autoref{sect:ISRF}). This interaction creates an electron-positron pair, as shown in \autoref{fig:pair production diagram}, with each particle obtaining half of the energy of the initial \grayn{} in the centre-of-momentum reference frame. The electron-positron pair may instantly annihilate with one another or diffuse away and become lower-energy secondary CRs. As pair absorption reduces the \graya{} flux that reaches Earth it is common to define the effect of pair absorption as an optical depth~($\tau_{\gamma\gamma}$) along the line of sight. The optical depth calculation used in \GP{} accounts for the full directionality of the ISRF~\citep{MoskalenkoI.2006,PorterT.2018} and is given by:

\begin{figure}
    \centering
    \begin{tikzpicture}[node distance=1cm and 1.5cm]

        \coordinate[label=left:$\gamma$] (tl);   % Top left
        \coordinate[below right=of tl] (centre); % centre
        \coordinate[below left=of centre, label=left:$\gamma$] (bl); % bottom left
        \coordinate[above right=of centre, label=right:$e^{-}$] (tr); % top right
        \coordinate[below right=of centre, label=right:$e^{+}$] (br); % bottom right

        \draw[photon] (tl) -- (centre); % top left photon
        \draw[photon] (bl) -- (centre); % top left photon
        \draw[particle] (centre) -- (br); % bottom right electron
        \draw[particle] (centre) -- (tr); % top right positron

    \end{tikzpicture}
    \captionsetup{type=figure,justification=centering,format=plain}
    \captionof{figure}{A pair production event, where two photons ($\gamma$) collide, creating an electron-positron pair ($e^{-}$ and $e^{+}$, respectively). The equation for the pair production interaction is given by: $\gamma + \gamma \rightarrow e^{-} + e^{+}$.}
    \label{fig:pair production diagram}
\end{figure}
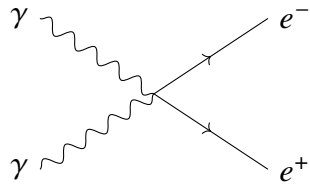

\begin{align}
    \tau_{\gamma\gamma}(\mathscr{l},\,\mathscr{b},\,E_{\gamma}) &= \iiint \left( \frac{\mathrm{d}\mathcal{F}(\Vec{r},\,E_{b},\,\Omega)}{\mathrm{d}E_{b}\,\mathrm{d}\Omega} \sigma_{\gamma\gamma}(E_{c}) \left( 1 - \cos\Theta \right) \right)\,\mathrm{d}x\,\mathrm{d}E_{b}\,\mathrm{d}\Omega \\
    E_{c} &= \sqrt{ \frac{E_{b}E_{\gamma} \left( 1-\cos\Theta \right) }{2} }
\end{align}

\noindent
where $E_{c}$ is the centre-of-momentum system energy of the two photons, $\Theta$ is the angle between the momenta of the two photons from the observer's perspective, $\Omega$ is the solid angle, $\mathrm{d}\mathcal{F}(\Vec{r},\,E_{b},\,\Omega)/\mathrm{d}E_{b}\mathrm{d}\Omega$ is the differential background photon flux at the point $\Vec{r}$, and $\sigma_{\gamma\gamma}$ is the KN cross section for the two photons interacting. The pair-absorption cross section in the relativistic regime from~\citet{JauchJ.1976} is given by:

\begin{align}
    \sigma_{\gamma\gamma}(E_{c}) &= r_{e}^{2} \pi \left( \frac{m_{e} c^{2}}{E_{c}} \right)^{2} \left( \ln \left( \frac{2 E_{c}}{m_{e} c^{2}} \right) - 1 \right)
\end{align}

\noindent
where all variables have been defined previously. Similarly as for IC scattering, there are two regimes for the pair absorption that depends on the KN suppression factor. For $E_{e} < 100$\,TeV the \grays{} will interact with the IR component of the ISRF, while for $E_{e} > 100$\,TeV the \grays{} will create electron-positron pairs on the CMB radiation.

\begin{figure}[t]
    \centering
    \includegraphics[width=\textwidth]{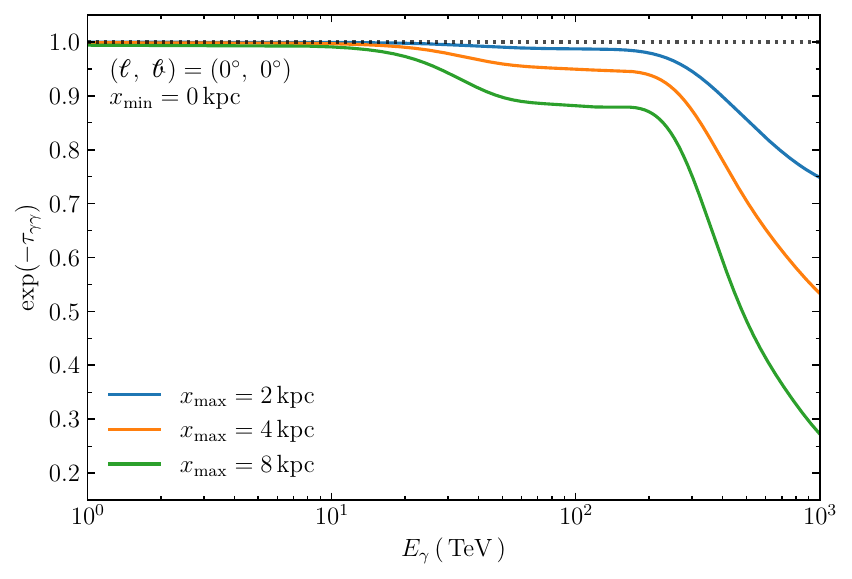}
    \caption{The transmittance along the line-of-sight due to pair-production as a function of the \graya{} energy. The line-of-sight integral is in the direction of the Galactic centre~(GC) and goes from 0\,kpc to 2\,kpc~(blue), 4\,kpc~(orange), and 8\,kpc~(green). The absorption was calculated for the R12 ISRF model~(see \autoref{ssect:R12 Model}).}
    \label{fig:pair absorption}
\end{figure}

The pair-production transmittance~(i.e.~the fraction of \graya{} flux not absorbed by the production of electron-positron pairs) along the line-of-sight is shown in \autoref{fig:pair absorption}. The observed absorption in the 10--200\,TeV energy range is due to \graya{} pair-production on the IR photons. For energies >200\,TeV the observed absorption is due to \graya{} pair-production on CMB photons.

For a \graya{} source located 4\,kpc towards the Galactic centre~(GC), which is approximately halfway between Earth and the GC, $\sim$5\% of \grays{} in the 10--200\,TeV energy range are absorbed due to pair-production. The fraction of absorbed \grays{} increases to $\sim$45\% for 1\,PeV \grays.
For a source located at the GC, approximately 15\% of \grays{} are absorbed due to pair-production in the 10--200\,TeV energy range, with the absorbed fraction increasing to almost 75\% for 1\,PeV \grays.

As the absorption for \graya{} energies in the 10--200\,TeV range depends on the ISRF density, the absorption is maximised towards the GC and minimised for the Galactic poles. For \graya{} energies >200\,TeV the absorption depends only on the isotropic CMB field. Therefore, the \graya{} pair-absorption on the CMB depends only on the distance that the \grays{} travel. On Galactic scales there is an effective horizon for $\sim$1\,PeV \grays{} beyond which the \grays{} are unable to reach Earth.

\subsection{Cooling Distances} \label{ssect:cooling lengths}

As CRs interact with the ISM and create \grays{} their kinetic energy is converted into photons. Due to their energy losses there will be maximum diffusion distance that a CR of a given energy will be able to diffuse.
For CR protons above 1\,GeV \pps~(as discussed in \autoref{ssect:hadronic emission}) are the dominant energy-loss~(i.e.~`cooling') process. For CR electrons the dominant cooling process varies between bremsstrahlung, synchrotron, and IC, depending on the ISM conditions~(e.g.~the ISM gas density and magnetic field strength) and kinetic energy.

The maximum distance that a CR can travel before losing its kinetic energy via a given process is known as its cooling distance, and the time it takes for the CR to lose that energy is known as the cooling time. Both the cooling distance and cooling time must be known to a first-order approximation to define a reasonable simulation with \GP.
To simulate CR diffusion, \GP{} creates a spatial grid and calculates the CRs travelling between the propagation cells~(discussed later in \autoref{ssect:grid}). For accurate results the propagation cells in a \GP{} simulation must be smaller than the shortest cooling distance of any particle being propagated such that the CRs can travel between the propagation cells.

The cooling distance for each process can be calculated to first order by substituting the relevant cooling times into the equation for the diffusion distance~(\autoref{eq:diffusion distance}). The cooling times for the four main interactions were derived in Sections \ref{ssect:hadronic emission} and \ref{ssect:leptonic emission} after taking a range of assumptions for the ISM conditions and CR energies. The cooling times are given by: \autoref{eq:cool pp}~(\pps), \autoref{eq:cool brem}~(bremsstrahlung), \autoref{eq:cool sync}~(synchrotron), \autoref{eq:cool ic}~(IC). The cooling distances for all four processes are then given by:

\begin{align}
    d_{\mathrm{pp}}(E_{p}) &= 3.25 \times 10^{-11} \sqrt{ \left( \frac{D_{0}}{\mathrm{cm}^{2}\,\mathrm{s}^{-1}} \right) \left( \frac{E_{p}}{10\,\mathrm{GeV}} \right)^{\delta} \left( \dfrac{n_{\mathrm{H}}}{\mathrm{cm}^{-3}} \right)^{-1} } \ \mathrm{pc} \label{eq:cool length pp} \\
    d_{\mathrm{brem}}(E_{e}) &= 2.82 \times 10^{-11} \sqrt{ \left( \frac{D_{0}}{\mathrm{cm}^{2}\,\mathrm{s}^{-1}} \right) \left( \frac{E_{e}}{10\,\mathrm{GeV}} \right)^{\delta} \left( \dfrac{n_{\mathrm{H}}}{\mathrm{cm}^{-3}} \right)^{-1} } \ \mathrm{pc} \label{eq:cool length brem} \\
    d_{\mathrm{syn}}(E_{e}) &= 1.54 \times 10^{-11} \sqrt{ \left( \frac{D_{0}}{\mathrm{cm}^{2}\,\mathrm{s}^{-1}} \right) \left( \frac{E_{e}}{10\,\mathrm{GeV}} \right)^{\delta} \left( \frac{B}{\mathrm{\mu G}} \right)^{-2} \left( \frac{E_{e}}{\mathrm{TeV}} \right)^{-1} } \ \mathrm{pc} \label{eq:cool length sync} \\
    d_{\mathrm{IC}}(E_{e}) &= 7.95 \times 10^{-11} \sqrt{ \left( \frac{D_{0}}{\mathrm{cm}^{2}\,\mathrm{s}^{-1}} \right) \left( \frac{E_{e}}{10\,\mathrm{GeV}} \right)^{\delta} \left( \frac{U_{\mathrm{rad}}}{\mathrm{eV} \ \mathrm{cm}^{-3}} \right)^{-1} \left( \frac{E_{e}}{\mathrm{GeV}} \right)^{-1} } \ \mathrm{pc} \label{eq:cool length ic}
\end{align}

\noindent
where $D_{0}$ is the normalisation of the diffusion coefficient, $\delta$ is the diffusion index, $E_{p}$ and $E_{e}$ are the kinetic energies of the proton and electron, respectively, $n_{\mathrm{H}}$ is the density of the ionised hydrogen gas in the ISM, $B$ is the magnetic field strength, and $U_{\mathrm{rad}}$ is the radiation energy density of the ISRF.
Note that the IC cooling distance~(\autoref{eq:cool length ic}) is an upper-bound that assumes the CR electron is not in the KN regime~(see \autoref{ssect:leptonic emission}). This assumption is valid for $E_{e} < 100$\,TeV, with \autoref{eq:cool length ic} being an overestimate for higher CR electron energies\footnote{As the purpose of these equations is to find the maximum energy losses of the CRs, and as the synchrotron losses dominate above $E_{e} < 100$\,TeV, the KN suppression of the IC scattering cross section at higher energies is not accounted for here.}.

Taking $\delta=0.5$, \autoref{eq:cool length pp} shows that the maximum distance CR protons can diffuse is proportional to $E_{p}^{0.25}$. As the diffusion distance has an increasing slope in kinetic energy, the CR protons travel further as energy increases.

There are many processes via which CR electrons convert their kinetic energy into other forms, and those processes compete depending on the energy of the electron. Hence, the cooling distances for CR electrons varies significantly with energy. From \autoref{eq:cool length brem}, the cooling distance for bremsstrahlung interactions is proportional to $E_{e}^{0.25}$. From Equations \ref{eq:cool length sync} and \ref{eq:cool length ic} the cooling distance for synchrotron and IC, respectively, is proportional to $E_{e}^{-0.25}$ and depends on either the magnetic field strength or the background IR radiation energy density. The dominant energy loss process changes with both kinetic energy and the ISM conditions in a given position. If either synchrotron or IC losses dominate, then the CR electrons diffuse shorter distances as their energy increases.

\begin{figure}
    \centering
    \includegraphics[width=\textwidth]{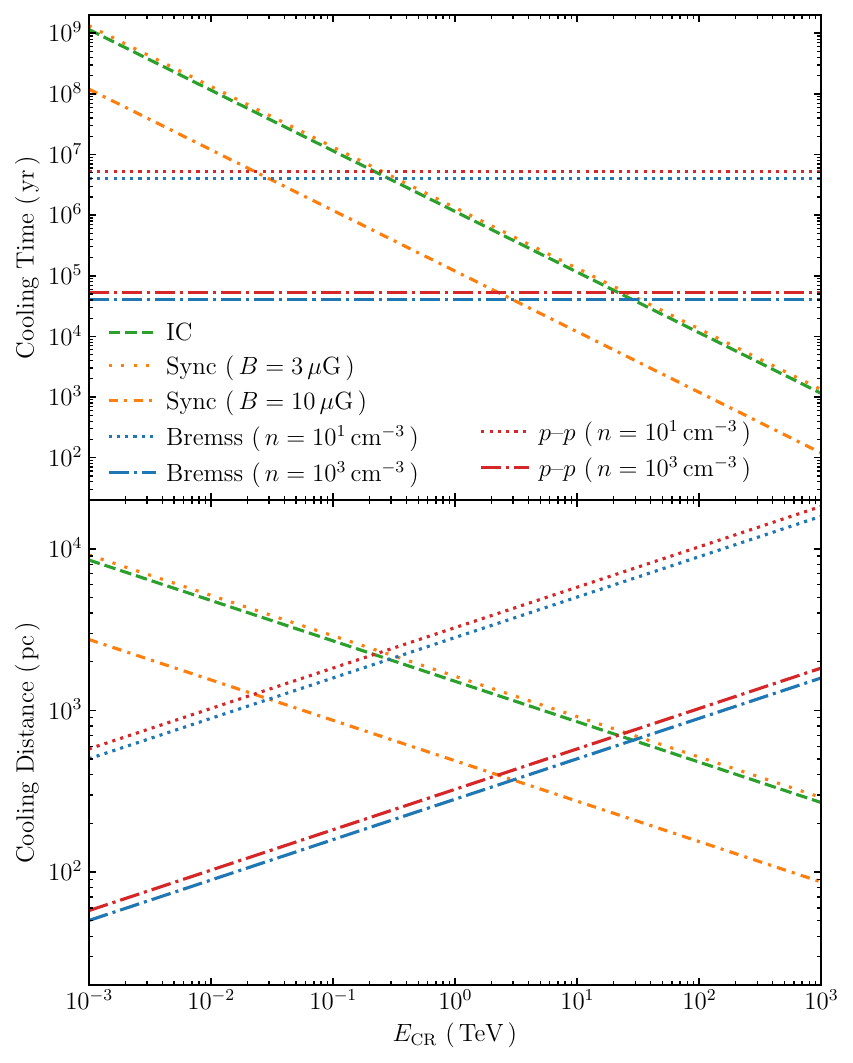}
    \caption{The cooling times~(top) and cooling distances~(bottom) as functions of the CR proton or electron kinetic energy. The cooling times for \pps, bremsstrahlung, synchrotron, and IC are given by Equations~\ref{eq:cool pp},~\ref{eq:cool brem},~\ref{eq:cool sync}, and~\ref{eq:cool ic}, respectively. The cooling distances are taken from Equations~\ref{eq:cool length pp}--\ref{eq:cool length ic}. The values for $n$ and $B$ were taken as Galactic averages, the IC cooling times/distances are calculated for the IR component of the ISRF, and the value $D_{0}=10^{28}$\,cm$^{2}$\,s$^{-1}$ was used.}
    \label{fig:cooling times/lengths}
\end{figure}

How the dominant cooling process changes with energy and the ISM conditions is shown in \autoref{fig:cooling times/lengths} For a strong magnetic field~(10\,$\mu$G) and low gas density~($10^{1}$\,cm$^{-3}$), bremsstrahlung losses dominate for CR electrons up to energies~$\sim$20\,GeV. For a weak magnetic field~(3\,$\mu$G) and a high gas density~($10^{3}$\,cm$^{-3}$), bremsstrahlung losses dominate up to energies~$\sim$5\,TeV.

For a weak Galactic magnetic field~(3\,$\mu$G) the IC losses and synchrotron losses compete for electron kinetic energies $E_{e^{-}} < 100$\,TeV. For this regime the ratio of the IC losses to the synchrotron losses is approximately equal~\citep{ModerskiR.2005}. For $E_{e^{-}} > 100$\,TeV the IC scattering cross section with the CMB field becomes KN suppressed, and the IC cooling times and cooling distances in \autoref{fig:cooling times/lengths} become an overestimate.
However, as the synchrotron energy losses begin to dominate for $E_{e^{-}} > 100$\,TeV, \autoref{fig:cooling times/lengths} is valid for finding the timestep size and spatial grid size for all shown kinetic energies to first order.
\section{CR Source Distributions} \label{sect:source dist.}

The transport equation~(\autoref{eq:Transport Equation}), includes a source term, $Q(\Vec{r},\,p)$ that captures all new CRs of momentum $p$ at location $\Vec{r}$ being added into the MW through various processes. The source term due to secondary particle creation was discussed in \autoref{sssect:secondary creation}. The source term for CRs injected into the MW by sources is defined as the source distribution~($\rho(\Vec{r})$) multiplied by the injection spectra, where the injection spectra can be a function of time. The source distribution defines where the CRs are being injected into the MW, and the source term represents the CR flux being injected at any given location. The time-independent source term\footnote{The time-dependent source term will be discussed later in \autoref{ssect:TDD source dist.}.} for injected CRs is given by:

\begin{align}
    Q(\Vec{r},\,p) &= \rho(\Vec{r}) \frac{\mathrm{d}\mathcal{N}_{\mathrm{CR}}}{\mathrm{d}p} \label{eq:source term}
\end{align}

\noindent
where $\mathcal{N}_{\mathrm{CR}}$ is the injected CR density.
Due to the short cooling times of the electrons, a 3D spatial distribution for the CR injection must be used to reproduce the local electron and positron spectra~\citep{ShenC.1970,ShenC.1971}.
Additionally, a 3D~source distribution must be utilised to accurately model the large-scale structure in the \graya{} sky~\citep{PorterT.2017}.

\GP{} originally used a source distribution that was Gaussian in both the $R$ and $Z$ coordinates~\citep{StrongA.1998} and was able to reproduce the EGRET \graya{} observations from~\citet{StrongA.1996a}.
Currently \GP{} uses a superposition of two source distributions, referred to as the spiral arm component and the Galactic disk component~\citep{WernerM.2015}\footnote{The CR propagation code \textsc{Dragon} models the CR source distribution as a two-dimensional disk~\citep{EvoliC.2017}, while \textsc{Picard}~\citep{KissmannR.2015} uses the same CR source distribution as \GP.}.
The source distributions are parameterised in cylindrical coordinates $(R,\,\phi,\,Z)$ with the origin at the GC.

When operating in a time-independent mode \GP{}  does not simulate individual CR acceleration sites such as SNRs and PWNe, with each cell injecting a constant flux of CRs.
For the time-dependent mode the source distribution represents a probability density function and will be discussed in more depth in \autoref{ssect:TDD source dist.}.
In both cases, the CR source distribution is unitless. The Galactic CR density is normalised at the end of the simulation such that the CR flux at the Solar location is equal to the observed CR flux at Earth~(discussed in more depth in \autoref{ssect:parameter optimisation}).

\subsection{Galactic Disk}

The Galactic disk component of the CR source distribution is modelled following~\citet{YusifovI.2004} and consists of a shifted radial exponential function, where the exponential is shifted such that the density at the GC is non-zero. The flat disk is then convolved with an exponential scale height above the plane~\citep{PorterT.2017}. The source density within the disk is given by:

\begin{align}
    \rho_{\mathrm{disk}}(R,\,Z) &= \left( \frac{R+R_{\mathrm{off}}}{R_{\odot}+R_{\mathrm{off}}} \right)^{\aleph_{1}} \exp \left[ -\aleph_{2} \left( \frac{R-R_{\odot}}{R_{\odot}+R_{\mathrm{off}}} \right) \right] \exp \left( \frac{|Z|}{H_{Z,\rho}} \right) \label{eq:source dist disk}
\end{align}

\noindent
where $\aleph_{1}=1.64$ and $\aleph_{2}=4.01$ are parameters found by fitting the function to the distribution of observed pulsars via a least-mean-squares method. $R_{\mathrm{off}}=0.55$\,kpc is the radial distance the exponential is shifted by, and $H_{Z,\rho}=0.2$\,kpc is the scale height of the disk. The distance to the GC is given by $R_{\odot}=8.5$\,kpc, which was the IAU recommended distance from the GC to Earth at the time of the model's creation~\citep{KerrF.1986}. The Galactic disk component is symmetric in the angle around the GC~($\phi$).

\subsection{Spiral Arms} \label{ssect:SA distribution}

The spiral arm component of the CR source density follows the model of the major arms from~\citet{RobitailleT.2012}~(R12; which is discussed in more depth in \autoref{ssect:R12 Model}). As the R12 model was constructed to follow the distribution of infrared photons in the MW, the CR spiral arms were altered to better follow the distribution of SNRs in the MW~\citep{PorterT.2017}. The scale height~($H_{Z,\rho}$) was set to be equal to that from the disk source distribution, and the four spiral arms are normalised such that all four arms inject an equal amount of total CR luminosity into the Galaxy. The four arms are denoted by $j=1$, 1$^{\prime}$, 2, and 2$^{\prime}$, with the total spiral arm distribution being the sum over all four. The spiral arm source density is given by the equations:

\begin{align}
    \rho_{\mathrm{arms}}(R,\,\phi,\,Z) &= \sum_{j} \exp \left( -\frac{R-R_{\odot}}{H_{R}} \right)
    \exp \left( -\frac{\left[\phi-\phi_{j}(R)\right]^{2}}{W_{\rho}^{2}} \right)
    \exp \left( -\frac{|Z|}{H_{Z,\rho}^{2}} \right) \label{eq:source dist arm} \\
    \phi_{j}(R) &= \alpha_{j} \ln \left( \frac{R} {R_{j,\mathrm{min}}} \right) + \phi_{j,\mathrm{min}} \label{eq:arm shape function}
\end{align}

\noindent
where the scale radius $H_{R}=3.5$\,kpc, the width of the spiral arms has been set to a constant value of $W_{\rho}=0.75$\,kpc, and the function $\phi_{j}$ is the spiral arm winding equation. The arm parameter $\alpha_{j}$ is the winding constant, $R_{j,\mathrm{min}}$ is the minimum radius for the arm, and $\phi_{j,\mathrm{min}}$ is the starting angle of the arm.
The maximum angular extent of the arms is given by $\Delta=6$\,rad~(i.e.~the arms extend from $\phi_{j,\mathrm{min}}$ to $\phi_{j,\mathrm{min}} + \Delta$).
The parameters for each of the four spiral arms can be found in \autoref{tab:source distro arm params}.

\begin{table}
    \centering
    \bgroup
    \def\arraystretch{1.25}
    \begin{tabular}{l c c c}
        \\ \hline
        Arm & $\alpha_{j}$ & $R_{j,\mathrm{min}}$ & $\phi_{j,\mathrm{min}}$ \\
         & & (kpc) & (rad) \\ \hline
        \\[-1em]
        1            & 4.18 & 3.8 & 0.234 \\
        1$^{\prime}$ & 4.18 & 3.8 & 3.376 \\
        2            & 4.19 & 4.5 & 5.425 \\
        2$^{\prime}$ & 4.19 & 4.5 & 2.283 \\ \hline
    \end{tabular}
    \egroup
    \caption{Parameters used for constructing the spiral arm source distribution~\citep[][shown in \autoref{fig:distribution demo}]{RobitailleT.2012}, where $\alpha_{j}$ is the spiral arm winding constant, $R_{j,\mathrm{min}}$ is the minimum radius of the spiral arm, and $\phi_{j,\mathrm{min}}$ is the starting angle of the arm.}
    \label{tab:source distro arm params}
\end{table}

\subsection{Disk and Spiral Arm Combinations}

The combined source distributions use the disk and spiral arm components~(\autoref{eq:source dist disk} and \autoref{eq:source dist arm}, respectively). The combined distribution is defined such that some fraction of the CR flux at the Solar position is due to the spiral arm component, with the remaining CR flux from the disk component. This normalisation condition ensures that all of the combined distributions return the same CR flux at the Solar position~\citep{PorterT.2017}.

\begin{figure}
    \centering
    \includegraphics[height=7.5cm]{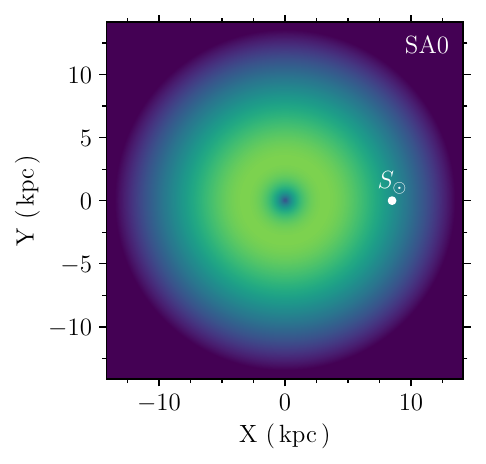}
    \includegraphics[height=7.5cm]{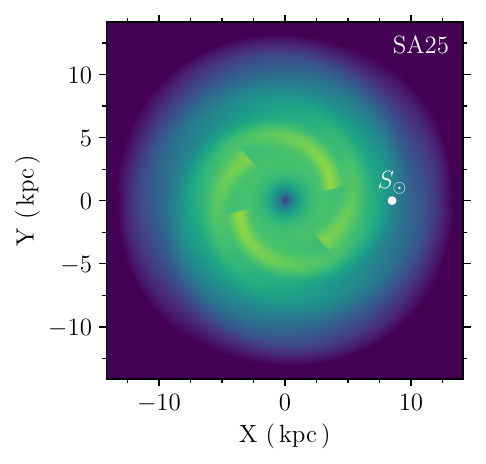}
    \includegraphics[height=7.5cm]{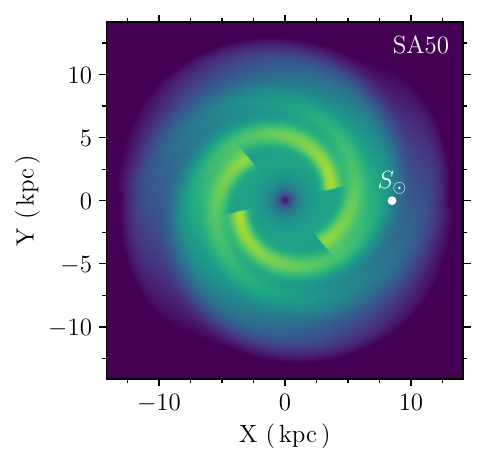}
    \includegraphics[height=7.5cm]{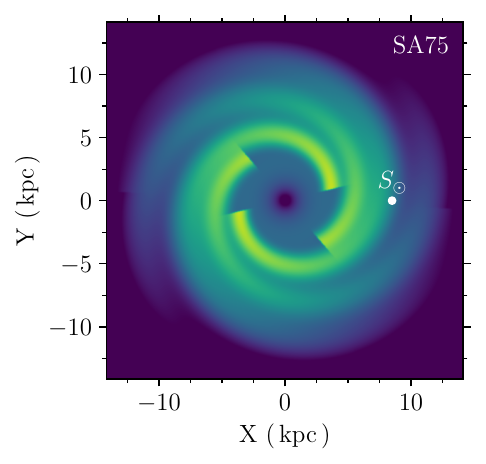}
    \includegraphics[height=7.5cm]{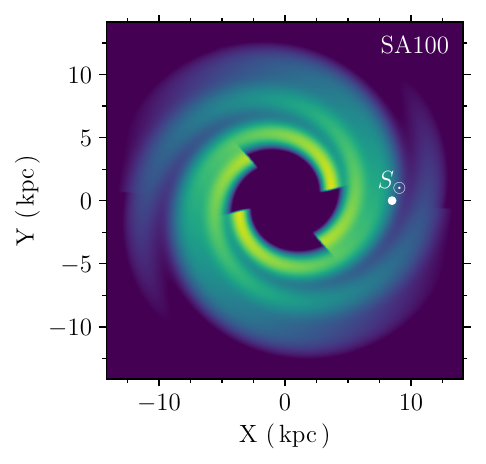}
    \caption{The CR energy density in arbitrary units for the SA0~(top left), SA25~(top right), SA50~(centre left), SA75~(centre right), and SA100~(bottom) integrated over the $Z$-axis. All panels have an identical colour scaling, and the Solar location is shown by $S_{\odot}$.}
    \label{fig:distribution demo}
\end{figure}

In this thesis, five different injection ratios between the Galactic disk and the spiral arms were chosen. The source distributions are denoted by the percentage of CR luminosity at the Solar position from the spiral arm component, with the remaining luminosity at the Solar position being injected from the Galactic disk. For example, SA25 denotes that 25\% of the CR source luminosity at the Solar position is injected by the spiral arms, with the remaining 75\% being injected by the Galactic disk. The chosen CR source distributions were as follows: SA0, SA25, SA50, SA75, and SA100 -- giving a wide range of possible combinations to enable an analysis of how the simulated diffuse emission changes with the different SA models. The five chosen CR source distributions can be seen in \autoref{fig:distribution demo}.
\section{Interstellar Gas} \label{sect:ISM gas}

The ISM gas is mostly comprised of hydrogen and helium with ratios of approximately 90\% and 10\%, respectively~\citep{FerriereK.2001}. The hydrogen can be found in the atomic~(\hi), molecular~(\htwo), or ionised~(\hii) states, while the helium is mostly neutral. For the hydrogen, \hi{} constitutes $\sim$60\% of the mass averaged over the MW, while \htwo{} and \hii{} contain 25\% and 15\%, respectively~\citep{FerriereK.2001}. The \hii{} gas has a low number density and scale height $\sim$few~100\,pc, while the \htwo{} gas forms high density molecular clouds.

Both the pion-decay emission~(from \pps, see \autoref{sssect:pp collisions}) and the flux from bremsstrahlung interactions~(see \autoref{sssect:Bremsstrahlung}) are proportional to the interstellar gas density.
Therefore, structures observed in the hadronic \grays{} and the leptonic bremsstrahlung \grays{} are tied to the structure in the ISM gas density.
The spatial density distribution of the interstellar gas also impacts the CR energy losses and the fragmentation of the CR nuclei.
As the ISM gas density is responsible for both the propagation of CRs and the related \graya{} emissions, Galactic CR propagation models must utilise 3D ISM observations.

\GP{} uses the gas column-density maps based on the HI4PI survey~\citep{BenBekhtiN.2016} and the composite CO survey~\citep{DameT.2001}, with a correction for the contribution of `dark gas'\footnote{Dark gas refers to cold, dense gas. The dark gas is optically thick to its own emissions, and so cannot be observed via emission line spectroscopy. The mass of the dark gas in the MW is similar to the mass of the molecular gas. For a detailed description, see~\citet{GrenierI.2005}.}~\citep{GrenierI.2005} using a map of optical depth at 353\,GHz based on {\it Planck} data~\citep{AghanimN.2016}. Both of these gas surveys have angular resolutions of approximately 0.1$^{\circ}$.

Quantifying the ISM gas spatial density relies on measuring doppler shifts in spectral lines to find the velocity towards or away from Earth. The gas column density maps for the atomic and molecular hydrogen are split into Galactocentric radial bins using a rotation curve~\citep{AckermannM.2012}. 
The spatial resolution of the gas surveys along the line of sight is limited by the broadening of the spectral lines. The spectral lines are broadened by the random kinematics of the gas caused by, for example, the gas temperature and turbulence within the gas.
To improve the resolution along the line of sight \GP{} smooths the gas column density between the velocity bins by performing a weighted average over the velocity axis, where the weighting factors are derived from gas density models. The smoothing procedure ensures that the gas density is spread over the propagation cells, preventing discontinuities in the gas density that can arise due to the cells being much smaller than the line-of-sight resolution of the gas surveys.

For the \hii{} gas, \GP{} employs the NE2001 model~\citep{CordesJ.2002, CordesJ.2003, CordesJ.2004} with updates from~\citet{GaenslerB.2008}. The NE2001 model is comprised of a thin disk, a thick disk, and spiral arms. Cavities and clumps in the known ISM gas distribution are also included in the NE2001 model.
For the \hi{} gas, \GP{} employs the gas density model from~\citet{JohannessonG.2018}, which is obtained using a maximum-likelihood fit between an idealised distribution and the LAB \hi-survey~\citep{KalberlaP.2005} and the composite CO survey~\citep{DameT.2001}. The~\citet{JohannessonG.2018} model is comprised of a warped and flaring disc, four spiral arms, and a central bar. Due to the complexity of the ISM gas model, which includes cavities and clumps, there is no simplistic functional form. The models are calculated with the \textsc{Galgas} code~\citep{JohannessonG.2018}, with the resulting files from~\citet{JohannessonG.2018} used as inputs for \GP{} in this thesis.

\section{Interstellar Radiation Fields} \label{sect:ISRF}

Both the IC emission~(see \autoref{sssect:IC}) and Galactic absorption of \grays{} via pair production of electrons and positrons~(see \autoref{sssect:pair absorption}) depend on the background radiation in the MW. The background interstellar radiation field, or ISRF, has an average energy density close to 1\,eV\,cm$^{-3}$, which is similar to the energy density of CRs, the GMF, and the turbulence in the interstellar gas~\citep{StrongA.2013}.

The ISRF encompasses all low-energy electromagnetic radiation within the Galaxy generated by both internal and external sources. The ISRF includes infrared, optical, and UV emission from stars, infrared light from interstellar dust radiating heat, and the CMB. Early models of the ISRF include,~e.g.~\citet{CowsikR.1974,BignamiG.1977,MathisJ.1983}. However, these models were not self-consistent -- i.e.~they did not couple the absorption of light by dust with the re-emission of infrared light. Later models of the ISRF that are fully self-consistent include,~e.g.~\citet{PorterT.2005,PopescuC.2017}.

Within \GP{} there are two 3D ISRF models in use. These are referred to as F98~\citep[from][]{FreudenreichH.1998} and R12~\citep[from][]{RobitailleT.2012}. The F98 model includes a Galactic bar and does not include any spiral arms, while the R12 model includes the spiral arms but does not include a Galactic bar. Although the R12 more accurately predicts the infrared emission for the Galactic longitudes $|\mathscr{l}| \leq 60^{\circ}$, both the F98 and R12 models are equivalent solutions and are considered as upper and lower limits on the Galactic infrared emissions. Both of the ISRF models are self-consistent and are parameterised in cylindrical coordinates $(R,\,\phi,\,Z)$ with the origin at the GC. For the remainder of this section the radiation energy density, $U_{\mathrm{rad}}$, is denoted by $U_{\mathrm{F98}}$ and $U_{\mathrm{R12}}$ for the F98 and R12 distributions, respectively, with additional subscripts defining the components of the respective radiation energy densities.

\subsection{Freudenreich Spatial Model}

The F98 ISRF model~\citep{FreudenreichH.1998} combines both the stellar emission and infrared emission from dust. It has 47~free parameters and is constructed to have good agreement with the diffuse infrared background experiment~\citep[DIBRE;][]{FreudenreichH.1996} survey taken with the \textit{cosmic background explorer~(COBE)} satellite.
The F98 model combines a stellar disk, a Galactic bar, and a dust layer.
The F98 model is parameterised in cylindrical coordinates $(R,\,\phi,\,Z)$; however, the F98 model also uses the transformed coordinates~($X^{\prime},\,Y^{\prime}$) for creating the hole in the Galactic disk, and the transformed coordinates~($X^{\prime\prime},\,Y^{\prime\prime},\,Z^{\prime\prime}$) for constructing the central Galactic bar.

\begin{table}
    \centering
    \bgroup
    \def\arraystretch{1.25}
    \begin{tabular}{l c c c c}
        \\ \hline
        Normalisation & \multicolumn{4}{c}{Wavelength Band} \\
         & J & K & L & M \\ \hline
        \\[-1em]
        $U_{\mathrm{F98,disk}, 0}$ & 8.144 & 6.648 & 3.511 & 1.782 \\
        $U_{\mathrm{F98,dust}, 0}$ & 4.642 & 1.152 & 2.180 & 3.236 \\
        $U_{\mathrm{F98,bar}, 0}$  & 10.52 & 8.817 & 4.538 & 2.255 \\ \hline
    \end{tabular}
    \egroup
    \caption{\label{tab:F98 norms} The normalisation parameters for each of the wavelengths present in the DIBRE survey~\citep{FreudenreichH.1996}. All units are in MJy\,sr$^{-1}$\,kpc$^{-1}$ and can be converted using 1\,MJy\,sr$^{-1}$\,kpc$^{-1} = 9.522 \times 10^{9}$\,W\,pc$^{-3}$\,Hz$^{-1}$\,sr$^{-1}$.}
\end{table}

\begin{figure}
    \centering
    \includegraphics[height=8cm]{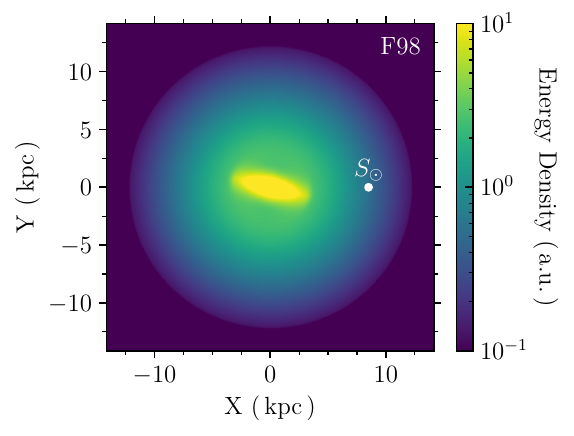}
    \caption{The F98 ISRF energy density in arbitrary units integrated over the $Z$-axis. The Solar location is shown by $S_{\odot}$.}
    \label{fig:F98 distribution}
\end{figure}

The disk and bar distributions are sums over the four wavelengths present in the DIBRE survey; namely $\nu=1.25$\,$\mu$m, 2.2\,$\mu$m, 3.5\,$\mu$m, and 4.9\,$\mu$m~(labelled as J, K, L, and M, respectively), with each wavelength having a separate normalisation. The normalisation values for each wavelength band and model component are shown in \autoref{tab:F98 norms}.
The total radiation energy density of the F98 ISRF is shown in \autoref{fig:F98 distribution}, and is given by the sum over all components:

\begin{equation}
    \begin{aligned}
        U_{\mathrm{F98}}(R,\,\phi,\,Z) ={} &U_{\mathrm{F98,disk}}(R,\,\phi,\,Z) + U_{\mathrm{F98,dust}}(R,\,\phi,\,Z)\,+ \\
        &U_{\mathrm{F98,bar}}(R,\,\phi,\,Z) \label{eq:F98 distribution}
    \end{aligned}
\end{equation}

\noindent
where $U_{\mathrm{F98}}$ denotes the total radiation energy density and the additional subscripts denote the individual components.

\subsubsection{F98 Stellar Disk Distribution}

The F98 stellar disk is described by an inner and outer truncated~(i.e.~set to zero) exponential in $R$, follows a $\sech^{2}$ in $Z$, and has a warped mid-point in the $Z$ axis. The distribution is a sum over the wavelength-band counting index, $\nu$, with each wavelength normalised by the parameter $U_{\mathrm{F98,disk},\,\nu,\,0}$. The normalisation values for the different frequency bands in the DIBRE data are given in \autoref{tab:F98 norms}. The radiation energy density of the disk is given by the equation:

\begin{equation}
    \begin{aligned}
        U_{\mathrm{F98,disk}}(R,\,\phi,\,Z) ={} &h_{\mathrm{disk}}(X^{\prime},\,Y^{\prime}) \sum_{\nu} U_{\mathrm{F98,disk}, \nu, 0} \\
        &\times \exp \left( -\frac{R}{H_{R, \mathrm{disk}}} \right) \\
        &\times \sech^{2} \left( \frac{Z-\Bar{Z}(R,\,\phi)}{H_{Z, \mathrm{disk}}} \right)
    \end{aligned}
\end{equation}

\noindent
where $H_{Z, \mathrm{disk}}=0.3457$\,kpc is the disk scale height, and the disk scale radius is given by:

\begin{align}
    H_{R, \mathrm{disk}} &=
    \begin{cases}
        2.6045\,\mathrm{kpc}, & R \leq R_{\mathrm{F98,max}} \\
        0.5\,\mathrm{kpc}, & R > R_{\mathrm{F98,max}}
    \end{cases} \label{eq:F98 disk scale height}
\end{align}

\noindent
where $R_{\mathrm{F98,max}}=12.18$\,kpc is the radius of the disk. The function $h_{\mathrm{disk}}(X^{\prime},\,Y^{\prime})$ creates a hole in the Galactic disk, where the~($X,\,Y$) coordinates are rotated clockwise by the angle $\phi_{0}=13.79^{\circ}$ to give the transformed coordinates~($X^{\prime},\,Y^{\prime}$). The functional form of the stellar hole is given by:

\begin{align}
    h_{\mathrm{disk}}(X^{\prime},\,Y^{\prime}) &= 1 - \exp \left[ -\left( \frac{R_{h}(X^{\prime},\,Y^{\prime})}{H_{R, \mathrm{disk-hole}}} \right)^{\omega_{H, \mathrm{disk}}} \right] \label{eq:F98 disk hole} \\
    \left[ R_{h}(X^{\prime},\,Y^{\prime}) \right]^{2} &= (X')^{2} + (e Y')^{2} \label{eq:F98 disk hole radius}
\end{align}

\noindent
where $\omega_{H, \mathrm{disk}}=1.711$ is the disk hole power, $H_{R, \mathrm{disk-hole}}=2.973$\,kpc is the scale radius of the stellar disk hole, and $e=0.8554$ is the eccentricity of the hole.

The mean height of the Galactic plane, $\Bar{Z}(R,\,\phi)$, is equal to zero for Galactic radii less than $R_{w}=4.34$\,kpc from the GC. After $R_{w}$, the mean height is modelled as a cubic polynomial in $R$ and is sinusoidal in $\phi$. The sinusoidal polynomial has the effect of warping the Galactic plane, where the warping increases in amplitude as the Galacto-centric radius increases. The warping of the Galactic disk is given by:

\begin{align}
    \Bar{Z}(R,\,\phi) &=
    \begin{cases}
        0, & u \leq 0 \\
        (c_{1} u + c_{2} u^{2} + c_{3} u^{3}) \sin(\phi - \phi_{w}), & u > 0
    \end{cases} \\
    u &= R - R_{w}
\end{align}

\noindent
where $\phi_{w}=0.44^{\circ}$ is the straight-line azimuthal node, and the polynomial coefficients are given by $c_{1}=0.01118$, $c_{2}=-0.00192$, and $c_{3}=0.000795$.

\subsubsection{F98 Dust Distribution}

The F98 dust layer follows a similar distribution to the F98 stellar disk, including a similar hole around the GC. The hyperbolic-sine function in the $Z$-axis is raised to a different power compared to the disk distribution, and the mean height is warped by an extra factor $\Pi=1.782$ to better reproduce the data from the DIBRE experiment. The radiation energy density of the dust distribution is given by the equation:

\begin{equation}
    \begin{aligned}
        U_{\mathrm{F98,dust}}(R,\,\phi,\,Z) ={} &h_{\mathrm{disk}}(X^{\prime},\,Y^{\prime}) \, h_{\mathrm{dust}}(X^{\prime},\,Y^{\prime}) \sum_{\nu} U_{\mathrm{F98,dust}, \nu, 0} \\
        &\times \exp \left( -\frac{R}{H_{R, \mathrm{disk}}} \right) \\
        &\times \exp \left( -\frac{2 R}{3 H_{R, \mathrm{dust}}} \right) \\
        &\times \sech^{10/3} \left( \frac{|Z-\Pi\Bar{Z}(R,\,\phi)|}{H_{Z, \mathrm{dust}}} \right)
    \end{aligned}
\end{equation}

\noindent
where the $H_{R, \mathrm{dust}}=3.066$\,kpc is the dust scale radius, $H_{Z, \mathrm{dust}}=0.1520$\,kpc is the dust scale height, the disk scale radius $H_{R, \mathrm{disk}}$ is given by \autoref{eq:F98 disk scale height}, and the disk hole function $h_{\mathrm{disk}}$ is given by \autoref{eq:F98 disk hole}. The dust hole function, $h_{\mathrm{dust}}$, is given by:

\begin{align}
    h_{\mathrm{dust}}(X^{\prime},\,Y^{\prime}) &= 1 - \exp \left[ - \left( \frac{R_{h}(X^{\prime},\,Y^{\prime})}{H_{R, \mathrm{dust-hole}}} \right)^{\omega_{H, \mathrm{dust}}} \right]
\end{align}

\noindent
where $H_{R, \mathrm{dust-hole}}=2.615$\,kpc is the dust scale radius, $\omega_{H, \mathrm{dust}}=2.150$ is the dust hole power, and the radius of the hole~($R_{h}$) is given by \autoref{eq:F98 disk hole radius}. Although \GP{} uses the dust distribution from the F98 model, some of the dust distribution parameters used in the above equations are not completely identical~\citep{PorterT.2017}. The scale lengths and scale heights, the central hole, and the elliptical warp are taken from the F98 model as above, while the dust properties and composition are changed such that they agree with~\citet{DraineB.2007}. The dust normalisation was then set to be identical to the R12 model. This was necessary as \GP{} requires a wider range of wavelengths down into the UV spectrum for its ISRF calculations, which are not present in the original F98 model. As the extinction and emissivity calculations for the F98 dust distribution in~\citet{FreudenreichH.1998} are unused by \GP, they are not included here. For completeness, the original F98 dust distribution normalisation values are included in \autoref{tab:F98 norms}.

\subsubsection{F98 Bar Distribution}

The F98 bar distribution\footnote{Shown here is the corrected form implemented in \GP{}, as per \citet{Troy_and_Macias_emails}} is described by ellipsoid in the coordinates $(X^{\prime \prime},\,Y^{\prime \prime},\,Z^{\prime \prime})$. These double-primed coordinates are found by rotating the primed coordinates $(X^{\prime},\,Y^{\prime})$ clockwise into the $Z$-axis by the best-fit pitch angle $\phi_{\mathrm{pitch}}=-0.023^{\circ}$. To represent the bar structures seen in other galaxies, the bar ellipsoid has `boxy' edges~(i.e.~the ellipse ends abruptly) along the $Z^{\prime \prime}$ axis and rounded edges along the $X^{\prime \prime}$ and $Y^{\prime \prime}$ axes. The radiation energy density of the ellipsoid decreases following a $\sech^{2}$ function with an exponential cut off at the Galacto-centric radius $R_{\mathrm{end}}=3.128$\,kpc. The radiation energy density of the bar is given by:

\begin{align}
    U_{\mathrm{F98,bar}}(R_{b}) &= \sum_{\nu} U_{\mathrm{F98,bar}, \nu, 0} \ \sech^{2} (R_{b}) \times
    \begin{cases}
        1, & R \leq R_{\mathrm{end}} \\
        \exp \left[ - \left(\dfrac{R-R_{\mathrm{end}}}{H_{\mathrm{end}}} \right)^{2} \right], & R > R_{\mathrm{end}}
    \end{cases}
\end{align}

\noindent
where $H_{\mathrm{end}}=0.461$\,kpc is the scale length of the bar cut off in the $(X^{\prime \prime},\,Y^{\prime \prime},\,Z^{\prime \prime})$ coordinate frame. The variable $R_{b}$ follows the equation for an ellipsoid in the $(X^{\prime \prime},\,Y^{\prime \prime},\,Z^{\prime \prime})$ coordinate frame, with the ellipsoid defined by the shape parameters $(\mathcal{S}_{X},\,\mathcal{S}_{Y},\,\mathcal{S}_{Z}) = (1.696,\,0.6426,\,0.4425)$. The bar is generated using ``face-on'' and ``edge-on'' parameters, denoted by $\perp$ and $\parallel$ respectively. The equation for the face-on radius~(i.e.~the radius perpendicular to the $Z^{\prime\prime}$ axis) is given by:

\begin{align}
    (R_{\perp})^{C_{\perp}} &= \left( \frac{|X^{\prime \prime}|}{\mathcal{S}_{X}} \right)^{C_{\perp}} + \left( \frac{|Y^{\prime \prime}|}{\mathcal{S}_{Y}} \right)^{C_{\perp}} \\
    \therefore R_{\perp} &= \left[ \left( \frac{|X^{\prime \prime}|}{\mathcal{S}_{X}} \right)^{C_{\perp}} + \left( \frac{|Y^{\prime \prime}|}{\mathcal{S}_{Y}} \right)^{C_{\perp}} \right]^{1/C_{\perp}}
\end{align}

\noindent
where $C_{\perp}=1.574$. The edge-on radius~(i.e.~the radius parallel to the $Z^{\prime\prime}$ axis) is then included to calculate the radius of the bar ellipsoid~($R_{b}$), which is given by:

\begin{align}
    (R_{b})^{C_{\parallel}} &= R_{\perp}^{C_{\parallel}} + \left( \frac{|Z^{\prime \prime}|}{\mathcal{S}_{Z}} \right)^{C_{\parallel}} \\
    \therefore (R_{b})^{C_{\parallel}} &= \left[ \left( \frac{|X^{\prime \prime}|}{\mathcal{S}_{X}} \right)^{C_{\perp}} + \left( \frac{|Y^{\prime \prime}|}{\mathcal{S}_{Y}} \right)^{C_{\perp}} \right]^{C_{\parallel}/C_{\perp}} + \left( \frac{|Z^{\prime \prime}|}{\mathcal{S}_{Z}} \right)^{C_{\parallel}}
\end{align}

\noindent
where the edge-on parameter is given by $C_{\parallel}=3.501$.
The bar ellipsoid is a rounded diamond in the $(X^{\prime \prime},\,Y^{\prime \prime})$ plane as the perpendicular shape parameter $C_{\perp}<2$. Conversely, the bar ellipsoid has more pronounced vertices in the $Z^{\prime \prime}$ axis as the parallel shape parameter $C_{\parallel}>2$.

\subsection{Robitaille Spatial Model} \label{ssect:R12 Model}

The R12 ISRF model~\citep{RobitailleT.2012} has good agreement with latitudinal and longitudinal profiles for the emission in the near-infrared to far-infrared wavelengths from the Galactic legacy midplane survey extraordinaire~\citep[GLIMPSE;][]{ChurchwellE.2009} and multiband infrared photometer for \textit{Spitzer}~\citep[MIPSGAL;][]{CareyS.2009} surveys taken with the \textit{Spitzer space telescope}. The R12 ISRF model also has good agreement with the improved reprocessing of the IRIS survey~\citep{Miville-DeschenesM.2005} survey taken with the \textit{infrared astronomical satellite}~(\textit{IRAS}) instrument. As the spiral arm source distribution~(\autoref{ssect:SA distribution}) is derived from the R12 spiral arm model, the two models share many parameters.

\begin{figure}
    \centering
    \includegraphics[height=8cm]{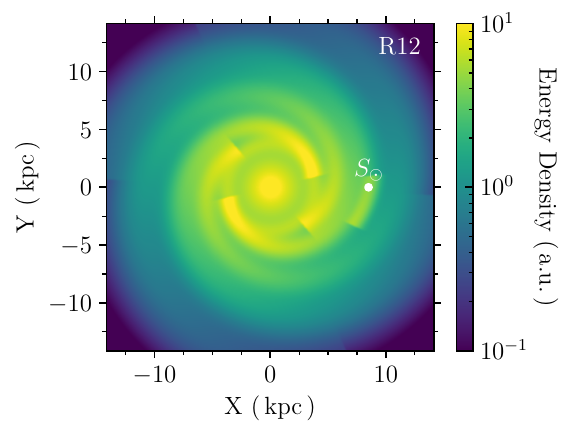}
    \caption{The R12 ISRF energy density in arbitrary units integrated over the $Z$-axis. The Solar location is shown by $S_{\odot}$.}
    \label{fig:R12 distribution}
\end{figure}

The R12 model components are constructed include an axisymmetric bulge, an exponential disk, a star forming ring, spiral arms, and a Galactic dust. The disk, bulge, and ring distributions are sums over the counting index $i=1,\ldots,87$~(hereafter referred to as the spectral type), which denote each of the different stellar types from the main sequence, pre-main sequence, giant stars, stars on the asymptotic giant branch, and planetary nebulae. These are named the SKY models, and are originally from~\citet{WainscoatR.1992}, being developed further by~\citet{CohenM.1993}, \citet{CohenM.1994}, and \citet{CohenM.1995}. Each stellar class has its own normalisation values~\citep[][see their Table~2]{WainscoatR.1992}.
The total radiation energy density of the R12 model is shown in \autoref{fig:R12 distribution}, and is given by the sum over all components:

\begin{equation}
    \begin{aligned}
        U_{\mathrm{R12}}(R,\,\phi,\,Z) ={} &U_{\mathrm{R12,disk}}(R,\,Z) + U_{\mathrm{R12,bulge}}(R,\,Z) + U_{\mathrm{R12,arms}}(R,\,\phi,\,Z)\,+ \\
        &U_{\mathrm{R12,ring}}(R,\,Z) + U_{\mathrm{R12,dust}}(R,\,Z) \label{eq:R12 total}
    \end{aligned}
\end{equation}

\noindent
where $U_{\mathrm{R12}}$ denotes the total radiation energy density and the additional subscripts denote the individual components.

\subsubsection{R12 Disk Distribution}

The radiation energy density within the Galactic disk is given by a decreasing exponential which is truncated beyond $R_{\mathrm{R12,max}}=15$\,kpc. The radiation energy density of the disk component is given by:

\begin{align}
    U_{\mathrm{R12,disk}}(R,\,Z) &=
    \begin{dcases}
        \sum_{i} U_{\mathrm{R12,disk}, i} \exp \left(-\dfrac{R-R_{\odot}}{H_{R}} \right) \exp \left(-\dfrac{|Z|}{H_{Z, i}} \right), & R < R_{\mathrm{R12,max}} \\
        0, & R \geq R_{\mathrm{R12,max}}
    \end{dcases} \label{eq:R12 disk}
\end{align}

\noindent
where $H_{R}=3.5$\,kpc is the radial scale length, $U_{\mathrm{R12,disk},i}$ is the spatial radiation energy density of the $i$\textsuperscript{th} spectral type, and $H_{Z,i}$ is the scale height perpendicular to the Galactic plane of the $i$\textsuperscript{th} spectral type. The distance to the GC is given by $R_{\odot}=8.5$\,kpc, which was the IAU recommended distance from the GC to Earth at the time of the model's creation~\citep{KerrF.1986}.

\subsubsection{R12 Bulge Distribution}

The central bulge of the MW represents a region where the distribution of stars extends further vertically out of the Galactic plane and is comprised mostly of stars that are older than those found within the Galactic plane~\citep{BinneyJ.1997}. The radiation energy density within the Galactic bulge is given by:

\begin{align}
    U_{\mathrm{R12,bulge}}(R,\,Z) &= U_{\mathrm{R12,bulge,0}} \left[ \kappa^{-1.8} \exp \left( -\kappa^{3} \right) \right] \sum_{i} U_{\mathrm{R12,bulge}, i} \\
    \kappa &= \frac{\sqrt{R^{2} + K_{\mathrm{bulge}}^{2} Z^{2}}}{R_{\mathrm{bulge}}}
\end{align}

\noindent
where $U_{\mathrm{R12,bulge,0}}=3.6$ is a unitless global normalisation factor, $U_{\mathrm{R12,bulge}, i}$ is the normalisation constant for the $i$\textsuperscript{th} spectral type, $\kappa$ is a unitless function of radial distance from the GC and height above the Galactic plane, $K_{\mathrm{bulge}}=1.6$ is the bulge axis ratio~(i.e.~major axis length over the minor axis length), and $R_{\mathrm{bulge}}=2$\,kpc is the central radius of the bulge.

\subsubsection{R12 Spiral Arm Distributions}

The radiation energy density within the spiral arms is split into two categories, with the counting variable $j=1,\,1^{\prime},\,2,\,2^{\prime}$ denoting the four major arms, and $j=L,\,L^{\prime}$ denoting the two local spurs~(also commonly referred to as the minor arms). All arms and spurs are given the same arm winding function as in the source spiral arms~(see \autoref{eq:arm shape function})., with the parameter values for all the R12 spiral arms given in \autoref{tab:local spur params}. The shape parameters for the four major arms are equal to those for the spiral arm source distribution given earlier in \autoref{sect:source dist.}. The radiation energy density within the arms is set to zero when the distance from the GC to the centre of the arms is less than $R_{\mathrm{min}}$.

\begin{table}
    \centering
    \bgroup
    \def\arraystretch{1.25}
    \begin{tabular}{l c c c c c}
        \\ \hline
        Arm & $\alpha_{j}$ & $R_{j,\mathrm{min}}$ & $\phi_{j,\mathrm{min}}$ & $W_{j}$ & $\Delta_{j}$ \\
         & & (kpc) & (rad) & (kpc) & (rad) \\ \hline
        \\[-1em]
        1            & 4.18 & 3.8   & 0.234 & 0.55 & 6    \\
        1$^{\prime}$ & 4.18 & 3.8   & 3.376 & 0.55 & 6    \\
        2            & 4.19 & 4.5   & 5.425 & 0.55 & 6    \\
        2$^{\prime}$ & 4.19 & 4.5   & 2.283 & 0.55 & 6    \\
        L            & 4.57 & 8.100 & 5.847 & 0.30 & 0.55 \\
        L$^{\prime}$ & 4.57 & 7.591 & 5.847 & 0.30 & 0.55 \\\hline
    \end{tabular}
    \egroup
    \caption{The spiral arm parameters used for constructing the R12 spiral arms~\citep{RobitailleT.2012}. The values for arms 1, 1$^{\prime}$, 2, and 2$^{\prime}$ are the same as for the spiral arm source distribution shown in \autoref{tab:source distro arm params}.}
    \label{tab:local spur params}
\end{table}

For the two local spurs~($j=L,\,L^{\prime}$) the radiation energy density within the arms is modelled as a step function. The local spur parameter values are given in \autoref{tab:local spur params}. The distance from the centre of the local spur to some position is given by $\varphi_{j}$. The radiation energy density is given by the same functional form as for the disk~(see \autoref{eq:R12 disk}) when $\varphi_{j} \leq W_{j}$, and the radiation energy density is set to zero when $\varphi_{j} > W_{j}$. The local spurs are described by the equation:

\begin{align}
    U_{\mathrm{R12},j=L,L^{\prime}}(R,\,\phi,\,Z) &=
    \begin{dcases}
        U_{\mathrm{R12,A}} \sum_{i} U_{\mathrm{R12,arms}, i} \exp \left(-\dfrac{R-R_{\odot}}{H_{R}} \right) \exp \left(-\dfrac{|Z|}{H_{Z, i}} \right), & \varphi_{j} \leq W_{j} \\
        0, & \varphi_{j} > W_{j}
    \end{dcases} \label{eq:R12 spurs}
\end{align}

\noindent
where $U_{\mathrm{R12,A}} = 5$ is a unitless global normalisation factor, and the other constants are the same as for the disk component given in \autoref{eq:R12 disk}. The distance from some position to the centre of an arm, $\varphi_{j}$, is given by:

\begin{align}
    \varphi_{j} &= \exp \left( \frac{\phi_{j}(R) - \phi_{j,\mathrm{min}}}{\alpha_{j}} \right) R_{j,\mathrm{min}}
\end{align}

\noindent
where the equation is derived by rearranging the arm winding function given by \autoref{eq:arm shape function}. For the four major arms $(j=1,\,1^{\prime},\,2,\,2^{\prime})$ the radiation energy density is given by the same functional form as for the disk; however, unlike for the minor arms which use a step function, the major arms are modelled with a Gaussian profile perpendicular to the arms. The major arms are then given by:

\begin{equation}
    \begin{aligned}
        U_{\mathrm{R12},j=1,1^{\prime},2,2^{\prime}}(R,\,\phi,\,Z) ={} &U_{\mathrm{R12,A}} \sum_{i} U_{\mathrm{R12,arms}, i} \\
        &\times \exp \left(-\frac{R-R_{\odot}}{H_{R}} \right) \\
        &\times \exp \left(-\frac{|Z|}{H_{Z, i}} \right) \\
        &\times \exp \left( - \frac{\left[\phi-\phi_{j}(R)\right]^{2}}{W_{j}} \right) \label{eq:R12 arms}
    \end{aligned}
\end{equation}

\noindent
where the arm-specific values are given in \autoref{tab:local spur params} and $\phi_{j}(R)$ is given by \autoref{eq:arm shape function}. The four major arms are described as Gaussian cylinders which are wound around the Galaxy as logarithmic spirals. In a similar fashion as for the source distribution arms, $U_{\mathrm{R12},j=1,1^{\prime},2,2^{\prime}}(R,\,\phi,\,Z) = 0$ for $\phi_{j} \geq \Delta_{j}$~(i.e.~the arms extend from $\phi_{j,\mathrm{min}}$ to $\phi_{j,\mathrm{min}} + \Delta_{j}$). The radiation energy density is also set to zero when the distance from the GC to the centre of the arm is less than $R_{j,\mathrm{min}}$~(i.e.~$U_{\mathrm{R12},j=1,1^{\prime},2,2^{\prime}}(R,\,\phi,\,Z) = 0$ for $R < R_{j,\mathrm{min}}$).

The total spiral arm radiation energy density is given by the sum of all the radiation energy densities across the arms and spurs, which are themselves sums over the spectral type $i$. The total component of the radiation energy density from the spiral arms for the R12 model is then given by:

\begin{align}
    U_{\mathrm{R12,arms}}(R,\,\phi,\,Z) = \sum_{j} U_{\mathrm{R12,j}}
\end{align}

\subsubsection{R12 Ring Distribution}

The Galactic ring represents a toroidal structure around the GC~\citep{ClemensD.1988}, though its existence is debated~\citep[e.g.~see][]{DobbsC.2012}. The Galactic ring is included in the R12 model, with the radiation energy density within the Galactic ring being given by a Gaussian distribution. The maximum of the Gaussian,~i.e.~the centre of the ring, is located at $R_{\mathrm{ring}}=6.75$\,kpc from the GC. The ring distribution has a Gaussian width of $H_{R,\mathrm{ring}}=0.96$\,kpc. The radiation energy density decreases as an exponential from the Galactic plane, using the same scale height as \autoref{eq:R12 disk} and \autoref{eq:R12 spurs}. Hence, the ring component is given by the equation:

\begin{align}
    U_{\mathrm{R12,ring}}(R,\,Z) &= U_{\mathrm{R12,ring},0} \sum_{i} U_{\mathrm{R12,ring}, i} \exp \left( -\frac{\left(R-R_{\mathrm{ring}}\right)^{2}}{2 H_{R,\mathrm{ring}}^{2}} \right) \exp \left( -\frac{|Z|}{H_{Z, i}} \right)
\end{align}

\noindent
where $U_{\mathrm{R12,ring},0}=25$ is a unitless global normalisation constant, and $U_{\mathrm{R12,ring}, i}$ is the normalisation constant for the $i$\textsuperscript{th} spectral type.

\subsubsection{R12 Dust Distribution}

The radiation energy density of IR emitted by dust within the Galaxy is given by a radial Gaussian up to the transition radius $R_{\mathrm{smooth}}$. The dust distribution is modelled as decreasing exponential for radii $R>R_{\mathrm{smooth}}$. Similarly to the previously discussed R12 components, the radiation energy density from the dust is modelled as a decreasing exponential in $Z$. It is worth noting that unlike the previous R12 components, the dust distribution does not depend on the stellar type. The dust model is adopted from~\citet{DraineB.2007}, which uses a mixture of dust grains from neutral and ionised polycyclic aromatic hydrocarbons, amorphous silicates, and carbonaceous dust grains which follow a size distribution from~\citet{WeingartnerJ.2001}. The radiation energy density from the dust component is then given by:

\begin{equation}
    \begin{aligned}
        U_{\mathrm{R12,dust}}(R,\,Z) ={} &U_{\mathrm{R12,dust}, 0} \exp \left( -\frac{|Z|}{H_{Z,R12,\mathrm{dust}}} \right) \\
        &\times
        \begin{cases}
            \varkappa_{0} \exp \left( -\dfrac{(R-R_{\mathrm{dust}})^{2}}{2 H_{R,R12,\mathrm{dust}}^{2}} \right), & R < R_{\mathrm{smooth}} \\
            \exp \left( -\dfrac{R}{H_{R}} \right), & R \geq R_{\mathrm{smooth}}
        \end{cases}
    \end{aligned}
\end{equation}

\noindent
where the dust normalisation constant is given by $U_{\mathrm{R12,dust}, 0}=10^{-25}$\,g\,cm$^{-3}$, the scale height of the dust is given by $H_{Z,R12,\mathrm{dust}}=0.1$\,kpc, the Gaussian peaks at a distance of $R_{\mathrm{dust}}=4.5$\,kpc, the Gaussian has a scale radius of $H_{R,R12,\mathrm{dust}}=1$\,kpc, and $H_{R}=3.5$\,kpc is the same as before.

The value $R_{\mathrm{smooth}}$ is defined as the transition radius, where the radial distribution goes from a Gaussian to a decreasing exponential. The Gaussian distribution is then normalised by the value $\varkappa_{0}$, which is defined such that the transition is smooth to the first order -- i.e.~both the function and its first derivative are equal at the transition point $R=R_{\mathrm{smooth}}$. The equations for $R_{\mathrm{smooth}}$ and $\varkappa_{0}$, as well as their approximate values, are given by:

\begin{align}
    R_{\mathrm{smooth}} &= \frac{H_{R,R12,\mathrm{dust}}^{2}}{H_{R}} + R_{\mathrm{dust}} \approx 4.79\,\mathrm{kpc} \\
    \varkappa_{0} &= \dfrac{\exp \left( -\dfrac{R_{\mathrm{smooth}}}{H_{R}} \right) }{\exp \left( -\dfrac{(R_{\mathrm{smooth}}-R_{\mathrm{dust}})^{2}}{2 H_{R,R12,\mathrm{dust}}^{2}} \right) } \approx 0.27
\end{align}

\noindent
where all parameters have been defined previously.
\section{Galactic Magnetic Fields} \label{sect:GMF}

The synchrotron cooling~(see \autoref{sssect:Synchrotron}), which is a major contributor to the electron energy losses at 1--10\,TeV energies~(as shown in \autoref{fig:cooling times/lengths}), depends on the magnetic fields that permeate the ISM. As VHE CRs travel through the MW their paths become twisted and bent around the magnetic field, where the degree of deflection depends on the CR energy and the magnetic field strength~(as discussed in \autoref{sect:CR diffusion}).
The Galactic CR density distribution will in part be determined by the GMF, as CRs~(especially electrons) will be slowed or even trapped in regions where the magnetic field is strong.

The GMF has two components that are approximately equal in strength. Charged particles flow along the field lines of the large-scale regular component~\citep{BeuermannK.1985}. There is also the small-scale irregular component~\citep[sometimes referred to as the `random' component, see][]{SunX.2008} which is randomly oriented by the ionised interstellar gas clouds and the turbulent motions within the ISM gas. The random component is produced by SNe and other powerful outflows in the Galaxy and are typically smaller than 100\,pc in size~\citep{GaenslerB.1995,HaverkornM.2008}. There is a further random anisotropic field~(also known as a `striated' component) that is produced by a large-scale compression and stretching of the irregular component of the GMF~\citep{BeckR.2001}. The striated field is aligned with the large-scale regular component, but with reversals in direction on smaller scales.

\GP{} has a variety of available GMF models from across the literature, two of which are utilised in this work. Although the \GP{} axisymmetric exponential~(GASE) model is a simple exponential disk that is not fit to observations, it is used commonly throughout the literature~\citep[e.g.][]{KorsmeierM.2022,QiaoB.2022}. The other GMF investigated here is the~\citet{PshirkovM.2011} bisymmetric spiral~(PBSS) model, which includes the spiral arm structure and is fit to rotation measures~(RMs) of extragalactic sources. Other notable GMF models within \GP{} that are not used in this thesis include,~e.g.~\citet{SunX.2008,JaffeT.2010,SunX.2010,JanssonR.2012}.
The GMF models are parameterised in cylindrical coordinates $(R,\,\phi,\,Z)$ with the origin at the GC.

\subsection{GALPROP Axisymmetric Exponential Model}

The \GP{} axisymmetric exponential~\citep[GASE;][]{StrongA.2000} model is a simple exponential disk in the radial direction and an absolute exponential in height. The GASE model does not include any spiral arms or other more complex Galactic structures. The magnetic field strength is given by:

\begin{align}
    B_{\mathrm{GASE}}(R,\,Z) &= B_{\mathrm{GASE},0} \exp \left( \frac{R - R_{\odot}}{H_{R,\mathrm{GASE}}} \right) \exp \left( -\frac{|Z|}{H_{Z,\mathrm{GASE}}} \right) \label{eq:GASE distribution}
\end{align}

\noindent
where $B_{\mathrm{GASE},0}=5$\,$\mu$G is the field strength at Earth, $H_{R,\mathrm{GASE}}=10$\,kpc is the radial scale-length, and $H_{Z,\mathrm{GASE}}=2$\,kpc is the scale height of the disk. The distance to the GC is given by $R_{\odot}=8.5$\,kpc, which was the IAU recommended distance from the GC to Earth at the time of the model's creation~\citep{KerrF.1986}.

\begin{figure}
    \centering
    \includegraphics[height=8cm]{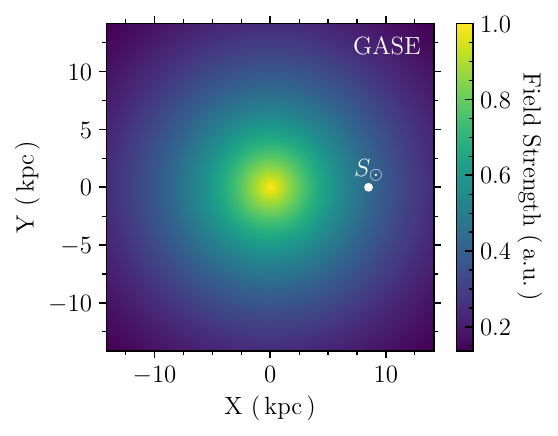}
    \caption{The GASE GMF field strength in arbitrary units integrated over the $Z$-axis. The Solar location is shown by $S_{\odot}$.}
    \label{fig:GASE distribution}
\end{figure}

The GASE distribution is shown in \autoref{fig:GASE distribution}. Due to the GASE GMF's incompatibility with rotation measures of extragalactic radio sources it is not considered an accurate representation of the magnetic field. However, it is still commonly used throughout the literature due to its simplicity~\citep[e.g.][]{KorsmeierM.2022,QiaoB.2022}. The GASE model is considered as a baseline distribution to compare against in \autoref{chap:paper 1}.

\subsection{Pshirkov Bi/Axisymmetric Spiral Models} \label{ssect:Pshirkov}

\begin{figure}
    \centering
    \includegraphics[height=8cm]{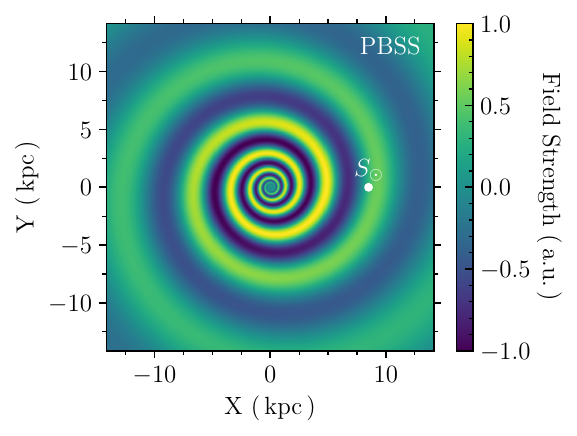}
    \caption{The PBSS GMF field strength in arbitrary units integrated over the $Z$-axis. The Solar location is shown by $S_{\odot}$. The halo component of the PBSS distribution has not been included to reproduce the distribution employed by \GP~\citep{PorterT.2017}.}
    \label{fig:PBSS distribution}
\end{figure}

There are two GMF models from~\citet{PshirkovM.2011}, both of which are constructed to agree with the Faraday RMs of extragalactic radio sources in the national radio astronomy observatory very large array~(NRAO~VLA) sky survey~\citep[named the NVSS;][]{CondonJ.1998} RM catalogue, as well as a compilation of RMs of sources from~\citet{KronbergP.2011}. The \citet{PshirkovM.2011} GMFs consist of two components: a Galactic halo, and the spiral arms. The two \citet{PshirkovM.2011} models differ only in the direction of the spiral arm magnetic field. The two models are referred to as the Pshirkov bisymmetric spiral~(PBSS) distribution and the Pshirkov axisymmetric spiral~(PASS) distribution. The PBSS GMF is shown in \autoref{fig:PBSS distribution}.

\subsubsection{Pshirkov Spiral Arm Distributions}

The PBSS and PASS spiral arms are constructed from a decreasing exponential in height with logarithmic spiral arms. Each arm is identical in size and rotated $\pi$ radians with respect to one another. The strength of the~\citet{PshirkovM.2011} spiral arm magnetic field is given by:

\begin{align}
    B_{\mathrm{PBSS}}(R,\,\phi,\,Z) &= B_{\mathrm{P,disk}}(R) \cos \left[ \phi - \phi_{b} \ln \left( \frac{R}{R_{\odot}} \right) + \phi_{a} \right] \exp \left( -\frac{|Z|}{H_{Z,\mathrm{P}}} \right) \label{eq:PBSS arm distribution}
\end{align}

\noindent
where $H_{Z,\mathrm{P}}=1$\,kpc is the scale height of the magnetic field. The only functional difference between the PBSS and PASS GMF models is that the absolute value is taken on the cosine term for the PASS model to ensure that the spiral arms point in the same direction. The disk term, $B_{\mathrm{P,disk}}$, is constant for the Galacto-centric radius $R<R_{\mathrm{P,disk}}$ and inversely proportional to the Galacto-centric radius for $R>R_{\mathrm{P,disk}}$. The central disk is given by:

\begin{align}
    B_{\mathrm{P,disk}}(R,\,\phi) &= B_{\mathrm{P,disk},0} \frac{R_{\odot}}{\cos \phi} \times
    \begin{cases}
        1 / R_{\mathrm{P,disk}}, & R < R_{\mathrm{P,disk}} \\
        1 / R, & R \geq R_{\mathrm{P,disk}}
    \end{cases}
\end{align}

\noindent
where $B_{\mathrm{P,disk},0}=2$\,$\mu$G is the magnetic field strength normalisation and $R_{\mathrm{P,disk}}=5$\,kpc is the distance from the GC where the field strength is assumed to be independent of the Galacto-centric radius. The cosine term from \autoref{eq:PBSS arm distribution}, which controls how the arms are wound around the Galaxy, depends on the angles $\phi_{a}$ and $\phi_{b}$ which are given by:

\begin{align}
    \phi_{a} &= \phi_{b} \ln \left( 1 + \frac{R_{\mathrm{rev}}}{R_{\odot}} \right) - \frac{\pi}{2} \label{eq:PBSS theta} \\
    \phi_{b} &= \frac{1}{\tan (\phi_{p})} \label{eq:PBSS b}
\end{align}

\noindent
where $\phi_{p}=-6^{\circ}$ is the pitch angle of the arms and $R_{\mathrm{rev}}=-0.6$\,kpc is the radial distance from the Solar position towards the first field reversal, with negative distances implying that the field reversal occurs in the direction towards the GC.

\subsubsection{Pshirkov Halo Distribution}

The halo component represents a large structure located above and below the Galactic plane. It is constructed from a decreasing radial exponential and a height function. The magnetic field strength of the Galactic halo is given by:

\begin{align}
    B_{\mathrm{halo}}(R,\,Z) &= B_{\mathrm{halo},0} \left[ 1 + \left( \frac{|Z| - Z_{\mathrm{halo}}}{H_{Z,\mathrm{halo}}} \right)^{2} \right]^{-1} \frac{R}{H_{R,\mathrm{halo}}} \exp \left( 1 - \frac{R}{H_{R,\mathrm{halo}}} \right) \label{eq:PBSS halo}
\end{align}

\noindent
where $B_{\mathrm{halo},0} = 4$\,$\mu$G is the halo magnetic field strength normalisation, $Z_{\mathrm{halo}} = 1.3$\,kpc is the vertical location of the halo, and $H_{R} = 8$\,kpc is the scale radius of the halo. The parameter $H_{Z}$ is the scale height of the halo and differs between the northern~(positive $Z$) and southern~(negative $Z$) hemispheres. For the northern hemisphere the scale height is given by $H_{Z,\mathrm{halo,north}} = 0.25$\,kpc, and for the southern hemisphere the scale height is given by $H_{Z,\mathrm{halo,south}} = 0.4$\,kpc. 

As the halo component of the~\citet{PshirkovM.2011} GMF lies 1.3\,kpc above/below the plane, it has a negligible impact on Galactic CR diffusion and \graya{} production in the MW~\citep{PorterT.2017}. Hence, it is not used within \GP. The halo component is included here for completeness.
\section{GALPROP Setup} \label{sect:GALPROP setup}

Within \GP{} the CR densities are normalised by a single condition -- that all CR spectra~(protons, electrons, helium, etc) at the Solar position reproduce their respective locally observed spectra within measurement systematics and uncertainties.
The local spectra are post propagation, with the CR fluxes depending on the chosen source distribution, the ISM gas, the ISRF, and the GMF. Therefore, the CR normalisation procedure in \GP{} must be performed post-propagation. Additionally, the propagation grid must be appropriately defined -- the cells must be smaller than the shortest cooling distance of any CR~(see \autoref{ssect:cooling lengths}). This section details how the CR normalisation is performed and how the propagation grid is defined.

\subsection{Diffusion and Injection Parameter Optimisation} \label{ssect:parameter optimisation}

The critical condition for any CR diffusion simulation is that all CR spectra, from electrons to protons to iron, reproduce the measured CR spectra at the Solar location. Reproduction is ensured by treating the diffusion and injection parameters for each CR species as free parameters. Optimising these parameters needs to be performed for each CR source distribution, as altering where the CRs are injected impacts the local CR spectra.

To normalise and optimise the propagation parameters and source spectra, the procedure in~\citet{PorterT.2017} and~\citet{JohannessonG.2018} are followed using the CR data from~\citet{JohannessonG.2019}~(see their Table~1) for each source distribution. An initial optimisation of the propagation models is made by fitting to the observed CR spectra using data from the alpha magnetic spectrometer~(AMS-02)~\citep{AguilarM.2014,AguilarM.2015b,AguilarM.2015a,AguilarM.2017,AguilarM.2018} and \textit{Voyager}~\citep{CummingsA.2016} in the GeV energy range where the diffuse CR sea is the dominant source of CRs. The fitting procedure is performed individually for the following CR spectra: Be, B, C, O, Mg, Ne, and Si. These seven spectra are then kept fixed. The next step in the optimisation procedure is to fit the spectra for electrons, protons, and He. The proton normalisation directly impacts the propagation and injection of the heavier elements, so the proton spectrum is optimised iteratively until all spectra converge to the local observations. As the optimisation method extrapolates outside the energy range of the data~(MeV to TeV), a best-fit model is then determined via a $\chi^{2}$ test on each of the spectra.

The optimisation procedure can be performed for any arbitrary source distribution and gas distribution and is included as a separate program in \GP{} version~57. Although the program automatically calculates all relevant parameters, it must be manually initiated by the user prior to any \GP{} propagation for each new source and ISM gas distribution. The optimisation program ensures that all of the source distributions used in \GP{} give the same local CR spectra, which reduces any inconsistences between the distributions that may arise due to the limited data statistics and coverage over the modelled energy range.

Currently, the normalisation procedure only accounts for the source distribution and the gas distribution. This assumption is valid for protons and heavier elements which are not heavily impacted by the ISRF or GMF. However, it may not be valid for electrons in the >10\,TeV regime which are significantly impacted by the GMF~\citep{MarinosP.2023}.

In \GP, and commonly elsewhere in the literature, the CR injection spectra are given by power laws. In \GP{} the injection spectra can be defined separately for each individual CR species. However, it is typical to define the spectra separately for electrons, protons, helium, with all CR species heavier than helium sharing the same spectrum with different normalisations. For this thesis, the proton and electron spectra are normalised at the Solar position to the CR normalisation constants given by $J_{p}$ and $J_{e^{-}}$ at the energies $E_{k, p}=100$\,GeV and $E_{k, e^{-}}=34.5$\,GeV respectively, with heavier elements normalised relative to the proton spectrum at 100\,GeV\,nuc$^{-1}$. The injection spectra are given by:

\begin{align}
    \frac{\mathrm{d}\mathcal{N}_{\mathrm{CR}}}{\mathrm{d}p} &= J_{\mathrm{CR}} \varrho_{\mathrm{CR}}^{-\eta_{\mathrm{CR}}} \label{eq:CR injection spectra}
\end{align}

\noindent
where $p$ is the momentum of the CR, $\mathcal{N}_{\mathrm{CR}}$ is the injected CR density, $J_{\mathrm{CR}}$ is the normalisation of the CR spectrum, $\varrho_{\mathrm{CR}}$ is the rigidity of the CR, and $\eta_{\mathrm{CR}}$ is the injection spectral index for the CR species. The spectral index in \GP{} can be broken into different regions, where the slope of the CR injection spectrum can be altered. The breaks used in this thesis are given by the rigidities $\varrho_{\mathrm{CR},1}$ and $\varrho_{\mathrm{CR},2}$, and the spectral indices are given by $\eta_{\mathrm{CR}}=\eta_{\mathrm{CR},0}$ for rigidities $\varrho_{\mathrm{CR}}<\varrho_{\mathrm{CR,1}}$, $\eta_{\mathrm{CR}}=\eta_{\mathrm{CR},1}$ for rigidities $\varrho_{\mathrm{CR},1}<\varrho_{\mathrm{CR}}<\varrho_{\mathrm{CR},2}$, and $\eta_{\mathrm{CR}}=\eta_{\mathrm{CR},2}$ for rigidities $\varrho_{\mathrm{CR}}>\varrho_{\mathrm{CR},2}$.

In \GP{} the injection spectra can have an arbitrary number of spectral breaks, and the transitions between each of breaks can be smoothed via the parameter $\mathscr{s}$. Although this functionality is not used in this thesis, the full form for the CR injection spectrum for each CR species is then given by:

\begin{align}
    \dfrac{\mathrm{d}\mathcal{N}}{\mathrm{d}p} &= J \left( \dfrac{\varrho}{\varrho_{0}} \right)^{-\eta_{0}} \prod_{\mathscr{i}} \left( 1 + \left( \dfrac{\varrho}{\varrho_{\mathscr{i}}} \right)^{\left[ \frac{\eta_{\mathscr{i}}-\eta_{\mathscr{i}+1}}{\mathscr{s_{\mathscr{i}}}} \right] } \right)^{\mathscr{s_{\mathscr{i}}}} \label{eq:full CR injection spectra}
\end{align}

\noindent
where $\varrho_{0}$ is the normalisation rigidity for the given CR species, $\mathscr{i}$ is the counting parameter for the spectral breaks, and $\varrho_{\mathscr{i}}$ is the rigidity of the $\mathscr{i}$\textsuperscript{th} spectral break. The parameter $\mathscr{s}_{\mathscr{i}}$ is the smoothing parameter for the $\mathscr{i}$\textsuperscript{th} spectral break, and is negative for $|\eta_{\mathscr{i}}|<|\eta_{\mathscr{i}+1}|$ and positive for $|\eta_{\mathscr{i}}|>|\eta_{\mathscr{i}+1}|$. The CR subscripts have been dropped for simplicity; however, each parameter in \autoref{eq:full CR injection spectra} can be controlled separately for each individual CR species.

The diffusion coefficient from \autoref{eq:diffusion coefficient}, is normalised to $D_{xx,0}$ at the rigidity $\varrho_{0}=4$\,GV, and has a spectral index given by $\delta$. The diffusion coefficient spectral index depends on the chosen source model. As of version~57, \GP{} assumes an isotropic and homogeneous spatial diffusion coefficient, with an anisotropic diffusion tensor planned for a future release.
The diffusion coefficient is sometimes linked to the magnetic field strength due to the CRs diffusing more slowly through stronger magnetic fields~\citep[e.g.~see][]{GabiciS.2007a}. Linking the magnetic field to the diffusion coefficient is important for regions with magnetic fields on the order of 10\,$\mu$G. 
Although the diffusion coefficient in \GP{} can be linked to the GMF strength~\citep[e.g.][]{AckermannM.2015}, this functionality is not used in this thesis as the average GMF strength is on the order of 3\,$\mu$G.
The CR spectral parameters and the diffusion parameters for all five source distributions used throughout this thesis are shown in \autoref{tab:SA pars}. The source distributions themselves, and the equations that define them, are detailed in \autoref{sect:source dist.}.

\begin{table}
    \centering
    \caption{The optimised \GP{} propagation parameters for each of the five source distributions.}
    \label{tab:SA pars}
    \bgroup
    \def\arraystretch{1.25}
    \begin{tabular}{lccccc}
        \hline
        Parameter & SA0 & SA25 & SA50 & SA75 & SA100 \\
        \hline
        $D_{xx,0}$~[$10^{28}$]              & 4.36  & 4.39  & 4.55  & 4.67  & 4.66  \\ % 3 s.f.
        $\delta$                            & 0.354 & 0.349 & 0.344 & 0.340 & 0.339 \\ % 3 s.f. D_g_1
        $v_{\mathrm{Alfven}}$               & 17.8  & 18.2  & 18.1  & 19.8  & 19.1  \\ % 3 s.f.
        $J_{p}$~[$10^{-9}$]                 & 4.096 & 4.404 & 4.113 & 4.329 & 4.394 \\ % 4 s.f.
        $J_{e^{-}}$~[$10^{-10}$]            & 3.925 & 4.444 & 3.994 & 4.428 & 4.502 \\ % 4 s.f.
        $\eta_{e^{-},0}$                    & 1.616 & 1.390 & 1.488 & 1.455 & 1.521 \\ % 4 s.f.
        $\eta_{e^{-},1}$                    & 2.843 & 2.756 & 2.766 & 2.763 & 2.753 \\ % 4 s.f.
        $\eta_{e^{-},2}$                    & 2.493 & 2.460 & 2.470 & 2.447 & 2.422 \\ % 4 s.f.
        $\varrho_{e^{-},1}$                 & 6.72  & 5.27  & 5.14  & 5.54  & 5.29  \\ % 3 s.f.
        $\varrho_{e^{-},2}$                 & 52.4  & 81.6  & 67.7  & 70.7  & 79.7  \\ % 3 s.f.
        $\eta_{\mathrm{p},0}$               & 1.958 & 1.928 & 1.990 & 1.965 & 2.009 \\ % 4 s.f.
        $\eta_{\mathrm{p},1}$               & 2.450 & 2.464 & 2.466 & 2.494 & 2.481 \\ % 4 s.f.
        $\eta_{\mathrm{p},2}$               & 2.391 & 2.411 & 2.355 & 2.374 & 2.414 \\ % 4 s.f.
        $\varrho_{\mathrm{p},1}$            & 12.0  & 12.3  & 12.2  & 14.5  & 13.5  \\ % 3 s.f.
        $\varrho_{\mathrm{p},2}$            & 202   & 157   & 266   & 108   & 125   \\ % 3 s.f.
        $\eta_{\mathrm{He},0}$              & 1.925 & 1.886 & 1.956 & 1.937 & 1.971 \\ % 4 s.f.
        $\eta_{\mathrm{He},1}$              & 2.417 & 2.421 & 2.432 & 2.467 & 2.443 \\ % 4 s.f.
        $\eta_{\mathrm{He},2}$              & 2.358 & 2.369 & 2.320 & 2.347 & 2.376 \\ % 4 s.f.
        $\varrho_{\mathrm{He},1}$           & 12.0  & 12.3  & 12.2  & 14.5  & 13.5  \\ % 3 s.f.
        $\varrho_{\mathrm{He},2}$           & 202   & 157   & 266   & 108   & 125   \\ % 3 s.f.
        $\eta_{\mathrm{Z},0}$               & 1.328 & 1.519 & 1.426 & 1.630 & 1.624 \\ % 4 s.f.
        $\eta_{\mathrm{Z},1}$               & 2.377 & 2.390 & 2.399 & 2.399 & 2.418 \\ % 4 s.f.
        $\eta_{\mathrm{Z},2}$               & 2.377 & 2.390 & 2.399 & 2.399 & 2.418 \\ % 4 s.f.
        $\varrho_{\mathrm{Z},1}$            & 3.16  & 4.21  & 3.44  & 4.61  & 4.50  \\ % 3 s.f.
        $\varrho_{\mathrm{Z},2}$~[$10^{3}$] & 5.00  & 5.00  & 5.00  & 5.00  & 5.00  \\ % 3 s.f.
        \hline
    \end{tabular}
    \egroup
    \\
    \begin{itemize}\setlength\itemsep{3.11pt}
        \item[--] $\eta_{\mathrm{CR},0}$ is the power-law index before the first break
        \item[--] $\eta_{\mathrm{CR},1}$ is the power-law index between the first and second breaks
        \item[--] $\eta_{\mathrm{CR},2}$ is the power-law index after the second break
        \item[--] $\varrho_{\mathrm{CR},1}$~[GV] is the rigidity of the first break
        \item[--] $\varrho_{\mathrm{CR},2}$~[GV] is the rigidity of the second break
        \item[--] The subscript $Z$ is used for all CR nuclei with atomic numbers $Z \ge 3$
        \item[--] The diffusion coefficient~($D_{xx,0}$) is measured in cm$^{2}$\,s$^{-1}$
        \item[--] The Alfven velocity~($v_{\mathrm{Alfven}}$) is measured in km\,s$^{-1}$
        \item[--] The CR normalisation constants~($J_{p}$ and $J_{e^{-}}$) are measured in MeV$^{-1}$\,cm$^{-2}$\,s$^{-1}$\,sr$^{-1}$
    \end{itemize}
\end{table}

\subsection{Timestep} \label{ssect:time-indep time grid}

The transport equation, given in \autoref{eq:Transport Equation}, is solved for some timestep, $\Delta t$.
Numerically solving the transport equation only provides information on processes that occur on timescale of a similar size to the chosen timestep.
However, the CRs diffusing in the MW lose energy at different rates depending on their energy~(as seen in \autoref{fig:cooling times/lengths}).
To capture the diffusion of all particles at all energies, the timestep used in the time-independent solution\footnote{The time-dependent timestep will be discussed later in \autoref{ssect:time-dep time grid}.} of the transport equation varies during the simulation.
The initial timestep, $\Delta t_{0}$, should be on the order of the longest cooling time of any particle being simulated.
The timestep is then decreased iteratively, with the $k$\textsuperscript{th} timestep being given by:

\begin{align}
    \Delta t_{k} = \mathscr{f}^{k} \Delta t_{0}
\end{align}

\noindent
where $\mathscr{f}$ is the timestep factor and must be less than one. The timestep is reduced iteratively until the final timestep~($\Delta t_{\mathrm{final}}$) is reached, which should be as short as the shortest cooling time of any particle being simulated~(see \autoref{fig:cooling times/lengths}). Solving the transport equation via this method allows changes across many orders of magnitude of time to be simulated accurately and efficiently. The transport equation is solved iteratively for each timestep $M$ times, or until the solution is stable.

\subsection{Propagation Grid} \label{ssect:grid}

\GP{} creates a virtual MW that consists of cells, where this grid of cells contains the ISM densities and is what the CRs diffuse through. The propagation grid is defined in the coordinates $(X,\,Y,\,Z)$, with three different grid functions available in \GP{} -- a linear function, a tan function, and a step function. A description of the linear and tan grid functions is provided below, and an explanation on setting up the propagation grid within \GP{} is provided in \autoref{sect:grid galdef}.

For the CR propagation results to be valid, the cell size at the observer's position must be small enough to accurately capture the CR densities. In order to capture the CR diffusion across the cell, the cell size must be smaller than the shortest cooling distance of any CR particle included in the simulation\footnote{Depending on the chosen differential equation solver, the cells may need to be an order of magnitude smaller than the shortest cooling distance~\citep[][Appendix D]{PorterT.2022}.}. As the cooling distances for protons and electrons have different energy dependencies~(see \autoref{ssect:cooling lengths}), it will be the lowest-energy proton or the highest-energy electron that dictates the cell size. See \autoref{fig:cooling times/lengths} for calculations on the diffusion and cooling distances.

\begin{figure}
    \centering
    \hspace*{\fill}%
    \subfloat[width=7.5cm][The gridding function for the linear and tan functions.]{\includegraphics[width=7.5cm, align=t]{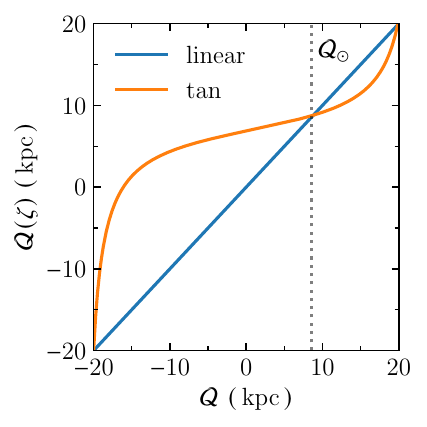}
    \label{fig:grid functions}}
    \hfil \hspace{0.2cm}
    \subfloat[width=7.5cm][The derivative of the gridding function, which is equivalent to the cell size.]{\includegraphics[width=7.5cm, align=t]{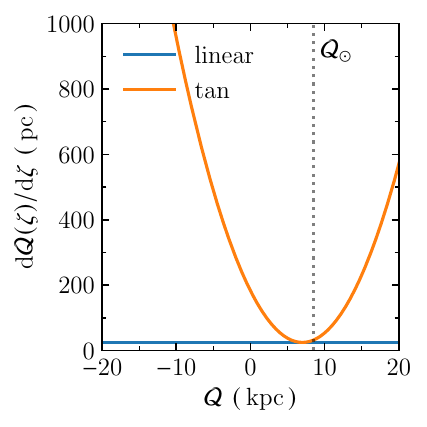}
    \label{fig:grid function derivative}}
    \hspace*{\fill}%
    \\
    \centering
    \hspace*{\fill}%
    \subfloat[width=7.5cm][The subtended angle of the cell size in arcminutes.]{\includegraphics[width=7.5cm, align=c]{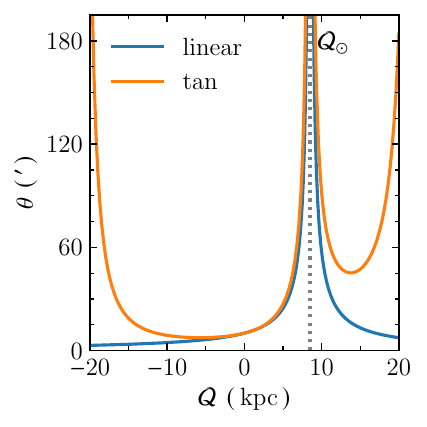}
    \label{fig:grid function subtended angle}}
    \hfil \hspace{1.9cm} % Difficult to align the figures horizontally, due to the table. Couldn't find a good solution, found some amount of space that looked good
    \subfloat[width=7.5cm][The parameter values used for Figures~\ref{fig:grid functions} through~\ref{fig:grid function subtended angle}, with the resulting number of cells given by $N_{\mathrm{cells}}$.]{
    \vspace{0pt}
    \bgroup
    \def\arraystretch{1.25}
    \begin{tabular}[c]{lrr}
        \hline
        Parameter & Linear & Tan \\ \hline
        $\Delta \mathcal{Q}$ \ [kpc] & 0.025 & 0.025 \\
        $\mathcal{Q}_{\mathrm{min}}$ [kpc] & $-20$ & $-20$ \\
        $\mathcal{Q}_{\mathrm{max}}$ [kpc] & 20 & 20 \\
        $\mathcal{Q}_{0}$ [kpc] & -- & 7 \\
        $\mathcal{Q}_{\mathrm{ref}}$ [kpc] & -- & 26.5 \\
        $\lambda$ & -- & 50 \\ \hline
        $N_{\mathrm{cells}}$ & 1601 & 316
    \end{tabular}
    \egroup
    \label{tab:grid function parameters}}
    \hspace*{\fill}%
    \caption{An example of the linear and tan grid functions that can be used in \GP. In all subfigures the Solar location is given by $\mathcal{Q}_{\odot}=8.5$\,kpc and shown by the vertical dashed line. The parameter values used~(d) are representative of a typical $X$ axis in \GP. For an example where the variable cell size of the tan gridding function is visible on the two-dimensional grid see \autoref{fig:GALPROP p vs e}.}
    \label{fig:grid function examples}
\end{figure}

For the descriptions of the grid functions provided below, the following definition is used: $\mathcal{Q}(\zeta)$ is the transformed coordinate at the $\zeta$\textsuperscript{th} point, and can refer to either the $X$, $Y$, or $Z$ axes.
The parameters $\mathcal{Q}_{\mathrm{min}}$ and $\mathcal{Q}_{\mathrm{max}}$ are the minimum and maximum extents of the axis, and the relationship $\mathcal{Q}(\zeta=0)=\mathcal{Q}_\mathrm{min}$ and $\mathcal{Q}(\zeta=N_{\mathrm{cells}}-1)=\mathcal{Q}_{\mathrm{max}}$ are used to define the bounds of the transformed axes. The linear and tan gridding functions, their first derivative~(i.e.~the sizes of the cells as a function of distance from the GC), and their subtended angles from Earth, are shown in \autoref{fig:grid function examples}. The impact of the tan gridding function on the 2D cell sizes can be observed in \autoref{fig:GALPROP p vs e}.

\subsubsection{Linear Grid Function} \label{sssect:linear grid function}

The linear grid function is the standard approach. It gives all cells the same size defined by the step size~($\Delta \mathcal{Q}$). The linear grid function is given by:

\begin{align}
    \mathcal{Q}_{\mathrm{lin}}(\zeta) &= \Delta \mathcal{Q} \zeta + \mathcal{Q}_{\mathrm{min}} \label{eq:linear grid function}
\end{align}

\noindent
where the physical size of the cells along the $\mathcal{Q}$ axis is given by the derivative, and has the functional form:

\begin{equation}
    \begin{aligned}
        \frac{\mathrm{d}\mathcal{Q}_{\mathrm{lin}}(\zeta)}{\mathrm{d}\zeta} &= \Delta \mathcal{Q} \\
        &= \frac{\mathcal{Q}_{\mathrm{max}} - \mathcal{Q}_{\mathrm{min}}}{N_{\mathrm{cells}} - 1}
    \end{aligned} \label{eq:linear grid size}
\end{equation}

\noindent
where the parameters $\mathcal{Q}_{\mathrm{min}}$, $\mathcal{Q}_{\mathrm{max}}$, and $\Delta \mathcal{Q}$ are given by the user for each coordinate axis~(i.e.~$X$, $Y$, and $Z$), and $N_{\mathrm{cells}}$ is the number of cells along the axis. The distance from the Solar position to a cell can be defined as~$\mathcal{Q}_{R}=\mathcal{Q}(\zeta)-\mathcal{Q}_{\odot}$. Typically, the coordinate system used by \GP{} sets $X_{\odot}=8.5$\,kpc, $Y_{\odot}=0$\,kpc, and $Z_{\odot}=0$\,kpc.

The angle subtended by the cells as a function of their distance from the Solar position is given by:

\begin{equation}
    \begin{aligned}
        \theta_{\mathrm{lin}} &= 2 \left| \arctan \left( \frac{\mathrm{d}\mathcal{Q}_{\mathrm{lin}}(\zeta) / \mathrm{d}\zeta}{2\mathcal{Q}_{R}} \right) \right| \\
        &= 2 \left| \arctan \left( \frac{\Delta \mathcal{Q}}{2\mathcal{Q}_{R}} \right) \right| \label{eq:linear grid angle}
    \end{aligned}
\end{equation}

\noindent
where all variables have been defined previously. As the cell size is a constant, the angle subtended by the cell asymptotes to zero as the distance to said cell increases~(as seen in \autoref{fig:grid function examples}). Cells far from the observer's location have a minimal impact on the line-of-sight integral for the \graya{} intensities due to their small angular size. Additionally, as the emission follows the $r$-squared law, distant cells do not contribute significantly to the emission integrated over the line of sight. Hence, the linear grid does not efficiently utilise computational resources for regions far from the Solar location.

As the CR protons diffuse large~(>200\,pc) distances, the cells can be large~($\Delta \mathcal{Q} \sim 100$\,pc), saving on computation time while still giving reasonable results. However, for CR electrons the grid size needs to be on the order of the cooling distances, which can be less than 100\,pc for >10\,TeV electrons~(see \autoref{fig:cooling times/lengths}).

\subsubsection{Tan Grid Function} \label{sssect:tan grid function}

An optimisation to the gridding functions can be made to save on computational resources without impacting the propagation of the CRs by increasing the size of the cells for larger distances from the observer. The increasing cell size is achieved by using a `tan' grid function, which maintains an approximately constant subtended angle for large distances from the observer. The tan grid function is defined such that the solution to the partial differential equations remains stable, and is given by:

\begin{align}
    \mathcal{Q}_{\mathrm{tan}}(\zeta) &= \frac{\Delta \mathcal{Q}}{a} \tan\left[ a (\zeta - \zeta_{0} ) \right] + \mathcal{Q}_{0} \label{eq:tan grid function} \\
    a &= \frac{\Delta \mathcal{Q}}{\mathcal{Q}_{\mathrm{ref}} - \mathcal{Q}_{0}} \tan\left[ \arccos\left( \frac{1}{\sqrt{\lambda}} \right) \right] \nonumber \\
    &= \frac{\Delta \mathcal{Q}}{\mathcal{Q}_{\mathrm{ref}} - \mathcal{Q}_{0}} \sqrt{\lambda - 1} \label{eq:a_grid} \\
    \zeta_{0} &= \frac{-1}{a} \arctan\left( \frac{a ( \mathcal{Q}_{\mathrm{min}} - \mathcal{Q}_{0} ) }{\Delta \mathcal{Q}} \right) \label{eq:z0} \\
    N_{\mathrm{cells}} - 1 &= \frac{1}{a} \left[ \arctan\left( \frac{a ( \mathcal{Q}_{\mathrm{max}} - \mathcal{Q}_{0} ) }{\Delta \mathcal{Q}} \right) - \arctan\left( \frac{a ( \mathcal{Q}_{\mathrm{min}} - \mathcal{Q}_{0} ) }{\Delta \mathcal{Q}} \right) \right] \label{eq:N-1}
\end{align}

\noindent
where the parameters $\mathcal{Q}_{\mathrm{min}}$, $\mathcal{Q}_{\mathrm{max}}$, $\mathcal{Q}_{0}$, $\mathcal{Q}_{\mathrm{ref}}$, and $\lambda$ are set by the user. $\mathcal{Q}_{0}$ is the `centre' of the axis~(where the grid has the highest resolution), $\mathcal{Q}_{\mathrm{ref}}$ is the location where the size of the grid is equal to $\lambda \Delta \mathcal{Q}$, $\lambda$ is a parameter defining the rate of growth of the grid and must be greater than one, and $N$ is the number of cells along the axis.
\autoref{eq:a_grid} arises from the condition that the size of the grid~(i.e.~first derivative) at the position $\mathcal{Q}_{\mathrm{ref}}$ must equal $\lambda \Delta \mathcal{Q}$.
\autoref{eq:z0} arises from the condition that $\mathcal{Q}(\zeta=0)=\mathcal{Q}_{\mathrm{min}}$, and \autoref{eq:N-1} gives the number of grid cells along the axis and arises from the condition that $\mathcal{Q}(\zeta=N_{\mathrm{cells}}-1)=\mathcal{Q}_{\mathrm{max}}$. The size of the cells is given by the first derivative:

\begin{align}
    \frac{\mathrm{d}\mathcal{Q}_{\mathrm{tan}}}{\mathrm{d}\zeta} &= \frac{\Delta \mathcal{Q}}{\cos^{2}\left[ a ( \zeta - \zeta_{0} )\right]} \label{eq:tan grid size}
\end{align}

\noindent
where all variables have been defined previously. When $\mathcal{Q}(\zeta)=\mathcal{Q}_{\mathrm{ref}}$, then the size of the grid is d$\mathcal{Q}$/d$\zeta=\lambda \Delta \mathcal{Q}$. In the transformed coordinates, this occurs at:

\begin{align}
     \zeta_{\mathrm{ref}} &= \frac{1}{a} \left[ \arctan\left( \frac{a ( \mathcal{Q}_{\mathrm{ref}} - \mathcal{Q}_{0} )}{\Delta \mathcal{Q}} \right) - \arctan\left( \frac{a ( \mathcal{Q}_{\mathrm{min}} - \mathcal{Q}_{0} )}{\Delta \mathcal{Q}} \right) \right] \label{eq:max X grid size}
\end{align}

\noindent
where all variables have been defined previously. For the case that the grid cell size is largest at the edges of the Galaxy~(i.e.~$\mathcal{Q}_{\mathrm{ref}}=\mathcal{Q}_{\mathrm{max}}$), \autoref{eq:max X grid size} is equal to the number of cells minus one~(i.e.~$N_{\mathrm{cells}}-1$; \autoref{eq:N-1}). Just as for the linear grid, it is useful to find the angle subtended by a cell, which is given by:

\begin{align}
    \theta_{\mathrm{tan}} &= 2 \left| \arctan\left( \frac{\mathrm{d}\mathcal{Q}_{\mathrm{tan}} / \mathrm{d}\zeta)}{2 \mathcal{Q}_{R}} \right) \right|
\end{align}

\noindent
where all variables have been defined previously. For the majority of the MW the cells have a near-constant subtended angle. The tan grid function, the cell size, and the subtended angle are shown in \autoref{fig:grid function examples}. The impacts of using a tan grid on the size of the cells can be seen in \autoref{fig:GALPROP p vs e}. There is a discontinuity at the Solar position as when Earth is located in a cell, that cell covers the entire $4\pi$ steradians of the sky.
Towards the Galactic anticentre the cells are nearby and larger. Hence, the subtended angles of the cells towards the Galactic anticentre are large~(>1$^{\circ}$). However, the impact on the calculations is minor as the CR density~(and the \graya{} flux) is low in the outer Galaxy. Near $\mathcal{Q}_{\mathrm{min}}$ and $\mathcal{Q}_{\mathrm{max}}$ the angle subtended by the cells increases rapidly -- as these locations are far from $\mathcal{Q}_{\odot}$, the \grays{} produced in these cells has little impact on the \graya{} flux measured by the observer due to the $r$-squared law. The tan grid function has only a negligible impact on the diffuse \graya{} calculations.

    \chapter{TeV Gamma-Ray Observations with \hess} \label{chap:hess}

CRs with energies below 10\,PeV do not travel in straight lines. Hence, VHE CRs cannot be traced back to their original sites of acceleration. Insights into the underlying acceleration physics must then be found through observations of the \grays{} emitted as the CRs interact with the ISM. In this thesis the TeV \graya{} data from the high energy stereoscopic system~(\hess) is utilised.

When a \grayn~(or CR) strikes Earth's upper atmosphere a shower of high-energy particles is created~\citet{AugerP.1939}. The particles in the shower will be travelling faster than the local speed of light~(i.e.~the speed of light through air), creating a cone of Cherenkov light~\citep{CherenkovP.1937} in the upper atmosphere.
Imaging atmospheric Cherenkov telescopes~(IACTs) observe \grays{} indirectly by measuring this Cherenkov light.

\hess{} is an IACT array consisting of five telescopes located in the Khomas~Region of Namibia at an altitude of 1800~metres.
Four of the telescopes have diameters of 12\,m and are arranged in a $120\times120$\,m square with the vertices aligned north/south and east/west. The fifth IACT has a diameter of 28\,m and is located in the centre of the square.
As the telescopes are separated by 85--170\,m they can observe the same atmospheric showers from different perspectives~(up to five, one for each telescope) allowing for a more accurate reconstruction of the original location of the \grayn{} on the sky. \hess{} is well suited to observing \grays{} in the 30\,GeV to 10\,TeV energy range and is able to achieve a sensitivity of one percent of the flux of the Crab nebula for an observation time of 50~hours~\citep{AharonianF.2006a}. \hess{} has performed one of the most comprehensive surveys of the TeV \graya{} emission along the Galactic plane, named the \hess{} Galactic plane survey~(HGPS). This survey is the main observational dataset used throughout this thesis.

This chapter begins with a brief overview of how IACTs operate and the background physics behind their observations. The specifics of the \hess{} telescope array will then be detailed -- from its performance, to how it collects observational data. A discussion on the TeV diffuse \graya{} emission is also included, including the challenges faced in extracting the diffuse \graya{} emission from the HGPS.

\section{Gamma-Ray Observations} \label{sect:gamma-ray observations}

Visible light photons are only mildly attenuated as they travel through the atmosphere. Conversely, the atmosphere completely blocks \graya{} photons from reaching the ground from space.
As \grays{} strike the atmosphere they create a shower of energetic, charged particles that rain down to the ground called an extensive air shower~(EAS). This effect can be exploited by IACTs to indirectly observe \grays.

\subsection{Extensive Air Showers (EASs)} \label{ssect:EAS}

Extensive air showers are created by both CRs and \grays{} -- when they strike the Earth's atmosphere their energy is converted into one or more secondary particles. These secondary particles are energetic enough to create tertiary particles, and so on -- repeating until no energy remains in the system. These particle cascades were first discovered by~\citet{AugerP.1939} and can be comprised of tens of thousands of energetic sub-atomic particles. EASs typically cover approximately~1\,km$^{2}$ by the time they reach the ground.

There are two types of EASs. Hadronic showers are initiated by CR nuclei, and electromagnetic showers are initiated by electrons and \grays.
Hadronic showers also contain electromagnetic sub-showers~\citep{GriederP.2010}.
For \graya{} observations, the interest is in selecting only the showers initiated by \grays; however, both types are explained below as the hadronic component is an important background that must be accounted for in observations.

\begin{figure}

    \centering

    \begin{tikzpicture}[node distance=1cm and 1cm]
        % \draw[help lines] (0,0) grid (16,8);

        % Initial
        \coordinate[label=above:$\gamma$] (l) at (8,8);
        \coordinate (l_0) at (8,6);

        % Generation one
        \coordinate (l_1_1) at (4,4);
        \coordinate (l_1_2) at (12,4);

        % Generation two
        \coordinate (l_2_1) at (2,2);
        \coordinate (l_2_2) at (6,2);
        \coordinate (l_2_3) at (10,2);
        \coordinate (l_2_4) at (14,2);

        % Generation three
        \coordinate (l_3_1) at (1,0);
        \coordinate (l_3_2) at (3,0);
        \coordinate (l_3_3) at (5,0);
        \coordinate (l_3_4) at (7,0);
        \coordinate (l_3_5) at (9,0);
        \coordinate (l_3_6) at (11,0);
        \coordinate (l_3_7) at (13,0);
        \coordinate (l_3_8) at (15,0);

        % Lines
        \draw[photon] (l) -- (l_0);

        \draw[particle] (l_0) -- (l_1_1);
        \draw[particle] (l_0) -- (l_1_2);

        \draw[particle] (l_1_1) -- (l_2_1);
        \draw[photon] (l_1_1) -- (l_2_2);
        \draw[photon] (l_1_2) -- (l_2_3);
        \draw[particle] (l_1_2) -- (l_2_4);

        \draw[particle] (l_2_1) -- (l_3_1);
        \draw[photon] (l_2_1) -- (l_3_2);
        \draw[particle] (l_2_2) -- (l_3_3);
        \draw[particle] (l_2_2) -- (l_3_4);
        \draw[particle] (l_2_3) -- (l_3_5);
        \draw[particle] (l_2_3) -- (l_3_6);
        \draw[photon] (l_2_4) -- (l_3_7);
        \draw[particle] (l_2_4) -- (l_3_8);

        % Labels
        \coordinate[label=above left:$e^{-}$] (a) at (6,5);
        \coordinate[label=above right:$e^{+}$] (b) at (10,5);

        \coordinate[label=above left:$e^{-}$] (c) at (3,3);
        \coordinate[label=above right:$\gamma$] (d) at (5,3);
        \coordinate[label=above left:$\gamma$] (e) at (11,3);
        \coordinate[label=above right:$e^{+}$] (f) at (13,3);
        
        \coordinate[label=above left:$e^{-}$] (g) at (1.5,1);
        \coordinate[label=above right:$\gamma$] (h) at (2.5,1);
        \coordinate[label=above left:$e^{-}$] (i) at (5.5,1);
        \coordinate[label=above right:$e^{+}$] (j) at (6.5,1);
        \coordinate[label=above left:$e^{-}$] (k) at (9.5,1);
        \coordinate[label=above right:$e^{+}$] (l) at (10.5,1);
        \coordinate[label=above left:$\gamma$] (m) at (13.5,1);
        \coordinate[label=above right:$e^{+}$] (n) at (14.5,1);

        % Generation
        \draw[dashed, draw=black!50] (0,6) -- (16,6);
        \draw[dashed, draw=black!50] (0,4) -- (16,4);
        \draw[dashed, draw=black!50] (0,2) -- (16,2);
        \draw[dashed, draw=black!50] (0,0) -- (16,0);
        \coordinate[label=above right:{$g=0$}] (o) at (0,6);
        \coordinate[label=below right:{$g=1$}] (p) at (0,6);
        \coordinate[label=below right:{$g=2$}] (q) at (0,4);
        \coordinate[label=below right:{$g=3$}] (r) at (0,2);

    \end{tikzpicture}

    \vspace{1cm}

    \begin{tikzpicture}[node distance=1cm and 1cm]
        % \draw[help lines] (0,0) grid (16,9);

        % Initial
        \coordinate[label=above:$N$] (h) at (8,9);
        \coordinate (h_0) at (8,7);

        % Generation one
        \coordinate (h_1_1) at (2,5); % pi0
        \coordinate (h_1_2) at (6,5); % pi-
        \coordinate (h_1_3) at (10,5); % pi+
        \coordinate (h_1_4) at (14,5); % N

        % Generation two
        \coordinate (h_2_1) at (1,3); % photon
        \coordinate (h_2_2) at (3,3); % photon
        \coordinate (h_2_3) at (5,3);
        \coordinate (h_2_4) at (7,3);
        \coordinate (h_2_5) at (9,3);
        \coordinate (h_2_6) at (11,3);

        % Generation three
        \coordinate (h_3_1) at (3,0);
        \coordinate (h_3_2) at (5,0);
        \coordinate (h_3_3) at (7,0);
        \coordinate (h_3_4) at (9,0);
        \coordinate (h_3_5) at (11,0);
        \coordinate (h_3_6) at (13,0);

        % Lines
        \draw[particle] (h) -- (h_0); % CR

        \draw[particle] (h_0) -- (h_1_1); % pi0
        \draw[particle] (h_0) -- (h_1_2); % pi-
        \draw[particle] (h_0) -- (h_1_3); % pi+
        \draw[particle] (h_0) -- (h_1_4); % N

        \draw[photon] (h_1_1) -- (h_2_1); % pi0 => gamma
        \draw[photon] (h_1_1) -- (h_2_2); % pi0 => gamma
        \draw[particle] (h_1_2) -- (h_2_3); % pi- => mu-
        \draw[particle] (h_1_2) -- (h_2_4); % pi- => numu
        \draw[particle] (h_1_3) -- (h_2_5); % pi+ => numu
        \draw[particle] (h_1_3) -- (h_2_6); % pi+ => mu+

        \draw[particle] (h_2_3) -- (h_3_1); % mu- => e-
        \draw[particle] (h_2_3) -- (h_3_2); % mu- => numu
        \draw[particle] (h_2_3) -- (h_3_3); % mu- => nue
        \draw[particle] (h_2_6) -- (h_3_4); % mu+ => e+
        \draw[particle] (h_2_6) -- (h_3_5); % mu+ => numu
        \draw[particle] (h_2_6) -- (h_3_6); % mu+ => nue

        % Labels
        \coordinate[label=above left:$\pi^{0}$] (a) at (5,6);
        \coordinate[label=below right:$\pi^{-}$] (b) at (7,6);
        \coordinate[label=below left:$\pi^{+}$] (c) at (9,6);
        \coordinate[label=above right:$\chi$] (d) at (11,6);

        \coordinate[label=above left:$\gamma$] (e) at (1.5,4);
        \coordinate[label=above right:$\gamma$] (f) at (2.5,4);
        \coordinate[label=above left:$\mu^{-}$] (g) at (5.5,4);
        \coordinate[label=above right:$\Bar{\nu}_{\mu}$] (h) at (6.5,4);
        \coordinate[label=above left:$\Bar{\nu}_{\mu}$] (i) at (9.5,4);
        \coordinate[label=above right:$\mu^{+}$] (j) at (10.5,4);

        \coordinate[label=above left:$e^{-}$] (m) at (4,1.5);
        \coordinate[label=below right:$\Bar{\nu}_{\mu}$] (n) at (5,1.5);
        \coordinate[label=above right:$\Bar{\nu}_{e}$] (o) at (6,1.5);
        \coordinate[label=above left:$e^{+}$] (p) at (10,1.5);
        \coordinate[label=below right:$\Bar{\nu}_{\mu}$] (q) at (11,1.5);
        \coordinate[label=above right:$\Bar{\nu}_{e}$] (r) at (12,1.5);

        % Generation
        \draw[dashed, draw=black!50] (0,7) -- (16,7);
        \draw[dashed, draw=black!50] (0,5) -- (16,5);
        \draw[dashed, draw=black!50] (0,3) -- (16,3);
        \draw[dashed, draw=black!50] (0,0) -- (16,0);
        \coordinate[label=above right:{$g=0$}] (s) at (0,7);
        \coordinate[label=below right:{$g=1$}] (t) at (0,7);
        \coordinate[label=below right:{$g=2$}] (u) at (0,5);
        \coordinate[label=below right:{$g=3$}] (v) at (0,3);

    \end{tikzpicture}

    \captionsetup{type=figure,justification=centering,format=plain}
    \vspace{0.5cm}
    \captionof{figure}{Diagram of the Heitler model~\citep{HeitlerW.1954} of a lepton-initiated shower~(top) and a hadron-initiated shower~(bottom). The interaction/radiation lengths are not to scale, and multiple products may be created at each intersection on the diagram. The particles represented by $\chi$ may be protons, neutrons, and other atomic nuclei. See text for descriptions.}
    \label{fig:airshower diagram}
\end{figure}
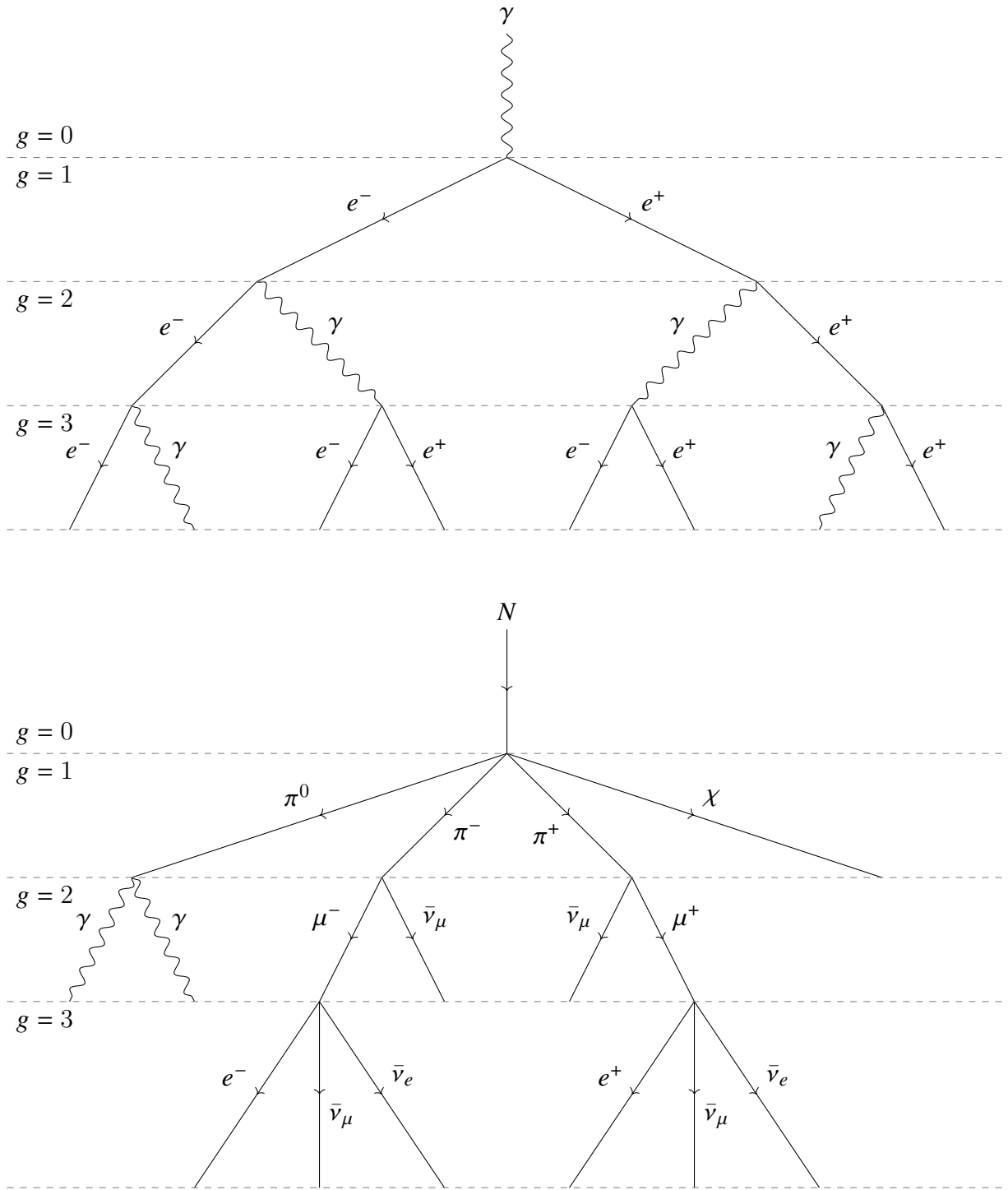

EASs are computed in generations~($g$), with the zeroth generation~($g=0$) being the initial particle, the first generation~($g=1$) being populated with the particles created from the initial interaction, the second generation~($g=2$) being all particles created by the first generation, and so on. A diagram of both leptonic and hadronic showers and how they evolve at each generation is shown in \autoref{fig:airshower diagram}.

\subsubsection{Electromagnetic Showers} \label{sssect:EM Showers}

Electromagnetic showers are initiated by either a \grayn{} or an electron striking the upper atmosphere. A cascade of \grays, electrons, and positrons travel towards the ground, creating additional particles and photons in the process. The Heitler shower model~\citep{HeitlerW.1954} assumes that the shower consists entirely of \grays{} and $e^{\pm}$ created via pair production and bremsstrahlung interactions. The equations describing the EM shower interactions are given by:

\begin{align}
    \gamma + N &\rightarrow e^{+} + e^{-} + N \label{eq:gamma shower} \\
    e^{\pm} + N &\rightarrow e^{\pm} + N + \gamma \label{eq:electron shower}
\end{align}

\noindent
where $\gamma$ represents a \grayn, $N$ represents a nucleus in Earth's atmosphere, and $e^{\pm}$ represent electrons/positrons. \autoref{eq:gamma shower} describes the \graya{} interactions~(pair production) and \autoref{eq:electron shower} describes the electron or positron interactions~(bremsstrahlung) as they travel through the atmosphere.
Although the ratio between the bremsstrahlung and pair-production interaction lengths is approximately~7/9, the Heitler model assumes that the interaction lengths are equal.
The Heitler assumption simplifies the situation, with the energy of the shower halving at each generation and each interaction process growing at the same rate. Therefore, the energy deposited by the shower as a function of the generation is given by:

\begin{align}
    E_{\mathrm{lep}}(g) &= \left( \frac{1}{2} \right)^{g} E_{0}
\end{align}

\noindent
where $E_{\mathrm{lep}}(g)$ is the energy of the leptonic shower at the $g$\textsuperscript{th} shower generation and $E_{0}$ is the initial energy of the lepton. The shower continues propagating downwards until ionisation losses with the atmosphere begin to dominate over the bremsstrahlung energy losses~\citep[$\sim$86\,MeV;][]{EngelR.2011}.

\subsubsection{Hadronic Showers}

Hadronic showers are initiated by CR hadrons,~i.e.~protons or other heavier nuclei, striking the upper atmosphere.
Hadronic showers in the TeV regime are considered a background component for \graya{} observations~(as will be discussed later in \autoref{ssect:CR background subtraction}).
A CR collision may create charged pions, neutral pions, or split the nuclei into smaller atoms. The charged pions continue to undergo further interactions with the atmospheric particles, while the neutral pions decay too quickly~\citep[$8.5 \times 10^{-17}$\,s;][]{ZylaP.2020} to interact with other nuclei. The charged pions may decay into charged muons and muon antineutrinos, and the charged muons decay into electrons, positrons, muon antineutrinos, and electron antineutrinos.
The decays of the neutral pions and charged muons create additional electromagnetic sub-showers within the hadronic shower.
Additionally, charged pions with energies $E_{\pi^{\pm}} \gtrsim 30$\,GeV are more likely to interact with the atoms and molecules in the atmosphere before decaying into charged muons~\citep{EngelR.2011}. The equations describing the hadronic shower interactions are given by:

\begin{align}
    N + N &\rightarrow \chi + n_{0}\pi^{0} + n_{1}\pi^{-} + n_{2}\pi^{+} \label{eq:pp collisions} \\
    \pi^{0} &\rightarrow \gamma + \gamma \label{eq:pi0 decay} \\
    \pi^{\pm} &\rightarrow \mu^{\pm} + \Bar{\nu}_{\mu} \label{eq:pi+- decay} \\
    \mu^{\pm} &\rightarrow e^{\pm} + \Bar{\nu}_{\mu} + \Bar{\nu}_{e} \label{eq:muon+- decay}
\end{align}

\noindent
where $\pi^{\pm}$ represents charged pions, $\pi^{0}$ represents a neutral pion, $\mu^{\pm}$ represents charged muons, $\Bar{\nu}_{\mu}$ represents a muon antineutrino, $\Bar{\nu}_{e}$ represents an electron antineutrino, and $\chi$ represents any additional particles created to conserve baryon number and electric charge~(e.g.~protons, neutrons, smaller nucleons, etc.) and create additional reaction chains within the shower. The values $n_{0}$, $n_{1}$, and $n_{2}$ are the numbers of neutral, negative, and positive pions, respectively.

Hadronic air showers quickly become a complex web of hadronic interactions, namely nucleon-nucleon~($N$--$N$), pion-nucleon~($\pi$--$N$), nucleon-nucleus~($N$--$A$), nucleus-nucleus~($A$--$A$), and pion-nucleus~($\pi^{\pm}$--$A$) interactions~\citep{GriederP.2010}. These interactions occur concurrently with the electromagnetic sub-showers from the decay of the neutral pions. At each generation, two thirds of the energy goes into the hadronic component of the shower and one third of the energy goes into the electromagnetic sub-showers. For the progenitor with initial energy $E_{0}$, the energy deposited into the hadronic and electromagnetic components is given by:

\begin{align}
    E_{\mathrm{had}}(g) &= \left( \frac{2}{3} \right)^{g} E_{0} \\
    E_{\mathrm{EM}}(g) &= \left[ 1 - \left( \frac{2}{3} \right)^{g} \right] E_{0}
\end{align}

\noindent
where $E_{\mathrm{had}}(g)$ is the energy of the hadronic shower component at the $g$\textsuperscript{th} shower generation, and $E_{\mathrm{EM}}(g)$ is the energy of the electromagnetic shower component at the $g$\textsuperscript{th} shower generation.

Due to the variety of interactions, some of which create particles with masses greater than that of an electron/positron, the hadronic showers appear more irregular than leptonic showers. Furthermore, the hadronic showers often have larger lateral extents due to the higher-mass particles being able to travel further from their interaction sites.
Hadronic showers are often discriminated apart from leptonic showers through the use of Monte~Carlo simulations of showers~\citep[e.g.~CORSIKA;][]{HeckD.1998}.

\subsection{Imaging Atmospheric Cherenkov Telescopes} \label{ssect:IACTs}

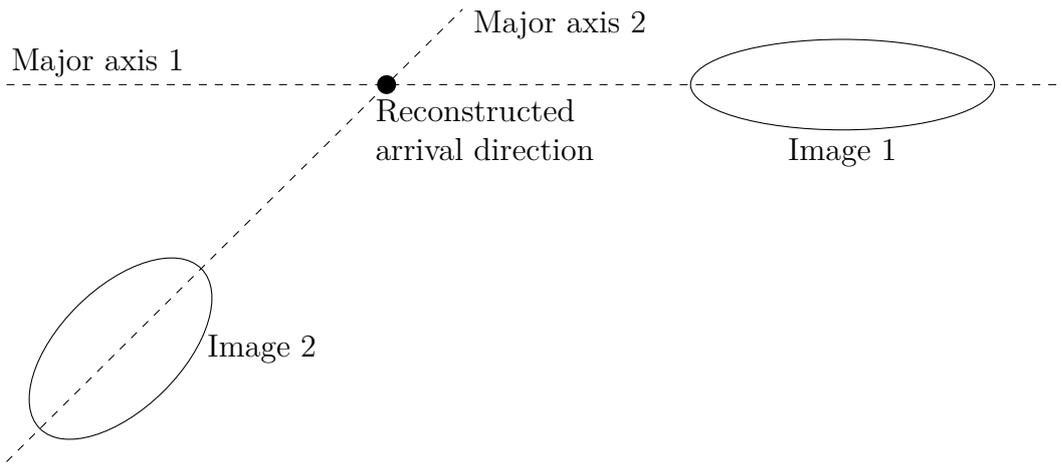
\begin{figure}
    \centering
    \begin{tikzpicture}
        
        % Draw the ellipses
        \draw[rotate around={45:(1.5,1.5)}] (1.5,1.5) ellipse (1.5cm and 0.8cm);
        \draw (11,5) ellipse (2cm and 0.6cm);

        % Draw the major axes
        \draw [dashed] (0,5) -- (14,5);
        \draw [dashed] (0,0) -- (6,6);

        % Draw the intersection point
        \node[circle,fill=black,inner sep=0pt, minimum size=0.25cm] (a) at (5,5) {};
        
        % Add text
        \node at (4.71,4.4) [anchor=west, text width=3.3cm] {Reconstructed arrival direction};
        \node at (1.2,5.3) {Major axis 1};
        \node at (6,5.8) [anchor=west, text width=4cm] {Major axis 2};
        \node at (11,4.1) {Image 1};
        \node at (2.5,1.5) [anchor=west, text width=4cm] {Image 2};
        
    \end{tikzpicture}
    \caption{A simplified diagram of a single EAS event observed by two telescopes. Each ellipse represents the image seen by an IACT, and the dashed lines represent the major axes of the ellipses. The intersection of the major axes of all the ellipses is the arrival direction of the initial \grayn.}
    \label{fig:hillas diagram}
\end{figure}

Unlike traditional telescopes that contend against atmospheric effects, IACTs focus on the light generated within the atmosphere. IACTs observe the Cherenkov light~\citep{CherenkovP.1937} emitted by charged particles travelling through the atmosphere.
This Cherenkov light is coherent and is emitted along the axis of the shower in a cone shape, where the shower axis is aligned with the progenitor's direction of travel.
When observing the shower directly along its axis, the Cherenkov cone appears as a near-perfect circle. In the much more likely case that the shower is observed off axis, the Cherenkov cone appears as an ellipse. The shower ellipse is parameterised with the `Hillas parameters'~\citep{HillasA.1985}, with the major axis of the ellipse pointing towards the shower axis. Two or more IACTs are then used to geometrically reconstruct the starting location of the shower in the sky. Viewing the shower with a larger number of IACTs provides a more accurate reconstruction of the arrival direction\footnote{See~\citet{BernlohrK.2013} for a summary of IACT observational techniques.}.
A simplified diagram showing an EAS viewed from two telescopes, along with the geometric reconstruction of the \grayn{} arrival direction, is provided in \autoref{fig:hillas diagram}.

IACTs observe EASs from both CRs and \graya{} photons. Unlike photons, which travel in straight lines and can be traced back to their origin, CRs are deflected by magnetic fields such as the Solar magnetic field and the Galactic magnetic field~(as discussed in \autoref{sect:CR diffusion}).
As the CRs below $100$\,PeV do not point to their origin, IACTs carefully select events to exclude as many CR showers as possible, with one of these methods discussed later in \autoref{ssect:CR background subtraction}.

As IACTs observe EASs within the atmosphere, their `effective' area is equal to the area of the Cherenkov light pool on the ground\footnote{In contrast, the area of a traditional optical telescope is equal to the area of the lens or primary mirror.}. The effective area of an IACT varies with the energy of the particle or photon that initiated the shower and various other observational parameters~(such as the zenith angle, etc.). For \hess{} the effective area can be on the order of~0.2\,km$^{2}$~\citep[$2 \times 10^{5}$\,m$^{2}$;][]{AharonianF.2006a}, and the future next-generation \graya{} observatory, the Cherenkov telescope array~(CTA), will reach an effective area on the order of~1\,km$^{2}$~\citep{BernlohrK.2013}\footnote{For further comparison, the detection areas of optical telescopes are generally on the order of~500\,cm$^{2}$. The largest optical telescopes currently in operation have detection areas on the order of~100\,m$^{2}$~\citep[e.g.~the large binocular telescope;][]{HillJ.2010}.}. These massive effective areas are integral for \graya{} astronomy, as for even the brightest sources the \graya{} flux is approximately~10\,photons\,km$^{-2}$\,s$^{-1}$ in the TeV regime~\citep[Crab flux;][]{AbdallaH.2018a}.

Examples of IACTs include \hess{} in the Khomas~Region~(Namibia), the major atmospheric gamma imaging Cherenkov telescopes~\citep[MAGIC;][]{BaixerasC.2003} on the Spanish island La~Palma, and the very energetic radiation imaging telescope array system~\citep[VERITAS;][]{WeekesT.2002} in Arizona~(United States of America). CTA sites are currently being constructed in both La~Palma~(Spain) and Paranal~(Chile).
\section{The \hess{} Galactic Plane Survey (HGPS)} \label{sect:HGPS}

The \hess{} Galactic plane survey~\citep[HGPS;][]{AharonianF.2005b,AharonianF.2006b,AbdallaH.2018a} was a decade-long observation program, with data being collected from~2004~to~2013.
The HGPS is constructed from many 28-minute-long observations, hereafter known as runs, with a total observation time of~2673~hours.
The observations span the Galactic latitudes $|\mathscr{b}| \leq 3^{\circ}$ and Galactic longitudes $65^{\circ} \leq \mathscr{l} \leq 250^{\circ}$, and has a resolution of $4.8^{\prime}$. The HGPS is the highest resolution and the most sensitive TeV \graya{} survey of the Galactic plane to date. The HGPS has a variable sensitivity as low as $0.4\%$ of the Crab nebula~(4\,mCrab)\footnote{Units of \%Crab are commonly used throughout VHE \graya{} astronomy, with the value used in this thesis given by $J_{\mathrm{Crab}} (E \ge 1\,\mathrm{TeV})=2.26\times10^{-11}\,\mathrm{cm}^{-2}\,\mathrm{s}^{-1}$~\citep{AharonianF.2006a} unless otherwise stated.} for the Galactic centre~(GC), worsening to~20\,mCrab at the edges of the survey where there was a lower exposure time.

\begin{figure}
    \centering
    \includegraphics[width=0.925\textheight, angle=90]{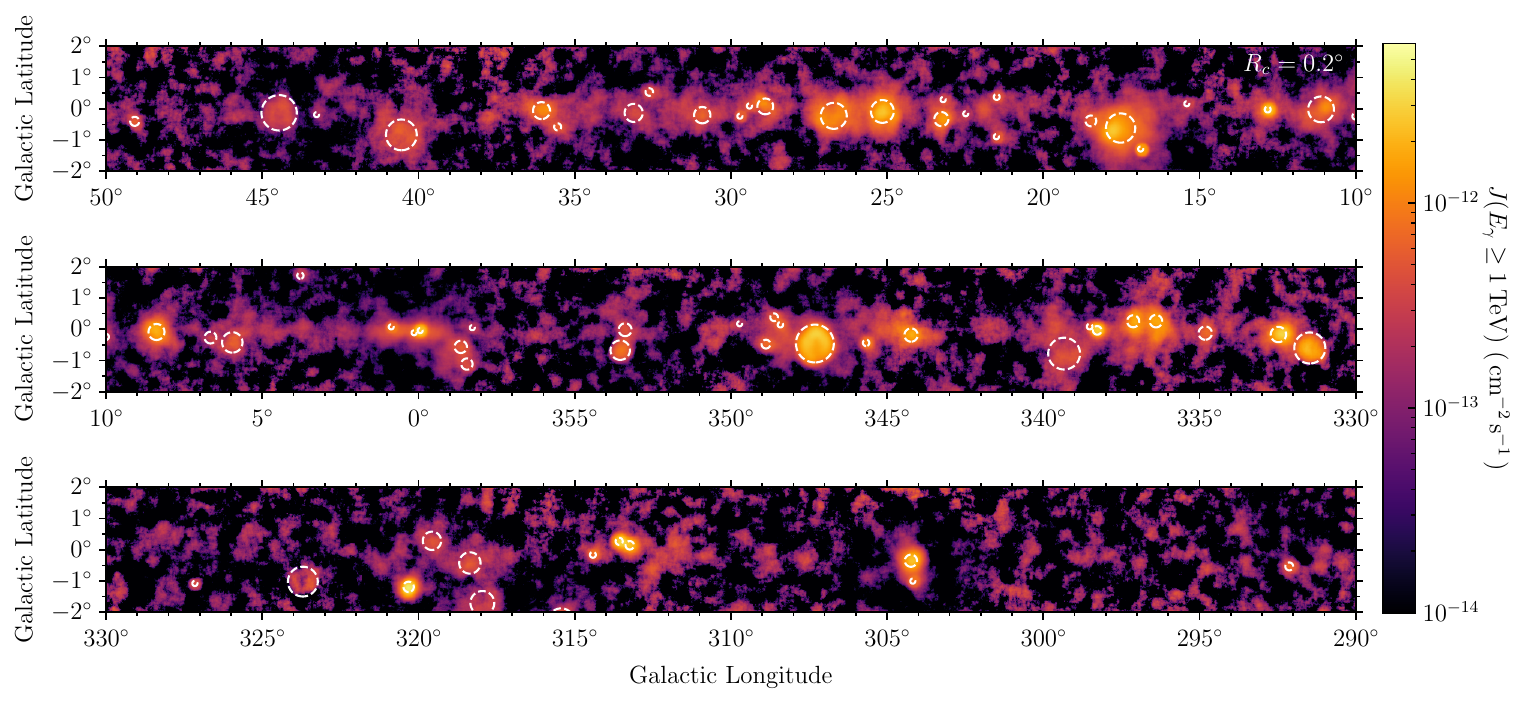}
    \caption{HGPS flux map integrated for energies $E_{\gamma} \geq 1$\,TeV for Galactic longitudes $\mathscr{l}=290^{\circ}$ to $\mathscr{l}=50^{\circ}$ and Galactic latitudes $|\mathscr{b}| \leq 2^{\circ}$ for the $R_{c}=0.2^{\circ}$ map~(described in \autoref{ssect:beam size}). The catalogued \graya{} sources are shown by the white dashed circles.}
    \label{fig:hgps}
\end{figure}

The HGPS \graya{} flux map\footnote{The HGPS~\citep{AbdallaH.2018a} \graya{} source catalogue, \graya{} flux, sensitivity, significance, \graya{} flux uncertainty, and \graya{} flux upper-limit data maps are available at \url{https://www.mpi-hd.mpg.de/hfm/HESS/hgps/}.} for Galactic longitudes $\mathscr{l}=290^{\circ}$ to $\mathscr{l}=50^{\circ}$ and Galactic latitudes $|\mathscr{b}| \leq 2^{\circ}$ is shown in \autoref{fig:hgps}, with catalogued \graya{} sources shown by the white circles.

\subsection{Gamma-Ray Flux} \label{ssect:gray flux}

IACTs, and other types of VHE instruments, record EASs from both CRs and \grays{} as `events'. The event list contains all properties of the observed air showers, such as the reconstructed arrival position, the reconstructed energy of the initial particle or photon, etc..
The number of observed EASs are then converted to a \graya{} flux.
For the HGPS two regions are defined, known as the \textit{on} and \textit{off} regions.
These two regions will be discussed later in Sections \ref{ssect:beam size} and \ref{ssect:CR background subtraction}, respectively.
The signal~(or `excess') number of \graya{} events, $N_{\gamma}$, is given by:

\begin{align}
    N_{\gamma} &= N_{\mathrm{on}} - \mathscr{a} N_{\mathrm{off}} \label{eq:N_gamma}
\end{align}

\noindent
where $N_{\mathrm{on}}$ and $N_{\mathrm{off}}$ are the number of events in the \textit{on} region and \textit{off} region, respectively, and $\mathscr{a}$ is the ratio of the exposure in each region. From the number of excess counts, the \graya{} flux in the HGPS was calculated via the equation:

\begin{align}
    F &= \frac{N_{\gamma}}{N_{\mathrm{exp}}} \int_{E_{1}}^{E_{2}} J_{\mathrm{ref}}(E_{\gamma})\,\mathrm{d}E_{\gamma} \label{eq:HGPS flux formula}
\end{align}

\noindent
where $F$ is the integral flux between the \graya{} energies $E_{1}$ and $E_{2}$, $E_{\gamma}$ is the \graya{} energy, $J_{\mathrm{ref}}(E_{\gamma})$ is the reference/assumed functional form of the \graya{} flux~(see below), and $N_{\mathrm{exp}}$ is the expected/predicted number of \graya{} counts. The expected number of \graya{} counts is given by:

\begin{align}
    N_{\mathrm{exp}} &= \sum_{R} t_{R} \int_{E_{\mathrm{min}}}^{\infty} J_{\mathrm{ref}}(E_{r}) A_{\mathrm{eff}}(E_{r}, \mathscr{q}_{R}) \, \mathrm{d}E_{r} \label{eq:N_exp}
\end{align}

\noindent
where $E_{r}$ is the reconstructed energy of the \grayn, the sum is performed over all observation runs~($R$) for a given source, $t_{R}$ is the live time of the given run, and $E_{\mathrm{min}}$ is the minimum threshold energy for an observation~(i.e.~the lowest \graya{} energy that could be observed in the run). $\mathscr{q}_{R}$ represents all observational parameters for the given run, including values such the zenith angle, which telescopes were included, and which cuts were applied to the data. $A_{\mathrm{eff}}(E_{r}, \mathscr{q}_{R})$ is the effective area of \hess{} for a given observation run as a function of the reconstructed energy and the observational parameters and is obtained through Monte~Carlo simulations.

The reference differential \graya{} source flux~($J_{\mathrm{ref}}$) is the assumed functional form of the \graya{} flux, and is used in calculating both the expected number of \graya{} counts~($N_{\mathrm{exp}}$) and the integrated \graya{} flux. The reference flux is taken as a power law and is given by:

\begin{align}
    J_{\mathrm{ref}}(E_{\gamma}) = J_{\gamma} \left( \frac{E_{\gamma}}{E_{0}} \right)^{-\eta} \label{eq:reference spectrum}
\end{align}

\noindent
where $J_{\gamma}$ is a reference normalisation differential flux, $E_{0}$ is the \graya{} energy at which the reference flux is normalised, and $\eta$ is a spectral index.
The integrated \graya{} flux given in \autoref{eq:HGPS flux formula} does not depend on $J_{\gamma}$~(which can be found by substituting in Equations \ref{eq:N_exp} and \ref{eq:reference spectrum}); therefore, the normalisation constant can be chosen arbitrarily.
The HGPS integrated \graya{} flux maps use a normalisation energy given by $E_{0}=1$\,TeV, and the \graya{} spectral index $\eta=2.3$ is taken as the average for the \graya{} sources in the Galactic plane.
The systematic uncertainty of $\eta$ is $\pm 0.2$ -- \citet{AbdallaH.2018a} found that when varying $\eta$ in this range the integrated flux varied by less than~5\%.

\subsection{Integration Beam} \label{ssect:beam size}

\begin{figure}
    \centering
    \includegraphics[width=0.905\textheight, angle=90]{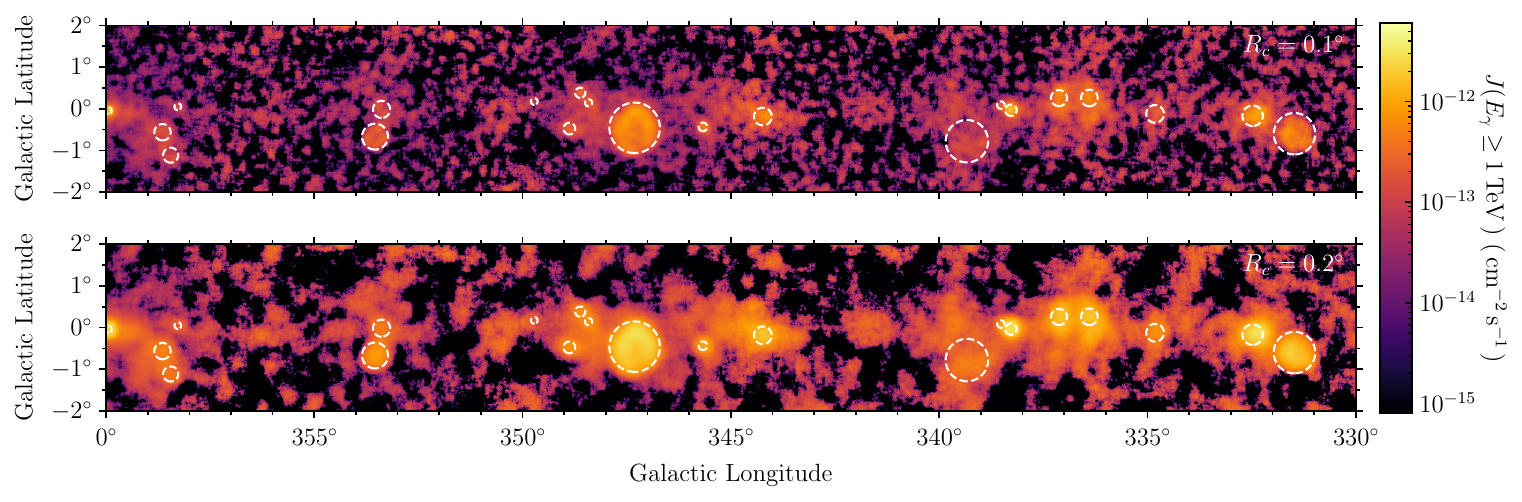}
    \caption{HGPS flux map for Galactic longitudes $\mathscr{l}=330^{\circ}$ to $\mathscr{l}=0^{\circ}$ and Galactic latitudes $|\mathscr{b}| \leq 2^{\circ}$ for the $R_{c}=0.1^{\circ}$ map~(top) and the $R_{c}=0.2^{\circ}$ map~(bottom). The catalogued \graya{} sources are shown by the white dashed circles, and both maps are shown with an identical colour scale.}
    \label{fig:hgps beams image}
\end{figure}

The number of excess counts~(\autoref{eq:N_gamma}) requires integrating the observed EAS events within an \textit{on}~region.
For the HGPS the \textit{on}~region is defined a circle with a constant radius, $R_{c}$, which is referred to as either the containment radius or the integration radius. The \textit{on}~region radius must be larger than the point spread function~(PSF) of the telescope~(i.e.~larger than~0.08$^{\circ}$ for \hess). The number of counts in a given region is the integration of all counts within said region, divided by the solid angle of the region. Hence, the number of EAS counts in the \textit{on}~region, $N_{\mathrm{on}}$, is given by:

\begin{align}
    N_{\mathrm{on}} &= \frac{1}{2\pi(1-\cos (R_{c}))} \int_{R_{c}} N_{\mathrm{events}} \,\mathrm{d}\Omega  \label{eq:on-region flux}
\end{align}

\noindent
where $\Omega$ is the solid angle of the \textit{on}~region with radius $R_{c}$ and $N_{\mathrm{events}}$ is the total number of events.
Utilising an \textit{on}~region that is larger than the PSF increases the significance of the observation, especially on scales of the same size as the integration radius. However, it reduces the ability to distinguish structure on scales smaller than $R_{c}$.
The HGPS flux maps uses integration radii of either $R_{c}=0.1^{\circ}$ or $R_{c}=0.2^{\circ}$, and so is well suited to resolve \graya{} structures that are approximately $0.1^{\circ}$ or $0.2^{\circ}$ in radius. A comparison between the $R_{c}=0.1^{\circ}$ and the $R_{c}=0.2^{\circ}$ \graya{} flux maps can be seen in \autoref{fig:hgps beams image}.

\subsection{CR Background Estimation} \label{ssect:CR background subtraction}

CR air showers, which typically outnumber \graya{} air showers by a factor of a thousand, are considered as background events for \graya{} observatories.
The CR EASs are suppressed by applying machine-learning algorithms to remove CR-like events~\citep[see][with the HGPS specifically using the analysis cuts in their Table~2a]{OhmS.2009}.

Some contamination due to the CR EASs will remain after the initial cuts are applied.
An estimate of the contamination due to CR air showers in the \textit{on}~region is calculated by defining an \textit{off}~region.
Ideally, the \textit{off}~region includes only CR EASs and not \graya{} showers. If this condition is satisfied, then the number of counts given in \autoref{eq:N_gamma} is due purely to \graya{} EASs.
If \graya{} emission is included in the \textit{off}~region, then the CR background in the \textit{off}~region will be over-estimated.

\hess{} employs multiple background estimation techniques, such as the `standard' ring method or the `reflected' ring method~\citep{BergeD.2007}. These background methods can introduce their own systematic uncertainties~\citep{AharonianF.2006a,BergeD.2007}, and are chosen based on the situation to reduce the systematics. However, neither of these two methods are suitable for the HGPS due to the large amount of statistically significant \graya{} emission in the Galactic plane.

To prevent statistically-significant \graya{} emission from being included in any background estimates, \citet{AbdallaH.2018a} defined `exclusion regions' where no background estimate would be taken.
The exclusion regions are defined from the statistical significance of an observation, $S$, where the HGPS uses the Li~and~Ma equation for the statistical significance~\citep{LiT.1983} given by:

\begin{align}
    S &= \sqrt{2 N_{\mathrm{on}} \ln \left[ \left( \frac{1 + \epsilon}{\epsilon} \right) \left( \frac{N_{\mathrm{on}}}{N_{\mathrm{on}} + N_{\mathrm{off}}} \right) \right] + 2 N_{\mathrm{off}} \ln \left[ (1 + \epsilon) \left( \frac{N_{\mathrm{off}}}{N_{\mathrm{on}} + N_{\mathrm{off}}} \right) \right] } \label{eq:Li and Ma}
\end{align}

\noindent
where $\epsilon$ is the ratio of time spent taking \textit{on}-region observations~($t_{\mathrm{on}}$) to the time spent taking \textit{off}-region observations~($t_{\mathrm{off}}$),~i.e.~$\epsilon = t_{\mathrm{on}} / t_{\mathrm{off}}$~\citep{LiT.1983}. \autoref{eq:Li and Ma} is utilised in the HGPS to create a significance map, where the \textit{off}~regions were defined as a ring with an inner radius of $0.7^{\circ}$ and a thickness of $0.44^{\circ}$. To avoid large systematic effects near the edges of the FoV, the only events included in the on and \textit{off}~regions are those reconstructed within a $2^{\circ}$ radius of the centre of the field of view~(FoV, which has a full radius of $5^{\circ}$). The $2^{\circ}$-radius FoV is referred to as the reduced FoV. This process identifies regions with a significance of $5\sigma$, which are then expanded by an additional $0.3^{\circ}$ beyond the $5\sigma$ contour.

When using the above ring-background method to define the exclusion regions, it was found that statistically significant emission was still included in the \textit{off}~regions~\citep{AbdallaH.2018a}.
Hence, the significance map is recreated iteratively. In each iteration the \textit{off}~regions cannot include emission in the previously calculated exclusion regions. Each iteration defines a new set of exclusion regions and is repeated until all of the exclusion regions are stable.

\begin{figure}
    \centering
    \includegraphics[width=\textwidth]{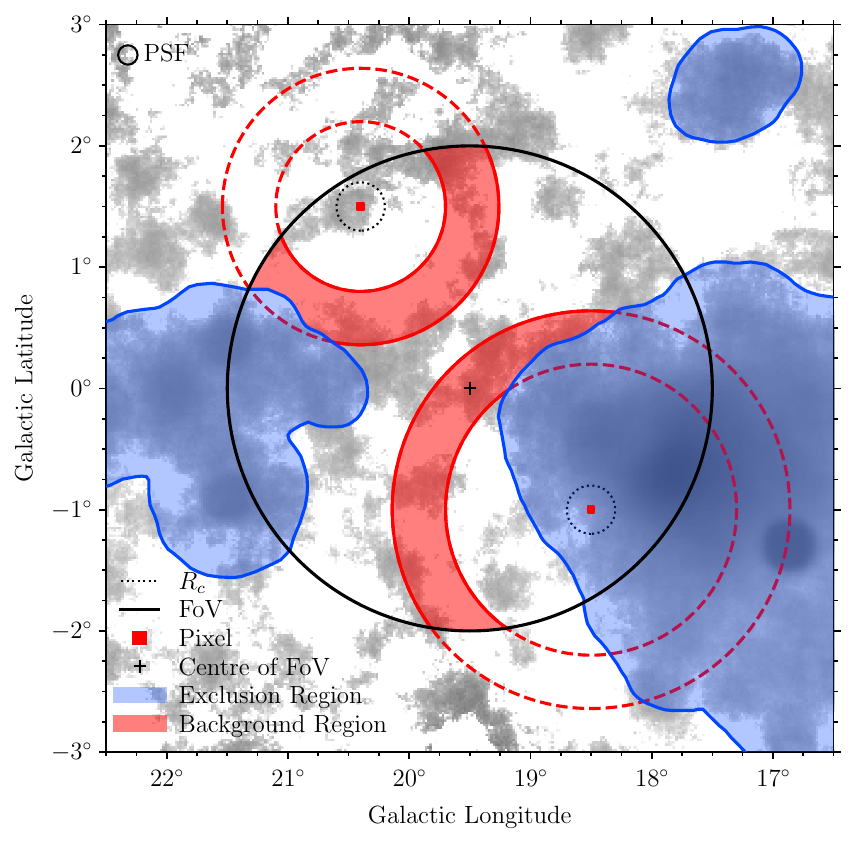}
    \caption{A demonstration of the adaptive ring background subtraction method, shown for an arbitrary location in the HGPS~\citep{AbdallaH.2018a}. The HGPS \graya{} flux is shown in reverse greyscale, with the PSF of \hess{} in the top-left corner. The reduced FoV has a radius of $2^{\circ}$ and is shown by the black circle, with the black cross representing the centre of the FoV.
    Two pixels were chosen arbitrarily and are shown by the red squares.
    The \graya{} flux for each pixel is integrated from the \textit{on}~regions shown by the dotted black circles. The background \graya{} flux is integrated from the \textit{off}~regions shown by the red rings.
    The background estimate is not taken from the exclusion regions~(shown in blue). The \textit{off}-region ring can be adaptively enlarged from an inner radius of $0.44^{\circ}$ to a maximum inner radius of $1.7^{\circ}$ to limit the statistical uncertainties, as demonstrated by the pixel highlighted in the lower-right. The process for calculating the exclusion regions and the radii of the rings is detailed in the text in \autoref{ssect:CR background subtraction}.}
    \label{fig:ring background}
\end{figure}

For the final flux map the HGPS uses an `adaptive' ring background. Much like for the construction of the exclusion regions, only events reconstructed within the central $2^{\circ}$ of the FoV are included. For a given pixel position, a ring is drawn with a constant thickness of $0.44^{\circ}$ and a minimum inner radius of $0.7^{\circ}$. If a large portion of the ring overlaps with an exclusion region then the inner radius is increased such that the acceptance integrated within the ring is greater than four times the acceptance at the given pixel position\footnote{The acceptance map is the expected number of observed CR events across all observations. The acceptance map is normalised such that outside the exclusion regions the expected number of counts is equal to the observed counts.}. The outer ring radius is limited to a maximum of $1.7^{\circ}$ to limit systematics associated with the outer region of the FoV. The number of EAS counts in the \textit{off}~region is given by:

\begin{align}
    N_{\mathrm{off}} &= \frac{1}{\Omega_{\mathrm{off}}} \int_{\mathrm{off}} N_{\mathrm{events}} \, \mathrm{d}\Omega
\end{align}

\noindent
where $\Omega_{\mathrm{off}}$ is the solid angle of the \textit{off}~region~(i.e.~background region). From \autoref{eq:N_gamma}, the number of counts for a given pixel is then given by:

\begin{equation}
    \begin{aligned}
        N_{\gamma} &= N_{\mathrm{on}} - \mathscr{a} N_{\mathrm{off}} \\
        &= \left( \frac{1}{2\pi(1-\cos (R_{c}))} \int_{R_{c}} N_{\mathrm{events}} \,\mathrm{d}\Omega \right) - \left( \frac{\mathscr{a}}{\Omega_{\mathrm{off}}} \int_{\mathrm{off}} N_{\mathrm{events}} \, \mathrm{d}\Omega \right) \label{eq:pixel flux}
    \end{aligned}
\end{equation}

\noindent
where all parameters have been defined previously, and the number of counts measured in the \textit{on}~region is given by \autoref{eq:on-region flux}. A demonstration of this method being applied to two arbitrarily-chosen pixels can be seen in \autoref{fig:ring background}.

\subsection{Flux Sensitivity} \label{ssect:hess sensitivity}

The flux sensitivity is defined as the minimum flux~($F_{\mathrm{min}}$) needed for a source fully contained within the \textit{on}~region to be detected above the CR background at the $5\sigma$ level. The flux sensitivity of the HGPS is calculated by taking \autoref{eq:HGPS flux formula} and substituting $N_{\gamma}$ with $\hat{N}_{\gamma}$, where $\hat{N}_{\gamma}$ is defined as the number of \graya{} events required for a $5\sigma$ detection above the background. The value of $\hat{N}_{\gamma}$ is determined by numerically solving \autoref{eq:Li and Ma} for $N_{\mathrm{on}}$,~i.e.~$\hat{N}_{\gamma}$ is found such that the condition $S(N_{\mathrm{on}}=\hat{N}_{\gamma})=5$ is satisfied for the number of observed background counts~($N_{\mathrm{off}}$).

\begin{figure}[t]
    \centering
    \includegraphics[width=\textwidth]{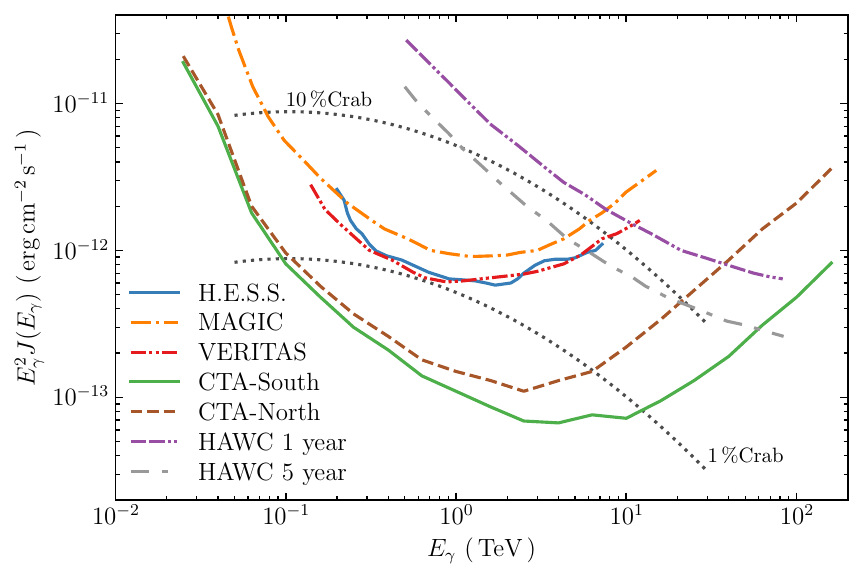}
    \caption{The~5$\sigma$ sensitivity for various \graya{} facilities. Sensitivities are shown for a~50-hour observation, unless stated otherwise. The sensitivity curves are from the following: \hess~\citep{HollerM.2015}, MAGIC~\citep{AleksicJ.2016}, VERITAS~\citep{HoranD.2007}, CTA-South/CTA-North~\citep{AcharyaB.2018}, and HAWC~\citep{AbeysekaraA.2017a}. The Crab flux~(dotted black line) has been calculated as a log-parabola following~\citet{AleksicJ.2015}.}
    \label{fig:iact sensitivities}
\end{figure}

The sensitivity curves for \hess, and many other \graya{} facilities, are shown in \autoref{fig:iact sensitivities}, where any flux above the sensitivity curves is considered to be a confident detection. The sensitivity curves are shown for points sources that are completely contained within the PSF of the observatory. The sensitivity of \hess{} and other IACTs can be simplified to a relation that depends on the radius of the source, $\sigma_{\mathrm{source}}$, and is given by:

\begin{align}
    F_{\mathrm{min}}(\sigma_{\mathrm{source}}) &\propto \sqrt{\dfrac{\sigma_{\mathrm{source}}^{2} + \sigma_{\mathrm{PSF}}^{2}}{\sigma_{\mathrm{PSF}}^{2}}} \label{eq:source size sensitivity}
\end{align}

\noindent
where $\sigma_{\mathrm{PSF}}=0.08^{\circ}$ is the PSF of \hess. For sources much smaller than the PSF~(i.e.~for point sources) the flux sensitivity is constant. For extended sources~(i.e.~for sources larger than the PSF) the flux sensitivity worsens linearly with source size~\citep{HintonJ.2009}.
Additionally, the flux sensitivity scales with the square-root of the livetime of an observation via the relationship given by:

\begin{align}
    F_{\mathrm{min}}(t_{\mathrm{on}}) &\propto \dfrac{1}{\sqrt{t_{\mathrm{on}}}} \label{eq:time sensitivity}
\end{align}

\subsection{\hess{} Gamma-Ray Source Horizon in the MW} \label{ssect:horizon}

\begin{figure}
    \centering
    \includegraphics[width=\textwidth]{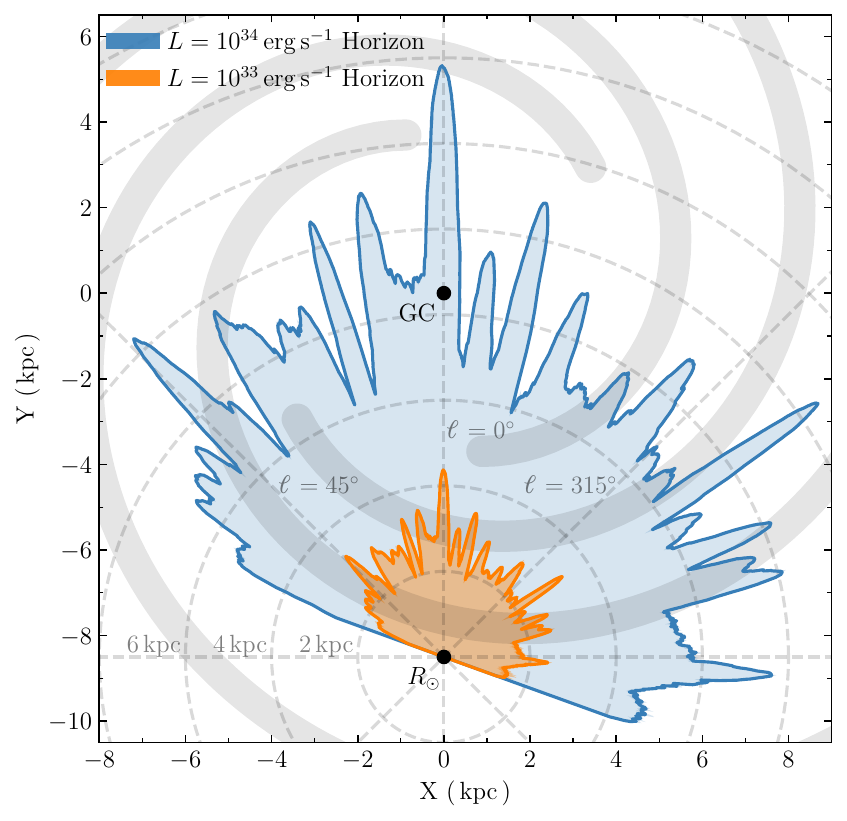}
    \caption{A top-down view of the MW showing the $5\sigma$ point-source detection horizon for the HGPS for \graya{} luminosities $10^{34}$\,erg\,s$^{-1}$~($\sim$100\%\,Crab;~blue) and $10^{33}$\,erg\,s$^{-1}$~($\sim$10\%\,Crab;~orange). The GC is located at the origin, with the Solar location shown by $R_{\odot}=8.5$\,kpc from the GC~\citep{KerrF.1986}. Galactic longitude and the radial distance from Earth are shown by the dashed grey lines. The solid grey lines represent the~\citet{RobitailleT.2012} spiral arms, which are discussed in more depth in \autoref{ssect:R12 Model}.}
    \label{fig:hgps horizon}
\end{figure}

The sensitivity can be utilised to calculate the astronomical horizon for a survey~($r_{\mathrm{max}}$). From the flux-luminosity relationship, the maximum distance is given by:

\begin{align}
    F_{\mathrm{min}} &= \frac{ L }{ 4 \pi r_{\mathrm{max}}^{2} } \\
    \therefore r_{\mathrm{max}} &= \sqrt{ \frac{ L }{ 4 \pi F_{\mathrm{min}} } }
\end{align}

\noindent
where $L$ is some \graya{} source luminosity. Taking the Crab nebula and using the flux defined earlier~\citep[$J_{\mathrm{Crab}}(E \geq 1\,\mathrm{TeV})=2.26\times10^{-11}$\,cm$^{-2}$\,s$^{-1}$;][]{AharonianF.2006a} at a distance of~2\,kpc~\citep{KaplanD.2008}, the luminosity of the Crab nebula is approximately $10^{34}$\,erg\,s$^{-1}$. \autoref{fig:hgps horizon} shows the \graya{} point-source source horizon for \graya{} sources with $L\approx100$\%\,Crab, which varies between~5--14\,kpc depending on the exposure along a given direction. For \graya{} sources with $L\approx10$\%\,Crab, the \graya{} point-source horizon decreases to between~1--4\,kpc.
As the \graya{} source horizon does not extend to the far side of the MW, estimates of the number of undetected \graya{} sources in the HGPS range from hundreds to thousands~\citep[e.g.~see][]{ActisM.2011,AcharyaB.2018}.

The horizon calculation above does not account for the angular size of the source. As shown in \autoref{eq:source size sensitivity}, the sensitivity of \hess{} worsens linearly with source size for extended \graya{} sources.
As~61 of the~78 catalogued \graya{} sources are extended, the depth of the HGPS is over-estimated in \autoref{fig:hgps horizon} for most known \graya{} sources.
\section{HGPS Source Characterisation and Classification}

CRs interact with the ISM to create \grays, with the various processes detailed in \autoref{sect:gamma-ray production}.
The CR flux is highest nearby to their sites of acceleration. Hence, the regions in close proximity to the CR accelerators will emit additional \grays. The regions of enhanced \graya{} emission are known as \graya{} sources.

The HGPS catalogued a total of~78~\graya{} sources along the Galactic plane.
However, many of the catalogued \graya{} sources are not located nearby to any known CR accelerator. This section will detail how the catalogued \graya{} sources in the HGPS are characterised and classified.

\subsection{HGPS Source Characterisation} \label{ssect:source characterisation}

The spatial profiles of most \graya{} sources in the HGPS can be modelled as one or more Gaussians. For a source at some Galactic coordinates~($\mathscr{l}_{0},\,\mathscr{b}_{0}$), the surface brightness of the source is modelled as:

\begin{align}
    S_{\mathrm{source}}(r,\,F_{\mathrm{source}},\,\sigma_{\mathrm{ang}}) &= F_{\mathrm{source}} \frac{1}{2 \pi \sigma_{\mathrm{ang}}^{2}} \exp \left( - \frac{r^{2}}{2 \sigma_{\mathrm{ang}}^{2}} \right) \label{eq:source subtraction} \\
    r &= \sqrt{ ( \mathscr{l} - \mathscr{l}_{0} )^{2} + ( \mathscr{b} - \mathscr{b}_{0} )^{2} } \label{eq:source radius}
\end{align}

\noindent
where $S_{\mathrm{source}}$ represents the surface brightness of the source, $F_{\mathrm{source}}$ is the totally spatially integrated flux of the source, and $r$ is the distance from the centre of the source. The angular size of the source region, $\sigma_{\mathrm{ang}}$, is calculated from the source radius~($\sigma_{\mathrm{source}}$) and the PSF of \hess~($\sigma_{\mathrm{PSF}} = 0.08^{\circ}$) and is given by:

\begin{align}
    \sigma_{\mathrm{ang}} &= \sqrt{ \sigma_{\mathrm{source}}^{2} + \sigma_{\mathrm{PSF}}^2 } \label{eq:total angular size}
\end{align}

\noindent
where all parameters have been defined previously.
For all point-like sources the radius is given by $\sigma_{\mathrm{source}}=0^{\circ}$; hence, the angular size of the source region is given by $\sigma_{\mathrm{ang}}=\sigma_{\mathrm{PSF}}$.
The modelling parameters for all~78~sources in the HGPS, as calculated by~\citet{AbdallaH.2018a}, can be found in \autoref{tab:all sources}.

Ten sources found in the HGPS are morphologically complex. Despite being characterised with a position, radius, and surface brightness, these ten sources cannot easily be modelled as a collection of Gaussians. The ten morphologically-complex \graya{} sources are listed as a `shell' spatial model in \autoref{tab:all sources}, or are a component of the GC region, and will be discussed in more depth in \autoref{sect:source cut}.

\subsection{HGPS Galactic TeV Source Classifications} \label{ssect:source classifications}

\begin{table}
    \centering
    \bgroup
    \def\arraystretch{1.25}
    \begin{tabular}{lr}
        \hline
        Classification        & Number of Sources \\ \hline
        Binary                &  3 \\
        Stellar Cluster       &  4 \\
        SNR                   &  8 \\
        Composite             &  8 \\
        Not Associated        & 11 \\
        PWN                   & 12 \\
        Not Firmly Identified & 32 \\ \hline
        Total                 & 78 \\
        \hline
    \end{tabular}
    \egroup
    \caption{A summary of the source classifications of the catalogued \graya{} sources observed in the HGPS~\citep{AbdallaH.2018a}. For details on the different source classifications, see text.}
    \label{tab:source types}
\end{table}

The~78~TeV \graya{} sources observed in the HGPS were originally classified into six categories, which are defined by the class of object responsible for the acceleration of the CRs. The original categories were titled: `supernova remnant~(SNR)', `pulsar wind nebula~(PWN)', `composite', `binary', `not firmly identified', and `not associated'. Four of the TeV \graya{} sources in the HGPS were later reclassified into a seventh category titled `stellar cluster'.
While \hess{} has recently observed the recurrent nova RS~Ophiuchi~\citep{AharonianF.2022a}, no TeV recurrent nova sources are present in the HGPS data. Additionally, while \hess{} is able to observe extra-Galactic \graya{} sources, no such sources are found in the HGPS data.
The seven \graya{} source categories represented in the HGPS are briefly discussed below, with the number of sources in each classification summarised in \autoref{tab:source types}.

\subsubsection{Supernova Remnants}

When a star undergoes a supernova explosion, a large amount of energy~($\sim$10$^{53}$\,erg) is released into the surrounding medium.
Of the total energy,~99\% is transferred into the creation of neutrinos. The remaining $10^{51}$\,erg goes into the kinetic energy of the expanding shock front~\citep{HillasA.2005}. The large shock is called a supernova remnant. CRs are accelerated within the SNR shock via a process known as diffusive shock acceleration~\citep[sometimes called first-order Fermi acceleration;][]{FermiE.1949,AharonianF.2004a}. Due to the large amount of \grays{} being produced within the SNR shock, and due the CRs escaping and creating \grays{} in the nearby surrounding ISM, SNRs can have complex morphologies that are difficult to model. The eight SNR \graya{} sources identified in the HGPS are given in \autoref{tab:SNR sources}.

\begin{table}
    \centering
    \bgroup
    \def\arraystretch{1.25}
    \begin{tabular}{lll}
        \hline
        Source Name & Accelerator & Classification Reference \\ \hline
        HESS~J0852--463 & Vela~Junior      & \citet{AharonianF.2005e}  \\
        HESS~J1442--624 & RCW~86           & \citet{AbramowskiA.2018}  \\
        HESS~J1534--571 & G323.7001.0      & \citet{AbdallaH.2018c}    \\
        HESS~J1713--397 & RX~J1713.7--3946 & \citet{AharonianF.2004a}  \\
        HESS~J1718--374 & G349.7+0.2       & \citet{AbramowskiA.2015b} \\
        HESS~J1731--347 & G353.6--0.7      & \citet{AbramowksiA.2011a} \\
        HESS~J1801--233 & W~28             & \citet{AharonianF.2008d}  \\
        HESS~J1911+090  & W~49B            & \citet{AbdallaH.2018d}    \\
        \hline
    \end{tabular}
    \egroup
    \caption{The eight SNR \graya{} sources in the HGPS, along with the name of the SNR that is accelerating the CRs and the reference that classified the source.}
    \label{tab:SNR sources}
\end{table}

\subsubsection{Pulsar Wind Nebulae}

Pulsars are created when a large star undergoes a supernova explosion.
The most likely outcome from the creation of the pulsar is that the rotational axis and magnetic axis will be misaligned. The misaligned, rotating magnetic field creates an electric field via Faraday's law~\citep{FaradayM.1832}, which in turn creates a drag force that slows the rotation of the pulsar.
Therefore, the pulsar's rotational energy is transferred into the electromagnetic fields, which then accelerate CRs. This process creates an energetic wind of CRs~\citep{GaenslerB.2006}, especially electrons, which interact with the ISM to create a nebula that emits \grays~(see \autoref{sect:gamma-ray production}). The~12~PWN \graya{} sources identified in the HGPS are given in \autoref{tab:PWN sources}.

\begin{table}
    \centering
    \bgroup
    \def\arraystretch{1.25}
    \begin{tabular}{lll}
        \hline
        Source Name & Accelerator & Classification Reference \\ \hline
        HESS~J0835--455 & Vela~X        & \citet{AharonianF.2006c}  \\
        HESS~J1303--631 & G304.10--0.24 & \citet{AbramowskiA.2012}  \\
        HESS~J1356--645 & G309.92--2.51 & \citet{AbramowskiA.2011b} \\
        HESS~J1418--609 & G313.32+0.13  & \citet{AharonianF.2006f}  \\
        HESS~J1420--607 & G313.54+0.23  & \citet{AharonianF.2006f}  \\
        HESS~J1514--591 & MSH~15--52    & \citet{AharonianF.2005d}  \\
        HESS~J1554--550 & G327.15--1.04 & \citet{AbdallaH.2018a}    \\
        HESS~J1747--281 & G0.87+0.08    & \citet{AharonianF.2005a}  \\
        HESS~J1818--154 & G15.4+0.1     & \citet{AbramowskiA.2014c} \\
        HESS~J1825--137 & G18.00--0.69  & \citet{AharonianF.2006g}  \\
        HESS~J1837--069 & G25.24--0.19  & \citet{MarandonV.2008}    \\
        HESS~J1849--000 & G32.64+0.53   & \citet{AbdallaH.2018a}    \\
        \hline
    \end{tabular}
    \egroup
    \caption{The twelve PWN \graya{} sources in the HGPS, along with the name of the pulsar that is accelerating the CRs and the reference that classified the source.}
    \label{tab:PWN sources}
\end{table}

\subsubsection{Composite Sources}

When a pulsar is created during a supernova it will be given some `kick' velocity. Depending on the magnitude of the kick velocity, the pulsar and surrounding PWN will remain inside the SNR shell for approximately~40\,kyr~\citep{GaenslerB.2006}. After the pulsar escapes the SNR the two accelerators can be easily differentiated. However, while the pulsar is within the SNR it can be difficult to distinguish which object is the dominant CR accelerator. Sources where the pulsar are within the SNR shell are classified as composite \graya{} sources. The eight composite \graya{} sources identified in the HGPS are given in \autoref{tab:composite sources}.

\begin{table}
    \centering
    \bgroup
    \def\arraystretch{1.25}
    \begin{tabular}{lll}
        \hline
        Source Name & Accelerator & Classification Reference(s) \\ \hline
        HESS~J1119--614 & G292.2--0.5 & \citet{AbdallaH.2018a}                            \\
        HESS~J1640--465 & G338.3--0.0 & \citet{AbramowskiA.2014a}, \citet{GotthelfE.2014} \\
        HESS~J1714--385 & CTB~37A     & \citet{AharonianF.2008c}                          \\
        HESS~J1813--178 & G12.8--0.0  & \citet{FunkS.2007}, \citet{GotthelfE.2009}        \\
        HESS~J1833--105 & G21.5--0.9  & \citet{AbdallaH.2018a}                            \\
        HESS~J1834--087 & W~41        & \citet{AbramowskiA.2015a}                         \\
        HESS~J1846--029 & G29.7--0.3  & \citet{AbdallaH.2018a}                            \\
        HESS~J1930+188  & G54.1+0.3   & \citet{AcciariV.2010}, \citet{AbdallaH.2018a}     \\
        \hline
    \end{tabular}
    \egroup
    \caption{The eight composite \graya{} sources in the HGPS, along with the name of the composite object that is accelerating the CRs and the reference(s) that classified the source.}
    \label{tab:composite sources}
\end{table}

\subsubsection{Binaries}

There are two types of binary \graya{} sources: accretion powered and rotation powered~\citep{MirabelI.2006}. For an accretion-powered binary \graya{} source a stellar-mass black hole or neutron star orbits close enough to accrete matter from the companion star, forming a relativistic jet. Photons are boosted into the TeV energy range via IC upscattering with the CRs in the jet~\citep[e.g.~LS~5039;][]{ParedesJ.2000}. For a rotation-powered binary \graya{} source the wind from a rapidly spinning pulsar interacts with the wind of the companion star. The two opposing winds create a shock that can accelerate CRs into the TeV regime. Additionally, the VHE CRs upscatter photons via IC interactions~\citep[e.g.~PSR~B1259--63;][]{DubusG.2006}. The three compact binary \graya{} sources identified in the HGPS are given in \autoref{tab:binary sources}.

\begin{table}
    \centering
    \bgroup
    \def\arraystretch{1.25}
    \begin{tabular}{lll}
        \hline
        Source Name & Accelerator & Classification Reference \\ \hline
        HESS~J1018--589~A & 1FGL~J1018.6--5856 & \citet{AbramowskiA.2015e} \\
        HESS~J1302--638   & PSR~B1259--63      & \citet{AharonianF.2005c}  \\
        HESS~J1826--148   & LS~5039            & \citet{AharonianF.2006e}  \\
        \hline
    \end{tabular}
    \egroup
    \caption{The three binary \graya{} sources in the HGPS, along with the name of the object accelerating the CRs and the reference that classified the source.}
    \label{tab:binary sources}
\end{table}

\subsubsection{Stellar Clusters}

There are two types of stellar clusters: open clusters and globular clusters. Both types of stellar clusters are groupings of tens to hundreds of massive stars.
The colliding stellar winds between the massive stars within the cluster creates many shock fronts which can accelerate CRs into the TeV regime via Fermi acceleration.
Stellar clusters have only become a confirmed accelerator of TeV CRs in recent years, after the original HGPS data was released.
The four \graya{} sources in the HGPS that were later reclassified as stellar clusters are given in \autoref{tab:cluster sources}.

\begin{table}
    \centering
    \bgroup
    \def\arraystretch{1.25}
    \begin{tabular}{lll}
        \hline
        Source Name & Accelerator & Classification Reference \\ \hline
        HESS~J1023--575 & Westerlund~2 & \citet{MestreE.2021} \\
        HESS~J1646--458 & Westerlund~1 & \citet{AharonianF.2022b} \\
        HESS~J1747--248 & Terzan~5     & \citet{AbramowskiA.2011c} \\
        HESS~J1848--018 & W~43         & \citet{AbdallaH.2018a} \\
        \hline
    \end{tabular}
    \egroup
    \caption{The four \graya{} sources in the HGPS that were later reclassified as stellar clusters, along with the name of the cluster accelerating the CRs and the reference that classified the source.}
    \label{tab:cluster sources}
\end{table}

\subsubsection{Not Firmly Identified}

The \hess{} sources that are not firmly identified have one or more nearby CR accelerators from the categories discussed above. However, it is not known which accelerator is responsible for the observed \graya{} emission.
The sources that are not firmly identified can be placed into three sub-categories:
\begin{itemize}
    \item There is a CR source observed nearby; however, the \graya{} emission can be explained by any of the categories discussed above,~i.e.~the class of the CR counterpart is not known. Examples include: HESS~J1018.6--589~B~\citep{AbramowskiA.2015f}, HESS~J1427--608~\citep{GuoX.2017}, and both HESS~J1843--033 and HESS~J1844--030~\citep{CaoZ.2021b}.
    \item The CR accelerator class responsible for the \graya{} emission is known; however, the CR counterpart is only detected with a low significance. An example includes HESS~J1848--018~\citep{AbdallaH.2018a}.
    \item Multiple CR accelerators are observed in close proximity to the \graya{} source. However, the \graya{} emission can be explained by any of the possible counterparts. For example, regions around HESS~J1825--137~\citep{CollinsT.2021} exhibit spectral characteristics indicative of both an SNR and a PWN contributing to the CR acceleration around the source. Other examples sources include HESS~J1804-216~\citep{FeijenK.2020} and HESS~J1841--055~\citep{AcciariV.2020}.
\end{itemize}
There are a total of~32~sources that are not firmly identified in the HGPS, all of which are included in \autoref{tab:all sources} along with the parameters to model their spatial profiles.

\subsubsection{Sources with No Association} \label{sssect:no association}

\begin{table}
    \centering
    \bgroup
    \def\arraystretch{1.25}
    \begin{tabular}{lrr}
        \hline
        Source Name & Longitude ($^{\circ}$) & Latitude ($^{\circ}$) \\ \hline
        HESS J1457--593 & 318.35 & --0.42 \\
        HESS J1503--582 & 319.57 &   0.29 \\
        HESS J1626--490 & 334.82 & --0.12 \\
        HESS J1702--420 & 344.23 & --0.19 \\
        HESS J1708--410 & 345.67 & --0.44 \\
        HESS J1729--345 & 353.39 & --0.02 \\
        HESS J1741--302 & 358.28 &   0.05 \\
        HESS J1745--303 & 358.64 & --0.56 \\
        HESS J1828--099 &  21.49 &   0.38 \\
        HESS J1832--085 &  23.21 &   0.29 \\
        HESS J1858+020  &  35.54 & --0.58 \\
        \hline
    \end{tabular}
    \egroup
    \caption{All eleven \graya{} sources in the HGPS where it is not known whether or not they are a component of the diffuse TeV \graya{} emission. These sources are referred to as \graya{} sources with no known possible association. Longitude and latitude are listed in Galactic coordinates.}
    \label{tab:noassoc sources}
\end{table}

The \hess{} sources with no association do not have a known CR counterpart,~i.e.~the CR source responsible for the \graya{} emission is unknown.
There are two possible types of unassociated \graya{} sources.
First, the \graya{} emission may be due to a dense molecular cloud being impacted by diffuse CRs~\citep[e.g.~HESS~J1457--593;][]{AbdallaH.2018b} and is therefore an enhanced region of the diffuse \graya{} emission.
Second, the \graya{} emission may be due to an undiscovered CR counterpart. The undetected accelerator could be a faint SNR or PWN in the nearby area~\citep[e.g.~HESS~J1741--302;][]{AbdallaH.2018b} or an ancient PWN~\citep[e.g.~HESS~J1708--410;][]{KaufmannS.2018}.
Hence, it is impossible to determine if the unassociated \graya{} sources are part of the diffuse TeV emission with current observations.
The sources with no association, and their relationship to estimates of the diffuse emission, will be discussed in more depth in \autoref{sssect:Chosen Sources}. All~11~\graya{} sources with no association are given alongside their Galactic coordinates in \autoref{tab:noassoc sources}.
\section{The HGPS Large-Scale Gamma-Ray Emission} \label{sect:HGPS diffuse emission}

The HGPS observations show two components to the \graya{} emission along the Galactic plane: the~78~discrete \graya{} sources discussed above and a large-scale component. However, the large-scale emission was not detected to the $5\sigma$ level due to the worsening sensitivity for larger source sizes~(discussed in \autoref{ssect:hess sensitivity}). Hence, no conclusion can be made on the origin of the large-scale TeV emission -- be it from a diffuse `sea' of CRs or a population of unresolved \graya{} sources~\citep{AbramowskiA.2014b}. As the large-scale emission was not detected to the~5$\sigma$ level, no further analysis was conducted during the HGPS.

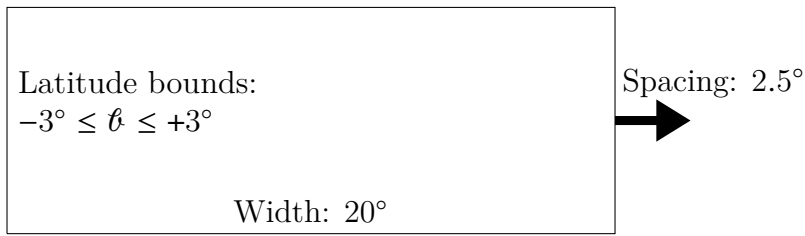
\begin{figure}
    \centering
    \begin{tikzpicture}
        
        % Draw the box
        \draw (0,0) -- (8,0) -- (8,3) -- (0,3) -- (0,0);
        
        % Draw the arrow
        \draw[-{Triangle}, line width=1.2mm] (8,1.5) -- (9,1.5);
        
        % Add text
        \node at (4,0.3) {Width: $20^{\circ}$};
        \node at (0,1.75) [anchor=west, text width=4cm] {Latitude bounds: $-3^{\circ} \leq \mathscr{b} \leq +3^{\circ}$};
        \node at (9.3,2) {Spacing: $2.5^{\circ}$};
        
    \end{tikzpicture}
    \caption{A diagram of the sliding window method used by~\citet{AbdallaH.2018a}. The window is defined to be $20^{\circ}$ wide and uses the Galactic latitude bounds $|\mathscr{b}| 3^{\circ}$. After the analysis is performed within the window it slides $2.5^{\circ}$ in Galactic longitude, and the process is repeated.}
    \label{fig:sliding window diagram}
\end{figure}

If the large-scale structure is not properly accounted for it will contaminate the all \graya{} source analyses~(such as morphological and spectral analyses) along the Galactic plane.
To prevent this \graya{} flux contamination \citet{AbdallaH.2018a} created an ad-hoc large-scale model. First, an analysis mask based on the exclusion regions~(see \autoref{ssect:CR background subtraction}) is defined to ensure that significant \graya{} emission is not included in the large-scale emission model. A sliding window is then defined to be $20^{\circ}$ wide in Galactic longitude with Galactic latitude bounds of $-3^{\circ} \leq \mathscr{b} \leq +3^{\circ}$. Within the window the emission is characterised as a Gaussian. The window then slides $2.5^{\circ}$ in Galactic longitude and the process is repeated. A diagram of the sliding window is shown in \autoref{fig:sliding window diagram}. The sliding window procedure is repeated for the entire plane and is interpolated to the longitudinal resolution of the HGPS~(0.02$^{\circ}$). The Gaussian within the window is constructed with the form:

\begin{align}
    F_{\mathrm{win}}(\mathscr{b}) &= F_{\mathrm{win,peak}} \exp \left( - \frac{( \mathscr{b} - \mu_{\mathrm{win}} )^{2}}{2 \sigma_{\mathrm{win}}^{2}} \right) \label{eq:window Gauss}
\end{align}

\noindent
where $F_{\mathrm{win}}$ is the flux of the large-scale emission in the window as a function of the Galactic latitude~($\mathscr{b}$), $F_{\mathrm{win,peak}}$ is the peak flux within the window~(not including the flux within the masked regions), $\mu_{\mathrm{win}}$ is the latitudinal location of the peak flux in the window, and $\sigma_{\mathrm{win}}$ is the width of the Gaussian in the window. All variables~($F_{\mathrm{win,peak}}$, $\mu_{\mathrm{win}}$, and $\sigma_{\mathrm{win}}$) are calculated for each window using a maximum-likelihood function. The sliding-window method results in a 2D~Gaussian as a function of both Galactic longitude and Galactic latitude.
The estimate of the large-scale \graya{} emission obtained via this model is subtracted from the HGPS flux maps before the \graya{} sources are characterised\footnote{Note that the model of the large-scale emission is not subtracted from the public maps and is only applied when characterising the \graya{} sources.}.

\subsection{The HGPS Diffuse Gamma-Ray Emission} \label{ssect:HGPS diffuse}

Although the sliding-window Gaussian provides a model of the large-scale emission, it does not distinguish between the unresolved sources or the truly diffuse components. 
Separating the unresolved source and diffuse components is challenging with the HGPS for three reasons: the integration beam size~(\autoref{ssect:beam size}), the adaptive ring background method~(\autoref{ssect:CR background subtraction}), and the limited \graya{} source horizon~(\autoref{ssect:horizon}).
As the diffuse emission is not detected to the~5$\sigma$ level, the \hess{} collaboration does not consider the 2D~Gaussian found via the sliding-window method as a formal estimate of the diffuse emission.

As discussed in \autoref{ssect:beam size}, the HGPS analysis is well suited to observing \graya{} emissions on the scale of the integration radius, $R_{c}$, where the provided public flux maps use $R_{c}=0.1^{\circ}$ and $R_{c}=0.2^{\circ}$. However, the diffuse \graya{} emission is expected to vary on scales greater than the maximum integration radius provided. The analysis by~\citet{AbramowskiA.2014b} found that the large-scale \graya{} emission for Galactic longitudes $-75^{\circ} < \mathscr{l} < 60^{\circ}$ is approximately a Gaussian profile across Galactic latitude. The large-scale Gaussian has a standard deviation of approximately $0.4^{\circ}$ in latitude. The analysis by~\citet{AbdallaH.2018a} further shows that the standard deviation of the latitudinal profile varies from~0.1$^{\circ}$--0.5$^{\circ}$ depending on the Galactic longitude, with an average Gaussian width of approximately $0.4^{\circ}$ in the GC region.
Utilising a larger integration radius would increase the significance of the observed large-scale structure. As the containment radius is used in the construction of the flux maps, it is not possible to increase $R_{c}$ without the original HGPS event lists.

The adaptive ring background method is used in the HGPS to subtract the CR background from the \graya{} observations~(see \autoref{ssect:CR background subtraction}). This background method assumes that there is no \graya{} emission in the \textit{off}~region.
Some \graya{} emission is prevented from being included in the CR background estimate due to the use of exclusion regions placed around areas of significant \graya{} emission. However, the chosen exclusion regions do not consider the large-scale Galactic \graya{} emission as significant emission, and do not take into account the low-significance unresolved source emission. Depending on the specific location of the pixel and the morphology of the nearby exclusion regions, the adaptive ring background can~(and often does) include the large-scale \graya{} emission as part of the CR background.
As the adaptive ring method over-estimates the CR background, the HGPS flux can only be considered as a lower limit on the true large-scale \graya{} emission.
The chosen background subtraction method has been shown to significantly alter the HGPS large-scale \graya{} analyses. The HGPS was compared to the HAWC GPS by~\citet{AbdallaH.2021}, subjecting the raw HPGS data to the same integration radius~($0.4^{\circ}$) and background method that HAWC uses~\citep[the FoV background technique;][]{BergeD.2007,AbdallaH.2021}. It was found that the HGPS flux was systematically increased by a median value of $1.9 \times 10^{-13}$\,cm$^{-2}$\,s$^{-1}$, or approximately~15\%, across the Galactic plane\footnote{The flux maps created by~\citet{AbdallaH.2021} are not publicly available. The estimate of the systematic difference between the adaptive ring background and the FoV background methods cannot be directly compared to the large-scale \graya{} estimates elsewhere in this thesis.}.

Although the HGPS has an unprecedented sensitivity for TeV \grays, it does not probe deep enough to reach the other side of the MW~(discussed in \autoref{ssect:horizon} and demonstrated in \autoref{fig:hgps horizon}). An estimate provided by~\citet{ActisM.2011} shows that there are hundreds~(if not thousands) of \graya{} sources in the MW that \hess{} is not sensitive enough to detect. Although unresolved, these \graya{} sources potentially contribute between~13\% and~60\% of the large-scale \graya{} flux~(\citealt{SteppaC.2020} and~\citealt{CataldoM.2020}, respectively). Detecting the unresolved \graya{} sources requires more sensitive observations, which requires a larger, more sensitive \graya{} facility.

    \chapter{Comparing GALPROP to the HGPS} \label{chap:comparing galprop and hess}

Comparing the \graya{} emission predicted by \GP{} to the \graya{} emission observed by \hess{} will aid in determining the contributions of the leptonic and hadronic components to the diffuse emission. Comparisons will also assist in constraining the Galactic structures such as the CR source distribution and the GMF.

Constructing a comparison between the models and observations is not trivial as the results from \GP{} and the HGPS differ in their construction.
The \GP{} emission is given as a differential flux in units of~MeV$^{-1}$\,cm$^{-2}$\,s$^{-1}$\,sr$^{-1}$, while the HGPS emission is given as a flux integrated above 1\,TeV in units of~cm$^{-2}$\,s$^{-1}$. Furthermore, \GP{} calculates the \graya{} flux by performing a line-of-sight integral for each pixel and emission type~(\autoref{ssect:GP flux calculation}), while the HGPS is built from an event list across many observation runs. Additionally, \GP{} is constructed such that the predicted \graya{} emission is truly diffuse. Conversely, the HGPS emission has contributions from unresolved \graya{} sources and a further~78~discrete \graya{} sources. The source components in the HGPS must be accounted for as they do not represent the diffuse emission.

Comparing the \GP{} predictions to the HGPS observations requires an analysis procedure that accurately represents both datasets simultaneously.
Following~\citet{AbdallaH.2018a} the analysis was chosen to be constructed from a sliding window.
However, the sliding window used in this thesis to compare \GP{} to the HGPS must be altered to ensure fair and consistent analyses across both the predictions and observations.
This chapter will detail the sliding window analysis method, and the specific parameter choices, that were made to ensure a fair comparison.
Additionally, the sensitivity of the sliding-window analysis to changes in the window parameters will be quantified. This chapter will also detail the transformations applied to ensure compatibility between the \GP{} and HGPS flux maps.
Comparisons between the \GP{} predictions and the HGPS observations are presented in \autoref{chap:paper 1}.

\section{Analysing the Large-Scale Emission} \label{sect:GALPROP vs HGPS}

The sliding-window analysis utilised by~\citet{AbdallaH.2018a} is discussed in \autoref{sect:HGPS diffuse emission}. A window is defined with some longitudinal and latitudinal extent, and the emission contained within is characterised as a 2D~Gaussian. The window then slides some distance in longitude and the analysis is repeated. This iterative process enabled the construction of an ad-hoc background component that could be subtracted from each \graya{} source analysis.
However, the results were not considered an estimate of the diffuse emission due to the low significance and the presence of unresolved sources. 

To improve the data statistics the sliding-window analysis used to compare the \GP{} predictions to the HGPS observations instead takes an average of the flux within the window.
The sliding window used here characterises the TeV large-scale emission as a longitudinal profile. 

\subsection{Sliding Window Parameters}

\begin{figure}
    \centering
    \begin{tikzpicture}
        
        % Draw the box
        \draw (0,0) -- (8,0) -- (8,3) -- (0,3) -- (0,0);
        
        % Draw the arrow
        \draw[-{Triangle}, line width=1.2mm] (8,1.5) -- (9,1.5);
        
        % Add text
        \node at (4,0.3) {$\Delta w$};
        \node at (0,1.5) [anchor=west, text width=4cm] {$\mathscr{b}_{\mathrm{min}} \leq \mathscr{b} \leq \mathscr{b}_{\mathrm{max}}$};
        \node at (8.5,2) {$\Delta s$};
        
    \end{tikzpicture}
    \caption{A diagram of the construction of the sliding window. The Galactic latitude bounds are given by $\mathscr{b}_{\mathrm{min}} \leq \mathscr{b} \leq \mathscr{b}_{\mathrm{max}}$, the width is given by $\Delta w$, and the distance that the window slides in Galactic longitude is given by $\Delta s$.}
    \label{fig:sliding window parameter diagram}
\end{figure}
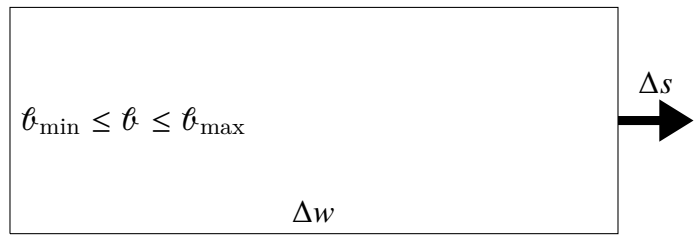

The sliding window is defined by the height/latitudinal extent given by $\mathscr{b}_{\mathrm{min}} \leq \mathscr{b} \leq \mathscr{b}_{\mathrm{max}}$, the width/longitudinal extent given by $\Delta w$, and the spacing between the windows given by $\Delta s$. A diagram of the sliding window and the parameter labels is shown in \autoref{fig:sliding window parameter diagram}.

As the sliding window analysis performed by~\citep{AbdallaH.2018a} was tuned for the creation of an ad-hoc background component, their analysis used the values $|\mathscr{b}| \leq 3^{\circ}$, $\Delta w = 20^{\circ}$, and $\Delta s = 2.5^{\circ}$.
As the analysis here is focused on characterising the large-scale emission the sliding window parameters are treated as free variables. The chosen parameters must be robust for both the \GP{} and HGPS analyses simultaneously,~i.e.~altering the sliding window parameters can have only a marginal impact on the analysis of either the \GP{} and the HGPS profiles. It should be noted that the HGPS results in this section use an integration radius of $R_{c} = 0.2^{\circ}$~(discussed in \autoref{ssect:beam size}) and have masked the catalogued \graya{} sources. Details of the source masking procedure is provided later in \autoref{ssect:Masking sources in the HGPS}. 

\subsubsection{Height of the Sliding Window}

\begin{figure}
    \centering
    \includegraphics[width=\textwidth]{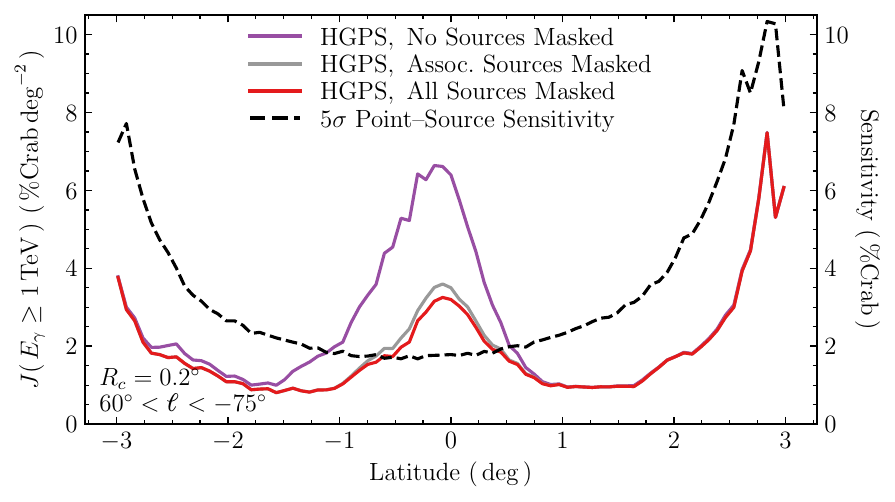}
    \caption{A latitudinal profile of the HGPS flux and sensitivity for the longitudes $-75^{\circ} < \mathscr{l} < 60^{\circ}$, with a beam size of $R_{c}=0.2^{\circ}$. The HGPS flux is shown in \%Crab\,deg$^{-2}$ for no sources masked~(purple), sources with an association masked~(grey), and all sources masked~(red). The sensitivity is shown by the black dashed line in units of \%Crab.}
    \label{fig:height test}
\end{figure}

The HGPS is built from many exposures focused on bright \graya{} sources, most of which lie within the central $2^{\circ}$ of the Galactic plane. Therefore, the sensitivity of the HGPS worsens towards regions above and below the Galactic plane.
\autoref{fig:height test} shows a latitudinal profile of the HGPS point-source sensitivity averaged across the longitudes $-75^{\circ} < \mathscr{l} < 60^{\circ}$. The \graya{} flux averaged across these longitudes is also shown for comparison.
The sensitivity is shown in units of \%Crab, where the flux of the Crab nebula is defined as $J_{\mathrm{Crab}} (E \ge 1\,\mathrm{TeV})=2.26\times10^{-11}\,\mathrm{cm}^{-2}\,\mathrm{s}^{-1}$~\citep{AharonianF.2006a}.
For latitudes $|\mathscr{b}| > 1^{\circ}$ the HGPS point-source sensitivity rapidly worsens with the minimum observable flux doubling by $|\mathscr{b}| = 2^{\circ}$ and quadrupling by $|\mathscr{b}| = 2.5^{\circ}$. There is an asymmetry in the HGPS exposure due to the focus on \graya{} sources at latitudes $|\mathscr{b}| < 0^{\circ}$.

The maximum latitude bound of the sliding window is set to $\mathscr{b}_{\mathrm{max}}=1^{\circ}$, where the average sensitivity at is approximately equal to 2.25\% of the Crab flux.
Due to the asymmetry in the HGPS exposure, the sensitivity worsens to the same level fir the negative latitudes around $b_{\mathrm{min}}=-1.5^{\circ}$. The latitudinal bounds of the sliding window are then given by $-1.5^{\circ} < \mathscr{b} < +1.0^{\circ}$, ensuring an approximately uniform exposure across the analysis region.

While the latitude bounds for the sliding window are an important consideration for the HGPS observations, the simulated \GP{} results have no effect analogous to the observational sensitivity worsening for higher latitudes.
To ensure a consistent and equal comparison the latitude bounds for the HGPS analysis are also applied to the \GP{} sliding-window analysis.

\subsubsection{Width of the Sliding Window}

\begin{figure}
    \centering
    \includegraphics[width=\textwidth]{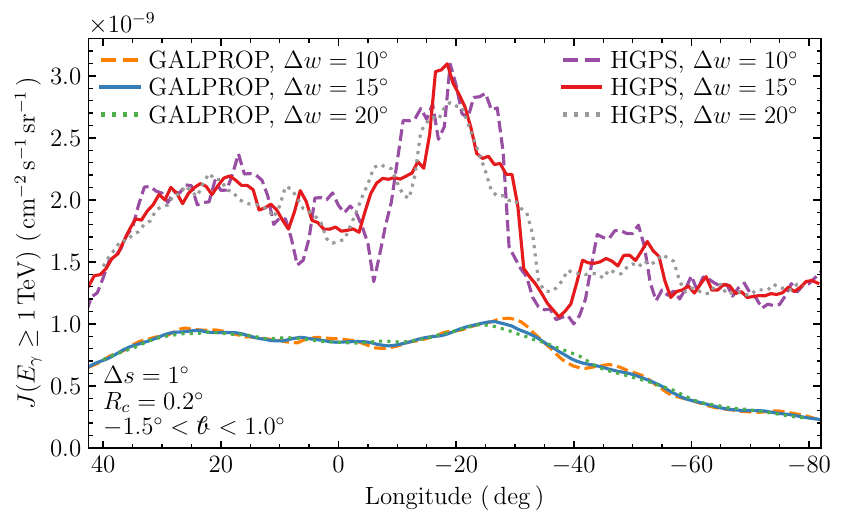}
    \caption{A longitudinal profile of both the HGPS and \GP{} flux, using a beam size of $R_{c}=0.2^{\circ}$ and the latitude range $-1.5^{\circ} < \mathscr{b} < +1.0^{\circ}$. The width of the sliding window, $\Delta w$, is varied from $10^{\circ}$~(dashed), $15^{\circ}$~(solid), and $20^{\circ}$~(dotted).}
    \label{fig:window width test}
\end{figure}

To improve the data statistics of the sliding-window analysis the window needs to be as wide as possible.
However, if $\Delta w$ is too large then the large-scale structure becomes averaged over Galactic longitude.
For example, the Galactic centre~(GC) and spiral arm tangents are approximately $20^{\circ}$ in longitude apart. Therefore, emission from the GC and spiral arm tangents can contaminate measurements of one another if the sliding window has a width $\Delta w > 20^{\circ}$.
Conversely, it was found that there were significant deviations in the HGPS longitudinal profile for sliding window widths $\Delta w < 10^{\circ}$ due to the lower data statistics for Galactic longitudes $-40^{\circ} \leq \mathscr{l} \leq -80^{\circ}$.
\autoref{fig:window width test} shows the results of the sliding window analysis on both the HGPS and \GP{} emission when varying the width of the sliding window from $10^{\circ}$ to $20^{\circ}$.

As the \GP{} results are smoothly varying along the Galactic plane, there is little variation in the longitudinal profile between $\Delta w = 10^{\circ}$, $\Delta w = 15^{\circ}$, and $\Delta w = 20^{\circ}$. However, some spatial information is lost in the $\Delta w = 20^{\circ}$ longitudinal profile in the GC region~($\mathscr{l}=0^{\circ}$). For the HGPS, there is little difference between the longitudinal profiles for $\Delta w = 15^{\circ}$ and $\Delta w = 20^{\circ}$. Larger deviations in the longitudinal profile due to lower data statistics begin to appear in the $\Delta w = 10^{\circ}$ profile.

The sliding window width was set to $\Delta w = 15^{\circ}$ for both \GP{} and the HGPS. This value for the window width ensures that the analysis robust to $\pm 30\%$ changes in $\Delta w$ for both the \GP{} predictions and the HGPS observations.

\subsubsection{Spacing Between Windows}

\begin{figure}
    \centering
    \includegraphics[width=\textwidth]{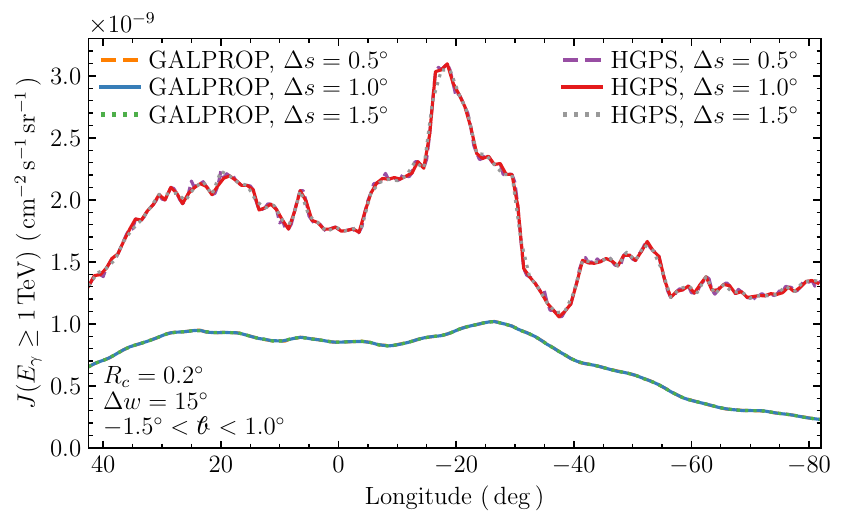}
    \caption{A longitudinal profile of both the HGPS and \GP{} flux, using a beam size of $R_{c}=0.2^{\circ}$, a latitude range of $-1.5^{\circ} < \mathscr{b} < +1.0^{\circ}$, and a window width of $\Delta w = 15^{\circ}$. The spacing between the sliding windows, $\Delta s$, is varied from $0.5^{\circ}$~(dashed), $1.0^{\circ}$~(solid), and $1.5^{\circ}$~(dotted).}
    \label{fig:window space test}
\end{figure}

The spacing between the sliding windows,~i.e.~the longitude that the window is moved between each iteration, must be defined such that changes in the longitudinal profile can be extracted. The Nyquist condition states that to observe variation on a given scale the sampling rate must be at least half of said scale.

It is possible for the large-scale emission to vary on the scale size of the \graya{} sources. The average diameter of a \graya{} source in the HGPS is $0.18^{\circ}$~\citep{AbdallaH.2018a}. Hence, to satisfy the Nyquist theorem, capturing variation on this scale size of the discrete \graya{} sources would require a sliding window spacing of $0.09^{\circ}$. Sampling on this scale is impossible as gas surveys used in the construction of the ISM models used by \GP{} have angular resolutions of approximately 0.1$^{\circ}$~(see \autoref{sect:ISM gas}).

The longitudinal structure of the large-scale emission may vary on the scale size of the angular separation between the \graya{} sources. The average angular separation between \graya{} sources is $\sim$2.5$^{\circ}$ in longitude~\citep{AbdallaH.2018a}. Approximately 20\% of the \graya{} sources lie within $1.0^{\circ}$ of another \graya{} source, with a particular pair being separated by only $0.03^{\circ}$ in longitude. Although the structure between all sources cannot be captured due to the limit resolution of the \GP{} maps, the majority of the structure between \graya{} sources could be captured with a window spacing between $0.5^{\circ}$ and $1.5^{\circ}$.

The sliding window analysis was performed on both the HGPS and \GP{} results, varying the sliding window spacing from $0.5^{\circ}$ to $1.5^{\circ}$. The results are shown in \autoref{fig:window space test}.
As the sliding window analysis is not sensitive to changes in $\Delta s$ for either the \GP{} predictions or HGPS observations for any of the chosen window spacings, the value $\Delta s=1^{\circ}$ is chosen for simplicity.
It should be noted that large deviations occur in the sliding-window analysis of the HGPS observations if the window spacing is larger than the average angular separation between \graya{} sources~(i.e.~for $\Delta s > 3^{\circ}$).
\section{Adapting the GALPROP Results for Comparison to the HGPS} \label{sect:changes to galprop}

\GP{} calculates the \graya{} emission for each process discussed in \autoref{sect:gamma-ray production} as a separate map.
The maps are calculated for the entire sky using the \hpx{} convention~\citep{GorskiK.2005}, which ensures an equal solid angle for each pixel. The highest resolution achievable by \GP{} is limited by the resolution of the gas maps used in the propagation and \graya{} emission calculations~(as discussed in \autoref{sect:ISM gas}). As of writing this thesis \GP{} is limited to ninth-order \hpx{} files,~i.e.~the minimum pixel dimensions are $0.1145^{\circ} \times 0.1145^{\circ}$.

\subsection{GALPROP Flux Integration}

The public data from the HGPS is a \graya{} flux map with units of~cm$^{-2}$\,s$^{-1}$ and has been integrated above 1\,TeV. For the comparison between the HGPS and \GP{} to be compatible, the differential \GP{} flux needs to be integrated over the same range of energies,~i.e.~$E_{\gamma}\geq1$\,TeV. However, the maximum simulated energy must be finite. For $\gtrsim$10\,GeV the differential \graya{} flux, $J(E_{\gamma})$, can be described by a power law. The differential flux, and the integrated flux between energies $E_{1}$ and $E_{2}$, are given by the equations:

\begin{align}
    J(E_{\gamma}) &\propto E_{\gamma}^{-\eta} \label{eq:flux} \\
    \Rightarrow \int_{E_{1}}^{E_{2}} J(E_{\gamma})\,\mathrm{d}E_{\gamma} &\propto \frac{1}{\eta-1} \left( E_{1}^{1-\eta} - E_{2}^{1-\eta} \right) \label{eq:integrated flux}
\end{align}

\noindent
where $E_{\gamma}$ is the \graya{} energy and $\eta$ is the spectral index. For discrete \graya{} sources, the spectral index typically takes values $2<\eta<3.5$, with the spectral index of the diffuse emission at TeV energies being in the range of $2.5<\eta<3$~\citep{AbdallaH.2018a}.
From \autoref{eq:integrated flux} it can be seen that setting $E_{1}=1$\,TeV and varying $E_{2}$ from 50\,TeV to 100\,TeV impacts the integral flux by less than one part in ten thousand.
The negligible impact was then confirmed numerically with the predicted emission from \GP.
As the error is well within the modelling uncertainties of \GP, the integrated \GP{} emission uses the energy bounds of $E_{1}=1$\,TeV and $E_{2}=50$\,TeV.
The \GP{} skymaps integrated for energies $E_{\gamma} \geq 1$\,TeV have units of~cm$^{-2}$\,s$^{-1}$\,sr$^{-1}$.

\subsection{Applying a Telescope Beam to GALPROP}

\GP{} calculates the \graya{} skymaps by performing a line-of-sight integral over the emissivity~(as discussed in \autoref{ssect:GP flux calculation}). The line-of-sight integral effectively draws a line from the observer in the direction of a given pixel and `counts' the photons along that line. The flux for a given \GP{} pixel is then divided by the solid angle of the pixel.
In contrast, the \graya{} emission given in the HGPS~\citep{AbdallaH.2018a} is an integral over a circular `beam' centred on each pixel~(as discussed in \autoref{ssect:beam size}).

To ensure compatibility between the \GP{} predictions and the HGPS observations, a telescope beam was applied to the \hpx{} skymaps.
To begin, the \GP{} predictions were converted to units of~cm$^{-2}$\,s$^{-1}$ by multiplying the flux by the solid angle of the pixels.
An integration beam with radius $R_{c}$ was then drawn around each pixel in the skymap.
However, integrating square pixels within a circular beam requires a correction factor, $C$, to be applied to the integral. The correction factor is equal to the solid angle of the beam divided by the total solid angle of all pixels within the beam, and is given by:

\begin{align}
    C &= \frac{A_{\mathrm{beam}}}{N_{\mathrm{beam}} A_{\mathrm{pixel}}}
\end{align}

\noindent
where $A_{\mathrm{beam}}=2\pi(1-\cos(R_{c}))$ is the solid angle of the applied beam, $A_{\mathrm{pixel}}$ is the area of a pixel, and $N_{\mathrm{beam}}$ is the number of pixels contained within the beam. It should be noted that the correction term approaches unity for $A_{\mathrm{beam}} \gg A_{\mathrm{pixel}}$. The values of $A_{\mathrm{pixel}}$ and $N_{\mathrm{beam}}$ depend on the resolution of the \hpx{} image.
After the integration beam is applied, each pixel is divided by the solid angle of the beam~($A_{\mathrm{beam}}$) such that the skymap has units of~cm$^{-2}$\,s$^{-1}$\,sr$^{-1}$. This algorithm can be expressed as a function, where the predicted \graya{} flux from \GP{} integrated over a telescope beam~($J_{\mathrm{beam}}$) is given by:

\begin{equation}
    \begin{aligned}
        J_{\mathrm{beam}}(\mathscr{l},\,\mathscr{b},\,E) &= \frac{1}{A_{\mathrm{beam}}} C \int_{R_{c}} A_{\mathrm{pixel}} J(\mathscr{l},\,\mathscr{b},\,E) \, \mathrm{d} \Omega \\
        &= \frac{1}{N_{\mathrm{beam}}} \int_{R_{c}} J(\mathscr{l},\,\mathscr{b},\,E) \, \mathrm{d} \Omega \label{eq:simplified galprop beam}
    \end{aligned}
\end{equation}

\noindent
where $(\mathscr{l},\,\mathscr{b})$ are the Galactic longitude and latitude, respectively, $\Omega$ is the solid angle, and the integral is performed over the beam with radius $R_{c}$. As can be seen from the simplified form of \autoref{eq:simplified galprop beam}, applying a telescope beam to the \GP{} output effectively averages the \graya{} flux over an angular scale size of $R_{c}$.

\begin{figure}
    \centering
    \includegraphics[width=\textwidth]{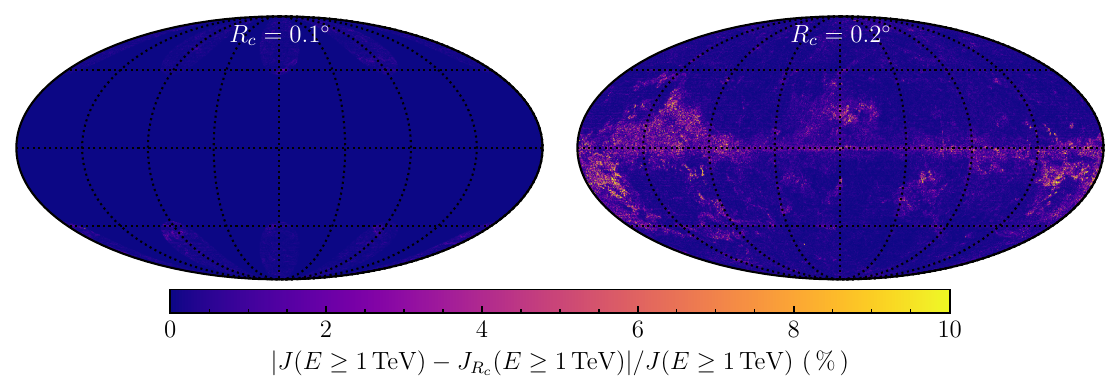}
    \caption{The absolute residuals, shown as a percentage, for the total \GP{} \graya{} flux integrated above 1\,TeV after subtracting the emission integrated over a beam with radius $R_{c}=0.1^{\circ}$~(left) and $R_{c}=0.2^{\circ}$~(right). For the $R_{c}=0.1^{\circ}$ beam the residuals are less than 1\%. For $R_{c}=0.2^{\circ}$, the residuals are less than 1\% for most of the sky, approximately 4--5\% for regions with \graya{} features with sizes $\sim$0.2$^{\circ}$, and can reach up to 10\% for the \graya{} features smaller than $0.2^{\circ}$.}
    \label{fig:GALPROP Rc subtraction}
\end{figure}

The \graya{} emission both before and after applying the telescope beam were compared to test if this procedure produces any systematic effects. This comparison was performed by calculating the residual emission between the original \GP{} predictions and the \graya{} flux after applying the telescope beam. The absolute residual emission, $J_{\mathrm{residual}}$, is given by:

\begin{align}
    J_{\mathrm{residual}}(\mathscr{l},\,\mathscr{b},\,E) &= \frac{|J(\mathscr{l},\,\mathscr{b},\,E) - J_{\mathrm{beam}}(\mathscr{l},\,\mathscr{b},\,E)|}{J(\mathscr{l},\,\mathscr{b},\,E)}
\end{align}

\noindent
where all parameters have been defined previously. The residual emission between the original \GP{} predictions and after applying the telescope beams can be seen in \autoref{fig:GALPROP Rc subtraction} for both the $R_{c}=0.1^{\circ}$ and $R_{c}=0.2^{\circ}$ beams.
The residual emission is calculated on the total \graya{} emission integrated above 1\,TeV.
The \GP{} outputs have pixels with a solid angle of $0.4 \times 10^{-5}$\,sr, which is similar in size to the solid angle of the $R_{c}=0.1^{\circ}$ beam~($10^{-5}$\,sr). Hence, the absolute residual emission for the $R_{c}=0.1^{\circ}$ beam is approximately zero across the entire 4$\pi$ steradians of the skymap.
At the polar regions the absolute residuals are larger due to the \hpx{} pixels being distorted\footnote{The \hpx{} pixels are distorted at the polar regions to ensure a constant solid angle for each pixel.}.
Also seen in \autoref{fig:GALPROP Rc subtraction} is that the absolute residuals for the $R_{c}=0.2^{\circ}$ beam can be as large as 10\%.
As shown by \autoref{eq:simplified galprop beam}, the \graya{} emission is averaged over the size of the beam. Hence, structures with radii smaller than the $R_{c} = 0.2^{\circ}$ beam are impacted. Such structures include the hadronic emission from nearby gas clouds.

\begin{figure}
    \centering
    \includegraphics[width=\textwidth]{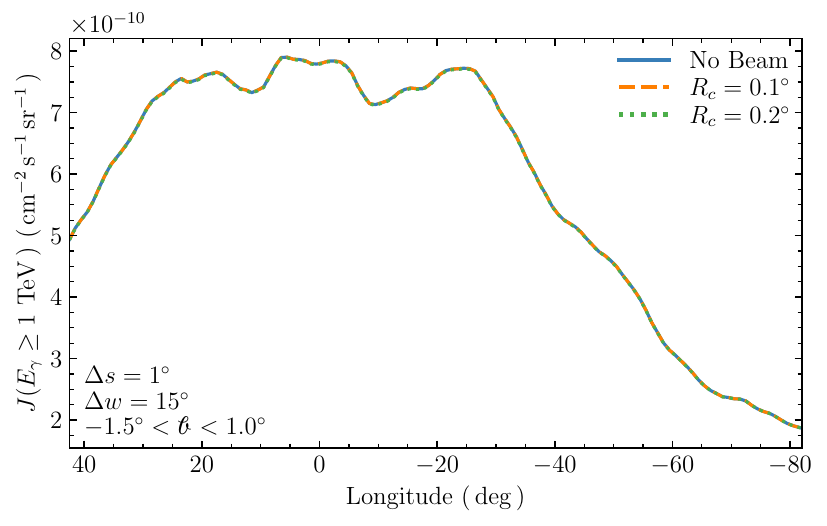}
    \caption{A longitudinal profile of the \GP{} emission with no telescope beam applied~(blue), and telescope beams with radii $R_{c}=0.1^{\circ}$~(orange dashed) and $R_{c}=0.2^{\circ}$~(green dotted). The analysis follows the sliding window recipe discussed in \autoref{sect:GALPROP vs HGPS}, with $\Delta s=1^{\circ}$ and $\Delta w=15^{\circ}$.}
    \label{fig:beam comparison}
\end{figure}

The large-scale \graya{} structure has a Gaussian width across Galactic latitude, varying between $0.1^{\circ}$--0.5$^{\circ}$, with an average width of $0.4^{\circ}$ around the GC region~\citep{AbramowskiA.2014b,AbdallaH.2018a}. The large-scale emission then varies on scales larger than either the $R_{c}=0.1^{\circ}$ or $R_{c}=0.2^{\circ}$ beams. Therefore, while the absolute residual emission after applying the $R_{c}=0.2^{\circ}$ can be as large as 10\%, it is expected that the large-scale structure would be unaffected.

To quantify any systematic effects that the telescope beam may have on the large-scale diffuse emission, a sliding-window analysis was performed~(see \autoref{sect:GALPROP vs HGPS}).
The results from the sliding-window analysis on the original \GP{} flux and after applying the $R_{c}=0.1^{\circ}$ and $R_{c}=0.2^{\circ}$ telescope beams are shown in \autoref{fig:beam comparison}.
Applying either of the two telescope beams had a negligible impact on the large-scale emission predicted by \GP, with the differences between the longitudinal profiles being on the order of one part in ten thousand.
Therefore, the \GP{} results are robust against the integration beam on the angular scales of interest~(i.e.~$0.1^{\circ}$ and $0.2^{\circ}$).

\section{Adapting the HGPS Results for Comparisons to GALPROP} \label{sect:changes to hgps}

The HGPS uses an equirectangular~(i.e.~cartesian) projection for the flux maps, with pixel dimensions of $0.02^{\circ} \times 0.02^{\circ}$ and units of~cm$^{-2}$\,s$^{-1}$. For compatibility with the \GP{} skymaps, the HGPS flux is divided by the solid angle of the integration beam~($2\pi[1-\cos(R_{c})]$), where $R_{c}=0.1^{\circ}$ or $R_{c}=0.2^{\circ}$.
Additionally, before the HGPS observations can be compared to the predicted diffuse TeV emission from \GP, all of the~78~catalogued \graya{} sources~(discussed in \autoref{ssect:source classifications}) need to be accounted for.

\subsection{Masking the Catalogued HGPS Sources} \label{ssect:Masking sources in the HGPS}

\begin{figure}
    \centering
    \includegraphics[width=0.905\textheight, angle=90]{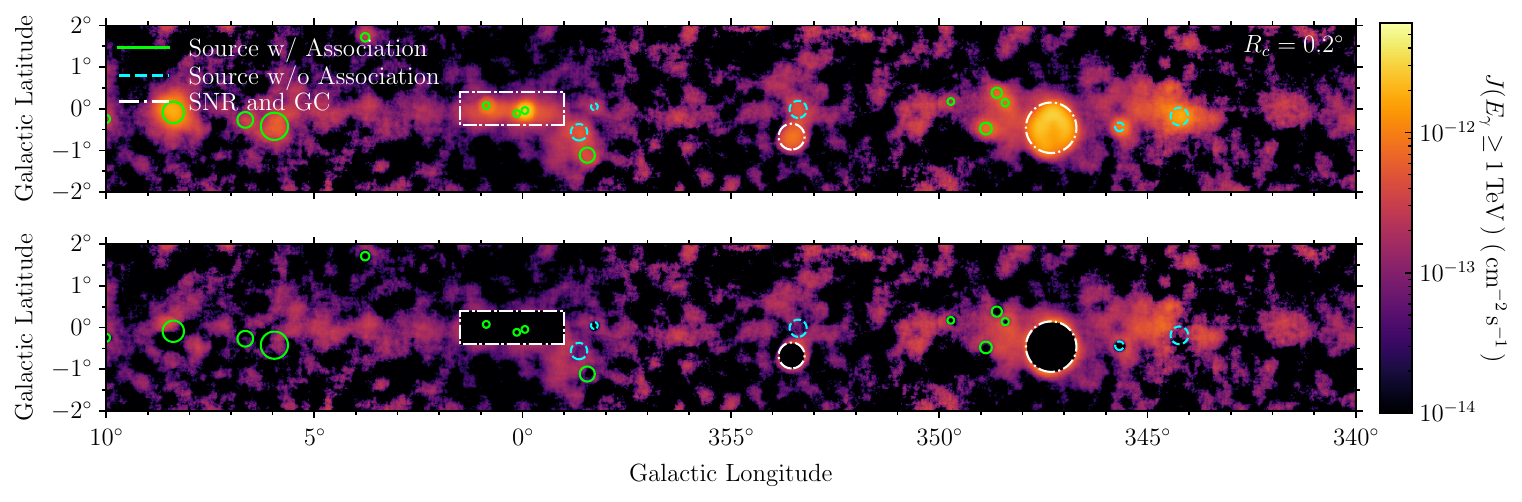}
    \caption{HGPS flux map for Galactic longitudes $10^{\circ} \leq \mathscr{l} \leq 340^{\circ}$ and Galactic latitudes $-2^{\circ} < \mathscr{b} < +2^{\circ}$ for the $R_{c}=0.2^{\circ}$ map. The flux is shown before the mask is applied~(top) and after the mask is applied~(bottom). For a description of sources with and without an association, see text.}
    \label{fig:mask demo}
\end{figure}

The largest contribution to the \graya{} emission observed in the HGPS is that from discrete, resolved sources. These sources are commonly referred to as catalogued \graya{} sources. To obtain an estimate of the large-scale \graya{} emission all of the catalogued \graya{} sources must be masked out.
To account for the catalogued source emission, the recipe from~\citet{AbdallaH.2018a} was followed. The morphological structure of the catalogued \graya{} sources is modelled and subtracted~(i.e.~masked) from the HGPS. A demonstration of the \graya{} source masking can be seen in \autoref{fig:mask demo}.

The catalogued \graya{} sources can be classified by their morphological structure into one of the following four categories:

\begin{enumerate}
    \item Shell: Sources in this category cannot be modelled in an easy, repeatable method. Shell-like sources show enhanced emission along their edges due to limb brightening and may have more complex structures within.
    \item GC: A complex morphological region filled with many bright \graya{} sources that cannot be described by a simple spatial model.
    \item Gaussian: Sources in this category can be modelled as one or more Gaussians. The majority of sources in the HGPS are in this category.
    \item Point: Sources in this category can be modelled as a single Gaussian, as the sources are intrinsically smaller than the point spread function~(PSF) of \hess.
\end{enumerate}

\subsubsection{Shell-type Sources and the Galactic Centre} \label{sect:source cut}

Shell-like sources are likely caused by SNRs accelerating CRs via diffusive shock acceleration as the SN shock expands into the surrounding ISM~\citep{FermiE.1949}. The collision between the SNR shock and ISM gas involves many complex interactions. The interactions between the two media results in a morphologically complex object. Shell-like sources are often asymmetric and contain detailed structures seen across many wavelengths. Due to the large number of poorly understood and unconstrained parameters, any systematic analysis applied to all shell-like sources performed by~\citet{AbdallaH.2018a} lead to unconstrained or diverging analyses.

As shell-like \graya{} sources cannot be modelled in a systematic, repeatable fashion, their emission is instead cut from the map. All pixels within the shell radius are excluded from any further analyses. The outline of shell-like sources is taken as a circle, where the radius of the HGPS spectral analysis region~(given in \autoref{tab:all sources}) is used for the radius of the cut-out region. An example of the exclusion procedure on a shell-type SNR can be seen in \autoref{fig:mask demo}. The excluded \graya{} source is HESS~J1713--397 at the coordinates $(\mathscr{l},\,\mathscr{b}) = (347.31^{\circ},\,-0.46^{\circ})$ with a radius of $0.60^{\circ}$.

The GC is a complex multi-source region. Although~\citet{AbdallaH.2018e} created a spatial model for the GC region, their model requires ISM gas data and multiple ad-hoc components. 
Similarly as for the shell-type sources, there exists no accurate, representative spatial model that can completely mask the GC without the inclusion of external data.
Following~\citet{AbdallaH.2018a} and the procedure for shell-like sources, the pixels within the GC region are excluded from any further analyses. The GC region is defined by the longitudinal bounds $-1.0^{\circ} \leq \mathscr{l} \leq +1.5^{\circ}$ and latitudinal bounds $|\mathscr{b}| \leq 0.4^{\circ}$~\citep{AbdallaH.2018a}, with the exclusion process shown in \autoref{fig:mask demo}.

\subsubsection{Point Sources and Gaussian Sources}

As discussed in \autoref{ssect:source characterisation}, the point-like \graya{} sources are modelled as a single three-dimensional Gaussian with a standard deviation width equal to the PSF of \hess~($0.08^{\circ}$). The Gaussian sources are modelled as one or more three-dimensional Gaussian(s) that have a standard deviation width larger than the PSF of \hess.

The surface brightness of a \graya{} source~($S_{\mathrm{source}}(\mathscr{l}, \mathscr{b})$; \autoref{eq:source subtraction}) is calculated individually for each point-like and Gaussian source as a function of the radial distance~($r$; \autoref{eq:source radius}) and the angular size~($\sigma_{\mathrm{ang}}$; \autoref{eq:total angular size}). \autoref{chap:HGPS source pars} details how the parameters were calculated for each source, and the parameter values for all catalogued sources are provided in \autoref{tab:all sources}.
The modelled source surface brightness is subtracted from the HGPS image via the equation:

\begin{align}
    F_{\mathrm{masked}}(\mathscr{l}, \mathscr{b}) &= F_{\mathrm{unmasked}}(\mathscr{l}, \mathscr{b})-S_{\mathrm{source}}(\mathscr{l}, \mathscr{b}) \label{eq:source mask}
\end{align}

\noindent
where $F_{\mathrm{unmasked}}$ is the original HGPS flux and $F_{\mathrm{masked}}$ is the residual flux after subtracting the catalogued source components.
For a given source, $S_{\mathrm{source}}$ is subtracted from the map for $r \leq 5 \sigma_{\mathrm{ang}}$ as $S_{\mathrm{source}}(r\geq5\sigma_{\mathrm{ang}})$ is always below the uncertainty in the HGPS flux map. Examples of the source masking procedure can be seen in \autoref{fig:mask demo}.

\subsubsection{Choosing the Sources to Mask} \label{sssect:Chosen Sources}

The \graya{} sources catalogued in the HGPS can be split into two categories. The sources with a known nearby accelerator of CRs~(or multiple possible accelerators) are labelled as an `associated source', and the \graya{} sources without any observed nearby CR accelerator are labelled as `not associated'~(or as `no known association').
The classifications for the CR accelerators are discussed in more depth in \autoref{ssect:source classifications}, with the binary, stellar cluster, SNR, composite, PWN, and `not firmly identified' classifications being considered associated sources.
The associated sources are not considered part of the large-scale \graya{} emission.
Conversely, it is not known if the unassociated \graya{} sources can be considered as part of the large-scale emission. 
The eleven \graya{} sources catalogued in the HGPS that have no known association are shown in \autoref{tab:noassoc sources}, with their masking parameters shown in \autoref{tab:all sources}.

\begin{figure}[t]
    \centering
    \includegraphics[width=\textwidth]{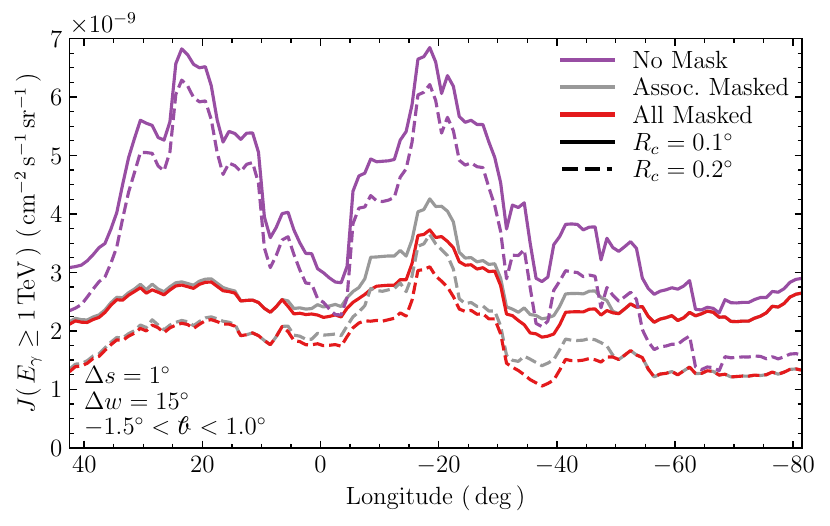}
    \caption{Longitudinal profile of the HGPS for $R_{c}=0.1^{\circ}$~(solid lines) and $R_{c}=0.2^{\circ}$~(dashed lines). The shown profiles are for no sources masked~(purple), only sources with an association masked~(grey), and all sources masked~(red). The grey and red lines represent the range of the estimates of the HGPS large-scale \graya{} emission. The analysis follows the sliding window recipe discussed in \autoref{sect:GALPROP vs HGPS} with $\Delta s=1^{\circ}$, $\Delta w=15^{\circ}$, and $-1.5^{\circ} < \mathscr{b} < 1.0^{\circ}$.}
    \label{fig:mask comparison}
\end{figure}

Two \graya{} flux maps were calculated from the HGPS -- one where all~78~catalogued \graya{} sources were masked, and another where only the sources with a known association were masked from the analyses. The two \graya{} flux maps represent a lower and upper limit, respectively, on the large-scale emission after accounting for the discrete sources observed in the HGPS. A comparison between the longitudinal profiles for the two masks~(as well as the un-masked HGPS) for both the $R_{c}=0.1^{\circ}$ and $R_{c}=0.2^{\circ}$ telescope beams is shown in \autoref{fig:mask comparison}.

The longitudinal profile of the HGPS with no sources masked~(shown in \autoref{fig:mask comparison}) represents the upper-bound on the TeV large-scale \graya{} emission along the Galactic plane. The difference between the `no sources masked' and `all sources masked' profiles represents the contribution of the catalogued \graya{} sources to the large-scale emission along the Galactic plane. Masking \graya{} sources decreases the large-scale emission by up to a factor of three; hence, the large-scale TeV \graya{} structure is dominated by \graya{} sources.
Additionally, the difference between the two longitudinal profiles demonstrates that there is a concentration of \graya{} sources along the spiral arm tangents~($\mathscr{l} \approx \pm 20^{\circ}$).

Comparing the large-scale emission between masking all sources and only masking sources with an association shows that most of the \graya{} sources with no known association lie between the longitudes $0^{\circ}<\mathscr{l}<310^{\circ}$, with the longitudinal profiles between the two masks differing by a maximum of 20\% in this region.
Masking all sources potentially overestimates the contribution of the \graya{} sources to the large-scale emission. Hence, the profile represents a lower limit on the large-scale emission after accounting for resolved sources.
Conversely, masking only the sources with a known association represents an upper limit as the \graya{} source component is potentially underestimated.
The difference between the two longitudinal profiles is a maximum of 20\% for the Galactic longitudes $10^{\circ}<\mathscr{l}<310^{\circ}$. The two profiles are within 1\% outside of this longitude range as only three of the eleven sources with no known association lie beyond the longitudes $10^{\circ}<\mathscr{l}<310^{\circ}$.

As discussed in \autoref{ssect:beam size} the HGPS emission is most sensitive to features with sizes similar to the integration radius, $R_{c}$.
However, the large-scale \graya{} structure varies on scales larger than either of the telescope beams used in the HGPS, with an average Gaussian width in the GC region of $0.4^{\circ}$~\citep{AbramowskiA.2014b,AbdallaH.2018a}.
Hence, the $R_{c}=0.2^{\circ}$ profile has a higher significance for the large-scale emission.
The difference between the longitudinal profiles taken on the $R_{c}=0.1^{\circ}$ and $R_{c}=0.2^{\circ}$ maps is independent of the applied source masks, varying between 0.5--1.0~$\times 10^{-9}$cm$^{-2}$\,s$^{-1}$\,sr$^{-1}$.
The $R_{c}=0.2^{\circ}$ HGPS map is used for all further estimates of the large-scale \graya{} emission in this thesis.

\subsection{Unresolved Sources in the HGPS} \label{ssect:Unresolved Sources}

As discussed in \autoref{ssect:horizon}, the HGPS has a limited source distance horizon beyond which \hess{} cannot individually detect point sources. The source horizon depends on the luminosity of the source, the distance to the source, and the angular extent of the source. As the horizon did not extend to the other side of the Galaxy~(\autoref{fig:hgps horizon}) there are many more \graya{} sources in the TeV energy range that remain undetected by \hess. The number of unresolved sources is estimated to be in the range of hundreds to thousands~\citep{ActisM.2011}. Although they are not individually detected, the unresolved sources contribute to the total observed large-scale \graya{} flux. While it is possible to mask the catalogued sources from the analysis, it is impossible to mask the unresolved sources. However, the flux that unresolved sources are likely to contribute to the large-scale emission can be estimated and subtracted from the observations.

An estimate from~\citet{SteppaC.2020} suggests that unresolved sources contribute 13\%--32\% to the large-scale emission. This estimate was obtained by comparing the total flux observed in the HGPS to the total TeV luminosity produced by a modelled distribution of sources, with the total luminosity of TeV \graya{} sources based on PWN distributions. Another estimate from~\citet{CataldoM.2020} suggests that the fraction of unresolved sources could be as large as 60\%. The~\citet{CataldoM.2020} estimate was obtained by using known SNR and PWN spatial distributions and luminosity distributions to calculate the number of detectable sources, as well as calculating the total \graya{} flux that all the sources would produce.

\begin{figure}
    \centering
    \includegraphics[width=\textwidth]{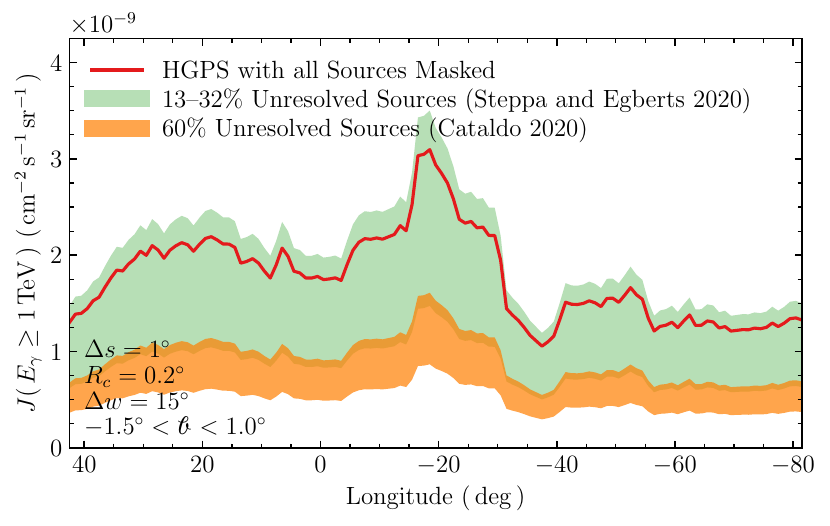}
    \caption{Longitudinal profile of the HGPS emission for a beam size of $R_{c}=0.2^{\circ}$ and catalogued sources masked~(red). Also shown is the HGPS after accounting for the systematic uncertainty~($\pm 30\%$), with unresolved source contributions of 13--32\%~\citep[green;][]{SteppaC.2020} and 60\%~\citep[orange;][]{CataldoM.2020}. The analysis follows the sliding window recipe discussed in \autoref{sect:GALPROP vs HGPS} with $\Delta s=1^{\circ}$, $\Delta w=15^{\circ}$, and $-1.5^{\circ} < \mathscr{b} < 1.0^{\circ}$.}
    \label{fig:HGPS unres masked}
\end{figure}

A combination of both the~\citet{SteppaC.2020} and~\citet{CataldoM.2020} unresolved source estimates with the $\pm 30\%$ systematic uncertainty in the \hess{} flux is shown in \autoref{fig:HGPS unres masked}. The HGPS profile is taken from the $R_{c}=0.2^{\circ}$ flux map after subtracting all catalogued \graya{} sources.
Subtracting the contribution from both the catalogued \graya{} sources and an estimate of the unresolved sources gives the residual emission of the HGPS on the scale-size of $R_{c}$.
The only remaining component in the residual emission is expected to be the large-scale diffuse \graya{} emission. Hence, the estimates shown in \autoref{fig:HGPS unres masked} can be considered estimates of the large-scale diffuse \graya{} emission. However, it should be noted that this estimate is below the~5$\sigma$ detection threshold.

As the longitudinal profile shown in \autoref{fig:HGPS unres masked} is taken after masking all~78~catalogued \graya{} sources, it is a lower bound on the residual large-scale emission.
As the unresolved source component from~\citet{CataldoM.2020} is an upper limit, the residual emission found after subtracting the \citet{CataldoM.2020} estimate is a lower bound of the TeV large-scale diffuse \graya{} emission.
    \chapter{First Paper: Diffuse TeV Gamma-Ray Predictions with GALPROP} \label{chap:paper 1}

This chapter contains the manuscript titled: ``The Steady-State Multi-TeV Diffuse Gamma-Ray Emission Predicted with \GP{} and Prospects for the Cherenkov Telescope Array'', which was published in the peer-reviewed journal \mnras{} in February~2023.

\vspace{2em}

As cosmic rays~(CRs) diffuse throughout the interstellar medium~(ISM) they interact with the ISM gas, the interstellar radiation field~(ISRF), and the Galactic magnetic field~(GMF). These interactions create broadband non-thermal emissions, from radio waves to ultra-high-energy gamma~rays~(\grays).
The \textit{Fermi} large area telescope~(\fermi) observed a large-scale diffuse \graya{} structure at GeV energies along the Galactic plane. More recently, the high-energy stereoscopic system~(\hess) in their Galactic plane survey~(named the HGPS) observed a large-scale structure at TeV \graya{} energies.
The emissions observed at GeV and TeV energies will be connected by a common origin of the CRs; however, the energy dependence of the emission, and the relative contribution from CRs that are diffuse or local to sources, is poorly understood.

In this manuscript it is shown that the model predictions from \GP{} broadly agree with the estimates of the diffuse TeV \graya{} emission from the HGPS after accounting for emission from catalogued \graya{} sources and the flux contribution from unresolved \graya{} sources.
It is also found that the \graya{} emissions from inverse Compton~(IC) scattering off CR electrons becomes an important component of the TeV \graya{} flux along the Galactic plane, and even dominates the total emissions above 10\,TeV. Due to the importance of the CR electrons, future \graya{} observations in the 100\,TeV energy range may be able to aid in constraining the GMF structure.
Given the lower bounds on the diffuse emission that was found from both the HGPS and \GP, the upcoming Cherenkov telescope array~(CTA) will likely be able to observe the large-scale diffuse structure at TeV \graya{} energies. The brightness of the \GP{} predictions indicate that more optimisation is necessary for applications to the more sensitive CTA~GPS.

\blindfootnote{Due to the large number of variables in this thesis, the notation in this published work is not necessarily consistent with the other chapters.}

% Include citations that have no real place in the thesis but are in the paper. This is to ensure that they still appear in the thesis bibliography
\nocite{CataldoM.2019,ChoiG.2022a,LipariP.2018,MaciasO.2019,SakoT.2009,ValleeJ.2020}
% CataldoM.2019 : Example of a model that doesn't agree with Tibet AS_gamma
% ChoiG.2022a   : Example that flux measurement uncertainties are large above 1 TeV
% LipariP.2018  : Example of a model that doesn't agree with Tibet AS_gamma
% MaciasO.2019  : A CR source distrution that is normalised to the inner Galaxy
% SakoT.2009    : Tibet AS_gamma reference
% ValleeJ.2020  : Different tracers show different spiral arm offsets

% Statement of authorship
\clearpage
\begin{figure}[ht]
    \centering
    \includegraphics[width=16cm]{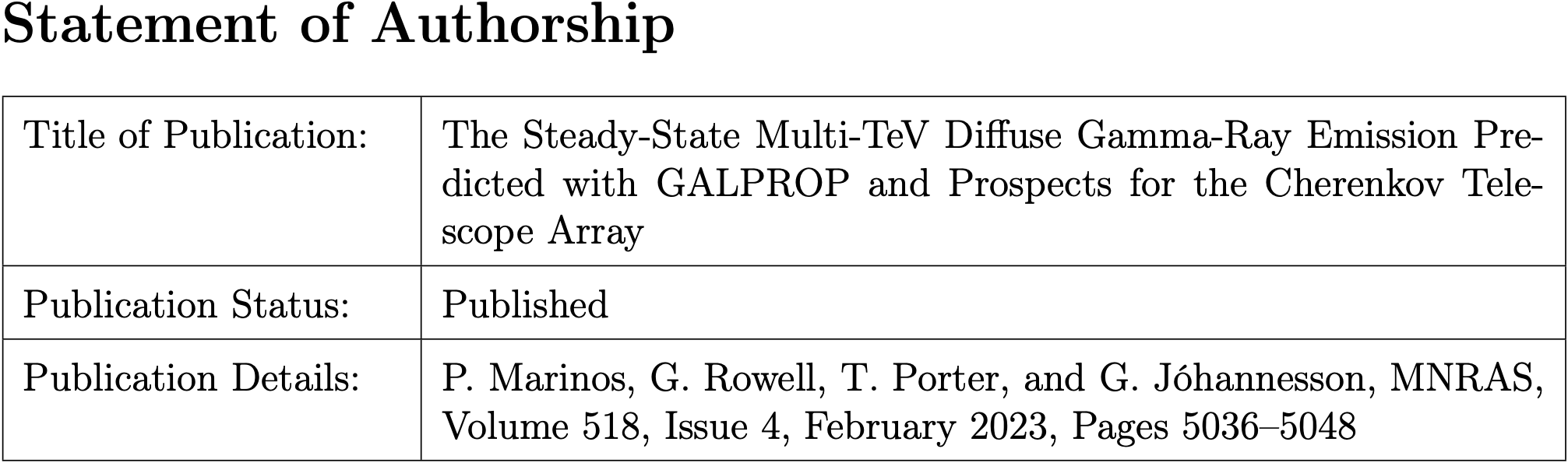}

    \vspace{2em}

    \includegraphics[width=16cm]{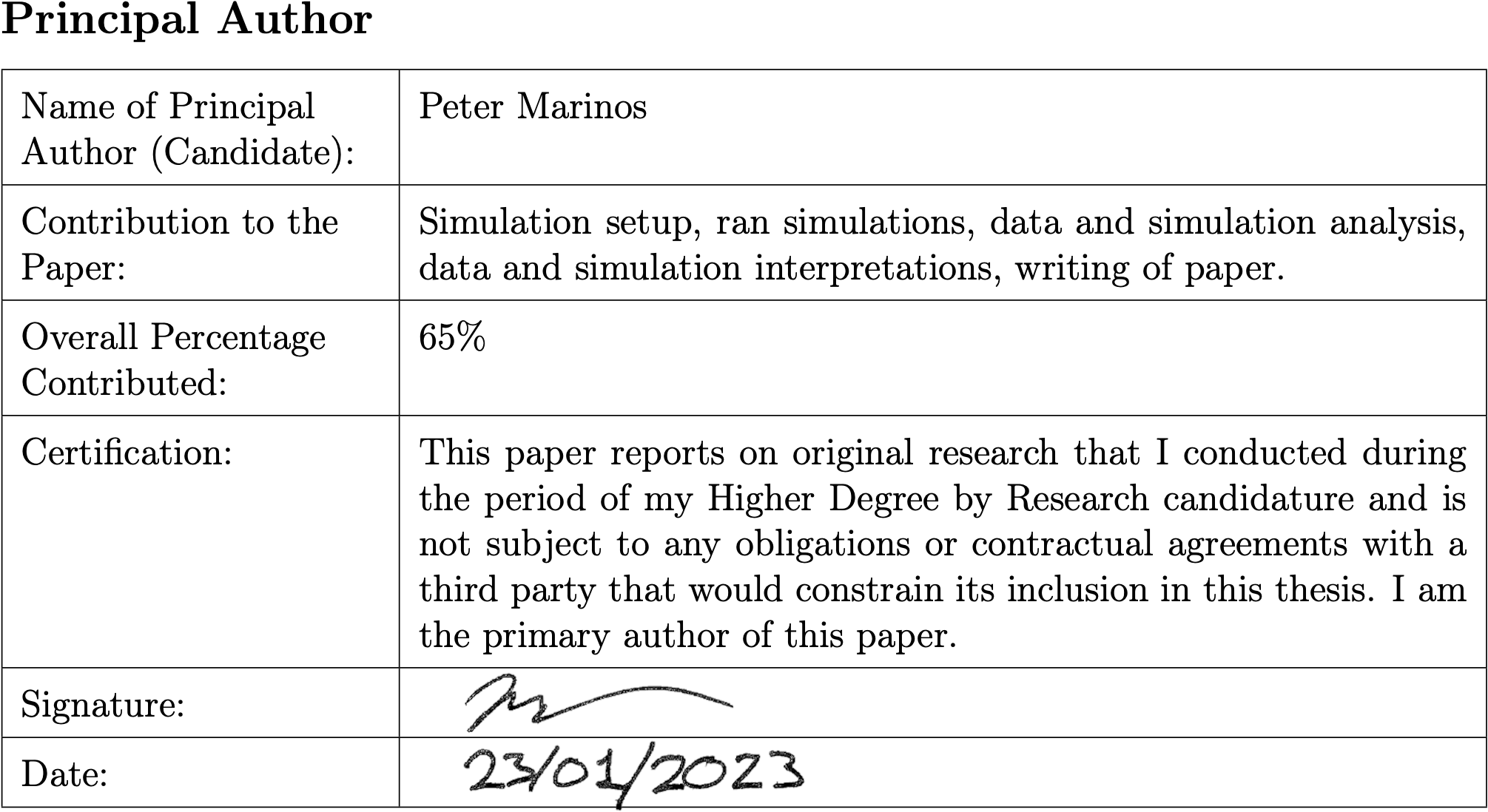}

    \vspace{2em}

    \includegraphics[width=16cm]{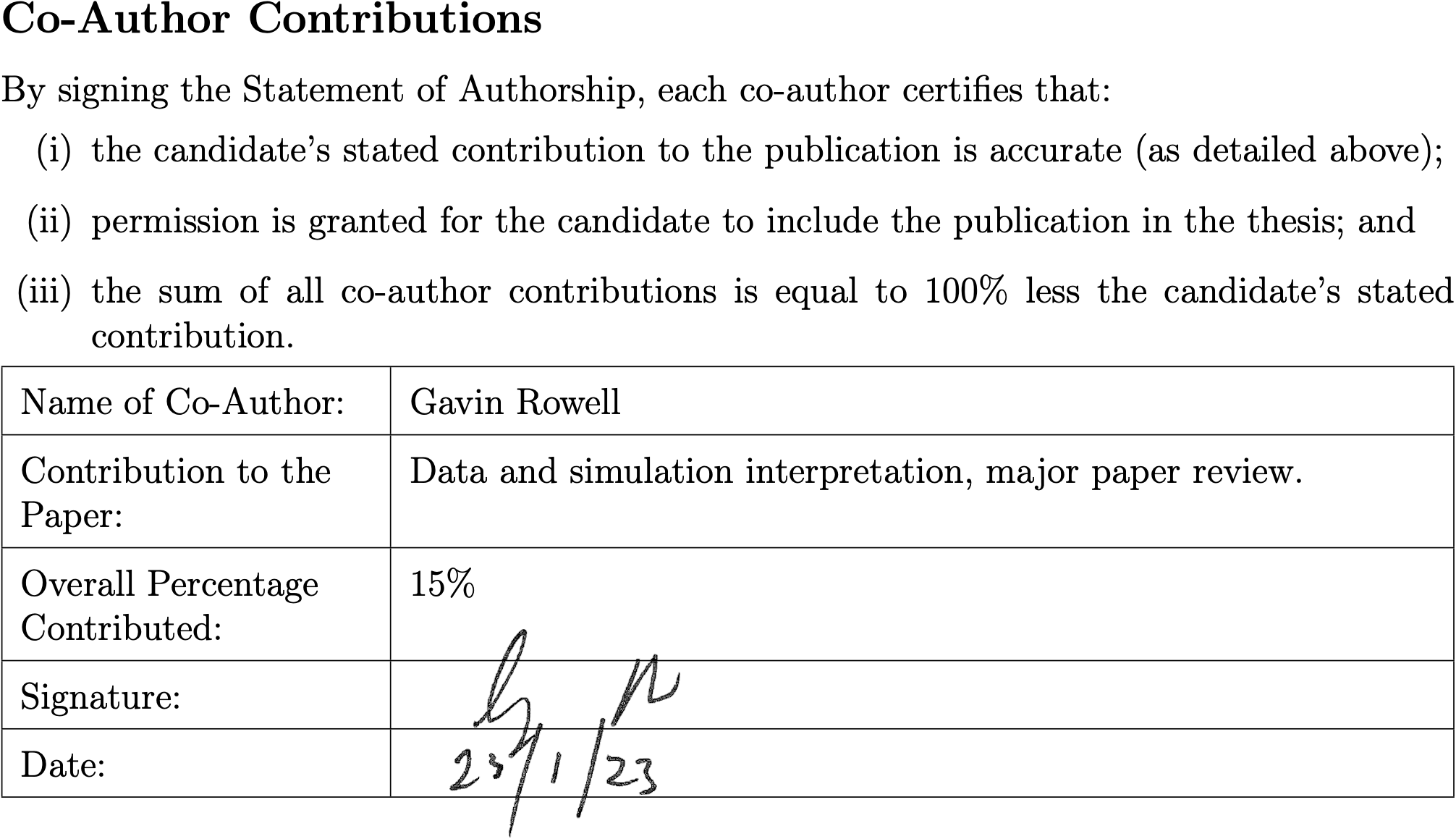}
\end{figure}
\clearpage
\begin{figure}[ht]
    \centering
    \includegraphics[width=16cm]{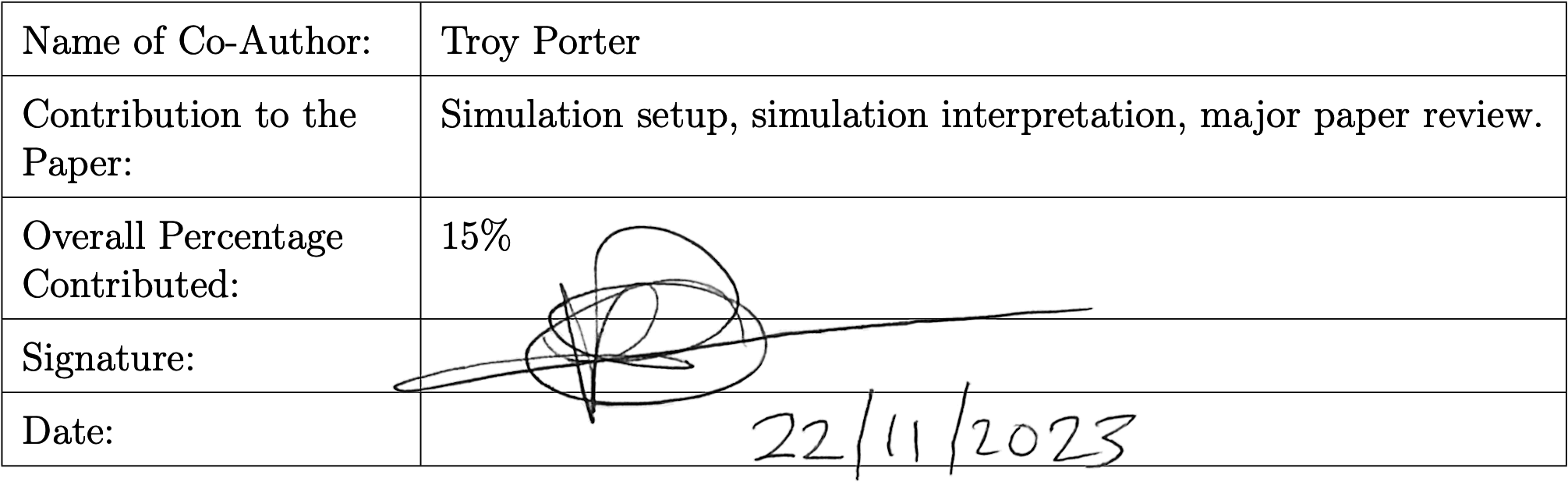}

    \vspace{1.5em}
    
    \includegraphics[width=16cm]{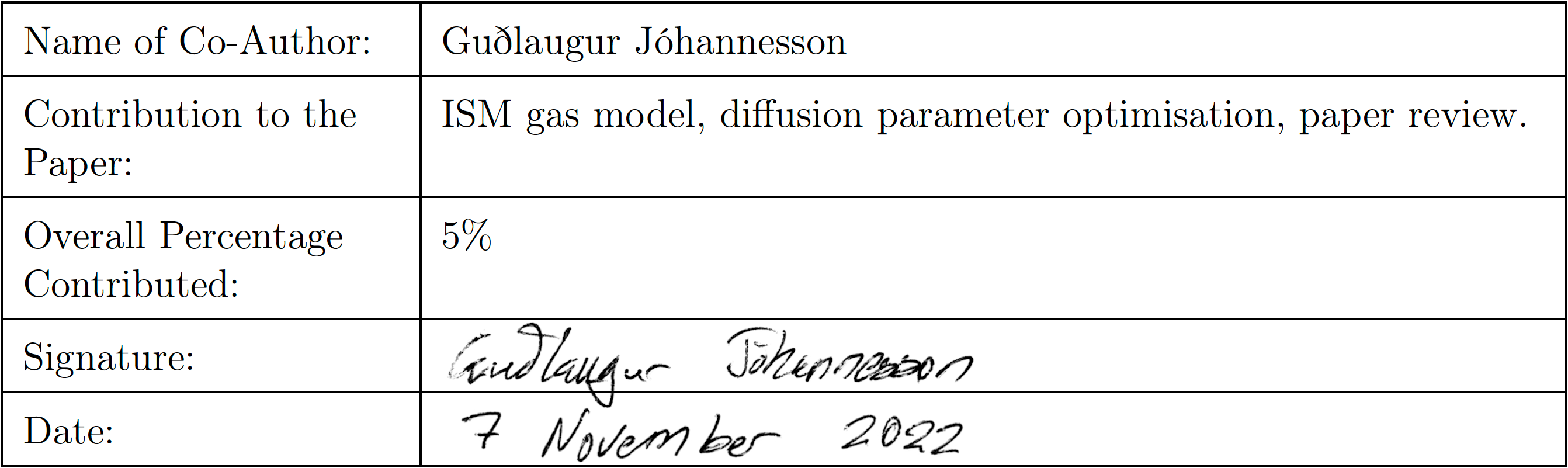}
\end{figure}

% Paper
% Scale = thesis textwidth / paper textwidth
%       = 160mm/7in
% Offset = binding offset - paper margin * scale
%        = 20mm - 0.65in * 0.842
%     Plus some extra fudging to get it perfect
% Offset is {horizontal offset}, {vertical offset}
\clearpage
\includepdf[pages=-, pagecommand={\thispagestyle{fancy}}, offset=9.3mm -10mm, noautoscale=true, scale=0.897]{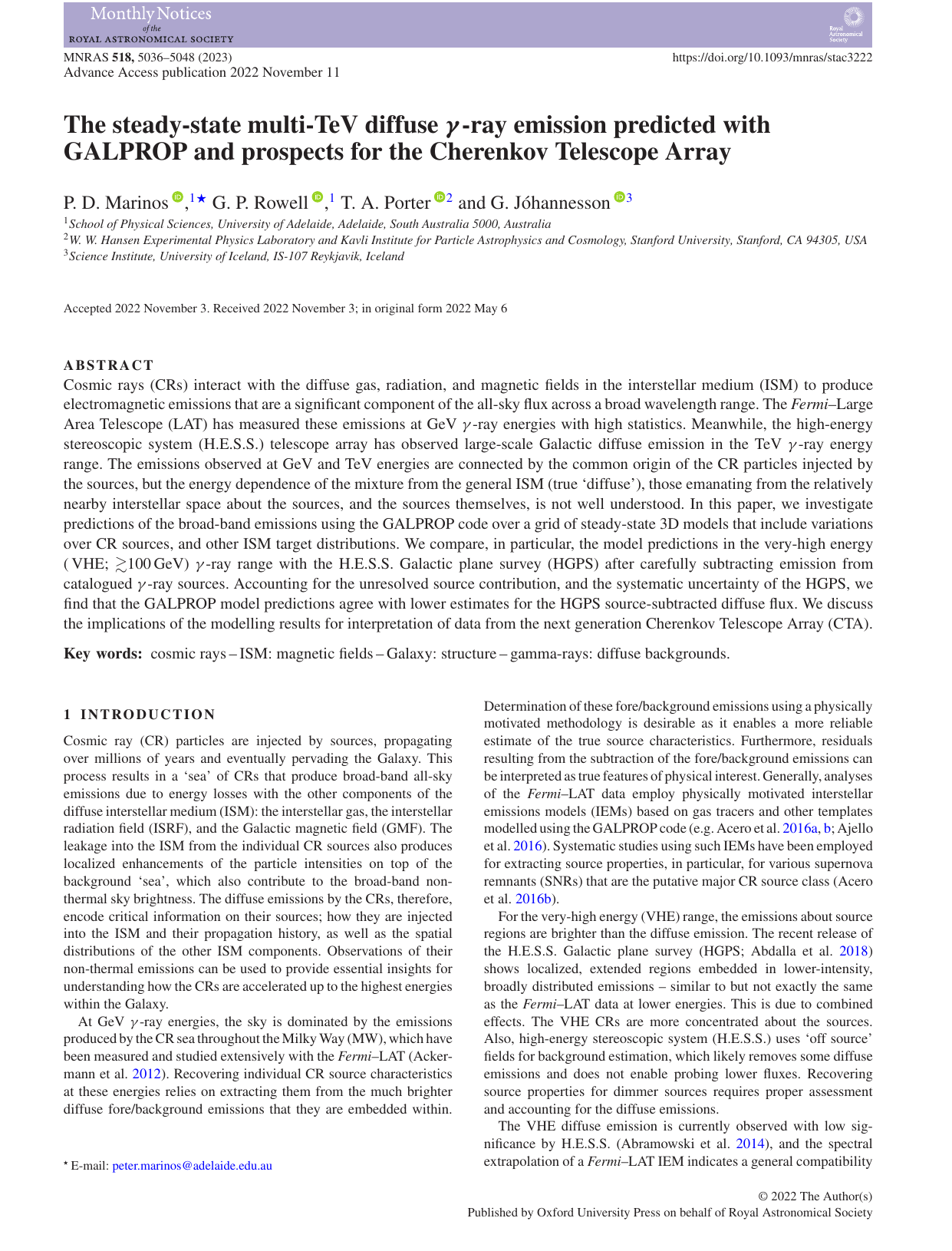}
    \chapter{Time-Dependent CR Injection} \label{chap:time dep}

The results in the previous two chapters utilise a steady-state~(time-independent) solution for the CR injection in the MW. The steady-state solution assumes that CRs are injected into the MW along a smoothly varying spatial distribution that does not evolve with time.

However, CRs are accelerated in localised, discrete region such as SNRs, PWNe, etc.~(see \autoref{ssect:source classifications} for discussions on the various source classes).
The steady-state assumption is valid for the >1\,GeV CR protons (and heavier nuclei) as they are able to diffuse >1\,kpc before losing their energy. Additionally, the cooling distances for the hadrons increases with their kinetic energy~(see \autoref{fig:cooling times/lengths}).
Conversely, CR electrons with energies greater than $\sim$10\,TeV have cooling times <1\,kyr and cooling distances <200\,pc, with the cooling times and distances decreasing with kinetic energy. Therefore, CR electrons in the TeV energy regime will not diffuse large distances away from their acceleration sites. It is not valid to assume that the CR electron density for kinetic energies $E_{e^{-}}>10$\,TeV is constant over time nor smoothly varying across the spatial axes.

This chapter will begin with a detailed discussion on the necessity of the time-dependent solution to CR injection. Also provided is a description of the methods \GP{} utilises to compute the time-dependent CR injection, and the computational resources required. Due to the additional complexity of the time-dependent models, tests are performed to ensure that \GP{} has been set up correctly and is providing accurate and reliable results.

\section{Motivation for a Time-Dependent Solution}

For the time-independent/steady-state solution in \GP, all CRs are injected into the MW by a smoothly varying source distribution~(as discussed in \autoref{sect:source dist.}). The source term for a given CR species, $Q(\vec{r},\,p)$, is the product of the source distribution and the CR injection spectra~(see \autoref{eq:source term}). The steady-state solution assumes that all CR species and energies are diffuse at all locations in the MW.

In reality, CRs are injected into the MW by a variety of objects~(see \autoref{ssect:source classifications}) with finite sizes and lifetimes. Considering the placement of CR sources in both time and space requires using a time-dependent simulation.
The injection and propagation of CRs in the ISM by individual, discrete sources with finite lifetimes has been investigated extensively~\citep[e.g.][]{LingenfelterR.1969,RamatyR.1970,LingenfelterR.1971,LingenfelterR.1973,LeeM.1979}. Additionally, \GP{} has been capable of injecting CRs from discrete sources using the full time-dependent solution over two decades~\citep[e.g.][]{StrongA.2001a,StrongA.2001b,SwordyS.2003}.
However, the limited data statistics electron and positron observations into the TeV regime, coupled with the significant increase in the required computational resources for the time-dependent solution, has prevented widespread use until recently.

As improved observational results~\citep[e.g.][]{AharonianF.2009,AmbrosiG.2017,KerszbergD.2017,RecchiaS.2019} have been acquired it has become increasingly apparent that a smoothly varying source distribution is unable to reproduce some features in the local CR flux above 1\,TeV. Recent studies that place individual CR accelerators within 1\,kpc of Earth have successfully reproduced the higher-energy features in the local CR electron flux~\citep[e.g.][]{KobayashiT.2004,PtuskinV.2006,MertschP.2011,LiuW.2015,MiyakeS.2015,MertschP.2018,EvoliC.2021,SudohT.2023}.
A time-dependent solution is also proving to be a requirement for accurate calculations of the \graya{} emission above 1\,TeV due to the increasing contribution of the IC emission~\citep[e.g.][]{MertschP.2018,PorterT.2019,MarinosP.2023}.

A variety of recent observational \graya{} results in the >10\,TeV energy range from HAWC, LHAASO, and \tibet{} further necessitate a time-dependent solution.
\begin{itemize}
    \item Results from HAWC~\citep{AbeysekaraA.2017b} and LHAASO~\citep{ZhaoS.2021} show that the localisation of \graya{} emissions nearby individual CR sources is stronger for energies $\gtrsim$10\,TeV, suggesting that the \graya{} emission may be partly leptonic in origin.
    \item HAWC results show that the observed \graya{} emission around 50\,TeV is not aligned with high-density gas clouds~\citep{AbeysekaraA.2020}, further suggesting that the \graya{} emission may not be purely hadronic in origin.
    \item LHAASO-KM2A observations of diffuse \grays{} in the 10\,TeV to 1\,PeV energy range for Galactic longitudes $15^\circ<\mathscr{l}<235^\circ$ have fluxes 2--3 times larger than expected from hadronic CRs interacting with the ISM gas~\citep{CaoZ.2023}.
    \item \tibet{} results indicate that many of the \grays{} detected within their $25^\circ<\mathscr{l}<100^\circ$, $|\mathscr{b}|<5^\circ$ window~\citep{AmenomoriM.2021} do not coincide with known VHE \graya{} sources, suggesting that the majority of >100\,TeV \graya{} emission is not from a diffuse collection of CRs~\citep[e.g.][]{DzhatdoevT.2021,FangK.2021}.
\end{itemize}
Hence, the HAWC, LHAASO, and \tibet{} results all suggest that either the hadronic or leptonic component of the \graya{} flux is underpredicted.
Recent results from IceCube~\citep{AbbasiR.2023} can be used to place limits on the hadronic component as neutrinos only arise from the hadronic interactions~(see \autoref{ssect:hadronic emission}). Comparisons between the expected neutrino flux and current steady-state models shows agreement~\citep[e.g.][]{AbbasiR.2023,VecchiottiV.2023}; however, the observational uncertainties are large and more modelling work is required.
Given the HAWC, LHAASO, \tibet, and IceCube results, it is likely that the >10\,TeV \graya{} emission is partly leptonic in origin. Therefore, a time-dependent solution to the CR transport equation is required to connect the \graya{} emissions from GeV to PeV energies.

Modelling the IC emission into the PeV regime requires sources of VHE electrons.
Electrons are likely accelerated to PeV energies by PWNe and PWN halos. PWN halos are suggested to be a major contributor to the unresolved source contribution to the \hess{} and HAWC \graya{} emission~\citep[e.g.][]{CataldoM.2020,MartinP.2022}, and a collection of PWNe could explain the discrepancy between models and the \tibet{} flux~\citep{VecchiottiV.2022}. As the sensitivity of HAWC, LHAASO, and \tibet{} improves, many sources that can accelerate electrons to energies above 100\,TeV are being discovered~\citep[e.g.][]{CaoZ.2021a,CaoZ.2021b,BurgessD.2022,AbeS.2023,ParkJ.2023a,ParkJ.2023b,WooJ.2023}.
\section{Description of the Time-Dependent Solution} \label{sect:time dep description}

The time-dependent solution in \GP{} has three major changes in its operation compared to the steady-state case: the source distribution, the source parameters, and the time-stepping method.

\subsection{Time-Dependent Source Distribution} \label{ssect:TDD source dist.}

The time-independent source term, $Q(\Vec{r},\,p)$, was given in \autoref{eq:source term}, with the source distributions, $\rho(\Vec{r})$, discussed in \autoref{sect:source dist.}. For the steady-state mode, the CR source term is the product of the source distribution and the CR injection spectra. Hence, the source distribution effectively defines the amplitude of the injection spectra for each location in the MW. The result is a smoothly varying spatial distribution injected CRs.

In the time-dependent mode the CRs are injected by individual sources that are created throughout the MW. The source distribution is altered such that it describes the probability of a source being created at a given position in space and time -- i.e.~the amplitude of the source distribution represents the likelihood of a source being placed at a given position. The probability of a source being created at~$\Vec{r}$ for a given timestep is described by:

\begin{align}
    P(\Vec{r},\,t) &\propto \frac{N_{\mathrm{sources}}}{t_{\mathrm{final}}} \Delta t \rho(\Vec{r}) \label{eq:tdep probability density} \\
    \int_{V}P(\Vec{r})\,\mathrm{d}V &= 1
\end{align}

\noindent
where the total probability is normalised such that the volume integral is equal to one, $N_{\mathrm{sources}}$ is the number of sources created throughout the simulation, and $t_{\mathrm{final}}$ is the length of time of the simulation. The factor $N_{\mathrm{sources}}/t_{\mathrm{final}}$ represents the average number of sources created per unit time. Once a source has been created, it will continue to inject CRs into the ISM throughout its lifetime given by $t_{\mathrm{life}}$.

Let $\mathcal{C}(\Vec{r},\,t)$ be defined such that if a source is placed at the location $(\Vec{r},\,t)$, then $\mathcal{C}(\Vec{r},\,t)=1$ for the lifetime of the source. If no source is located at $(\Vec{r},\,t)$, then $\mathcal{C}(\Vec{r},\,t)=0$. The time-dependent source term is then given by:

\begin{align}
    Q(\Vec{r},\,p,\,t) &=
    \begin{cases}
        \dfrac{\mathrm{d}\mathcal{N}_{\mathrm{CR}}}{\mathrm{d}p}, & \mathrm{if} \ \mathcal{C}(\Vec{r},\,t)=1 \\
        0, & \mathrm{if} \ \mathcal{C}(\Vec{r},\,t)=0
    \end{cases}
\end{align}

\noindent
where the subscript `CR' denotes that the variable can describe any CR particle. The CR spectra~($\mathrm{d}\mathcal{N}_{\mathrm{CR}}/\mathrm{d}p$; described in \autoref{ssect:parameter optimisation}) are given by broken power laws as functions of the CR momentum, with the spectral parameters for protons, electrons, and helium given in \autoref{tab:SA pars}.

\subsection{Time-Dependent Source Parameters} \label{ssect:tdep source pars}

Parameters defining the lifetimes and number of sources must also be defined for a time-dependent model.
The lifetime of the CR sources is given by $t_{\mathrm{life}}$, and the average time between sources being created is given by $t_{\mathrm{interval}} = t_{\mathrm{final}} / N_{\mathrm{sources}}$.

Estimates on the lifetime of CR accelerators vary significantly depending on the observational technique and the source type being considered. For SNRs, which make up the majority of the assumed CR accelerators, the estimated lifetimes vary between 30--300\,kyr~\citep{BlasiP.2012}. PWNe can accelerate CRs for longer than 100\,kyr, though the older PWNe contribute much less to the Galactic CRs~\citep{GiacintiG.2020}. Star clusters are also a candidate for the acceleration of CRs up to PeV energies~\citep{BykovA.2014} and may have lifetimes up to 10\,Myr~\citep{AharonianF.2019,BykovA.2019}.

Estimates on the rate of SNe vary depending on the observational technique. For example, estimates from the radioactive decay of \textsuperscript{26}Al range between one SNe every 35--125\,yr~\citep{DiehlR.2006}. Averaging over many observational techniques provides a tighter constraint of one SNe every 45--85\,yr~\citep{RozwadowskaK.2021}. Statistical modelling from~\citet{MertschP.2018} suggests a lower rate of one SN every 500\,yr to explain the CR electron spectrum above 1\,TeV. As pulsars~(and any associated PWNe) are formed by a subset of SNe, the creation rate of pulsars must be less than that for SNRs. However, the pulsar creation rate has no constraint tighter than that for SNe~\citep[e.g.][]{KeaneE.2008}.

The CR normalisation condition~(see~\autoref{ssect:parameter optimisation}) is not impacted by either the source lifetime or the source creation rate. Hence, $t_{\mathrm{life}}$ and $t_{\mathrm{interval}}$ are free parameters. Currently, all time-dependent sources in \GP{} have the same lifetime, creation rate, and injection spectra -- impulsive and continuous sources are not modelled separately, with no distinction between SNRs, PWNe,~etc.~being defined.
Hence, \GP{} only considers a single `average' source class.
Previous time-dependent \GP{} results use the values $t_{\mathrm{life}}=100$\,kyr and $t_{\mathrm{interval}}=100$\,yr~\citep{PorterT.2019}, which are representative of SNR-class CR accelerators. The effects of altering $t_{\mathrm{life}}$ and $t_{\mathrm{interval}}$ will be explored later in \autoref{chap:paper 2}.

\subsection{Time-Dependent Timesteps} \label{ssect:time-dep time grid}

To capture diffusion across many differing timescales the steady-state solution used a variable timestep~(see \autoref{ssect:time-indep time grid}).
However, the variable timestep method can only be applied if the CR injection is constant throughout time.
Hence, the time-dependent mode of \GP{} utilises a constant timestep, $\Delta t$, where the timestep must be less than the shortest cooling time of any particle being simulated. Capturing diffusion on longer scales then requires repeatedly solving the transport equation for the given timestep while simultaneously evolving the source distribution.

\GP{} will propagate CRs from the initial time, $t_{\mathrm{initial}}$, until the final time, $t_{\mathrm{final}}$. The number of timesteps is then given by $(t_{\mathrm{final}} - t_{\mathrm{initial}}) / \Delta t$.
The CR density in the MW is set to zero, with CR accelerators being placed via the method described in \autoref{ssect:TDD source dist.}.
The CR density at a given location will increase until the CR injection and energy-loss rates are balanced, after which the CR density will vary around some equilibrium value~(i.e.~steady-state value).
For 1\,GeV protons, reaching the steady-state values requires 500\,Myr of simulation time~\citep{PorterT.2019}.
Accurately capturing the cooling of the electrons up to 1\,PeV requires a timestep on the order of 100\,yr. Therefore, the time-dependent solution requires a total of $5 \times 10^{6}$~timesteps. For comparison, a similar steady-state simulation requires only 500~timesteps.

To reduce computation time, \GP{} has implemented a ``warm-start'' functionality\footnote{The warm-start parameters, and how to use them to improve the computational efficiency, is explained in \autoref{sect:time coordinate setup}.}. \GP{} can be operated with a coarse timestep~(e.g.~$\Delta t > 1$\,kyr) to allow all CRs to reach their steady-state values, after which the timestep can be reduced to the size required to accurately capture the cooling of the CR electrons~(e.g.~$\Delta t = 100$\,yr).
\section{Time-Dependent GALPROP Simulations: Trials and Optimisations}

Optimisations and considerations can be made to improve the runtime and storage requirements of the time-dependent \GP{} simulations. This section details these optimisations, as well as any additional considerations required for the time-dependent solution. Tests are performed to ensure the validity of the results.

\subsection{Length of Simulation}

Before interpreting any results from the time-dependent simulations, it is important to ensure that all CRs within the MW have reached the steady-state values in all locations of interest and for all included CR species and kinetic energies. 
The time taken to reach the steady-state flux will depend on the energy of the CR, the source creation rate, the source lifetime, the ISM distributions~(such as the ISM gas, the ISRF, and the GMF), and the source distribution.

\begin{figure}
    \centering
    \includegraphics[width=\textwidth]{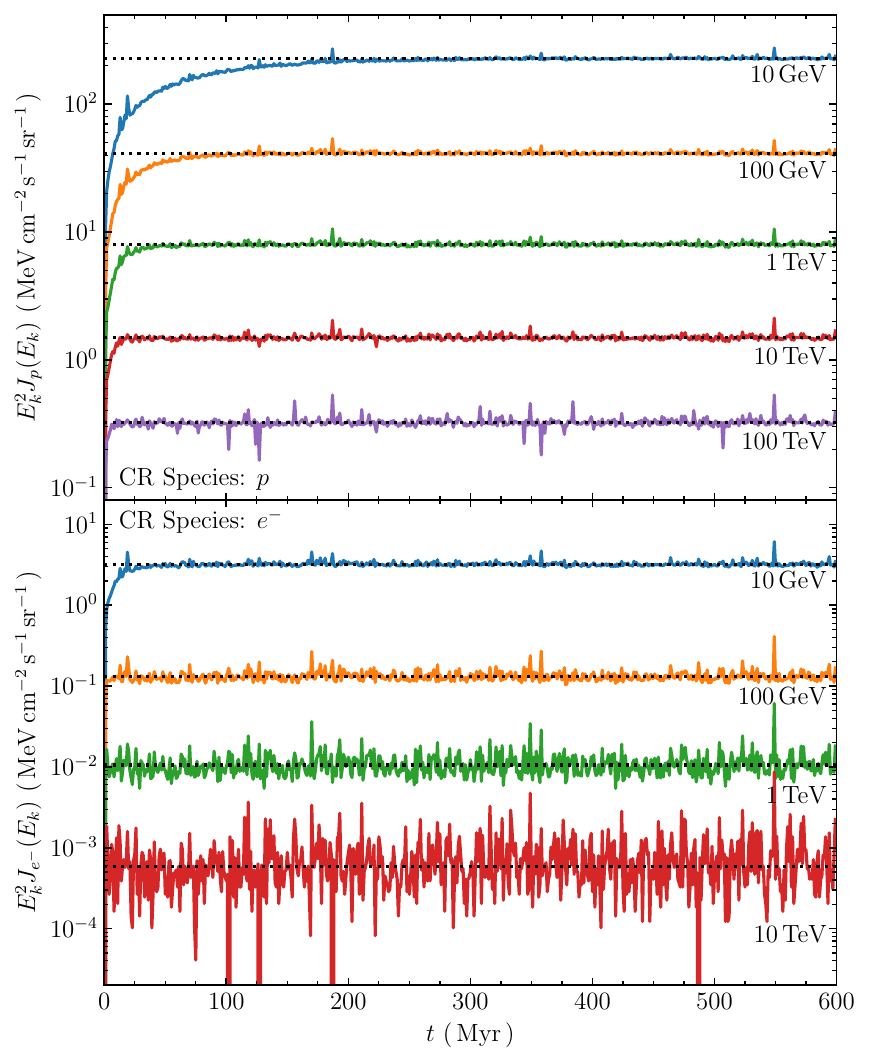}
    \caption{The CR proton~(top) and electron~(bottom) flux as a function of time at the Solar location for 10\,GeV~(blue), 100\,GeV~(orange), 1\,TeV~(green), 10\,TeV~(red), 100\,TeV~(purple). The time-independent~(steady-state) fluxes are shown by the dotted black lines. }
    \label{fig:simulation length test}
\end{figure}

\autoref{fig:simulation length test} shows the CR flux at the Solar location as a function of time for protons and electrons at various energies between 10\,GeV and 100\,TeV. The steady-state/time-independent fluxes are included for comparison. As the CR protons cool significantly more slowly than the CR electrons~(e.g.~see \autoref{fig:cooling times/lengths}) they take longer to reach their steady-state values. Additionally, as the lower-energy protons propagate more slowly than the higher-energy protons, they require more time to reach their steady-state values.

A sliding window analysis can be utilised to calculate the time taken for the time-dependent values to be considered as varying around the steady-state values. For a given CR species and energy, the flux is considered to have reached equilibrium if the average over a 10\,Myr period is greater than or equal to the steady-state value.
For the given example, the 100\,GeV CR protons require $\sim$300\,Myr to reach their steady-state values, while the 10\,GeV protons require $\sim$400\,Myr. The 10\,GeV CR electrons require only 50\,Myr to reach their steady-state values.

Once the time-dependent results have reached steady state, the observed fluctuations are then due entirely to the assumed spatial and temporal variation of the CR sources. \autoref{fig:simulation length test} also shows that the relative variation in the CR density increases with the kinetic energy of the CRs, with the 10\,TeV electrons varying by over a factor of ten at the Solar location. The relative variation in the CR flux, and the impacts on the Galactic diffuse \graya{} flux, will be explored further in \autoref{chap:paper 2}.

\subsection{Sampling Interval}

\GP{} will output the CR flux for each CR particle found in each of the propagation cells, with these outputs being used to calculate the \graya{} skymaps. In the time-independent mode, a single output is created at the end of the simulation. For the time-dependent mode the output is created every $t_{\mathrm{out}}$ timesteps~(referred to as the output interval or sampling interval). The sampling interval controls how much data is written during a time-dependent simulation and must be defined such that the variations on the scales of interest are captured.

The Nyquist condition states that to observe variation of a given frequency, $f_{N}$, the sampling frequency~($f_{S}$) must be at least double $f_{N}$,~i.e.~$f_{S} \geq 2 f_{N}$.
For the time-dependent solution there are two timescales of interest: how quickly the CR flux at the Solar location changes due to a CR accelerator being placed within a few hundred parsecs of Earth, and the variation on the timescale of the source lifetime.

\begin{figure}[t]
    \centering
    \includegraphics[width=\textwidth]{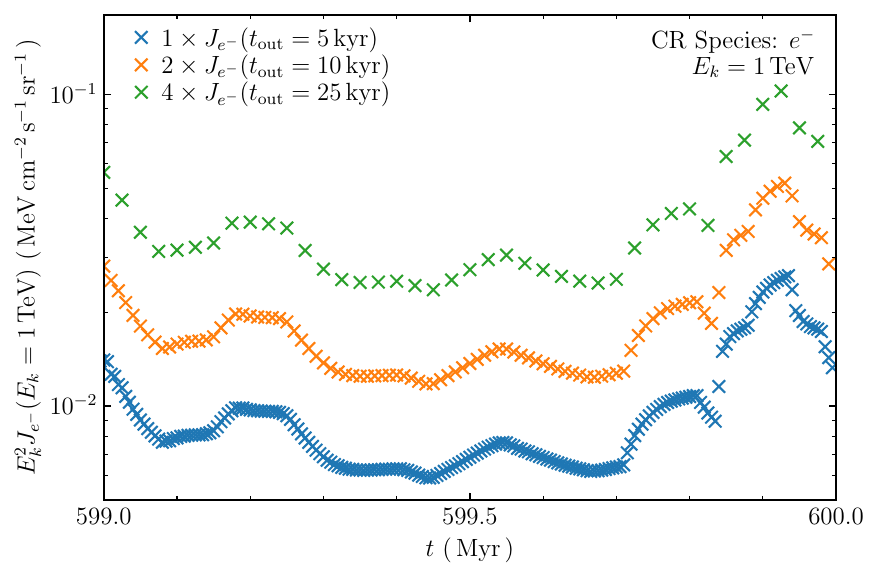}
    \caption{The local CR electron flux at 1\,TeV as a function of time for the output intervals $t_{\mathrm{out}}=5$\,kyr~(blue crosses), $t_{\mathrm{out}}=10$\,kyr~(orange crosses), and $t_{\mathrm{out}}=25$\,kyr~(green crosses). The time-series values have been multiplied by the given factors to increase contrast.}
    \label{fig:Sample test}
\end{figure}

From \autoref{fig:Sample test}, it can be seen that a single CR source being created near the Solar location increases the CR electron flux at 1\,TeV by a factor of two over a timeframe of 10--100\,kyr.
Therefore, capturing the rise and fall of the CR flux caused by the individual sources requires a sampling interval $\leq$5\,kyr. A sampling interval of 5\,kyr more accurately captures the details~(i.e.~the peaks and troughs) in the time series compared to the 10\,kyr sampling interval.

\autoref{fig:Sample test} also shows that any single CR source increases the local CR flux for approximately 100--200\,kyr. For the example shown in \autoref{fig:Sample test}, the source lifetime was defined as $t_{\mathrm{life}}=100$\,kyr. Therefore, capturing individual sources contributing to the local CR flux requires a sampling interval approximately half of the source lifetime. 

\subsection{Simulation Timestep} \label{ssect:timestep tests}

The computational resources required by \GP{} is proportional to the number of timesteps. However, as discussed in \autoref{ssect:time-dep time grid}, the timestep~($\Delta t$) must be shorter than the shortest cooling time~(see \autoref{fig:cooling times/lengths}) of any CR particle being propagated such that the cooling is accurately captured in the solution. The energy-loss calculations become approximated if the timestep is too large, with the impacts most significant nearby to the CR sources.

\begin{figure}
    \centering
    \includegraphics[width=\textwidth]{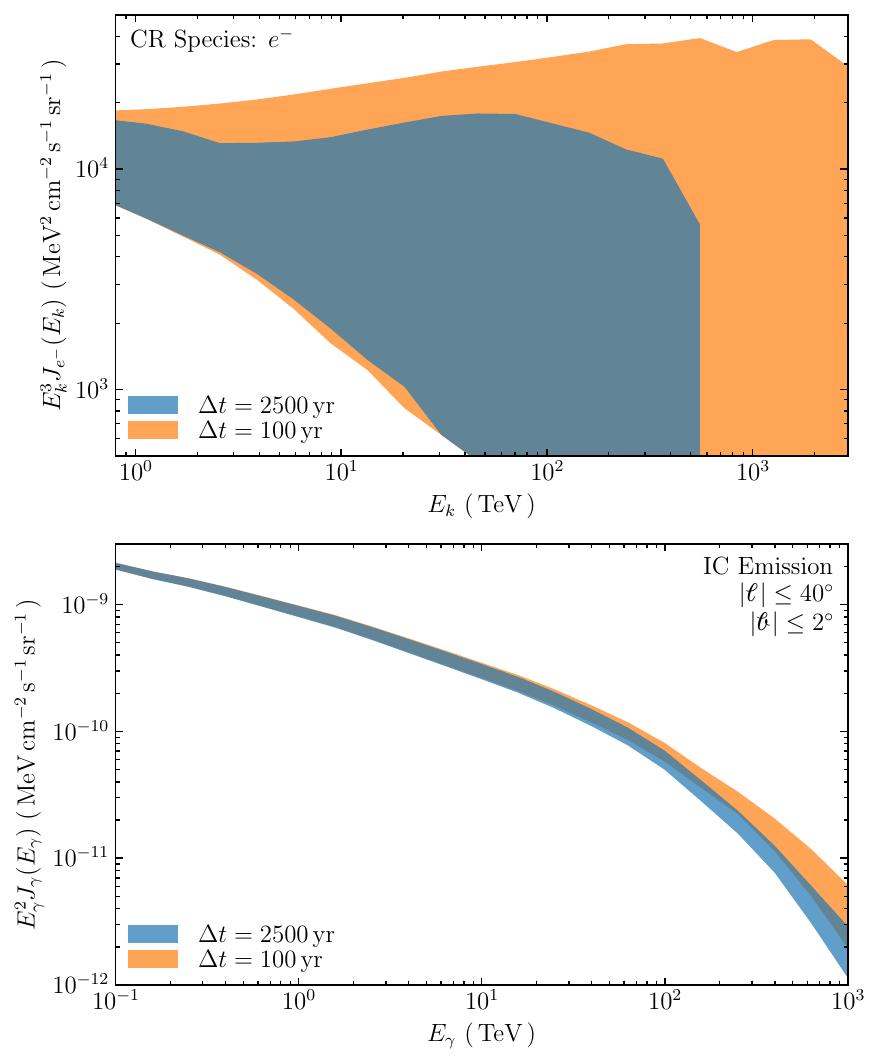}
    \caption{CR electron spectral envelopes at the Solar location~(top) and the IC \graya{} emission spectral envelopes towards the GC region~($|\mathscr{l}| \leq 40^{\circ}$, $|\mathscr{b}|<2^{\circ}$; bottom) for $\Delta t=2500$\,yr~(blue) and $\Delta t=100$\,yr~(orange). All envelopes use $t_{\mathrm{out}} = 25$\,kyr and are computed over a 5\,Myr simulation period.}
    \label{fig:timestep test}
\end{figure}

CR electron spectral envelopes are shown in \autoref{fig:timestep test} for $\Delta t=2500$\,yr and $\Delta t=100$\,yr, which are the approximate cooling times for a $\sim$1\,TeV and $\sim$1\,PeV electron, respectively. Both envelopes use $t_{\mathrm{out}} = 25$\,kyr and are calculated over the entire range of results,~i.e.~from the minimum to the maximum values.
CRs are injected up to 10\,PeV and use the following ISM distributions: the SA100 source distribution, the R12 ISRF model, and the PBSS GMF model~(see Sections \ref{sect:source dist.}, \ref{sect:ISRF}, and \ref{sect:GMF}, respectively).

Above 1\,TeV the CR flux envelopes begin to diverge. 
If $\Delta t > t_{\mathrm{cool}}$ then the energy losses will be approximated. For ISM regions with low density, the cooling of the CR electrons will be under-estimated. Conversely, the energy losses will be over-estimated in regions with high ISM density. As the sources are placed in regions of high ISM density more often~(see Sections \ref{sect:source dist.} to \ref{sect:GMF}), the energy losses are typically overestimated.

At 200\,TeV the difference in the upper bound of the electron spectra can be as large as a factor of two.
Above $\sim$500\,TeV no CR electrons reach the Solar location for $\Delta t = 2500$\,kyr.
For $\Delta t = 100$\,yr the local electron flux is computed accurately up to $E_{e^{-}}=1$\,PeV as the cooling time of a 1\,PeV electron is on the order of 100--500\,yr.
Decreasing the timestep further to $\Delta t=10$\,yr was found to have no significant impact on the CR electron spectra below 1\,PeV. For kinetic energies $E_{e^{-}}>1$\,PeV the uncertainty in the injection spectrum is larger than the variation between various timestep sizes.

\autoref{fig:timestep test} also shows the spectral envelopes for the IC \graya{} emission for the Galactic centre~(GC) region~($\mathscr{l}<40^{\circ}$, $|\mathscr{b}|<2^{\circ}$).
While the local electron spectra begin to diverge at $E_{e^{-}} = 1$\,TeV, the IC \graya{} emission towards the GC region for $\Delta t=2500$\,yr and $\Delta t=100$\,yr are approximately equal for \graya{} energies $E_{\gamma} < 100$\,TeV.
The IC \graya{} emission is less sensitive to changes in $\Delta t$ compared to the local electron spectrum as the \graya{} flux is calculated via a line-of-sight integral~(see \autoref{ssect:GP flux calculation}).
However, for the $\Delta t = 2500$\,yr timestep and energies $E_{e^{-}} > 100$\,TeV the over-estimated cooling often approximates the electron flux to zero.
Hence, there are systematically fewer >100\,TeV CR electrons in the MW, and the resulting IC \graya{} emission produced by the $E_{e^{-}} > 100$\,TeV electrons is underestimated.
For the 1\,PeV IC \graya{} emission, the flux is underestimated by a factor of~$\sim$2 for $\Delta t=2500$\,yr.

The cooling time of the CR protons is $\gtrsim$100\,kyr and independent of their kinetic energies~(see \autoref{fig:cooling times/lengths}). Hence, decreasing the timestep beyond 10\,kyr does not impact the local CR proton spectrum. Furthermore, as heavier nuclei have longer cooling times than the CR protons, all hadronic spectra are simulated correctly for all timestep sizes $\Delta t \lesssim 100$\,kyr.

\subsection{Spatial Grid}

Both the computation time and storage requirements for each individual timestep are proportional to the number of propagation cells along each axis. As discussed in \autoref{ssect:grid}, the propagation cells must be smaller than the shortest cooling distance of any included CR particle. Additionally, the size of the propagation cells may limit the resolution of the final \graya{} skymaps.

The CR hadrons have larger cooling distances than the CR electrons across most kinetic energies and Galactic conditions~(see \autoref{fig:cooling times/lengths}). The minimum cooling distances of CR electrons is approximately 10--300\,pc at 1\,PeV depending on the ISM conditions. For CR protons and other hadrons, the minimum cooling distances are approximately 60--$10^{3}$\,pc at 10\,GeV.

\begin{figure}
    \centering
    \includegraphics[width=\textwidth]{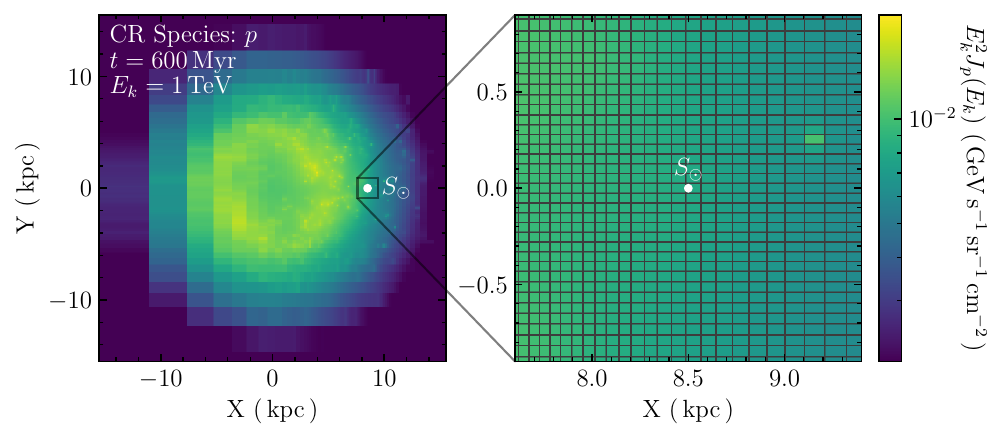}
    \includegraphics[width=\textwidth]{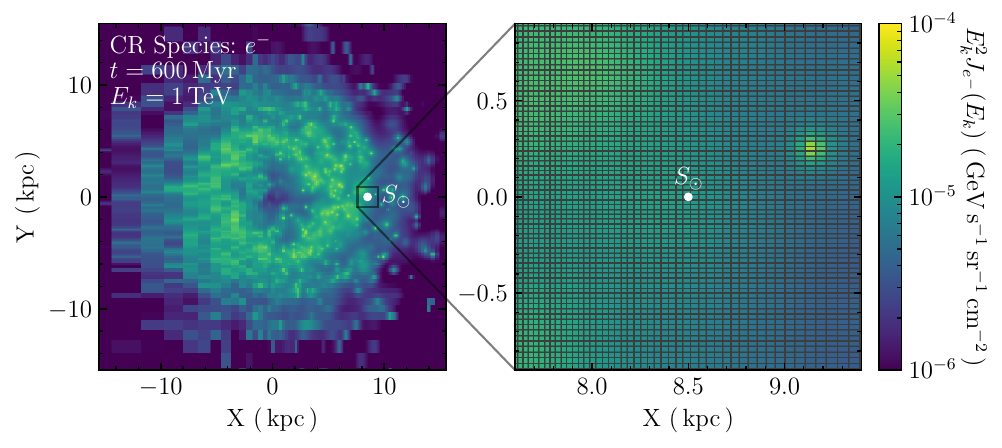}
    \caption{The CR energy density along the XY plane~(i.e.~$Z=0$), at the simulation epoch $t=600$\,Myr, and kinetic energy $E_{k}=1$\,TeV. The CR protons~(top) and electrons~(bottom) are shown for the whole MW~(left) and around the Solar location~(right). The CR energy density is shown in units of GeV\,s$^{-1}$\,sr$^{-1}$\,cm$^{-2}$ with the proton and electron energy densities shown on a separate scale. The electrons were propagated at a finer spatial resolution to capture their short cooling distances at high energies~(see text). The cell size increases with distance from the Solar location~(see \autoref{ssect:grid}), with the cell size for the proton and electron simulations increasing at different rates. A CR source can be seen in the right-hand panels at $(X,\,Y) \approx (9.1,\,0.3)$\,kpc for both protons and electrons.}
    \label{fig:Grid test}
\end{figure}

For time-dependent simulations the computation speed and the required storage can be improved by separating the CR hadrons and electrons into separate propagation grids, with the leptons having a finer spatial resolution.
As the hadrons diffuse further, they can be simulated at half the resolution across each of the $X$, $Y$, and $Z$ axes. Hence, the hadronic simulation will see a storage/speed improvement of a factor of $2^{3}$ when compared to simulating all particles together.
As the hadrons are no longer being simulated at a high spatial resolution, the simulation with a high spatial resolution will see a runtime and storage improvement by a factor equal to the number of unique species of CR hadrons being propagated.
\autoref{fig:Grid test} shows the hadrons and electrons being propagated on differing propagation grids.

\subsection{CR Source Placement in Space and Time}

To ensure consistency between \GP{} runs, propagating hadrons and leptons on separate spatial grids and altering the timestep cannot impact the placement of the CR sources. The placement of the CR sources must be identical in both the spatial and temporal axes such that comparisons can be made between the simulations.

In \autoref{fig:Grid test}, a single CR source can be seen near the Solar location. The CR source is located at $(X,\,Y) \approx (9.1,\,0.3)$\,kpc for both the CR hadron and CR electron simulations, demonstrating that the sources are placed consistently independent of the spatial axes.

\begin{figure}
    \centering
    \includegraphics[width=\textwidth]{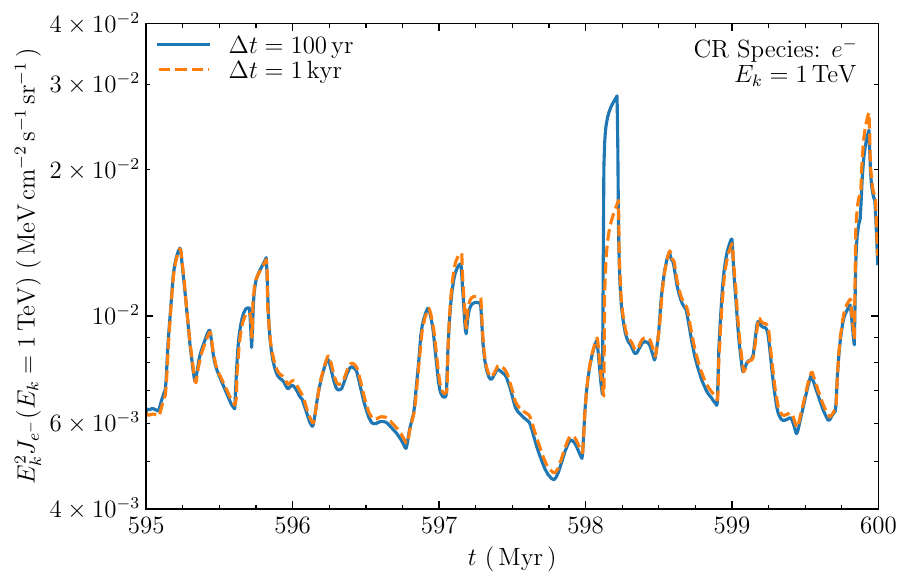}
    \caption{The CR electron flux at 1\,TeV shown at the Solar location as a function of time for the timestep sizes $\Delta t=100$\,yr~(solid blue) and $\Delta t=1$\,kyr~(dashed orange).}
    \label{fig:RNG test}
\end{figure}

\autoref{fig:RNG test} shows the local CR electron flux at 1\,TeV for both $\Delta t=100$\,yr and $\Delta t=1$\,kyr. The peaks in the time series are due to CR sources being placed <400\,pc from the Solar location. Importantly, the peaks and troughs are located at identical times between the two simulations, demonstrating that the sources are created at the same times independent of the timestep size.
Together, Figures \ref{fig:Grid test} and \ref{fig:RNG test} confirm that the placement of the CR sources is independent of the chosen spatial and temporal axes.

It should be noted that \autoref{fig:RNG test} shows a large increase in the CR electron flux at approximately 598.25\,Myr due to a CR source being placed within 100\,pc of the Solar location. There is a significant disagreement between the $\Delta t=100$\,yr and the $\Delta t=1$\,kyr time-series fluxes.
The close proximity of the source highlights the impact of approximating the cooling calculations with too large a timestep, as discussed in \autoref{ssect:timestep tests}.
The CR electron flux, and any temporal detail in the time-series flux, is more accurately calculated for $\Delta t=100$\,yr due to the cooling of the electrons being more accurately captured by the shorter timestep.
    \chapter{Second Paper: Time Variability of Galactic CRs and Gamma Rays} \label{chap:paper 2}

This chapter contains the manuscript titled: ``On the Temporal Variability of the Galactic Multi-TeV Interstellar Emissions'', which was published in the peer-reviewed journal \apj{} in March~2025.

\vspace{2em}

Cosmic rays~(CRs) propagate throughout the Milky~Way~(MW) and interact with the various components of the interstellar medium~(ISM) to produce broadband non-thermal emissions from radio waves to \grays.
As discussed in \autoref{chap:time dep}, calculations of CR propagation and the resulting \graya{} emission throughout the MW typically assume that particle injection is steady across time and space.
In reality, CRs are accelerated by discrete sources with finite lifetimes, such as: supernovae~(SNe), pulsar wind nebulae~(PWNe), and stellar clusters.
As the locations and spectra of all CR sources in the MW are not precisely known, considering injection by individual and discrete regions introduces additional uncertainties to models of CR propagation and \graya{} production.

\GP{} is used in this manuscript to model CR propagation and \graya{} emission over a range of CR source lifetimes and creation rates.
It was found that the variations in the CR and \graya{} fluxes that arise from uncertainties in the source parameters are similar in magnitude to the modelling uncertainty over an ensemble of steady-state simulations found in \autoref{chap:paper 1}.
Despite the influence of the underlying source properties, it was found that the parameters cannot be recovered from observations of the local CR flux or the diffuse \graya{} emission.
When comparing the \GP{} predictions to measurements of the diffuse \graya{} emission, it was found that the models and observations agree within their respective uncertainties up to 1\,PeV.
Additionally, the diffuse \graya{} predictions presented here can be used to estimate the unresolved source fraction in the LHAASO observations.
For future observations by CTA the modelling variation presented here must be considered when using CR propagation codes to estimate the diffuse \graya{} emission.

\blindfootnote{Due to the large number of variables in this thesis, the notation in this published work is not necessarily consistent with the other chapters.}

% Include citations that have no real place in the thesis but are in the paper. This is to ensure that they still appear in the thesis bibliography
\nocite{AbdollahiS.2017,AdrianiO.2023b,AguilarM.2021,AharonianF.2024,AlbertA.2019,BoschiniM.2020,CaoZ.2024,ChenE.2024,FahertyJ.2007,FangK.2018,FujitaY.2010,GabiciS.2007b,LiuY.2024,PorterT.1997,YanK.2024}
% AbdollahiS.2017 : Electron spectrum
% AdrianiO.2023b  : Electron spectrum
% AguilarM.2021   : Electron spectrum
% AharonianF.2024 : Updated measurements during the review process
% AlbertA.2019    : SWGO sensitivity
% BoschiniM.2020  : GALPROP agreement with various datasets
% CaoZ.2024       : Updated measurements during the review process
% ChenE.2024      : LHAASO sources not being completely masked
% Faherty.2007    : Geminga distance
% FangK.2018      : Two-zone models
% FujitaY.2010    : Particles being trapped for some time after source creation
% GabiciS.2007b   : CR accelerator injection spectra changing over time
% LiuY.2024       : More evidence on >PeV electrons that came out after thesis submission
% PorterT.1997    : Local electron flux is not representative of the entire MW
% YanK.2024       : More evidence on >PeV electrons that came out after thesis submission

% Others
\nocite{BoschiniM.2020,GabiciS.2007b}

% Statement of authorship
\clearpage
% \includepdf[pages=-, pagecommand={\thispagestyle{fancy}}, offset=20mm 0mm]{mainmatter/chapter_5/Statement_of_Authorship.pdf}
\begin{figure}[ht]
    \centering
    \includegraphics[width=16cm]{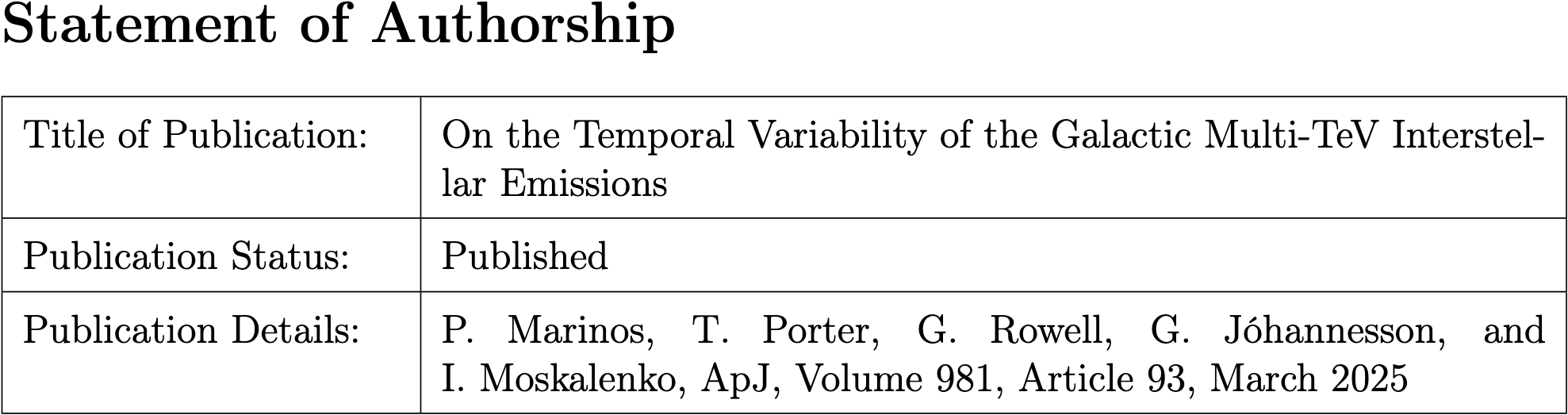}

    \vspace{2em}

    \includegraphics[width=16cm]{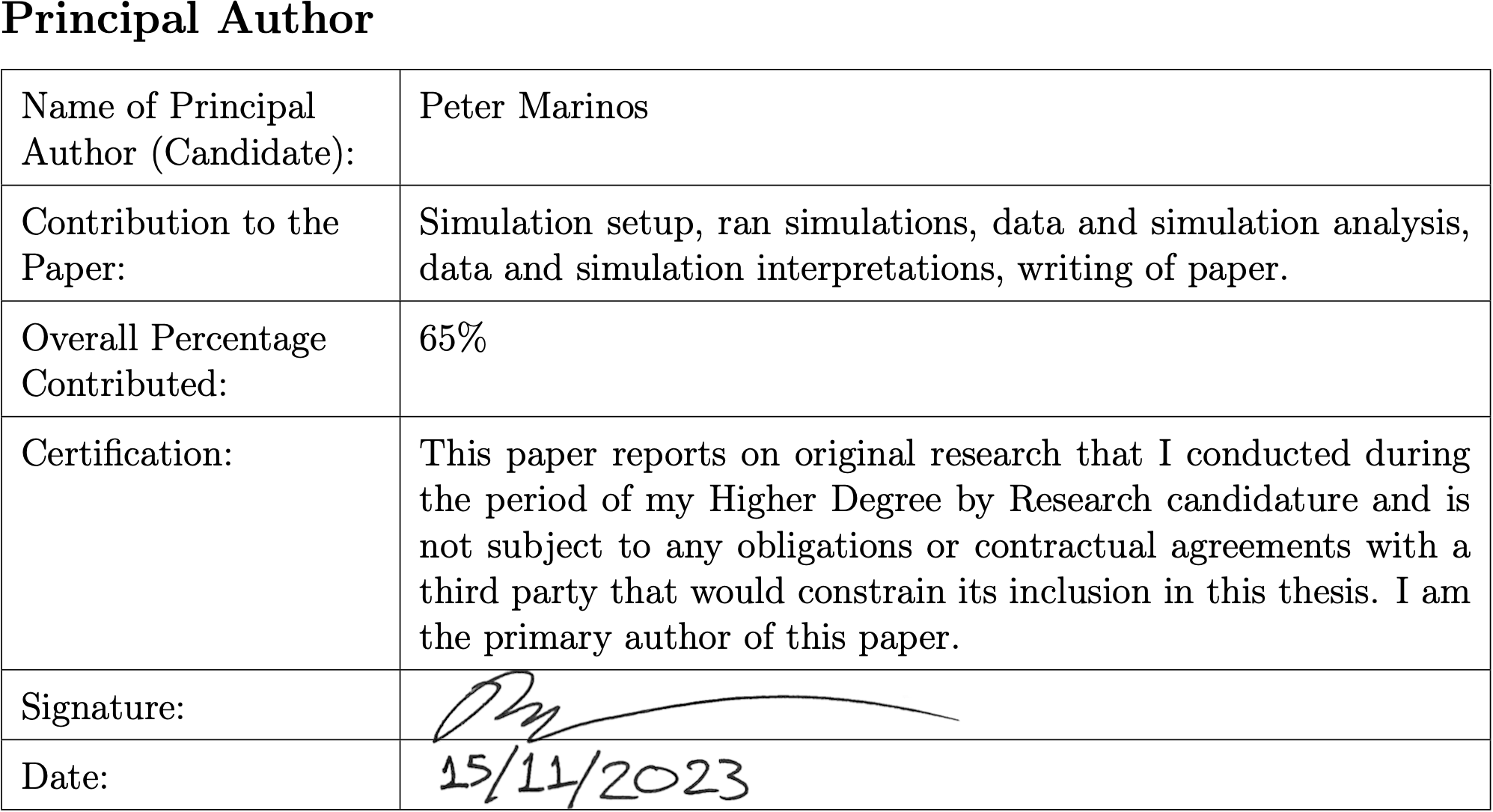}

    \vspace{2em}

    \includegraphics[width=16cm]{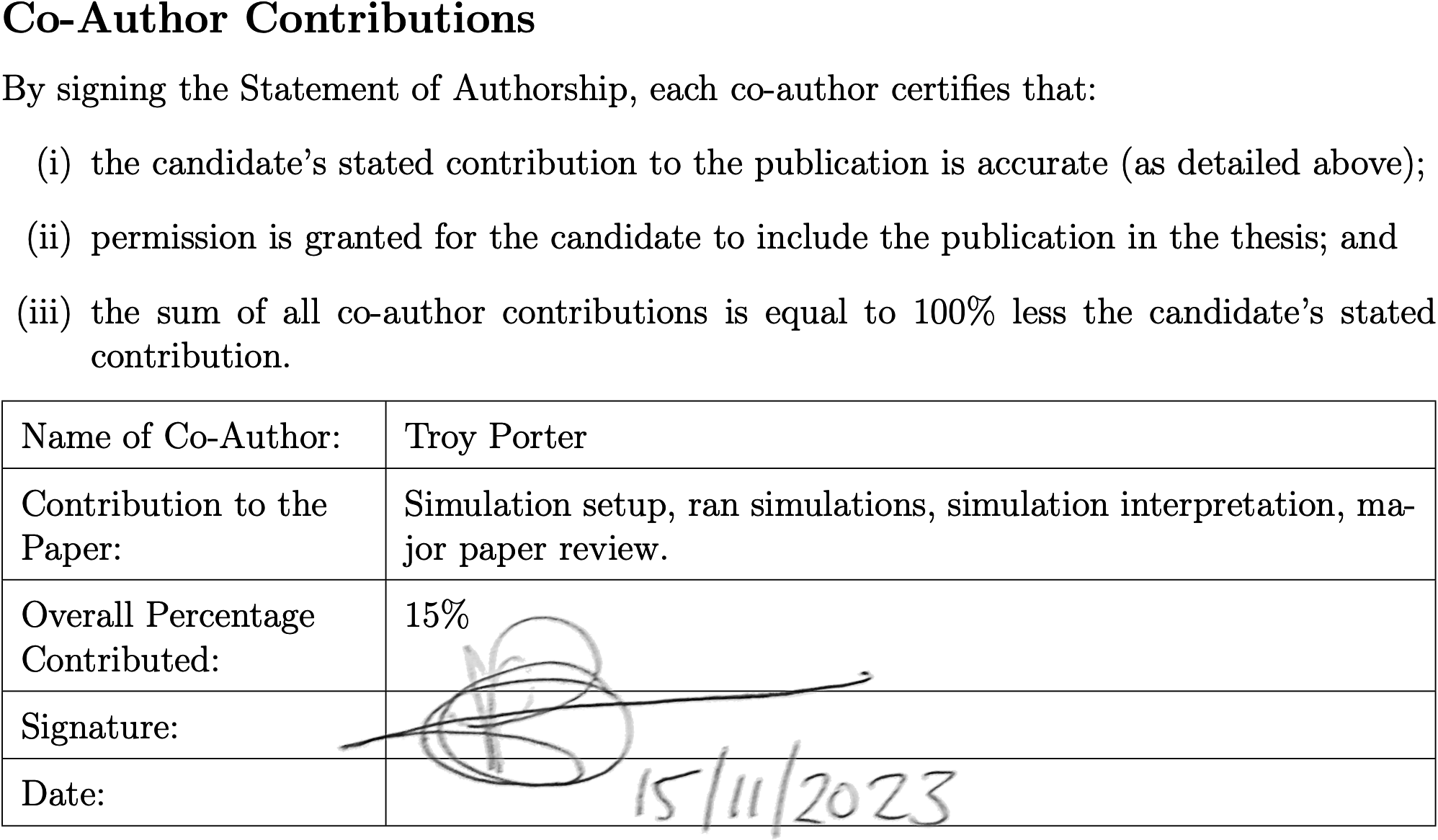}
\end{figure}
\clearpage
\begin{figure}[ht]
    \centering
    \includegraphics[width=16cm]{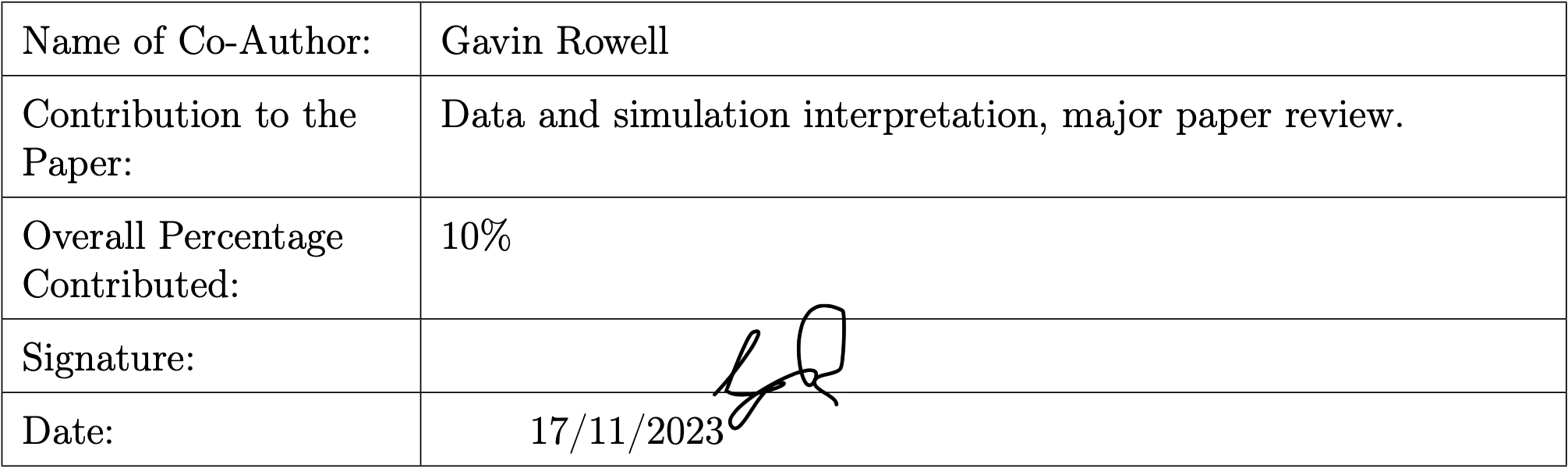}

    \vspace{2em}

    \includegraphics[width=16cm]{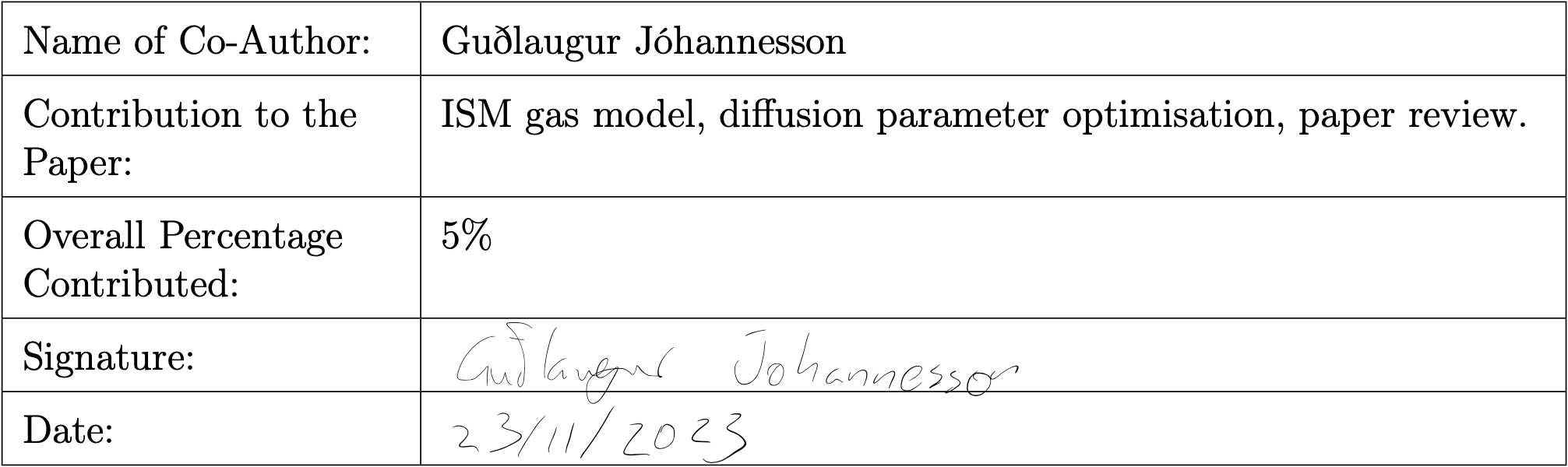}

    \vspace{2em}
    \includegraphics[width=16cm]{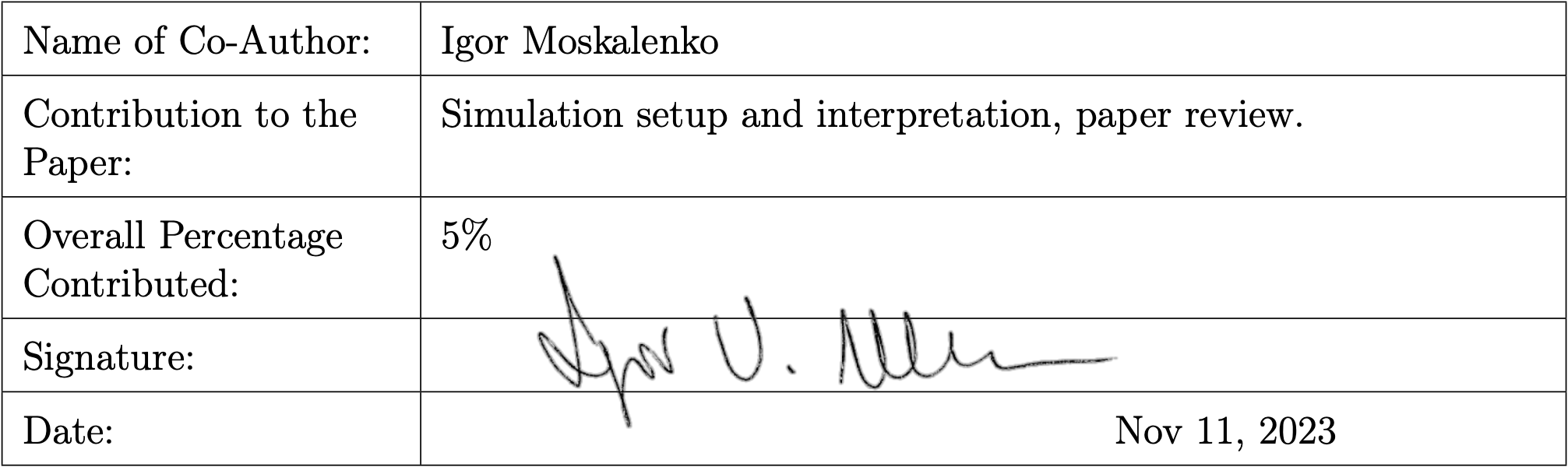}
\end{figure}

% Paper
% Scale = thesis textwidth / paper textwidth
%       = 160mm/7.30in
%       = 8.629
% Offset = binding offset - paper margin * scale
%        = 20mm - 0.7in * 0.888
%     Plus some extra fudging to get it perfect
% Offset is {horizontal offset}, {vertical offset}
\clearpage
\includepdf[pages=-, pagecommand={\thispagestyle{fancy}}, offset=9.8mm -10mm, noautoscale=true, scale=0.8628]{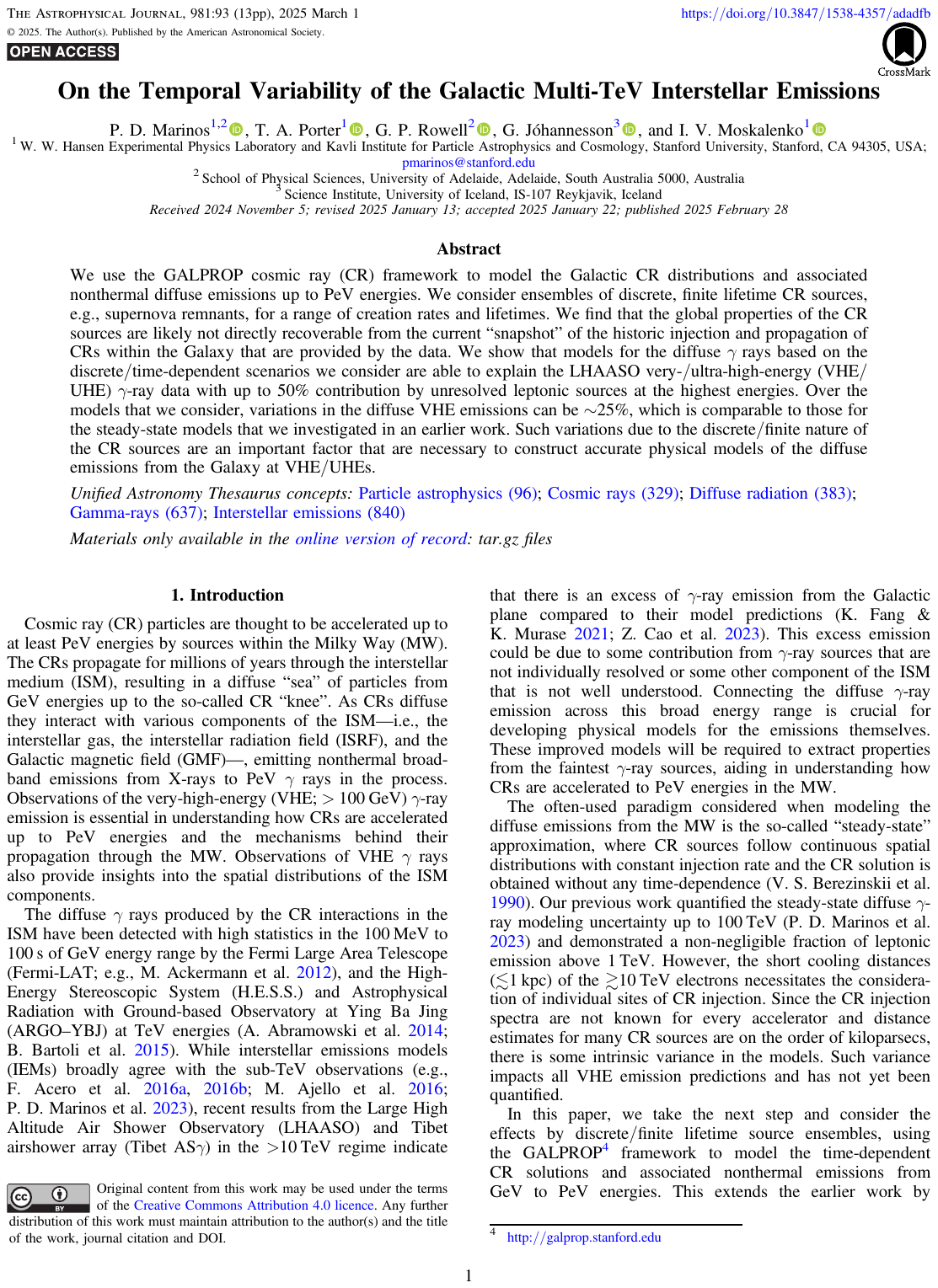}
    \chapter{CTA, the Galactic Plane Survey, and the Diffuse Emission} \label{chap:CTA and gammapy}

The Cherenkov telescope array~\citep[CTA;][]{AcharyaB.2018} will be comprised of two imaging atmospheric Cherenkov telescope~(IACT) arrays that are currently being constructed in both Paranal~(Chile) and La~Palma~(Spain). Situating the observatory across two locations will provide CTA with access to both the southern and northern hemispheres.
To cover a broad range of \graya{} energies the arrays will consist of three telescope sizes: large-sized telescopes~(LSTs; approximately~2\,GeV to~200\,GeV), medium-sized telescopes~(MSTs; approximately~100\,GeV to~10\,TeV), and small-sized telescopes~(SSTs; approximately~1\,TeV to~300\,TeV).

The SSTs and MSTs will have a field-of-view~(FoV) diameter of $8^{\circ}$~(almost double the \hess{} FoV) and the LSTs will have a FoV diameter of $5^{\circ}$. Additionally, CTA will have an angular resolution down to $\sim$1$^{\prime}$~(approximately five times better than \hess) and will be approximately ten times more sensitive than \hess~\citep{ActisM.2011,AcharyaB.2018}. The sensitivity for both the southern and northern sites of CTA are shown in \autoref{fig:iact sensitivities} with comparisons to various other \graya{} facilities. 
Much like \hess, one of the CTA consortium's key science goals is to perform a Galactic plane survey. 
CTA will be extremely sensitive to the details of the Galactic diffuse \graya{} emission~\citep{AceroF.2013}, and the improved FoV compared to \hess{} will enable more robust CR background subtraction methods. Additionally, the angular resolution and sensitivity improvements will aid in resolving fainter and smaller \graya{} sources that were not detected in the HGPS.

CTA's angular resolution of $\sim$1$^{\prime}$ will greatly improve studies of the \graya{} source morphology of the extended emission coming from the $\gtrsim$10--100\,pc region surrounding CR accelerators, allowing more accurate estimates of the underlying source properties~\citep{AceroF.2013,DubusG.2013,AmbrogiL.2016}.
CTA may be able to determine if the \graya{} sources with no association~(discussed in \autoref{sssect:no association}) are part of the diffuse emission or are located near CR accelerators~\citep{MartinezM.2013,MitchellA.2021}.

As CTA is expected to observe hundreds of new \graya{} sources in the Galactic plane the challenge of source confusion will need to be addressed~\citep{MestreE.2022}. Correctly resolving and identifying two or more individual low surface-brightness sources that are potentially overlapping one another necessitates an accurate and precise model of the diffuse background of \graya{} emission. The sliding window method employed in the HGPS~(see \autoref{sect:HGPS diffuse emission}) will be unable to prevent source confusion.
Using a simulated diffuse model, such as \GP, will aid the morphological studies around \graya{} sources, especially if the sources are overlapping. Using \GP{} will also inform all spectral studies, especially below $\sim$10\,TeV where the diffuse emission can be a significant component for faint~(i.e.~$\lesssim$1\,\%Crab) \graya{} sources.

This chapter begins with a discussion of the Python package \verb|gammapy| and the included instrument response functions~(IRFs) for CTA. A first look into the detectability of the diffuse \graya{} emission with CTA will be performed.

\section{The `gammapy' Python Package}

The Python package named \verb|gammapy|~\citep{DonathA.2015,DeilC.2017,DonathA.2023} was chosen as the ``CTA science analysis tools'' that will be used the scientific analysis of all CTA data.
The~\verb|gammapy| package is an open-source \verb|Python| package that has been developed to analyse \graya{} data from IACTs such as \hess, MAGIC, VERITAS, and CTA. Recently, \verb|gammapy| has been successfully used to analyse \graya{} data the water-tank Cherenkov detector HAWC~\citep{AlbertA.2022b}, and the \fermi{} analysis tool \verb|fermipy|~\citep{WoodM.2017} uses some elements of \verb|gammapy|.
This chapter makes use of \verb|gammapy| version~1.1 for all calculations\footnote{The most recent version can be found at \url{https://github.com/gammapy/gammapy}.}.

The~\verb|gammapy| package simplifies the computation of the so-called `data products' from an observational event list. The data products include flux images, significance images, spectra, light curves, etc.. Additionally, the future performance of CTA can be estimated by using the IRFs and \verb|gammapy| to predict the \graya{} flux that CTA would observe given some model \graya{} flux and assumed observational parameters.

\subsection{Instrument Response Functions}

IACTs observe airshowers and reconstruct the information on the particle or photon that initiated the airshower~(such as the arrival direction, energy, etc; see \autoref{sect:gamma-ray observations}).
The performance of the IACTs is characterised by the IRFs.
The CTA IRFs are calculated by performing Monte~Carlo simulations of airshowers using CORSIKA~\citep{HeckD.1998}, and then by modelling how the IACT array responds to CR and \graya{} airshowers for a range of observational parameters using \verb|sim_telarray|~\citep{BernlohrK.2008}.
The IRFs that are commonly used for IACT observations are: the energy dispersion, the angular point spread function, the effective area, and the background rate.

\begin{itemize}
    \item The true energy of a \grayn, $E_{\mathrm{true}}$, is not directly observed and must be computed from the airshower parameters. The energy that is reconstructed, $E_{\mathrm{reco}}$, is not necessarily equal to $E_{\mathrm{true}}$. The energy dispersion IRF characterises this difference and is defined as the ratio of the reconstructed energy on the true energy~($E_{\mathrm{reco}}/E_{\mathrm{true}}$). For CTA, $E_{\mathrm{reco}}$ is typically within $\sim$8\% of $E_{\mathrm{true}}$ above 1\,TeV.
    \item The arrival direction of the CR or \grayn{} that initiated the airshower must be reconstructed~(e.g.~see \autoref{fig:hillas diagram}). The point spread function~(PSF) represents the angle between the true and reconstructed arrival positions within which 68\% of \grays{} of a given true energy will fall. The PSF of CTA will be $\sim$3$^{\prime}$ at 1\,TeV and $\sim$1$^{\prime}$ at 100\,TeV.
    \item The effective area~($A_{\textrm{eff}}$) is the collection area of the IACT during the observation.
    The effective area is a function of the CR or \graya{} energy and the zenith angle of the observation. As the sensitivity of the telescope decreases towards the edge of the FoV, the effective area is also a function of the offset angle~(where the offset angle is defined as the angle between the centre of the FoV and the arrival position of the \grayn). The effective area for CTA is approximately 1\,km$^{2}$ for energies above 1\,TeV. 
    \item The background rate is the number of non-signal events per unit of time, per reconstructed energy, per solid angle. The background events include CR airshowers that have been misidentified as \grays{} and false triggers from the night-sky background. For observations the CR background is typically calculated via the standard ring background, reflected ring background, or the adaptive ring background~(see \autoref{ssect:CR background subtraction}). For this work the CR background~($J_{\mathrm{CR}}$) is given as the number of events per square degree following the simulated curves from~\citet{BernlohrK.2013}, and is approximately a powerlaw in energy.
\end{itemize}

The IRFs used here are similar to those from~\citet{BernlohrK.2013} but are calculated with a more recent version of CORSIKA, an updated detector model, and a more recent telescope layout.
The IRFs are based on the `alpha-configuration' of the CTA-South site, which has 14~MSTs and 37~SSTs, and have been optimised for a 20$^{\circ}$ zenith angle and a 50\,hr livetime\footnote{The alpha-configuration IRFs for both the southern and northern sites for various zenith angles, livetimes, and sub-arrays can be found at \url{https://zenodo.org/record/5499840\#.YUya5WYzbUI}.}.

\subsection{Reconstructing the GALPROP Diffuse Flux}

For some input \graya{} flux~($J_{\gamma}$) taken from \GP, \verb|gammapy| is used to calculate the number of \graya{} counts that CTA would be expected to observe~(given by $N_{\gamma}$). The expected number of \graya{} counts are then be converted to a reconstructed \graya{} flux assuming the background CR flux is given by $J_{\mathrm{CR}}$.
The predicted number of \graya{} events is given by:

\begin{align}
    N_{\gamma} &= N_{\mathrm{on}} - \mathscr{a} N_{\mathrm{off}} \label{eq:Ngamma}\\
    N_{\mathrm{on}} &= J_{\gamma}(E_{\mathrm{reco}},\,t_{\mathrm{on}},\,\Omega_{\mathrm{on}}) A_{\mathrm{eff}}(E_{\mathrm{reco}}) t_{\mathrm{on}} \Omega_{\mathrm{on}} \label{eq:Non2} \\
    N_{\mathrm{off}} &= J_{\mathrm{CR}}(E_{\mathrm{reco}},\,t_{\mathrm{off}},\,\Omega_{\mathrm{off}}) A_{\mathrm{eff}}(E_{\mathrm{reco}}) t_{\mathrm{off}} \Omega_{\mathrm{off}} \label{eq:Noff2} \\
    \mathscr{a} &= \frac{t_{\mathrm{on}} \Omega_{\mathrm{on}}}{t_{\mathrm{off}} \Omega_{\mathrm{off}}}
\end{align}

\noindent
where $E_{\mathrm{reco}}$ is the reconstructed energy of the particle or photon that initiated the air shower, $t_{\mathrm{on}}$ and $t_{\mathrm{off}}$ are the livetimes for the \textit{on} and \textit{off} regions, respectively, $\Omega_{\mathrm{on}}$ and $\Omega_{\mathrm{off}}$ are the solid angles of the \textit{on} and \textit{off} regions, respectively, and $\mathscr{a}$ is the ratio of exposure between the \textit{on} and \textit{off} regions.
The number of background counts~($N_{\mathrm{off}}$) is sampled from the CR flux, $J_{\mathrm{CR}}$, which is provided by the IRFs~\citep{BernlohrK.2013,DonathA.2023}. The number of on-region counts~($N_{\mathrm{on}}$) is sampled from the input \graya{} flux, $J_{\gamma}$.
The reconstructed fluxes in this chapter utilise an offset angle of $0.5^{\circ}$, the \textit{on} and \textit{off} regions have solid angles defined by $\Omega_{\mathrm{on}} = \Omega_{\mathrm{off}} = 2\pi[1 - \cos(0.2^{\circ})]$, and the livetimes are assumed to be equal~(i.e.~$t_{\mathrm{on}} = t_{\mathrm{off}}$). Hence, the ratio of exposures simplifies to $\mathscr{a}=1$.
As `\textit{on}' and `\textit{off}' are only symbolic terms in the current application of \verb|gammapy|, these assumptions were found to have only a minor impact on the results.

Assuming $N_{\mathrm{on}}$ and $N_{\mathrm{off}}$ follow a Poisson distribution, the significance of the excess counts as a function of the reconstructed energy is given by $S(E_{\mathrm{reco}})$. Taking the sum of the significances in quadrature gives the total significance of the reconstruction, $S_{\mathrm{total}}$, across all energies. Therefore, the significances are given by:

\begin{align}
    S(E_{\mathrm{reco}}) &= \frac{N_{\gamma}(E_{\mathrm{reco}})}{\sqrt{N_{\mathrm{off}}(E_{\mathrm{reco}})}} \\
    S_{\mathrm{total}} &= \sqrt{\sum_{E_{\mathrm{reco}}} \left[ S(E_{\mathrm{reco}}) \right]^{2}} \label{eq:S_total}
\end{align}

\noindent
where all parameters have been defined previously. The reconstructed flux points are shown as upper limits if $S(E_{\mathrm{reco}})<2$.

\section{Detectability of the Diffuse Gamma-Ray Emission with CTA}

For spectral and morphological studies performed by CTA, the diffuse \graya{} emission will contaminate estimates of \graya{} sources.
This contamination will be strongest in locations where the diffuse \graya{} emission is brightest -- e.g.~along the spiral-arm tangents~($20^{\circ} \leq |\mathscr{l}| \leq 40^{\circ}$; see \autoref{chap:paper 1}).
Accounting for this contamination will be particularly important for the $\gtrsim$10--100\,pc region around SNRs, as CRs that may have escaped the SNR will interact with the surrounding ISM and create \graya{} structures~\citep[e.g.][]{AharonianF.1996,GabiciS.2009,CasanovaS.2010,MalkovM.2013,EineckeS.2023,RowellG.2023}.
Hence, the analyses presented here are restricted to regions around SNRs detected by \hess{} towards Galactic longitudes $20^{\circ} \leq |\mathscr{l}| \leq 40^{\circ}$. The two chosen source regions are HESS~J1614--518~\citep{AharonianF.2005b} and RX~J1713.7--3946~\citep{PfeffermannE.1996}.

In this section, the variations over all \GP{} predictions will be used.
For the time-independent envelope, the variation is across the SA0, SA25, SA50, SA75, and SA100 source ditributions~(see \autoref{sect:source dist.}), the F98 and R12 interstellar radiation field~(ISRF) models~(see \autoref{sect:ISRF}), and the GASE and PBSS Galactic magnetic field~(GMF) models~(see \autoref{sect:GMF}). The time-independent variations are presented in \autoref{chap:paper 1}.
For the time-dependent variations, the envelope is calculated over six combinations of source lifetimes and source creation rates~(see \autoref{ssect:tdep source pars}). The source lifetimes range from 10\,kyr to 200\,kyr, and the source creation intervals range from one source every 50\,yr to one source every 500\,yr. The time-dependent variations are presented in \autoref{chap:paper 2}.

In this section the \GP{} diffuse \graya{} fluxes will be processed through the CTA-South IRFs using \verb|gammapy|~(i.e.~$J_{\gamma}$ used in \autoref{eq:Non2} will be taken from \GP).

\subsection{Diffuse Gamma-Ray Emission Near HESS~J1614--518}

\begin{figure}
    \centering
    \includegraphics[width=\textwidth]{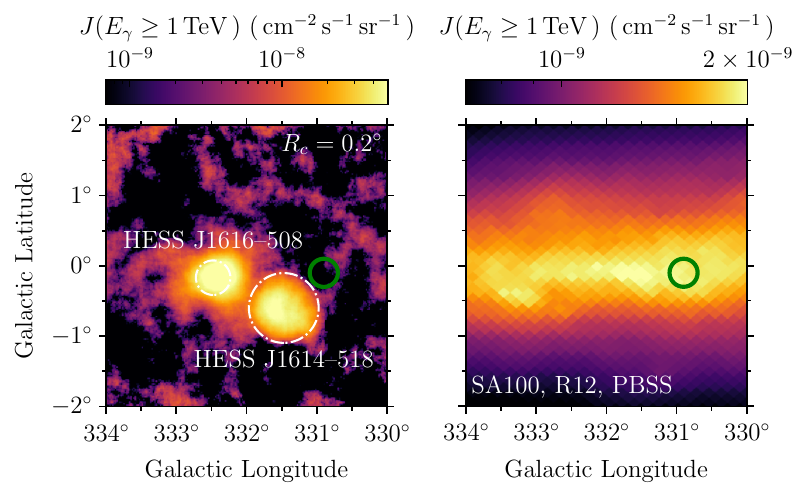}
    \caption{Integrated \graya{} flux above 1\,TeV around HESS~J1614--518 from the HGPS~(left) and the steady-state SA100/R12/PBSS combination from \GP~(right). The analysis region used for extracting the \GP{} emission is shown by the green circle. The HGPS sources HESS~J1614--518 and HESS~J1616--508 are shown by the white circles.}
    \label{fig:GALPROP and HGPS J1616}
\end{figure}

HESS~J1614--518~\citep{AharonianF.2005b} is an SNR~\citep{LauJ.2017,AbdallaH.2018c} located at $(\mathscr{l},\,\mathscr{b})=(331.47^{\circ},\,-0.60^{\circ})$ with a radius of $0.49^{\circ}$~\citep{AbdallaH.2018a}. The \graya{} emission observed in the HGPS around HESS~J1614--518 is shown in \autoref{fig:GALPROP and HGPS J1616} alongside the predicted diffuse emission from the steady-state SA100/R12/PBSS combination from \GP{} for the same region. Both the SNR HESS~J1614--518 and the unidentified source HESS~J1616--508 are shown by white circles.

The CTA reconstruction is computed from the midpoint of the \GP{} flux envelope, i.e.~the average between the upper and lower bound. The envelope represents the variation over the grid of twenty steady-state models from \autoref{chap:paper 1} as well as the six time-dependent models from \autoref{chap:paper 2}.
The \GP{} envelope is extracted from the analysis region shown by the green circle placed at the coordinates $(\mathscr{l},\,\mathscr{b})=(330.9^{\circ},\,-0.1^{\circ})$ with a radius of $0.2^{\circ}$.
The HGPS shows no significant emission in this region.
However, as the chosen analysis region lies on the edge of the SNR shell it may contain some \graya{} emission from recently escaped CRs. Thefore, the \GP{} emission in the region represents a lower bound on the expected \graya{} flux. 

\begin{figure}[t]
    \centering
    \includegraphics[width=\textwidth]{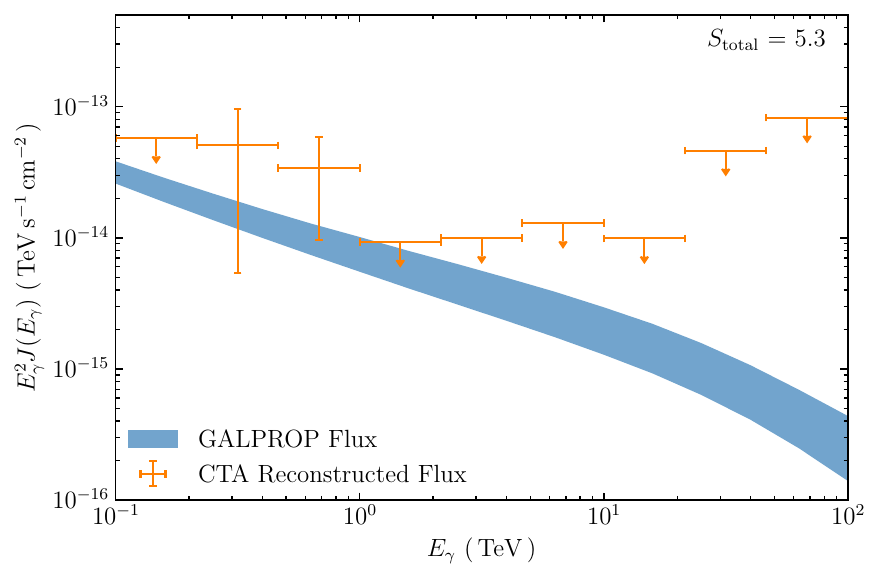}
    \caption{Differential photon flux envelope from \GP~(blue shaded band) and the CTA reconstruction~(orange points). Energy bins with $S<2$ are shown as upper limits, with the total significance shown in the top right corner. The reconstruction is calculated with three bins per decade and a livetime of 50\,hr.}
    \label{fig:reco livetime test}
\end{figure}

Both the \GP{} envelope and the CTA reconstruction are shown in \autoref{fig:reco livetime test}.
The computation was performed with three energy bins per decade. Additionally, the reconstruction uses a livetime of 50\,hr, which is the minimum expected observation time along the Galactic plane in the first ten years of the CTA GPS~\citep{AcharyaB.2018}.
The diffuse emission predicted by \GP{} below 1\,TeV is on the lower edge of CTA's sensitivity. 
The \GP{} flux is reconstructed for 0.2--1\,TeV with the predicted flux envelope contained within the uncertainties.
For energies above 1\,TeV the reconstructed points are shown as upper bounds as $S(E_{\mathrm{reco}}) < 2$.
Hence, CTA is expected to observe the diffuse emission around HESS~J1614--518 below 1\,TeV given a 50\,hr livetime. For \graya{} energies in the range 1--10\,TeV CTA will need to utilise, for example, more complex background subtraction methods or longer livetimes, for any potential detections the diffuse emission.

\subsection{Diffuse Gamma-Ray Emission Near RX~J1713.7--3946}

The SNR RX~J1713.7--3946~\citep{PfeffermannE.1996} is one of the brightest and most studied TeV \graya{} sources~\citep[e.g.][]{KoyamaK.1997,AharonianF.2004a,MoriguchiY.2005,AceroF.2009,MaxtedN.2012,TsujiN.2019,LeikeR.2021}.
Previous studies have predicted regions around RX~J1713.7--3946 to be emitting \grays{} due to the escaping CRs colliding with the surrounding gas~\citep[e.g.][]{CasanovaS.2010,RowellG.2023}. Some of the emission predicted to be around RX~J1713.7--3946 was later observed by \hess~\citep{AbdallaH.2018f}.
CTA observations around RX~J1713.7--3946 have the potential to confirm or deny the SNR as an accelerator of PeV hadrons -- which is unconfirmed for SNRs in general~\citep[e.g.][]{LagageP.1983,BellA.2004}.

\begin{figure}[t]
    \centering
    \includegraphics[width=\textwidth]{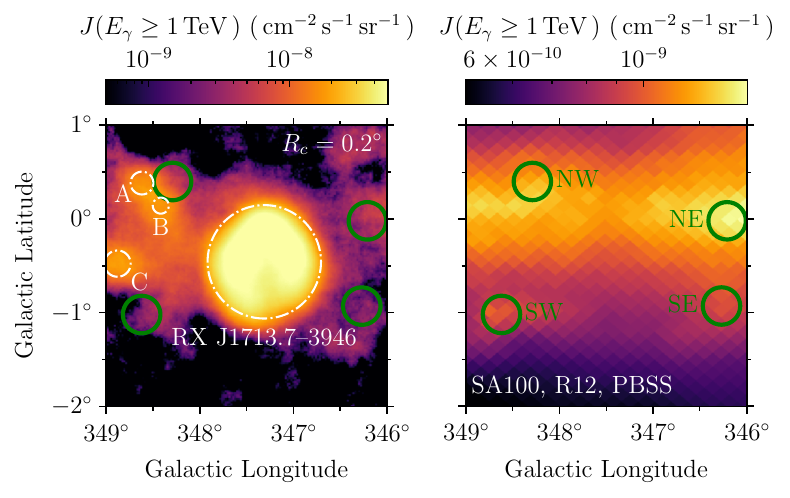}
    \caption{Integrated \graya{} flux above 1\,TeV around RX~J1713.7--3946 from the HGPS~(left) and the steady-state SA100/R12/PBSS combination from \GP~(right). The four analysis regions used for extracting the \GP{} emission are shown by green circles~(labelled as NW, NE, SW, and SE). The HGPS sources RX~J1713.7--3946, HESS~J1713--381~(labelled A), HESS~J1714--385~(labelled B), and HESS~J1718--385~(labelled C) are shown by white circles.}
    \label{fig:GALPROP and RXJ J1713}
\end{figure}

RX~J1713.7--3946 is located at $(\mathscr{l},\,\mathscr{b})=(347.31^{\circ},\,-0.46^{\circ})$ with a radius of $0.6^{\circ}$~\citep{AbdallaH.2018a}. The \graya{} emission observed in the HGPS around the SNR is shown in \autoref{fig:GALPROP and RXJ J1713} alongside the predicted diffuse emission from the steady-state SA100/R12/PBSS combination from \GP{} for the same region. RX~J1713.7--3946, along with the three nearby \graya{} sources HESS~J1713--381, HESS~J1714--385, and HESS~J1718--385~(labelled A, B, and C, respectively) are shown by the white circles.

Similarly as to the calculation around HESS~J1614--518, the CTA reconstruction is computed from the midpoint of the \GP{} flux envelope. The variation in the \GP{} flux over the grid of twenty steady-state models from \autoref{chap:paper 1} as well as the six time-dependent models from \autoref{chap:paper 2}.
The \GP{} envelope is extracted from the analysis regions shown by the green circles with radii of $0.2^{\circ}$ located at: NW; $(\mathscr{l},\,\mathscr{b})=(348.29^{\circ},\,0.40^{\circ})$, NE; $(346.21^{\circ},\,-0.02^{\circ})$, SW; $(346.27^{\circ},\,-0.93^{\circ})$, and SE; $(348.62^{\circ},\,-1.02^{\circ})$.
These four regions contain ISM clouds that are located at approximately the same line-of-sight distance from Earth and are predicted to emit additional \graya{} emission above the diffuse background~\citep{RowellG.2023}.

\begin{figure}[t]
    \centering
    \includegraphics[width=\textwidth]{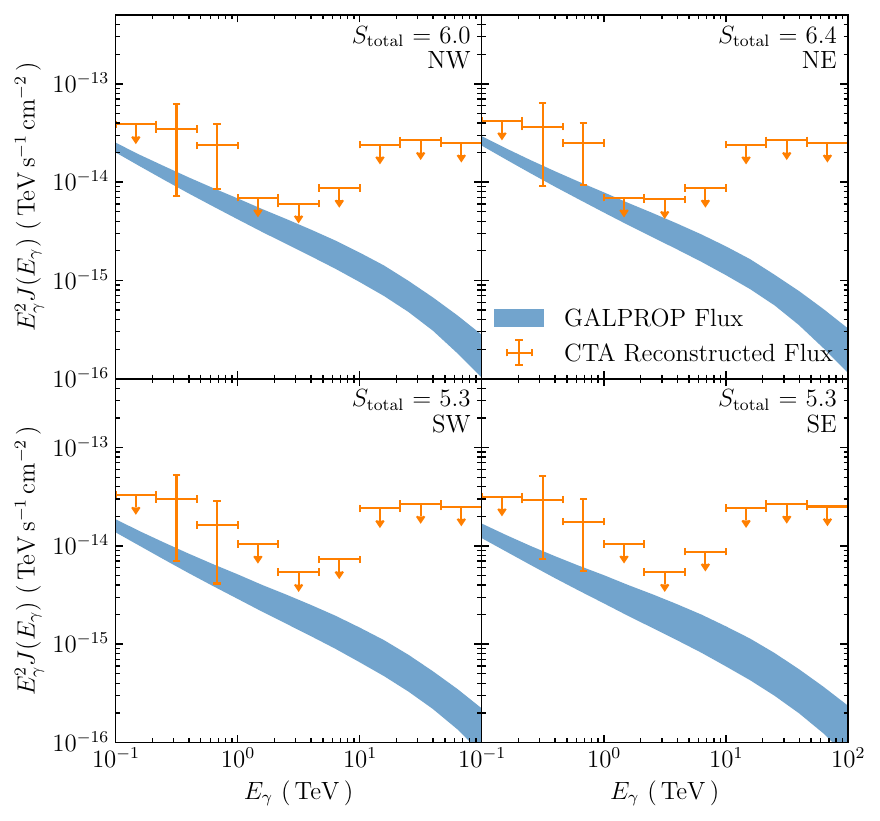}
    \caption{Differential photon flux envelopes from \GP~(blue shaded band) and the CTA reconstructions~(orange points). Energy bins with $S<2$ are shown as upper limits, with the total significance shown in the top right corner of each panel. The reconstruction is calculated with five bins per decade and a livetime of 100\,hr for the NW region~(top left), NE~(top left), SW~(bottom left), and SE~(bottom right).}
    \label{fig:RXJ1713 reco}
\end{figure}

The \GP{} envelopes and the CTA reconstructions are shown for the four regions in \autoref{fig:RXJ1713 reco}.
As additional observation time is planned for RX~K1713.7-3946~\citep{AcharyaB.2018} the computation was performed with a livetime of 100\,hr.
Just as for the HESS~J1614--518 reconstruction, the diffuse emission predicted by \GP{} is on the lower edge of the sensitivity of CTA. Below 1\,TeV CTA should be able to observe the diffuse emission, while for \graya{} energies 1--10\,TeV the reconstructed flux points are only presented as upper limits.

The \graya{} brightening due to CRs escaping the SNR was modelled by~\citet{RowellG.2023} for the same analysis regions presented here.
By comparing the models from \citet{RowellG.2023} to the diffuse \graya{} predictions in \autoref{fig:RXJ1713 reco}, it can be seen that the diffuse \graya{} emission below 10\,TeV could contribute up to half of the observed \graya{} flux. Hence, future observations by CTA for the regions around RX~J1713.7--3946 must account for the diffuse \graya{} emission before performing spectral and morphological analyses.

The calculations presented here demonstrate that CTA should detect the diffuse \graya{} emission. However, it should be noted that the \textit{on} and \textit{off} regions are purely symbolic representations.
A more accurate representation of the reconstructed fluxes would involve utilising the entire flux maps and computing the background counts from, e.g.~the ring background method~(see \autoref{ssect:CR background subtraction}).
Defining spatially separate \textit{on} and \textit{off} regions would allow more robust calculations of the significance, e.g.~\citet{LiT.1983}~(see \autoref{eq:Li and Ma}).
These calculations are left for future studies.
    \chapter{Summary and Future Work} \label{chap:summary}

Galactic CR transport models, such as \GP, have previously reproduced the Galactic diffuse \graya{} emission successfully up to 100\,GeV~\citep{AckermannM.2012}.
However, the diffuse \graya{} emission predicted by \GP{} had never been compared in detail to \graya{} observations at TeV~(or higher) energies.
This thesis presents the first comparison between the simulated diffuse TeV \graya{} emission from \GP{} to the TeV \graya{} observations from \hess. \GP{} was demonstrated to agree to the HGPS within modelling and experimental uncertainties. This thesis also presents the first demonstration that \GP{} agrees with the recent diffuse emission estimates from LHAASO up to 1\,PeV~\citep{CaoZ.2023}. Hence, the work in this thesis extended the demonstrated energy range that \GP{} can reproduce the diffuse \graya{} emission by an additional four orders of magnitude beyond previous comparisons.

For the CR transport models there will be some variation in the TeV predictions due to degeneracies in the Galactic distributions, such as in the CR source distribution, the interstellar radiation field~(ISRF), and Galactic magnetic field~(GMF).
Additionally, due to the short cooling times of the electrons above 10\,TeV, it becomes important to consider the injection of the CRs by discrete sources. The placement of the sources leads to another component of the modelling variation above 1\,TeV.
This thesis presents the first quantification of all components of the modelling variation in the TeV \graya{} predictions.

The steady-state modelling variation was quantified by simulating the diffuse \graya{} emission over a grid of reasonable Galactic distributions (see \autoref{chap:paper 1}). The choice of ISRF altered the simulated diffuse TeV emission by less than 5\%, while the choice of source distribution and GMF altered the simulated diffuse TeV emission by $\sim$40\%.
The time-dependent variation was quantified by simulating over multiple possible CR source lifetimes and CR source creation rates (see \autoref{chap:paper 2}). The time-dependent variation at 1\,TeV was similar to the full variation over all steady-state models,~i.e.~$\sim$40\%.

For the TeV observations from the \hess{} Galactic plane survey~\citep[HGPS;][]{AbdallaH.2018a}, there are multiple components that need to be considered before finding an estimate of the diffuse TeV \graya{} emission.
These components are the \graya{} emission from catalogued \graya{} sources, the emission from unresolved \graya{} sources, and the large-scale diffuse \graya{} emission.
The residual emission after subtracting the two \graya{} source components is considered to be an estimate of the observed diffuse emission.

After accounting for all of the uncertainties in the HGPS diffuse \graya{} estimate and the simulated diffuse \graya{} emission from \GP, it was found that the predictions are in good agreement with the observations.
However, the uncertainty band on the HGPS diffuse emission estimates found in this thesis can be as large as a factor of five. Firm conclusions on the diffuse \graya{} emission in the TeV energy regime will require the observations from the future TeV \graya{} observatory, the Cherenkov telescope array~\citep[CTA;][]{AcharyaB.2018}.

At 1\,TeV the diffuse \graya{} emission was found to have a significant leptonic component, especially above and below the Galactic plane. For Galactic longitudes $\mathscr{l} \leq |90^{\circ}|$ the leptonic component of the diffuse emission is approximately equal to the hadronic component of the flux.
For the steady-state models the leptonic contribution to the diffuse \graya{} emission increased with \graya{} energy, accounting for up to 90\% of the total flux at 100\,TeV for $\mathscr{l} \leq |90^{\circ}|$.
Extending to the time-dependent simulations, the increasing leptonic component of the diffuse emission can be seen as an increasing contribution due to unresolved leptonic CR sources.
Additionally, it was found that the local observed CR electron spectrum is entirely consistent with no discrete CR electron sources being located within 200\,pc of Earth, agreeing with various other types of models~\citep[e.g.][]{PorterT.1997,MertschP.2018}. It was also found that the electron flux above 10\,TeV can vary by over a factor of ten over a few million years.
This result shows that the injected CR electron spectrum used by any CR transport simulation cannot be normalised to the local CR electron spectrum above $\sim$1\,TeV, as is commonly performed in the literature.

This thesis also provided an initial look into the detectability of the diffuse TeV \graya{} emission by CTA.
The uncertainty band for both the HGPS diffuse \graya{} emission estimate and the variations in the \GP{} predictions are almost entirely above the proposed sensitivity for the CTA Galactic plane survey for Galactic longitudes $|\mathscr{l}| \lesssim 45^{\circ}$.
Therefore, CTA will detail the TeV diffuse emission, allowing the origin of the diffuse emission to be identified in the TeV regime and complementing the analysis of discrete \graya{} sources.

\section{Future Work}

The \GP{} modelling variation along the Galactic plane that was calculated in this thesis is an upper bound.
For example, the GASE GMF~\citep[][see \autoref{sect:GMF}]{StrongA.2000} was included in the estimate of the \graya{} modelling variation as it is commonly used throughout the literature.
However, the GASE GMF is not a complete representation of the MW. The modelling variation can be reduced by using other GMF models that are consistent with observations of extra-Galactic rotation measures, such as \citet{JanssonR.2012}.
Similarly, the SA0 and SA100 source distributions do not necessarily reproduce the data well but were included in the estimate of the modelling variation as they are commonly used throughout the literature.
For the time-dependent modelling variation the CR sources were approximated into a single average source class, with a wide range of values for the CR source lifetimes and CR source creation rates being taken.
Adding more recent, physically motivated Galactic distributions, finding tighter constraints on the CR source parameters, and accounting for the different classes of sources~(e.g.~SNRs, PWNe, etc.), will allow a more representative estimate of the modelling variation to be calculated.

Currently, \GP{} uses the LAB \hi-survey~\citep{KalberlaP.2005} and the CfA composite CO survey~\citep{DameT.2001} in the construction of the interstellar medium~(ISM) gas models~(see \autoref{sect:ISM gas}). These gas models are used for both the CR transport calculations and the \graya{} emission calculations. The gas surveys used in the construction of the ISM models have angular resolutions of approximately 0.1$^{\circ}$, limiting the maximum angular resolution for the \graya{} emission maps calculated by \GP{} to~$6.9^{\prime}=0.1145^{\circ}$.
The maximum resolution of \GP{} is currently adequate for comparisons to \hess, which has a point spread function~(PSF) full-width half-maximum~(FWHM) reaching $4.8^{\prime}=0.08^{\circ}$.
However, CTA will have a PSF FWHM reaching $1^{\prime} \approx 0.017^{\circ}$.
More recent, higher-resolution maps of the ISM gas would allow \GP{} to simulate the diffuse emission to a greater resolution and will be required before \GP{} can be used as a diffuse model by the CTA observatory.
Potential higher-resolution surveys that could be integrated into \GP{} include: Mopra~CO~\citep{BraidingC.2015,BraidingC.2018,CubukK.2023} and GASKAP~\hi/OH~\citep{DickeyJ.2013} for the southern hemisphere, and Nobeyama~45m~CO~\citep{MinamidaniT.2015,UmemotoT.2017} and THOR~\hi/OH~\citep{BeutherH.2016,WangY.2020} for the northern hemisphere.

\subsection{Applications to TeV Gamma-Ray Observations}

Currently \hess{} uses an ad-hoc model of the large-scale emission that is calculated from the observations.
With the large amount of significant emission that CTA is expected to observe along the Galactic plane a more robust method of subtracting the diffuse \graya{} background will be required.
By integrating the \GP{} predictions with the Python package \verb|gammapy| it will be possible to create a robust diffuse emission template that can be subtracted from analyses. This template could be applied retroactively to the HGPS, allowing for additional faint sources such as PWNe and transient sources to be discovered.
\autoref{chap:CTA and gammapy} presented a first look at the detectability of the diffuse emission with \GP{} using \verb|gammapy| and the provided IRFs. It was found that CTA should be able to directly measure the diffuse emission below 1\,TeV. Above 1\,TeV the diffuse emission will still be a significant component for faint~(i.e.~<1\,\%Crab) sources and will need to be considered for spectral and morphological studies.
Implementing a diffuse model will be integral for CTA as it is expected to detect hundreds of faint sources, making source confusion a future challenge for the observatory~\citep{MestreE.2022}.

CR air shower events typically outnumber \graya{} air showers by a factor of a thousand.
Throughout the literature there are many techniques to subtract the CR background from \graya{} observations, such as: the on-off method~\citep{WeekesT.1989}, the template method~\citep{RowellG.2003}, the standard ring and reflected ring methods~\citep{BergeD.2007}, and the field-of-view method~\citep{BergeD.2007}.
The technique used in the HGPS was the adaptive ring background method~(see \autoref{ssect:CR background subtraction}).
The adaptive ring background assumes that there is no \graya{} background contributing to the estimate of the CR flux.
However, the diffuse \graya{} emission will contaminate estimates of the CR background for all regions along the Galactic plane.
For the HGPS, the diffuse \graya{} emission and its morphology was not considered in the calculation of the CR background estimates. Hence, the CR background is overestimated in the HGPS.
The degree to which the CR background is overestimated will depend on the Galactic coordinates of the observation, as well as the pre-defined exclusion regions~(see \autoref{ssect:CR background subtraction}).
A diffuse emission model, such as \GP, could be used to estimate the \graya{} flux that would be contaminating the CR background estimate. This estimate could then be used to apply a correction, improving the significance of the observations.
Integrating a diffuse \graya{} model into the CR background pipeline could also be applied to CTA,~e.g.~to aid in identifying two overlapping, low-surface-brightness \graya{} sources.
    
    % Appendices
    \begin{appendix}
        \chapter{Running GALPROP} \label{chap:running galprop}

Producing physically meaningful and accurate results with \GP{} requires an appropriate setup, especially in regards to the size of the timestep~(see \autoref{ssect:time-indep time grid}) and the size of the of the propagation cells~(see \autoref{ssect:grid}). Most of the \GP{} run parameters are defined within the \galdef{} file, with some CR injection parameters defined in the \sclass{} file. Example \galdef{} and \sclass{} files are included in the installation, with many more examples available on the \GP{} website\footnote{\url{https://galprop.stanford.edu/}}. Details on the parameters within the \galdef{} file can be found in either the \GP{} explanatory supplement\footnote{The latest version of the explanatory supplement as of the writing of this thesis is the development version from April~2013~\citep{StrongA.2013}. All versions of the explanatory supplement are available at \url{https://galprop.stanford.edu/code.php?option=manual}.} or the example files included in the \GP{} installation. The information in this appendix supplements the information given in the two aforementioned sources and is current as of \GP{} version~57. This appendix ends with visualisations of some \GP{} outputs.

\section{Grid Coordinate Setup} \label{sect:grid galdef}

\GP{} numerically solves the CR energy density flowing from one volume element, or cell, into adjacent cells. There are three propagation grids that \GP{} can use: `linear' and `tan', `step'. The latter is not discussed in this thesis, while the linear and tan grids and their functional definitions are discussed in \autoref{ssect:grid}.

\begin{table}
    \centering
    \caption{The \GP{} coordinate parameters discussed in \autoref{ssect:grid} with how they are referred to in the \galdef{} file.} \label{tab:coordinates}
    \bgroup
    \def\arraystretch{1.25}
    \begin{tabular}{ll}
        \galdef{} Parameter & Symbolic Representation \\ \hline
        \verb|x_min|       & $X_{\mathrm{min}}$ \\
        \verb|x_max|       & $X_{\mathrm{max}}$ \\
        \verb|dx|          & $\Delta X$ \\
        \verb|x_grid_pars| & $X_{0}$, \ $X_{\mathrm{ref}}$, \ $\lambda_{X}$ \\
        \verb|y_min|       & $Y_{\mathrm{min}}$ \\
        \verb|y_max|       & $Y_{\mathrm{max}}$ \\
        \verb|dy|          & $\Delta Y$ \\
        \verb|y_grid_pars| & $Y_{0}$, \ $Y_{\mathrm{ref}}$, \ $\lambda_{Y}$ \\
        \verb|z_min|       & $Z_{\mathrm{min}}$ \\
        \verb|z_max|       & $Z_{\mathrm{max}}$ \\
        \verb|dz|          & $\Delta Z$ \\
        \verb|z_grid_pars| & $Z_{0}$, \ $Z_{\mathrm{ref}}$, \ $\lambda_{Z}$ \\
    \end{tabular}
    \egroup
\end{table}

The grid function is defined in the \galdef{} file, and each axis is specified independent of one another. The grid function is chosen via the parameters \verb|x_grid|, \verb|y_grid|, and \verb|z_grid|, which can be set to \verb|linear|, \verb|tan|, or \verb|step|. The parameters discussed in \autoref{ssect:grid} are shown alongside their \galdef{} names in \autoref{tab:coordinates}.

Appropriate values for the grid parameters need to be calculated before running \GP. To save computation time and RAM/storage requirements, the parameters should be adjusted for the goals of the given simulation. Calculating values for the grid size requires an analysis of the cooling distances of all CR species and energies that are being simulated. To accurately capture the CRs moving from one cell to an adjacent cell, the cell size along any given axis~($\Delta X$, $\Delta Y$, $\Delta Z$) should be smaller than the shortest cooling distance of any included CR. 
The cooling distances for the various energy-loss processes are given by Equations \ref{eq:cool length pp}--\ref{eq:cool length ic} and shown in \autoref{fig:cooling times/lengths}.
Note that if using the time-dependent solution, and depending on which solver is chosen, the grid size typically needs to be less than half of the shortest cooling distance~\citep[see][Appendix~D]{PorterT.2022}.

To further optimise computation time, memory, and storage, the tan grid~(\autoref{eq:tan grid function}) can be utilised. The tan grid function uses the scaling parameter, $\lambda$, to increase the cell size from $\Delta X$ at some `central' point, $X_{0}$, to the size $\lambda \Delta X$ at some reference point, $X_{\mathrm{ref}}$. By making some further assumptions, the tan grid parameters can be optimised to provide accurate diffusion calculations in the most important regions while sacrificing accuracy from regions far away from the central point. Reasonable values will change depending on the specifics of each individual simulation. Ideally, the grid parameters should be chosen such that the particle transport will be accurately modelled at the Solar position.
To accurately model the CR transport and \graya{} emission within and across the spiral arms in the MW, the region between $X \approx 4$\,kpc and the Solar position ($X = S_{\odot}$) should also be similar to the shortest cooling time.

To capture CR transport accurately for the entire MW, all ISM structures must be included within the propagation grid. For the $XY$-plane the maximum extent must be set to at least $\pm 20$\,kpc to capture all of the ISM structures~(e.g.~ISM  gas, GMF, etc.).
Increasing the extent of the $XY$-plane beyond $\pm 20$\,kpc does not significantly alter the \graya{} results as there is no large-scale structure beyond these limits. For the $Z$-axis there is a large-scale structure named the \textit{Fermi} bubbles~\citep{DoblerG.2010,SuM.2010} which extends to heights of $\pm 15$\,kpc~\citep{AckermannM.2014}. However, the mechanisms behind the \textit{Fermi} bubbles are still debated\footnote{See,~e.g.~\citet{AckermannM.2014,CrockerR.2015,Bland-HawthornJ.2019}, and references therein, for discussions on possible mechanisms behind the \textit{Fermi} bubbles.}, and \GP{} is not currently designed to reproduce them. Reproducing the local observations of radioactive secondary CR ratios requires that the $Z$-axis limits are set to at least $\pm 6$\,kpc~\citep{MoskalenkoI.2005,OrlandoE.2013,JohannessonG.2016}, and increasing the $Z$-axis limit beyond $\pm 6$\,kpc does not significantly change the \graya{} results~\citep{JohannessonG.2016}.
\section{Time Coordinate Setup} \label{sect:time coordinate setup}

\GP{} solves the transport equation~(\autoref{eq:Transport Equation}) numerically with a variety of finite differencing solvers that can be chosen by the user.
For the `Operator Splitting' solver the solution method choices are: `Explicit', `Implicit', and `Crank-Nicholson'.
In version~57 the `Eigen BiGCStab' solver was added, with the solution method choices `Diagonal' and `IncompleteLUT'.
All available solution methods operate under the same principle -- the CR density is calculated at some given time, and then the solution steps forward some length in time. 

\GP{} can solve the transport equation using either a `time-independent'~(also known as `steady-state') solution or a `time-dependent' solution. The time-independent solution uses a variable timestep length and a static distribution of sources, and the time-dependent solution uses a constant timestep length and a probabilistic source distribution.
The time-independent timestep is discussed in more depth in \autoref{ssect:time-indep time grid} and the time-dependent timestep is discussed in more depth in \autoref{ssect:time-dep time grid}.

\begin{table}
    \centering
    \caption{The \GP{} time parameters discussed in \autoref{ssect:time-indep time grid} and \autoref{ssect:time-dep time grid} with how they are referred to in the \galdef{} file.} \label{tab:time pars}
    \bgroup
    \def\arraystretch{1.25}
    \begin{tabular}{lcc}
        \multirow{2}{*}{\galdef{} Parameter} & \multicolumn{2}{c}{Symbolic Representation} \\
         & Time Independent & Time Dependent \\ \hline
        \verb|end_timestep|    & $\Delta t_{\mathrm{final}}$   & $\Delta t$ \\
        \verb|start_timestep|  & $\Delta t_{\mathrm{initial}}$ & $t_{\mathrm{initial}}$ \\
        \verb|timestep_output| & --                               & $t_{\mathrm{out}}$ \\
        \verb|timestep_init|   & --                               & $t_{\mathrm{warm}}$ \\
        \verb|timestep_factor| & $\mathscr{f}$                    & -- \\
        \verb|timestep_repeat| & $M$                              & $t_{\mathrm{final}}$ \\
    \end{tabular}
    \egroup
\end{table}

The time-independent and time-dependent solution methods are chosen in the \galdef{} file, where \verb|time_dependent_solution = 0| specifies the use of the time-independent solution and \verb|time_dependent_solution = 1| specifies the use of the time-dependent solution. The time-independent grid is defined by the parameters discussed in \autoref{ssect:time-indep time grid} and the time-dependent grid is defined by the parameters discussed in \autoref{ssect:time-dep time grid}. All the parameters for both the time-independent and time-dependent solutions are shown in \autoref{tab:time pars} alongside their names in the \galdef{} file.

Similar to the propagation grid, reasonable values for the time parameters need to be calculated before running \GP{} and will change depending on the specifics of each individual run. To accurately model the cooling of the CRs the timestep needs to be shorter than the most rapid cooling time of the CR species and energies included in the simulation and is critical for accurate modelling of the CR and \graya{} spectra. For time-independent solutions, the smallest timestep~($\Delta t_{\mathrm{final}}$) needs to be smaller than the fastest cooling time. For time-dependent solutions, all timesteps~($\Delta t$) need to be smaller than the fastest cooling time.
The timestep size is especially important for time-dependent simulations as the computation time can be on the order of weeks for timesteps $\Delta t < 10$\,kyr.

\begin{table}
    \centering
    \caption{The \GP{} parameters that control the number of sources in the MW and how they are referred to in the \sclass{} file.} \label{tab:source class pars}
    \bgroup
    \def\arraystretch{1.25}
    \begin{tabular}{ll}
        \sclass{} Parameter & Symbolic Representation \\ \hline
        \verb|epoch|             & $t_{\mathrm{epoch}}$  \\
        \verb|number_of_sources| & $N_{\mathrm{source}}$ \\
        \verb|source_on_time|    & $t_{\mathrm{life}}$ \\
        \verb|rng_seed|          & -- \\
    \end{tabular}
    \egroup
\end{table}

Time-dependent models have additional parameters to control the characteristics of the individual sources. These parameters are discussed in \autoref{ssect:tdep source pars} and are defined in the \sclass{} file. The number of sources created in the MW is given by $N_{\mathrm{source}}$, and each source will inject CRs into the MW for the length of time given by $t_{\mathrm{life}}$. The time $t_{\mathrm{epoch}}$ defines the length of time that \GP{} will create new CR accelerators.
The seed for the random number generator~(RNG) is also defined in the \sclass{} file. The RNG seed is the first number fed into the RNG algorithm. For ease of reproducibility the RNG seed is set to some known value such as zero.
The source parameters defined in the \sclass{} file are shown alongside their symbolic representation in \autoref{tab:source class pars}.

When using a time-dependent solution the required computation time for VHE CRs can be on the order of weeks. It can be useful to operate \GP{} in a `warm start' mode, where the simulation can be paused and restarted. This is especially useful on a high-performance computing cluster, where the \graya{} skymap creation can be split over many nodes. The start mode is set in the \galdef{} file with the parameter \verb|warm_start|, where \verb|warm_start = 0| specifies that the calculation will start at the time $t_{\mathrm{initial}}$. For \verb|warm_start = 1| and \verb|warm_start = 2| the calculation will start at the warm-start initialisation time $t_{\mathrm{warm}}$, with the latter only calculating the \graya{} skymaps. Note that to operate \GP{} in a warm-start mode, the runtime developer flag, \verb|-c|, must be specified in the command line.

Note that in the time-dependent mode \GP{} uses the timestep number~($t_{\mathrm{initial}}/\Delta t$) for reading and writing files. If performing a warm start with an altered timestep~(e.g.~to improve runtime efficiency, see \autoref{ssect:time-dep time grid}), the nuclei files will need to be renamed.
\section{GALPROP Output Examples}

Results from \GP{} are output as either \verb|.fits| or compressed \verb|.fits|~(i.e.~\verb|.fits.gz|) files. The CR nuclei outputs are stored as \verb|.fits| tables in units of~MeV\,cm$^{-2}$\,s$^{-1}$\,sr$^{-1}$ with array axes~($\mathrm{CR},\,E_{k},\,Z,\,Y,\,X$), where $E_{k}$ is the kinetic energy of the CR species. The \graya{} skymaps are stored as 3D \hpx{} files in units of~MeV$^{-1}$\,cm$^{-2}$\,s$^{-1}$\,sr$^{-1}$ with an axis for the \graya{} energy, $E_{\gamma}$. Each emission type is stored as a separate \hpx{} file.

This section contains visualisations for the \GP{} outputs. The model used in the creation of the visualisations was the SA50 source distribution~(see \autoref{sect:source dist.}), the R12 ISRF~(see \autoref{ssect:R12 Model}), and the PBSS GMF~(see \autoref{ssect:Pshirkov}).

\subsection{Time-Independent Output Examples}

\begin{figure}
    \centering
    \includegraphics[width=16cm]{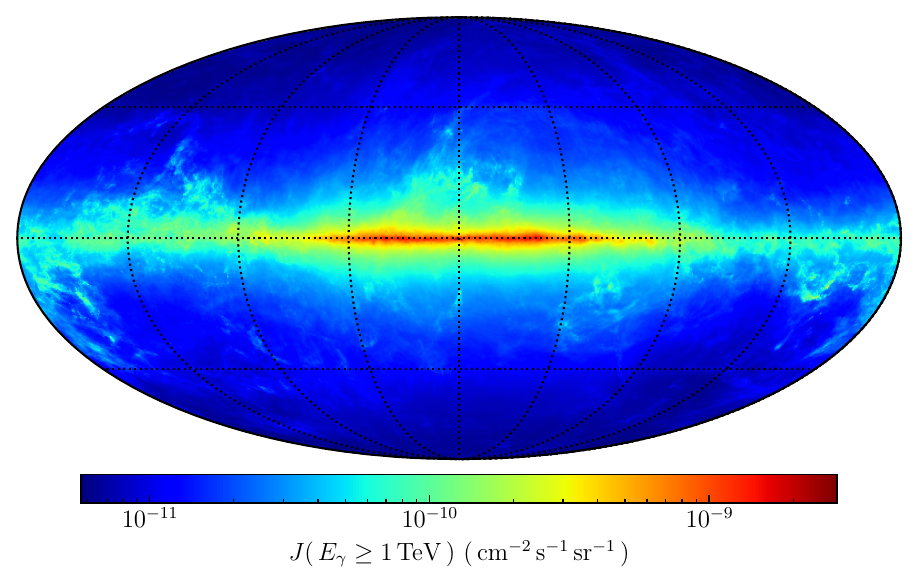}
    \caption{The total, integrated, time-independent \graya{} emission skymap from \GP{} in units of~cm$^{-2}$\,s$^{-1}$\,sr$^{-1}$. All components~(pion-decay emission, bremsstrahlung, and IC emission) have been summed together, and the flux has been integrated above 1\,TeV.}
    \label{fig:GALPROP skymap}
\end{figure}

\begin{figure}
    \centering
    \includegraphics[width=7.5cm]{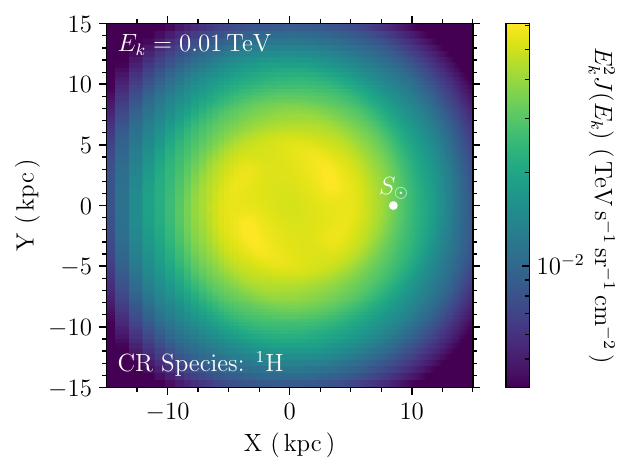}
    \includegraphics[width=7.5cm]{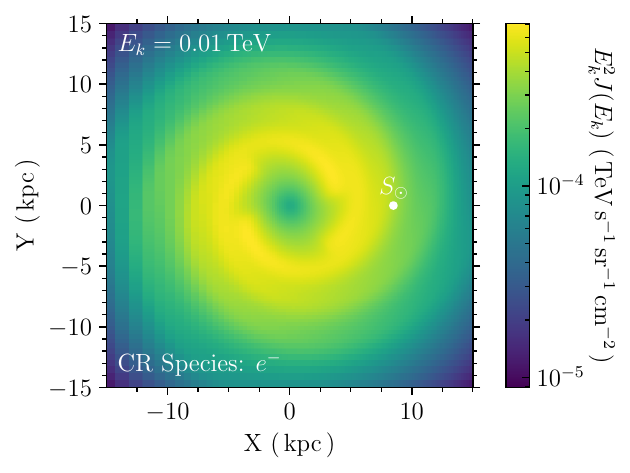}
    \includegraphics[width=7.5cm]{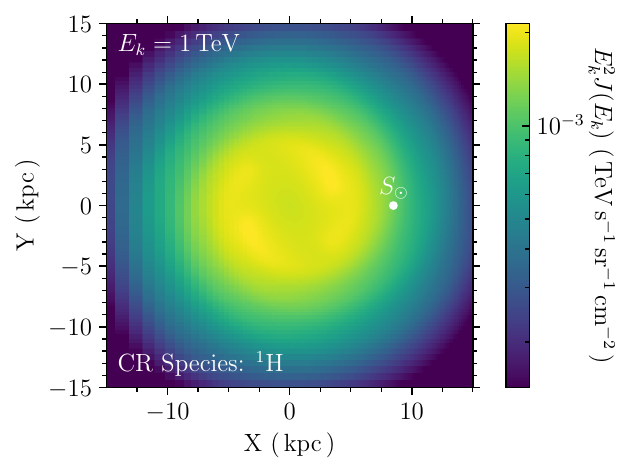}
    \includegraphics[width=7.5cm]{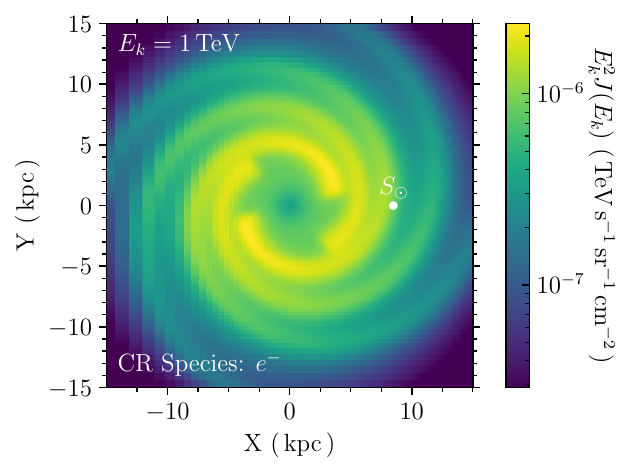}
    \includegraphics[width=7.5cm]{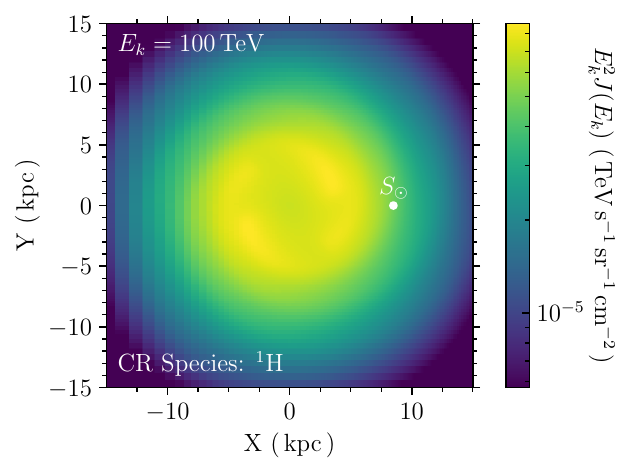}
    \includegraphics[width=7.5cm]{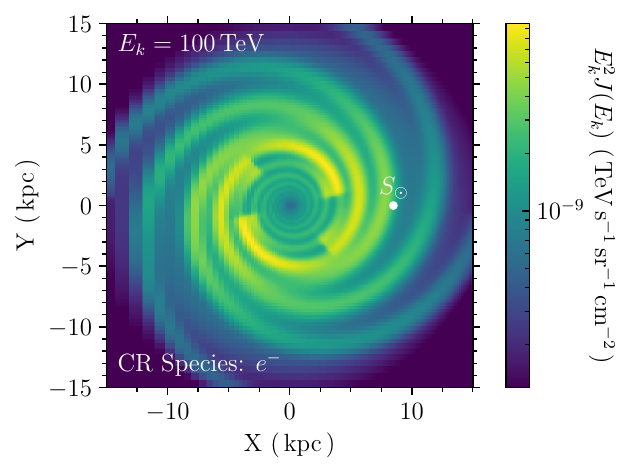}
    \caption{The time-independent CR proton energy density~(left) and time-independent CR electron energy density~(right) for the kinetic energies 10\,GeV~(top), 1\,TeV~(centre), and 100\,TeV~(bottom), with the Solar position shown by $S_{\odot}$. The density has been summed over the $Z$-axis, and the colourbar represents the CR energy density in units of~TeV\,cm$^{-2}$\,s$^{-1}$\,sr$^{-1}$.}
    \label{fig:GALPROP p vs e}
\end{figure}

\begin{figure}
    \centering
    \includegraphics[width=16cm]{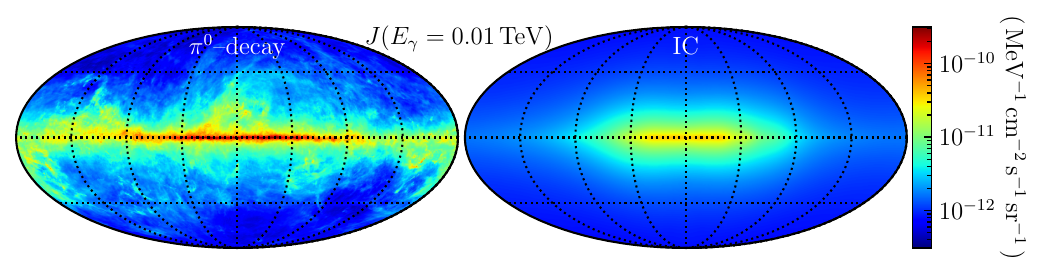}
    \includegraphics[width=16cm]{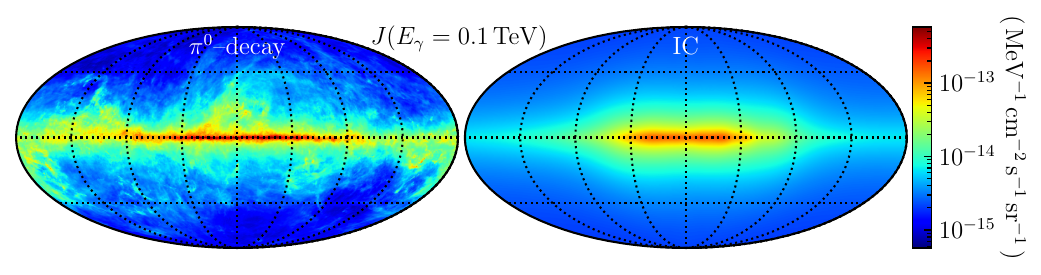}
    \includegraphics[width=16cm]{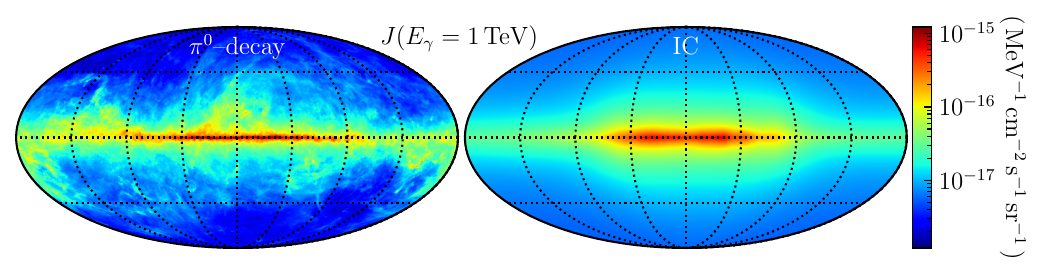}
    \includegraphics[width=16cm]{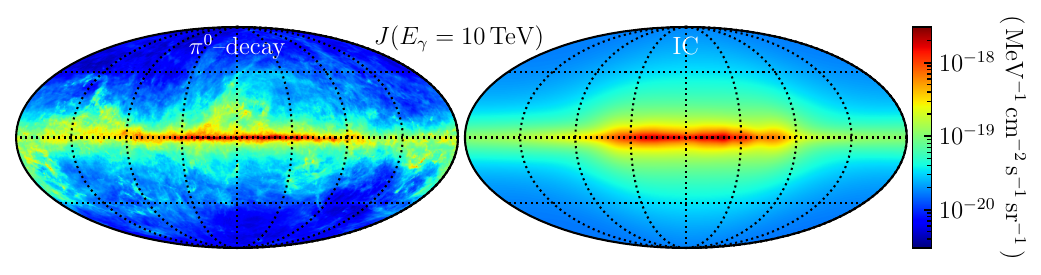}
    \includegraphics[width=16cm]{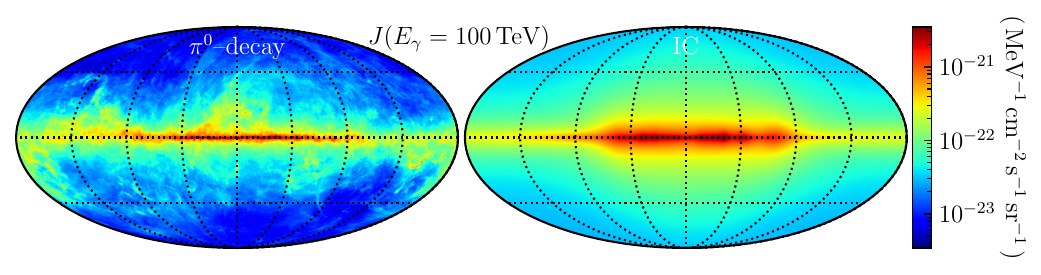}
    \caption{The time-independent pion-decay emission~(left) and time-independent IC emission~(right) for various \graya{} energies between 10\,GeV~(top) and 100\,TeV~(bottom). The flux is shown in units of~MeV$^{-1}$\,cm$^{-2}$\,s$^{-1}$\,sr$^{-1}$, with the scale applied to both emission types for a given energy.}
    \label{fig:GALPROP pi vs ic}
\end{figure}

For time-independent solution methods, all outputs are written to files after the propagation calculations have finished. An example visualisation of the total~(i.e.~sum over all \graya{} components) \graya{} skymap integrated above 1\,TeV is shown in \autoref{fig:GALPROP skymap}. Although it is not a direct output from \GP, the total, integrated skymap is used for the comparisons between \GP{} and \hess{} throughout this thesis, and is included here for reference.

An example visualisation of the CR proton and CR electron energy densities for various kinetic energies can be found in \autoref{fig:GALPROP p vs e}.
The CR proton density has a near-constant morphology across all energies as they travel distances on the order of 1\,kpc~(i.e.~they are diffuse).
Conversely, as the CR electrons travel shorter distances with increasing energy, their morphology has a strong dependence on energy. The structure in the CR electron density is strongly correlated with the source distribution~(SA50; \autoref{fig:distribution demo}) and strongly anti-correlated with the GMF~(PBSS; \autoref{fig:PBSS distribution}), with both structures apparent at 100\,TeV. The impact of the tan gridding function increasing the cell size on the other side of the MW can be seen in \autoref{fig:GALPROP p vs e} as a reduced resolution for negative $X$ values.

An example visualisation of the pion-decay and IC \graya{} emission skymaps can be found in \autoref{fig:GALPROP pi vs ic} for various energies in the range 0.01--100\,TeV. The pion-decay emission morphology is strongly correlated with the ISM gas density and independent of the \graya{} energy~(see \autoref{eq:pion intensity from emissivity}), while the IC emission morphology is strongly anti-correlated with the GMF and depends on the \graya{} energy.

\subsection{Time-Dependent Output Examples}

\begin{figure}
    \centering
    \includegraphics[width=16cm]{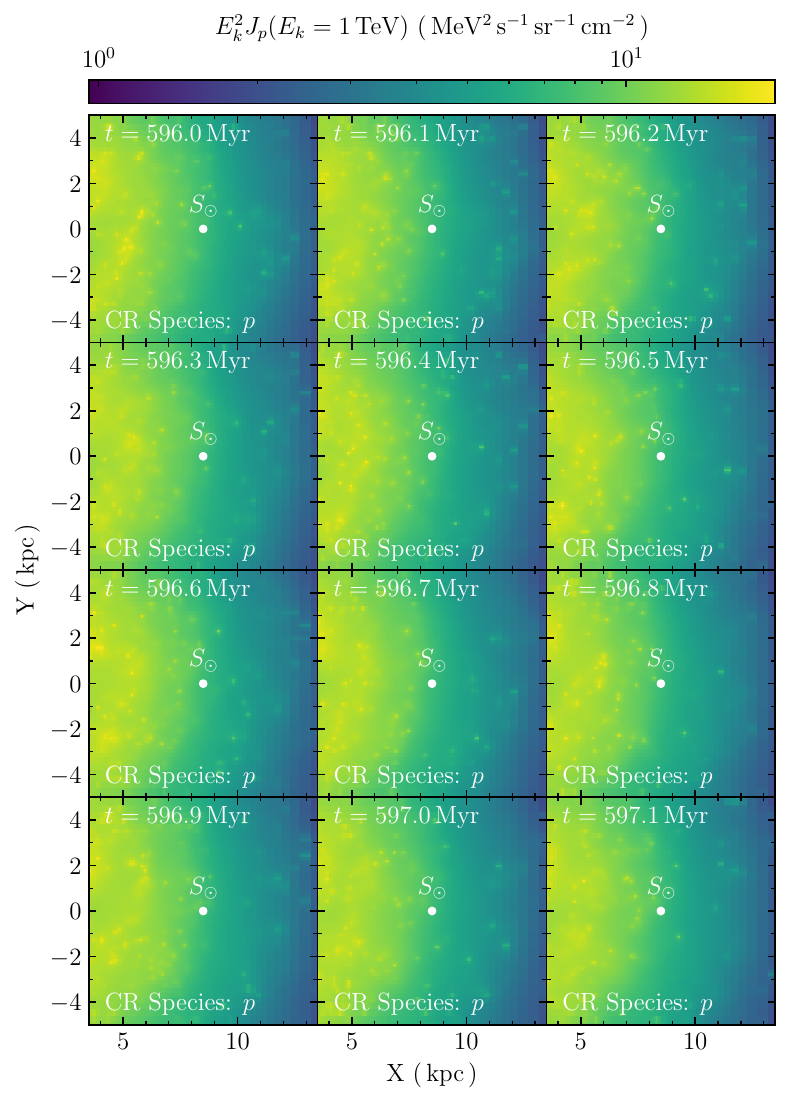}
    \caption{The time-dependent CR proton energy density at 1\,TeV, with the Solar position shown by $S_{\odot}$. The CR energy density is shown in units of~MeV$^{2}$\,cm$^{-2}$\,s$^{-1}$\,sr$^{-1}$. Each frame represents a 0.1\,Myr step in time, which is approximately equal to the lifetime of the sources.}
    \label{fig:GALPROP p density over time}
\end{figure}

\begin{figure}
    \centering
    \includegraphics[width=16cm]{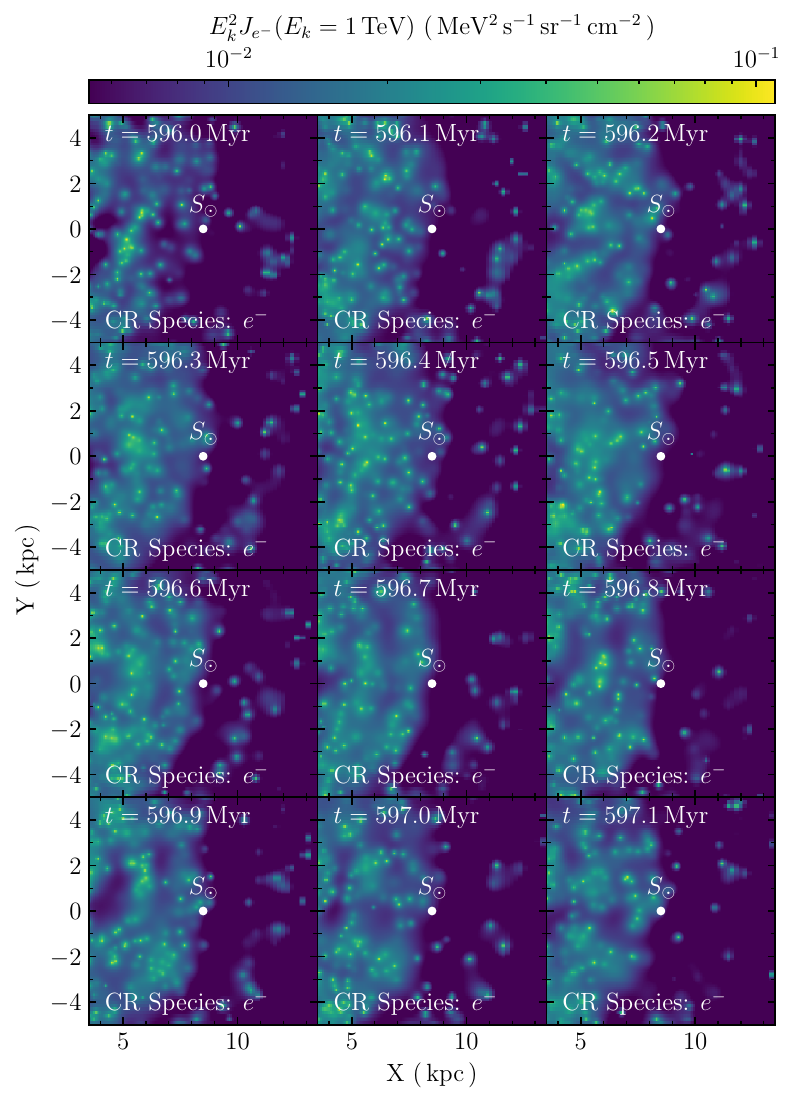}
    \caption{The time-dependent CR electron energy density at 1\,TeV, with the Solar position shown by $S_{\odot}$. The CR energy density is shown in units of~MeV$^{2}$\,cm$^{-2}$\,s$^{-1}$\,sr$^{-1}$. Each frame represents a 0.1\,Myr step in time, which is approximately equal to the lifetime of the sources.}
    \label{fig:GALPROP e density over time}
\end{figure}

\begin{figure}
    \centering
    \includegraphics[width=16cm]{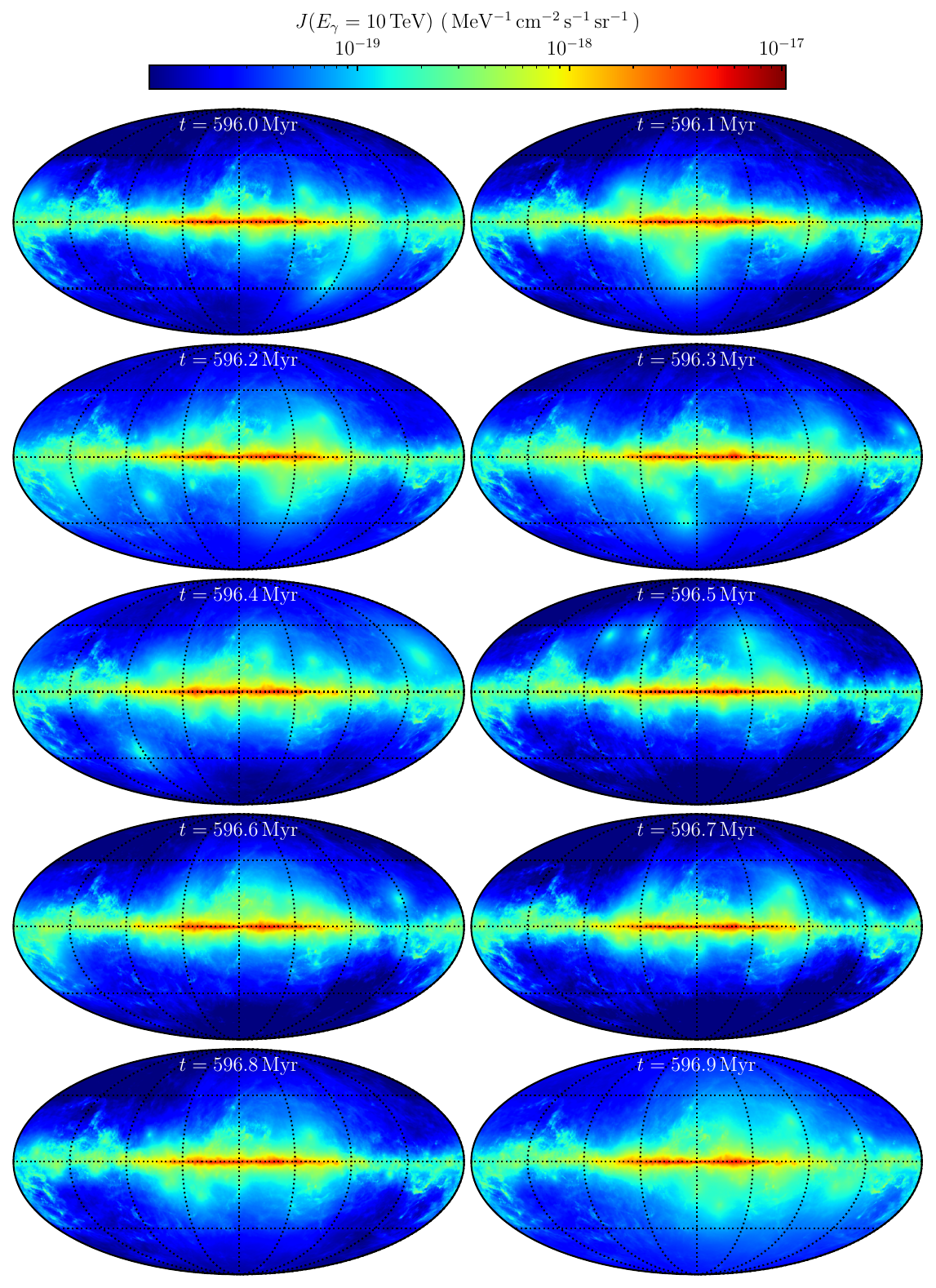}
    \caption{The \GP{} skymaps of the time-dependent total \graya{} emission at $E_{\gamma} = 10$\,TeV in units of~MeV$^{-1}$\,cm$^{-2}$\,s$^{-1}$\,sr$^{-1}$. All components~(pion-decay, bremsstrahlung, and IC emissions) have been summed together. Each frame represents a 0.1\,Myr step in time, which is approximately equal to the lifetime of the sources.}
    \label{fig:GALPROP total emission over time}
\end{figure}

The time-dependent outputs are formatted in an identical manner as their time-independent counterparts. The nuclei files are written at each output timestep, $t_{\mathrm{out}}$, with the timestep number appended to the end of the filename.
After all nuclei files have been created, \GP{} normalises the CR energy densities to the measured values at the Solar location for all CR species using the method described in \autoref{ssect:parameter optimisation}.
The normalisation constants are calculated for the final timestep and are then applied retroactively. After the nuclei have been normalised, \GP{} then begins creation of the \graya{} skymaps.

An example visualisation of the CR proton and CR electron energy densities at 1\,TeV around the Solar neighbourhood for various timesteps are shown in \autoref{fig:GALPROP p density over time} and \autoref{fig:GALPROP e density over time}, respectively.
The CR proton density does not vary significantly with time as the CR protons are diffuse at 1\,TeV; however, some CR sources are visible above the diffuse CR sea.
For the 1\,TeV CR electrons the individual sources are easily visible -- they do not travel far enough from their injection sites to form a diffuse sea. A cloud of VHE electrons continues to exist for some time after the CR accelerator ceases injecting CRs into the ISM. The length of time that a CR source will inject CRs into the ISM for, and how often sources are placed, is discussed in more depth in \autoref{ssect:TDD source dist.}. The length of time that the VHE electrons remain after the sources disappear depends on the kinetic energy of the CRs.

An example visualisation of the total~(i.e.~sum over all \graya{} components) \graya{} skymap integrated above 1\,TeV for various timesteps is shown in \autoref{fig:GALPROP total emission over time}. Similarly as for the CR electron density over time, the \graya{} emission shows bright spots caused by CR sources accelerating CRs into the ISM.
As discussed in \autoref{ssect:tdep source pars}, only a single, `average' source type is defined. Hence, all CR sources in the time-dependent model injects both hadrons and leptons. Therefore, a diffuse sphere of IC emission can be observed around each of the CR sources. Additionally, increased pion-decay emission can be observed from the ISM gas nearby each CR source.
        
\chapter{HGPS Source Parameters} \label{chap:HGPS source pars}

The procedure to mask a source from the HGPS is detailed in \autoref{ssect:Masking sources in the HGPS}. In particular, the equations that model the \graya{} source emission are given by Equations \ref{eq:source subtraction} to \ref{eq:total angular size}. Modelling each source requires the position of the source (given in Galactic longitude and latitude), the radius of the source~(given as a Gaussian standard deviation; $\sigma_{\mathrm{source}}$), and the totally spatially integrated flux of the source~($F_{\mathrm{source}}$). The source surface brightness is modelled as a function of the radial distance from the centre of the source~($S_{\mathrm{source}}(r)$; \autoref{eq:source subtraction}). 

To calculate the source parameters the H.E.S.S. collaboration first subtracted their model of the large-scale emission~(\autoref{sect:HGPS diffuse emission}) from the HGPS flux map. The large-scale model implicitly includes the diffuse \graya{} emission as well as emission from unresolved \graya{} sources.
The residual emission then represents only the \graya{} emission from catalogued sources.
The source parameters were approximated from the residual emission, and the surface brightness for each source was calculated by using Equations \ref{eq:source subtraction} to \ref{eq:total angular size}. The modelled surface brightness of each of the source was then subtracted from the flux map, with the source parameters being adjusted iteratively until the residual emission was minimised. For more depth, see Section~4.8 of~\citet{AbdallaH.2018a}.

All sources in the HGPS are listed in \autoref{tab:all sources} along with their spatial model, association, and modelling parameters. The association of each identified source is discussed in \autoref{ssect:source classifications}.

For the source-masking procedure in \autoref{ssect:Masking sources in the HGPS}, the sources with the `shell' spatial model were excluded from the map such that they did not contribute to any estimate of the diffuse emission. The Galactic centre was excluded in a similar fashion, and is defined by the bounds: $-1.0^{\circ} \leq l \leq +1.5^{\circ}$ and $|b| \leq 0.4^{\circ}$~\citep{AbdallaH.2018a}.

\clearpage

\bgroup
\def\arraystretch{1.25}
{\scriptsize
\begin{longtable}{lcccccc}
    \caption{\normalsize{A list of all~78~catalogued \graya{} sources in the HGPS, with the parameter values from~\citet{AbdallaH.2018a} and the HGPS catalogue. Longitude and latitude are listed in Galactic coordinates. For the `shell' spatial models, the value of $\sigma_{\mathrm{source}}$ is taken from the radius of the spectral analysis region used by the \hess{} collaboration. For all other sources the value of $\sigma_{\mathrm{source}}$ is listed as the radius of the source.}}
    \label{tab:all sources} \\
    \hline
    Source Name       & Spatial Model &           Association & Longitude & Latitude & $\sigma_{\mathrm{source}}$ & $F_{\mathrm{source}}$             \\
                      &               &                       &     (deg) &    (deg) &                      (deg) & ($10^{-12}$\,cm$^{-2}$\,s$^{-1}$) \\ \hline
    HESS~J0835--455   &    3-Gaussian &                   PWN &   263.96  & $-3.05$  &            $0.58 \pm 0.05$ &                  $15.36 \pm 0.53$ \\
    HESS~J0852--463   &         Shell &                   SNR &   266.29  & $-1.24$  &                     $1.00$ &                  $23.39 \pm 2.35$ \\
    HESS~J1018--589~A &      Gaussian &                Binary &   284.35  & $-1.67$  &                     $0.00$ &                   $0.30 \pm 0.05$ \\
    HESS~J1018--589~B &      Gaussian & Not Firmly Identified &   284.22  & $-1.77$  &            $0.15 \pm 0.03$ &                   $0.83 \pm 0.17$ \\
    HESS~J1023--575   &      Gaussian &       Stellar Cluster &   284.19  & $-0.40$  &            $0.17 \pm 0.01$ &                   $2.56 \pm 0.17$ \\
    HESS~J1026--582   &      Gaussian & Not Firmly Identified &   284.85  & $-0.52$  &            $0.13 \pm 0.04$ &                   $0.69 \pm 0.19$ \\
    HESS~J1119--614   &      Gaussian &             Composite &   292.13  & $-0.53$  &            $0.10 \pm 0.01$ &                   $0.87 \pm 0.13$ \\
    HESS~J1302--638   &      Gaussian &                Binary &   304.18  & $-1.00$  &                     $0.01$ &                   $0.40 \pm 0.05$ \\
    HESS~J1303--631   &    2-Gaussian &                   PWN &   304.24  & $-0.35$  &            $0.18 \pm 0.01$ &                   $5.26 \pm 0.27$ \\
    HESS~J1356--645   &      Gaussian &                   PWN &   309.79  & $-2.50$  &            $0.23 \pm 0.02$ &                   $5.53 \pm 0.53$ \\
    HESS~J1418--609   &      Gaussian &                   PWN &   313.24  & $ 0.14$  &            $0.11 \pm 0.01$ &                   $3.01 \pm 0.31$ \\
    HESS~J1420--607   &      Gaussian &                   PWN &   313.58  & $ 0.27$  &            $0.08 \pm 0.01$ &                   $3.28 \pm 0.24$ \\
    HESS~J1427--608   &      Gaussian & Not Firmly Identified &   314.42  & $-0.16$  &                     $0.05$ &                   $0.74 \pm 0.10$ \\
    HESS~J1442--624   &         Shell &                   SNR &   315.43  & $-2.29$  &            $0.30 \pm 0.02$ &                   $2.44 \pm 0.67$ \\
    HESS~J1457--593   &      Gaussian &                   N/A &   318.35  & $-0.42$  &            $0.33 \pm 0.04$ &                   $2.50 \pm 0.40$ \\
    HESS~J1458--608   &      Gaussian & Not Firmly Identified &   317.95  & $-1.70$  &            $0.37 \pm 0.03$ &                   $2.44 \pm 0.30$ \\
    HESS~J1503--582   &      Gaussian &                   N/A &   319.57  & $ 0.29$  &            $0.28 \pm 0.03$ &                   $1.89 \pm 0.28$ \\
    HESS~J1507--622   &      Gaussian & Not Firmly Identified &   317.97  & $-3.48$  &            $0.18 \pm 0.02$ &                   $2.99 \pm 0.31$ \\
    HESS~J1514--591   &    3-Gaussian &                   PWN &   320.32  & $-1.19$  &            $0.14 \pm 0.03$ &                   $6.43 \pm 0.21$ \\
    HESS~J1534--571   &         Shell &                   SNR &   323.70  & $-1.02$  &            $0.40 \pm 0.04$ &                   $1.98 \pm 0.23$ \\
    HESS~J1554--550   &      Gaussian &                   PWN &   327.16  & $-1.08$  &                     $0.02$ &                   $0.36 \pm 0.06$ \\
    HESS~J1614--518   &         Shell & Not Firmly Identified &   331.47  & $-0.60$  &            $0.42 \pm 0.01$ &                   $5.87 \pm 0.42$ \\
    HESS~J1616--508   &    2-Gaussian & Not Firmly Identified &   332.48  & $-0.17$  &            $0.23 \pm 0.03$ &                   $8.48 \pm 0.44$ \\
    HESS~J1626--490   &      Gaussian &                  N/A  &   334.82  & $-0.12$  &            $0.20 \pm 0.03$ &                   $1.65 \pm 0.33$ \\
    HESS~J1632--478   &      Gaussian & Not Firmly Identified &   336.39  & $ 0.26$  &            $0.18 \pm 0.02$ &                   $2.93 \pm 0.51$ \\
    HESS~J1634--472   &      Gaussian & Not Firmly Identified &   337.12  & $ 0.26$  &            $0.17 \pm 0.01$ &                   $2.90 \pm 0.37$ \\
    HESS~J1640--465   &    2-Gaussian &             Composite &   338.28  & $-0.04$  &            $0.11 \pm 0.03$ &                   $3.33 \pm 0.19$ \\
    HESS~J1641--463   &      Gaussian & Not Firmly Identified &   338.52  & $ 0.08$  &                     $0.04$ &                   $0.27 \pm 0.06$ \\
    HESS~J1646--458   &      Gaussian &       Stellar Cluster &   339.33  & $-0.78$  &            $0.50 \pm 0.03$ &                   $5.48 \pm 0.46$ \\
    HESS~J1702--420   &      Gaussian &                  N/A  &   344.23  & $-0.19$  &            $0.20 \pm 0.02$ &                   $3.91 \pm 0.65$ \\
    HESS~J1708--410   &      Gaussian &                  N/A  &   345.67  & $-0.44$  &            $0.06 \pm 0.01$ &                   $0.88 \pm 0.09$ \\
    HESS~J1708--443   &      Gaussian & Not Firmly Identified &   343.07  & $-2.32$  &            $0.28 \pm 0.03$ &                   $2.28 \pm 0.32$ \\
    HESS~J1713--381   &      Gaussian & Not Firmly Identified &   348.62  & $ 0.38$  &            $0.09 \pm 0.02$ &                   $0.65 \pm 0.13$ \\
    HESS~J1713--397   &         Shell &                   SNR &   347.31  & $-0.46$  &                     $0.50$ &                  $16.88 \pm 0.82$ \\
    HESS~J1714--385   &      Gaussian &             Composite &   348.42  & $ 0.14$  &                     $0.03$ &                   $0.25 \pm 0.05$ \\
    HESS~J1718--374   &    Point-Like &                   SNR &   349.72  & $ 0.17$  &                     $0.00$ &                   $0.12 \pm 0.04$ \\
    HESS~J1718--385   &      Gaussian & Not Firmly Identified &   348.88  & $-0.48$  &            $0.12 \pm 0.01$ &                   $0.80 \pm 0.14$ \\
    HESS~J1729--345   &      Gaussian &                   N/A &   353.39  & $-0.02$  &            $0.19 \pm 0.03$ &                   $0.86 \pm 0.17$ \\
    HESS~J1731--347   &         Shell &                   SNR &   353.54  & $-0.67$  &            $0.27 \pm 0.02$ &                   $2.01 \pm 0.15$ \\
    HESS~J1741--302   &    Point-Like &                   N/A &   358.28  & $ 0.05$  &                     $0.00$ &                   $0.16 \pm 0.04$ \\
    HESS~J1745--290   &    Point-Like & Not Firmly Identified &   359.94  & $-0.04$  &                     $0.00$ &                   $1.70 \pm 0.08$ \\
    HESS~J1745--303   &      Gaussian &                   N/A &   358.64  & $-0.56$  &            $0.18 \pm 0.02$ &                   $0.94 \pm 0.21$ \\
    HESS~J1746--285   &    Point-Like & Not Firmly Identified &     0.14  & $-0.11$  &                     $0.00$ &                   $0.15 \pm 0.05$ \\
    HESS~J1746--308   &      Gaussian & Not Firmly Identified &   358.45  & $-1.11$  &            $0.16 \pm 0.04$ &                   $0.68 \pm 0.22$ \\
    HESS~J1747--248   &      Gaussian &       Stellar Cluster &     3.78  & $ 1.71$  &            $0.06 \pm 0.01$ &                   $0.29 \pm 0.05$ \\
    HESS~J1747--281   &    Point-Like &                   PWN &     0.87  & $ 0.08$  &                     $0.00$ &                   $0.60 \pm 0.13$ \\
    HESS~J1800--240   &      Gaussian & Not Firmly Identified &     5.96  & $-0.42$  &            $0.32 \pm 0.04$ &                   $2.44 \pm 0.35$ \\
    HESS~J1801--233   &      Gaussian &                   SNR &     6.66  & $-0.27$  &            $0.17 \pm 0.03$ &                   $0.45 \pm 0.10$ \\
    HESS~J1804--216   &    2-Gaussian & Not Firmly Identified &     8.38  & $-0.09$  &            $0.24 \pm 0.03$ &                   $5.88 \pm 0.27$ \\
    HESS~J1808--204   &      Gaussian & Not Firmly Identified &    10.01  & $-0.24$  &                     $0.06$ &                   $0.19 \pm 0.04$ \\
    HESS~J1809--193   &    3-Gaussian & Not Firmly Identified &    11.11  & $-0.02$  &            $0.40 \pm 0.05$ &                   $5.27 \pm 0.29$ \\
    HESS~J1813--126   &      Gaussian & Not Firmly Identified &    17.31  & $ 2.49$  &            $0.21 \pm 0.03$ &                   $1.08 \pm 0.24$ \\
    HESS~J1813--178   &      Gaussian &             Composite &    12.82  & $-0.03$  &            $0.05 \pm 0.00$ &                   $1.98 \pm 0.15$ \\
    HESS~J1818--154   &      Gaussian &                   PWN &    15.41  & $ 0.16$  &                     $0.00$ &                   $0.17 \pm 0.04$ \\
    HESS~J1825--137   &    3-Gaussian &                   PWN &    17.53  & $-0.62$  &            $0.46 \pm 0.03$ &                  $18.41 \pm 0.56$ \\
    HESS~J1826--130   &      Gaussian & Not Firmly Identified &    18.48  & $-0.39$  &            $0.15 \pm 0.02$ &                   $0.86 \pm 0.17$ \\
    HESS~J1826--148   &      Gaussian &                Binary &    16.88  & $-1.29$  &                     $0.01$ &                   $1.28 \pm 0.04$ \\
    HESS~J1828--099   &      Gaussian &                   N/A &    21.49  & $ 0.38$  &                     $0.05$ &                   $0.43 \pm 0.07$ \\
    HESS~J1832--085   &      Gaussian &                   N/A &    23.21  & $ 0.29$  &                     $0.02$ &                   $0.21 \pm 0.05$ \\
    HESS~J1832--093   &      Gaussian & Not Firmly Identified &    22.48  & $-0.16$  &                     $0.00$ &                   $0.17 \pm 0.03$ \\
    HESS~J1833--105   &      Gaussian &             Composite &    21.50  & $-0.90$  &                     $0.02$ &                   $0.39 \pm 0.07$ \\
    HESS~J1834--087   &    2-Gaussian &             Composite &    23.26  & $-0.33$  &            $0.21 \pm 0.04$ &                   $3.34 \pm 0.24$ \\
    HESS~J1837--069   &    3-Gaussian &                   PWN &    25.15  & $-0.09$  &            $0.36 \pm 0.03$ &                  $12.05 \pm 0.45$ \\
    HESS~J1841--055   &    2-Gaussian & Not Firmly Identified &    26.71  & $-0.23$  &            $0.41 \pm 0.03$ &                  $10.16 \pm 0.42$ \\
    HESS~J1843--033   &    2-Gaussian & Not Firmly Identified &    28.90  & $ 0.07$  &            $0.24 \pm 0.06$ &                   $2.88 \pm 0.30$ \\
    HESS~J1844--030   &      Gaussian & Not Firmly Identified &    29.41  & $ 0.09$  &                     $0.02$ &                   $0.26 \pm 0.05$ \\
    HESS~J1846--029   &      Gaussian &             Composite &    29.71  & $-0.24$  &                     $0.01$ &                   $0.45 \pm 0.05$ \\
    HESS~J1848--018   &      Gaussian &       Stellar Cluster &    30.92  & $-0.21$  &            $0.25 \pm 0.03$ &                   $1.74 \pm 0.35$ \\
    HESS~J1849--000   &      Gaussian &                   PWN &    32.61  & $ 0.53$  &            $0.09 \pm 0.02$ &                   $0.53 \pm 0.09$ \\
    HESS~J1852--000   &      Gaussian & Not Firmly Identified &    33.11  & $-0.13$  &            $0.28 \pm 0.04$ &                   $1.30 \pm 0.25$ \\
    HESS~J1857+026    &    2-Gaussian & Not Firmly Identified &    36.06  & $-0.06$  &            $0.26 \pm 0.06$ &                   $3.77 \pm 0.40$ \\
    HESS~J1858+020    &      Gaussian &                   N/A &    35.54  & $-0.58$  &            $0.08 \pm 0.02$ &                   $0.53 \pm 0.11$ \\
    HESS~J1908+063    &      Gaussian & Not Firmly Identified &    40.55  & $-0.84$  &            $0.49 \pm 0.03$ &                   $6.53 \pm 0.50$ \\
    HESS~J1911+090    &    Point-Like &                   SNR &    43.26  & $-0.19$  &                     $0.00$ &                   $0.15 \pm 0.03$ \\
    HESS~J1912+101    &         Shell & Not Firmly Identified &    44.46  & $-0.13$  &            $0.49 \pm 0.04$ &                   $2.49 \pm 0.35$ \\
    HESS~J1923+141    &      Gaussian & Not Firmly Identified &    49.08  & $-0.40$  &            $0.12 \pm 0.02$ &                   $0.78 \pm 0.15$ \\
    HESS~J1930+188    &      Gaussian &             Composite &    54.06  & $ 0.27$  &                     $0.02$ &                   $0.29 \pm 0.09$ \\
    HESS~J1943+213    &      Gaussian & Not Firmly Identified &    57.78  & $-1.30$  &                     $0.03$ &                   $0.32 \pm 0.10$ \\
    \hline
\end{longtable}
}
\egroup
    \end{appendix}
    
    % References/Bibliography
    \newpage
    \phantomsection
    \addcontentsline{toc}{chapter}{Bibliography}
    \printbibliography
    
\end{document}